\documentclass[preprint]{aastex631}

\submitjournal{AAS Journals}

\shorttitle{Dippers 2026}
\shortauthors{Sitko et al.}

\begin{document}

\title{Gas Flows and Mass Accretion Rates in Eight Dippers: HD 142666, HD 143006, HD 145718,  V935 Sco, DoAr 25, EPIC 204638512, EPIC 205151387, and EPIC 203850058}

\correspondingauthor{Michael L Sitko}
\email{sitko@spacescience.org}

\author[0000-0003-1799-1755]{Michael L. Sitko}
\affiliation{Center for Exoplanetary Systems, Space Science Institute, 4765 Walnut Street, Suite B, Boulder, CO 80301, USA}

\author[0000-0002-7818-2305]{Ray W. Russell}
\affiliation{The Aerospace Corporation, Los Angeles, CA, USA}

\author[0000-0002-2131-4346]{Korash D. Assani}
\affiliation{Department of Astronomy, University of Virginia, Charlottesville, VA22903,USA}

\author[0000-0002-6328-6099]{Lucia A. Villanueva}
\affiliation{Universidad Nacional Autonoma de Mexico, Instituto de Astronomia, AP 106, Ensenada 22800, BC, Mexico}

\author[0000-0001-9797-5661]{Jesus Hernandez}
\affiliation{Universidad Nacional Autonoma de Mexico, Instituto de Astronomia, AP 106, Ensenada 22800, BC, Mexico}

\begin{abstract}
One of the scenarios used to explain the dipper phenomenon in stars is having the innermost disk regions being close to the line of sight to the star - inclined greater than 60 degrees - regardless of the inclination of the outermost disk (``mis-aligned'' or ``broken'' disks). If the dust and gas in these disks sample the same material, edge-on disks will be viewed through larger column densities of gas and dust than the average disk. We have examined inter-night variability of eight Kepler-discovered dipper stars using the SpeX spectrograph on NASA's Infrared Telescope facility at a spectral resolving power of R$\sim$750.  The He I line at 1.083 $\mu$m exhibits a wide variety of profiles, which change from night to night. In the majority of cases, an inverse P Cygni profile, indicative of inflowing gas, is present. The line profile (and continuum flux level) often change on time scales of 1 day. The Paschen and Brackett lines also change on similar time scales. In one object, V935 Sco, Pa$\beta$ and Br$\gamma$  went from being in emission to being in absorption over the course of 5 weeks. In another, EPIC 203850058, Pa$\beta$ went from being in emission in 2017 to vanishing altogether in 2018. The accretion rates determined using Pa$\beta$ for these stars tend to be smaller than those using Br$\gamma$, indicating that the former line is more susceptible to self-absorption than the latter, as would be expected for a highly inclined disk. \\
\end{abstract}


\section{Introduction} \label{sec:intro}

Dipper stars are objects that exhibit large non-periodic changes in brightness over time scales of days or even hours. They became important targets for the Kepler extended (K2) mission, and \citet{ansdell16} reported finding many in the Upper Sco and $\rho$ Oph stellar associations.  Most of these would be considered Weak-line T Tauri stars (WTTs, \citet{ansdell16}). Similar dipper activity was observed in stars in the young star-formation region NGC 2264 \citep{stauffer16}. While disks observed edge-on would be likely sources of such behavior,  \citet{ansdell20} examined a sample of 24 dippers in their sample, and found that the outer ALMA-detected disks had an isotropic distribution of inclinations. Thus the dipper phenomenon was not linked to the inclination of the outer disk. This implied that the inner disks were likely ``misaligned''  from the outer disk detected in ALMA images (they used the term ``broken'' disks).  \citet{ansdell20} also deduced that these are also not ``freak'' objects, but may be quite common in the evolution of pre-main sequence disks. In an early compilation of results of transiting hot jupiters, \citet{winn10}  found that 5 of their sample of 15 stars had inclinations greater than 90 degrees, placing them on retrograde orbits (one had an inclination of 180 degrees). \citet{bn12} found that 20\% of their sample of hot Jupiters had high inclinations (including retrograde orbits). Should gravitational scattering of young planets occur while the inner disk is still present, torquing of that inner disk material into orbits with inclinations out of the plane of the distant outer portions of the disk will also be expected to be common. Given that the protostar itself formed  prior to these scattering events, it is likely that they share the same rotation axis as the \textit{outer disk}, and not the often highly mis-inclined \textit{inner disk}. \citet{sitko26} have found that among the Herbig Ae/Fe stars, the inclinations of the inner disks tend to be systematically greater than the inclinations of the outer disks, suggesting a bias toward mis-inclinations, that may not be real.  \\

%\citet{ansdell16a} found a number of dippers in a survey of Kepler K2/C2 observations  Upper Sco and Oph Associations. They noticed that three objects, EPIC 204638513, often referred to as 2MASS J16042165-2130284 (or even ``J1604'' for short), EPIC 205151387, and EPIC 203850058, were found to have outer disks, based on images using ALMA, that had low inclinations presented to the observer \citep{ansdell16b}. That of EPIC 204638512 had an inclination - almost face-on - of $\sim$ 6$^{\circ}$. This prompted the study of a more extensive sample of 24 dippers in that region. Remarkably, among this sample of dippers, the inclination of the outer disks was \textit{isotropic}, meaning that the dipper phenomenon had nothing to do with the outer disk. One possible culprit was a mis-aligned inner disk (``broken disk'' using their terminology). For EPIC 204638513, there is little doubt that this inner disk is the source of the dipper phenomenon. If mis-inclined inner disks are common, they would provide a natural source for the dipper phenomenon. \\

Recent computational models suggest that the inner disk regions, as well as the star itself, might be mis-aligned, and the inner disk may be warped \citep{pelkonen25}, due to the structure of the pre-existing gas that the star and disk accrete from. Without added information, such as that which exists for EPIC 204638512 and HD 143006, our working framework will be that rotational axis of the star is probably close to that of the orbital axis of the outer disk, with the inner disk being mis-aligned to both. \\

Recent imaging data for  the dipper EPIC 204638512 (= 2MASS J16042165-2130284) using ALMA \citep{mayama18} and VLT/SPHERE \citep{pinilla18} provide evidence for such a misaligned inner disk.  With ALMA observations, a twisted  ``butterfly'' radial velocity map in CO (3-2) clearly shows a highly inclined inner disk, despite the outer disk being seen nearly face-on. Shadows of the inner disk cast on the outer disk are also visible in the dust continuum images \citep{pinilla18,sa20}. They found that the outer disk's darker regions are nearly 180$^{\circ}$ apart, indicating shadowing by a highly inclined inner disk. Furthermore, the shadowed regions varied in brightness. Intense photometric monitoring  suggested a variable scale height in the orbiting inner disk. The conclusion is that this object has a misaligned inner disk with surface irregularities. These irregularities could be the results of nearby planet (turbulence or a pressure bump), or the launch points of the accretion columns. \citet{zhong24} reported that the shadows disappeared between June and August, 2015. \\

Using the Michigan Infrared Combiner Exeter (MIRC-X) at the Center for High Angular Resolution Astronomy (CHARA) and the Precision Integrated-Optics Near-infrared Imaging ExpeRiment (PIONIER) at the Very Large Telescope Interferometer (VLTI), \citet{codron25} found that the inner disk has an inclination of i=22$^{\circ}$$\pm$3$^{\circ}$, that was mis-aligned from the outer disk by 39$^{\circ}$$\pm$4$^{\circ}$. This firmly establishes the disk mis-alignment scenario for this dipper star. It also indicates that the dipper phenomenon can occur when the observer is within $\sim$20$^{\circ}$ of the inner disk mid-plane ($\sim$66\% probability).  \\

Among the atomic species used for determining the gas dynamics of the inner regions of the star+disk region  in T Tauri  stars is that of He I. \citet{beristain01} and \citet{edwards03} have analyzed the profiles of the 0.5876 $\mu$m line and the 1.083 $\mu$m lines respectively. Along with \citet{kwan07}, they found evidence of both stellar and disk winds in the majority of the stars, producing a ``P Cygni'' line profile. The lower energy level of the 1.083 $\mu$m is metastable, with a lifetime for radiative de-excitation of $\sim$2.2 \textit{hours} \citep{lp01}. In this sense, it behaves like Ly$\alpha$ where the line is strongly scattered where the He I is present. The transition that produced the 0.5876 $\mu$m line feeds directly into the upper level of the 1.083 $\mu$m line transition, thus adding added de-excitation transitions beyond the resonant scattering transitions of that line. \citet{erkal22} has summarized the plethora of line profiles, and has expanded the profile  types to include infalling accreting gas with a red-shifted absorption component {``inverse P Cygni'' profle)  as well as a combination of wind and infalling gas. \\

Multi-epoch medium-resolution (R$\sim$750) spectroscopic observations of three dippers (EPIC 204638512, EPIC 205151387, and EPIC 203850058) were reported by \citet{sitko24}, who found that the line profiles that varied significantly with time. Those of EPIC 205151387 were perhaps the most telling, in that the He I line at 1.083 $\mu$m was in emission when the surrounding continuum was highest - the observation least attenuated by the dust in the inner disk. At somewhat lower continuum levels, with more attenuation by dust, the line took on an inverse P Cygni (hereafter IPC) profile, indicative of gas inflow. At the highest dust attenuation, the red-shifted absorption core of the profile completely obliterated all but a tiny part of the emission portion of the line. \citet{sitko24} interpreted these correlated behaviors of the dust extinction and gas flows within the framework of models of \citet{vinkovic21}, wherein the gas and dust react to the radiative pressure of the star \textit{and the disk}, along with a gas accretion model where gas and dust flow inward toward the star along the surface of the disk. Slightly above the surface of the disk (defined where the dust optical depth $\sim$1) an outflow in a disk wind occurs. Deeper down into the disk, the dynamics are dominated by the inflowing gas, that feeds the mass accretion onto the star. Such inflowing gas can reach velocities of  hundreds of km/s \citep{vinkovic24}. These describe the dynamics of the disk surface regions, not the accretion columns. An added complication may come to play: gas turbulence.  \citet{zhu24} and \citet{zhang24} have shown that when a magnetic field threads the disk, density and velocity fluctuations near the (observable) disk surface may occur. \\

A further complication arises near the innermost regions of the stars of later spectral types - the orientation of the stellar magnetic field to the disk at its inner edge. There is no guarantee that the axis of the stellar magnetic field aligns with the rotation axis of the inner disk. Thus, even if the stellar magnetic field is a simple bipolar one, aligned with the stellar rotation axis, these may be highly inclined to the mis-aligned inner disk. In principle,  the ``simple'' correlation of the dust extinction and He I gas velocities could, with more observations, be rather complex. \\

In this paper, we present new observations of a number of dippers including the earlier data from \citet{sitko24}. Our sample include HD 142666, HD 143006, HD 145718, V935 Sco (EPIC 204176565), DoAr25 (EPIC 203843911), EPIC 204638512, EPIC 205151387, and EPIC 203850058. The light curves obtained by \textit{Kepler }are illustrated in \citet{ansdell20}.

%Variability due to disk ripples. How? \textit{Late type stars - stellar multipolar magnetic fields attached to the disk at the co-rotation radius}. These produce \textit{multiple accretion columns} and some dust levitation.\\

%What are the rotation rates of TTs? Have they already be slowed down by braking?  \\
%What about the A stars? F stars? SAO 206462 may be rotating close to breakup speed! Some show broad absorption in H I lines, but metal lines not broadened much. \\

%SAO 206462 is a ``broken disk'' - Muller 2011\\

%This is basically the same scenario as that envisioned by \citet{ellerbroek14} for HD 163296, which itself could b e accused of being a dipper. . Of the dippers in the current study, EPIC 20461358 would seem to PC and occasional IPC.

\section{Observations}

We observed the eight dippers on multiple nights at an effective spectral resolution using the 0.8" slit from 0.7 - 2.4 $\mu$m using the SXD grating of the SpeX spectrograph \citep{rayner03} on NASA's Infrared Telescope Facility (IRTF). With that slit width, the spectral resolving power is R$\sim$750. In order to determine the absolute fluxes of these slit spectra, which are subject to time-variable throughput due to the effects of astronomical seeing and telescope guiding, we also observed them in the photometric  K-band using the SpeX guide camera with a 9-point dither pattern. Both spectra and images were flux-calibrated using the nearby A0V star HD 145127 at nearly the same airmass. The spectral data reduction was carried out using the Spextool software, running under IDL. \citep{vacca03,cushing04}. The photometric accuracy was determined using the combined standard deviations of the science target, its photometric comparison star, and the uncertainty in the magnitude of the standard star, as listed in Simbad. \\

With the exception of HD 142666, all of the stars were observed only once per night. For HD 142666 we observed the star twice per night on two nights - 240611 and 240612 UT, where the dates are given as \textit{yymmdd}. \\

\clearpage

\begin{deluxetable}{lcc}[!ht]
\tablecolumns{3}
\tablewidth{0pc}
%\rotate
\tabletypesize{\scriptsize}
\tablecaption{SpeX Observations of 8 Dipper Stars}
\tablehead{
\colhead{Object} & \colhead{Date \textit{yymmdd} (UT)}  & \colhead{Photometric  Accuracy}}
\startdata 
HD 142666 & 240611 \#1 & 4.4\% \\
                   & 240611 \#2 & 2.9\% \\
                   & 240612 \#1 & 3.5\% \\                  
                   & 240612 \#2 & 4.4\% \\     
                   & 240716  & 4.5\% \\  
                   & 240717 & 6.2\% \\  
                   & 240719  & 3.3\% \\    
HD 143006 & 240611 & 5.5\% \\  
                   & 240612 & 4.1\% \\ 
                    & 240619 & 4.3\% \\ 
HD 145718 & 240610 & 2.4\% \\  
                   & 240611 & 3.0\% \\ 
                    & 240612 & 4.0\% \\ 
                    & 240719 & 4.4\% \\ 
V935 Sco    & 240610 & 2.1\% \\  
                    & 240611 & 3.4\% \\ 
                    & 240612 & 4.0\% \\    
                    & 240716 & 5.6\% \\  
                    & 240717& 3.3\% \\ 
                    & 240718 & 3.6\% \\ 
 DoAr25       & 240611 & 5.5\% \\  
                    & 240612 & 4.5\% \\  
EPIC 204638512 &  170810 & 5.0\% \\  
                             &  170811 & 4.1\% \\  
                             &  180518 & 4.1\% \\  
                             &  180519 & 3.7\% \\  
                             &  240612& 4.5\% \\  
                             &  240717  & 3.3\% \\  
                             &  240718 & 2.2\% \\  
 EPIC 205151387 &  170810 & 6.2\% \\  
                             &  170811 & 4.8\% \\  
                             &  180518 & 4.3\% \\  
                             &  180519 & 4.0\% \\  
                             &  240423 & 7.3\% \\  
                             &  240524 & 8.6\% \\  
                             &  240601 & 3.3\% \\  
                             &  240615 & 3.5\% \\  
                             &  240621 & 3.4\% \\  
                             &  240716& 5.5\% \\  
                             &  240718 & 3.6\% \\  
 EPIC 203850058 & 170810 & 17\% \\  
                             &  170811 & 9\% \\  
                             &  180518 & 6.5\% \\  
                             &  180519 & 4.5\% \\  
  \enddata
\end{deluxetable}

\clearpage

\section{Results} 

\subsection{Gas Flows}

In all of the dippers studied here, the emission lines of  H I are quite weak, as seen by the Pa$\beta$ and Pa$\gamma$ lines - the latter lying adjacent to the He I lines in Figures 1-8. In this sense, they are similar to the weak-lined TT stars (WTTs), such as those recently reported by \citet{tb22} and \citet{tb23}. They first identified late-type stars with He I emission that is normally present in accreting systems. Within this sample, they observed the H$\alpha$ line at  high spectral resolution (most with R$\sim$32,500), resulting in a velocity resolution of 7-16 km/s. Their WTTs include systems including both weakly accreting and ``non-accreting'' systems. \citet{tb22} modeled their observed spectra using a gas inflow model, plus chromospheric emission. They extracted mass accretion rates, and found that the derived rates depended on the inclination of the disk feeding the accretion onto the star, suggesting the observed rates were biased, with those in the most highly-inclined disks likely being smaller than lower-inclination objects. This can occur when the higher inclination disks have disk dust that obscures part of the line-emitting regions, or where the column density of the gas inducers self-absorption (saturation). \\

\begin{deluxetable}{lcc}[!ht]
\tablecolumns{3}
\tablewidth{0pc}
%\rotate
\tabletypesize{\scriptsize}
\tablecaption{Disk Inclinations}
\tablehead{
\colhead{Object} & \colhead{Outer Disk i$^{\circ}$  \tablenotemark{a}} & \colhead{Inner Disk i$^{\circ}$}  
}
\startdata
HD 142666 & 61 & 78$\pm$2 \tablenotemark{b} \\
HD 143006 &19 & 27-28\tablenotemark{c} \\
HD 145718 & 70 & 62-63\tablenotemark{c}  \\
V935 Sco & 47 & \\
DoAr 25 & 66 & \\
EPIC 203850058 & 84 & \\
EPIC 204638512 & 8  & nearly edge-on\tablenotemark{d}  \\
EPIC 205151387 & 54  & \\
\enddata
\tablenotetext{a}{\citet{ansdell20}} 
\tablenotetext{b}{\citet{gravity19}}
\tablenotetext{c}{\citet{lazareff17}}
\tablenotetext{d}{\citet{sa20}}
\end{deluxetable}

\begin{figure}[!h]
\includegraphics[width=9.5cm, height=7cm]{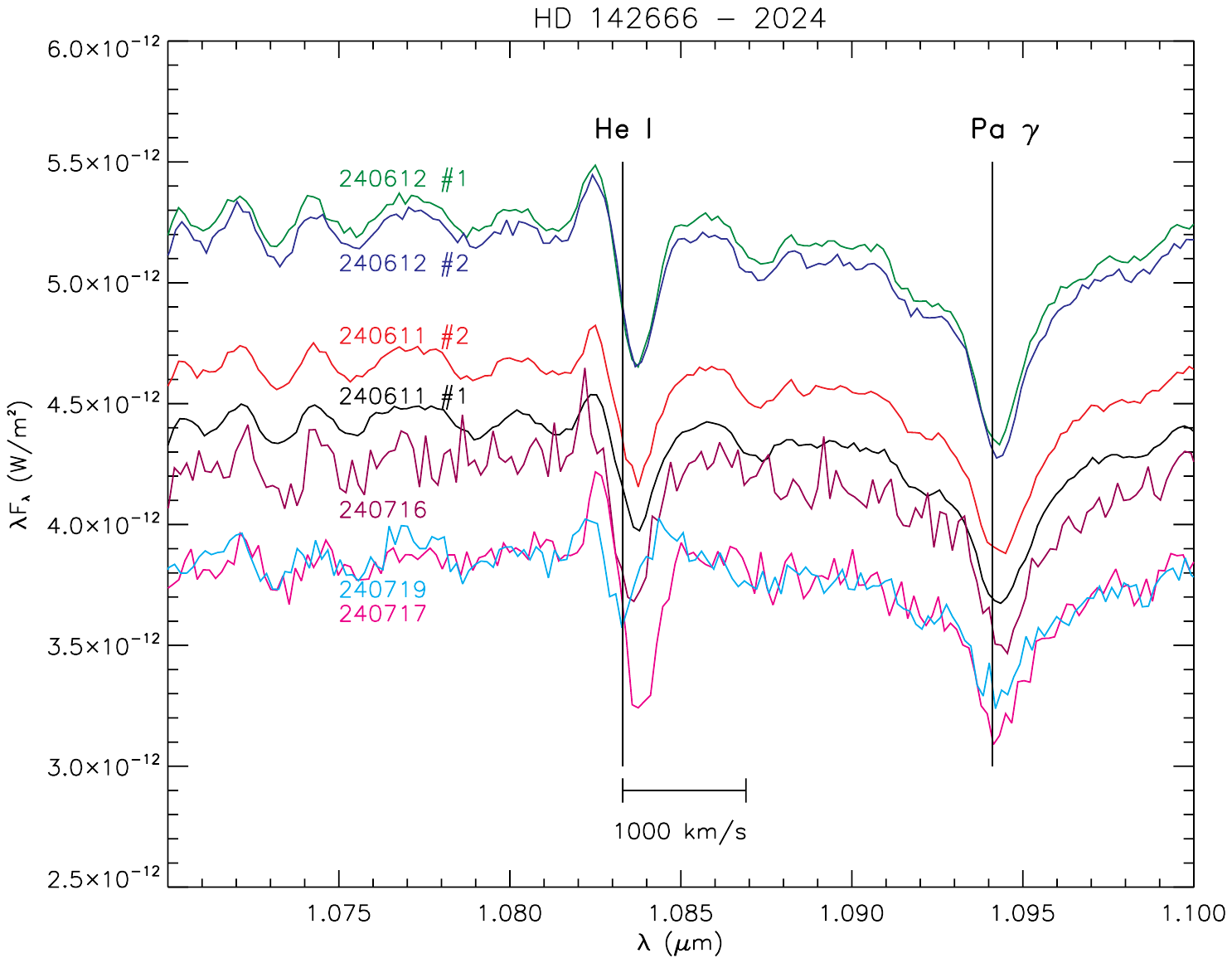}
\includegraphics[width=9.5cm, height=7cm]{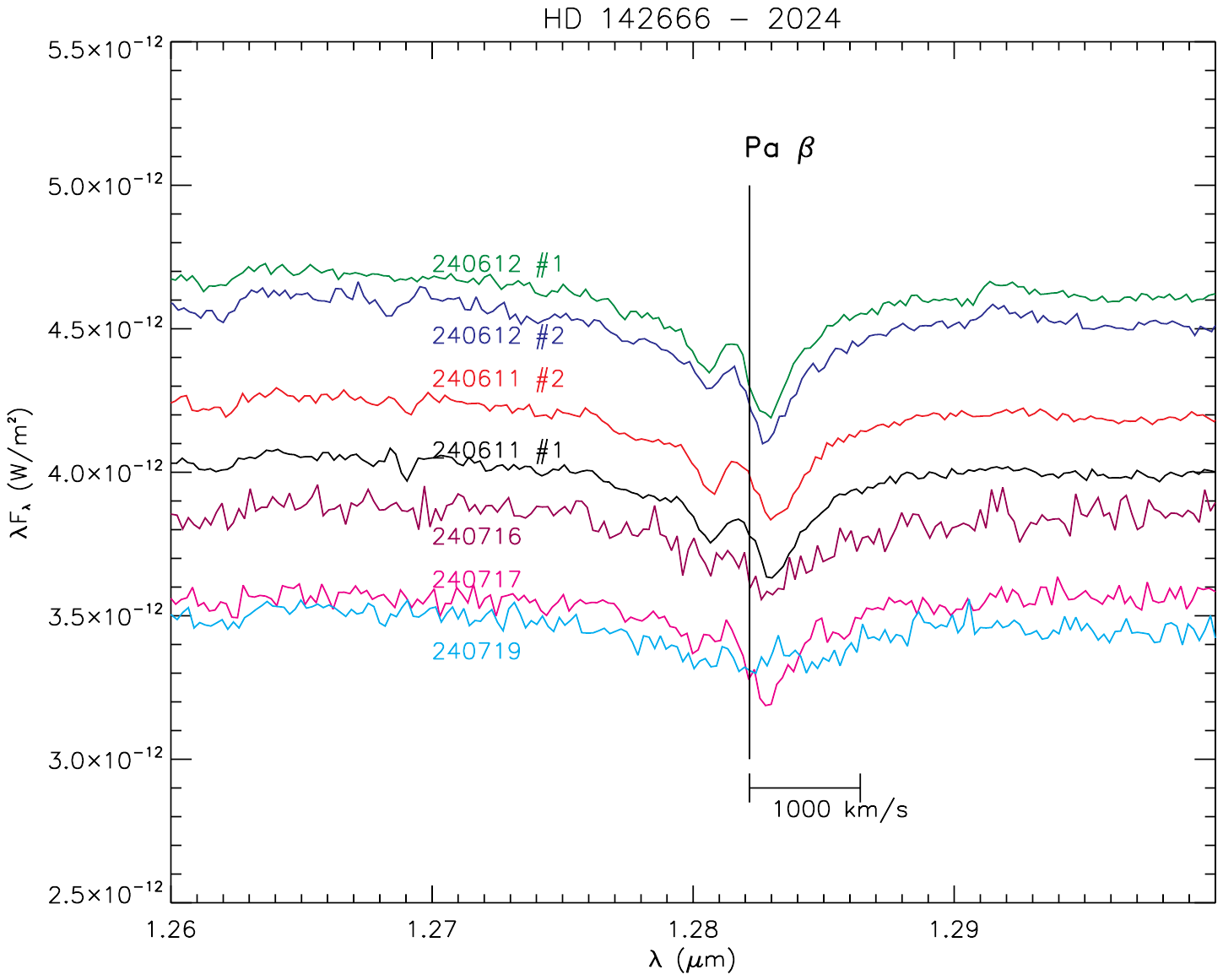}
\caption{In Figure 1 we show the SpeX He I and Pa$\beta$ spectra of HD 142666, normalized to their independently-determined K-band fluxes. The epoch color-coding is the same for both figures.\label{fig:f1}}
\end{figure}

\begin{figure}
\includegraphics[width=9.5cm, height=7cm]{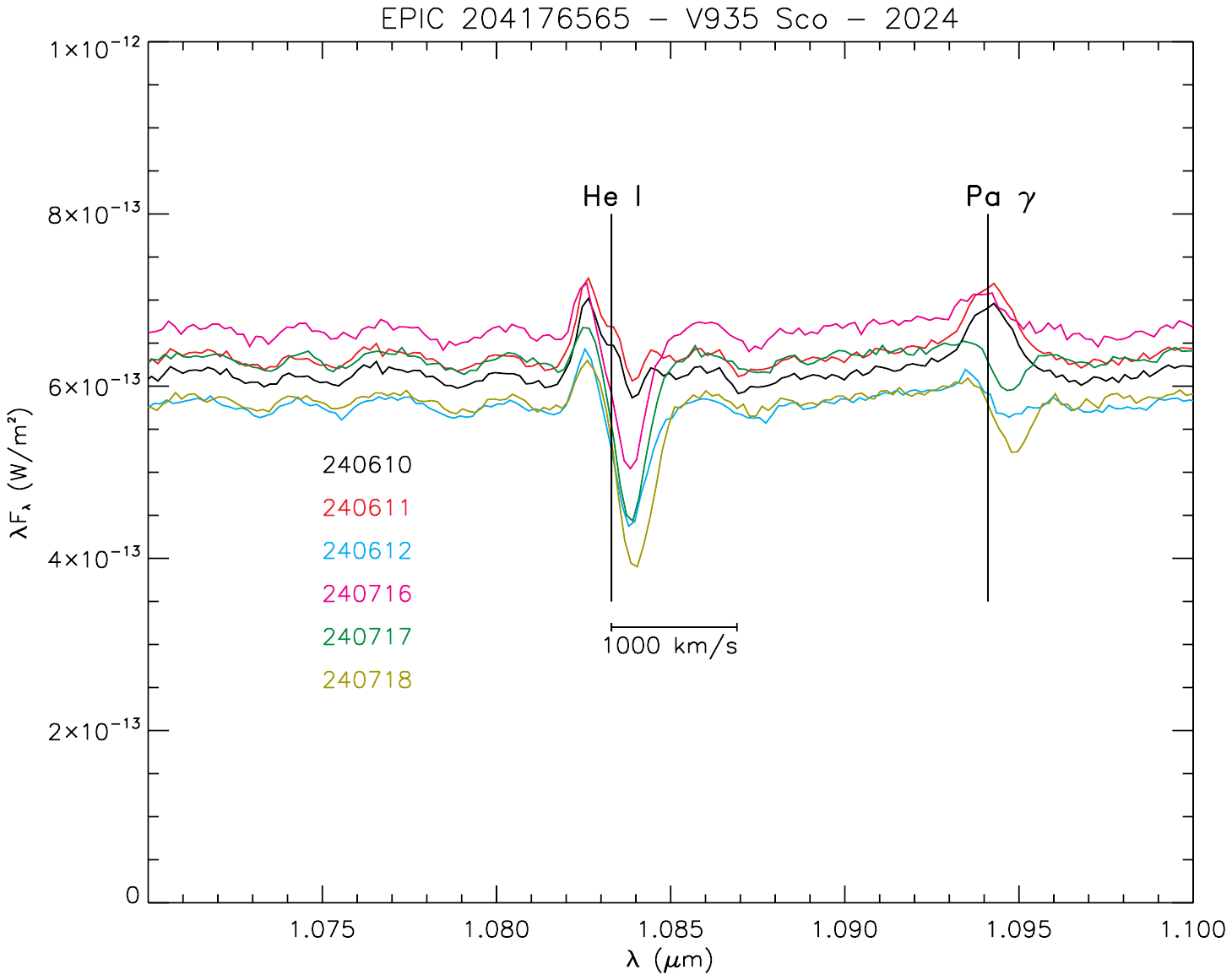}
\includegraphics[width=9.5cm, height=7cm]{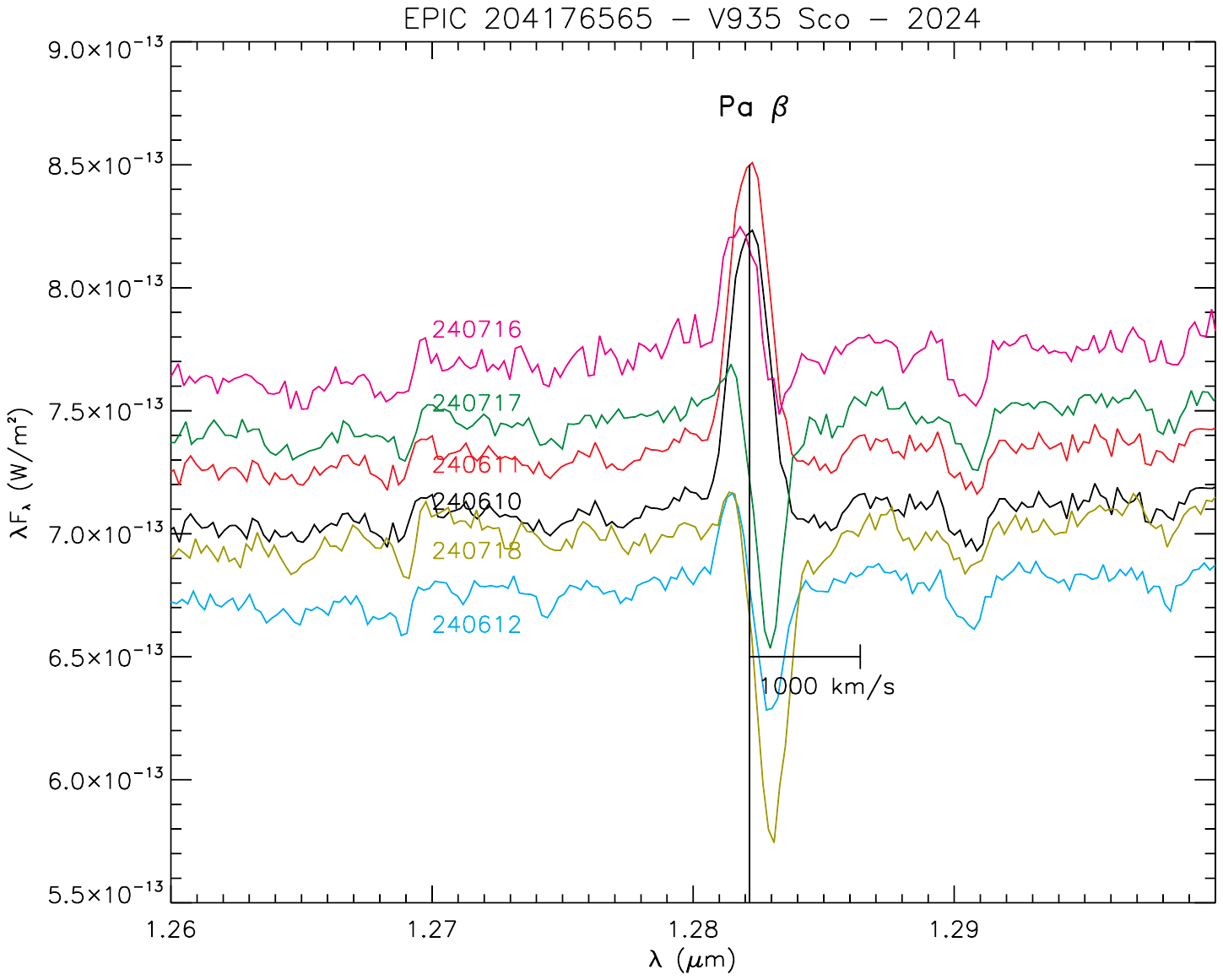}
\caption{In Figure 2 we show the SpeX He I and Pa$\beta$ spectra of V935 Sco - EPIC 204176565, normalized to their K-band fluxes. The epoch color-coding is the same for both figures.\label{fig:f2}}
\end{figure}

\begin{figure}
\includegraphics[width=9.5cm, height=7cm]{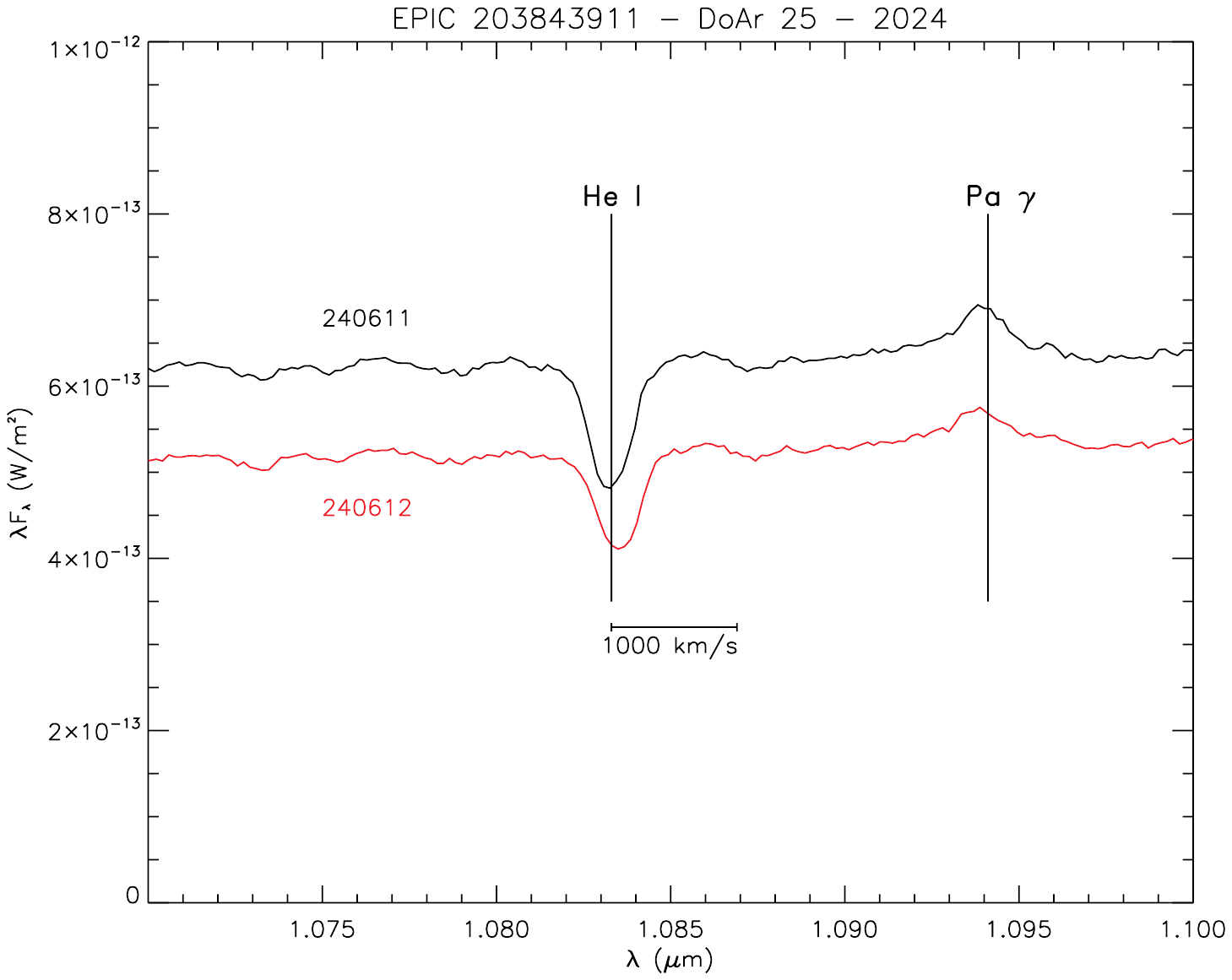}
\includegraphics[width=9.5cm, height=7cm]{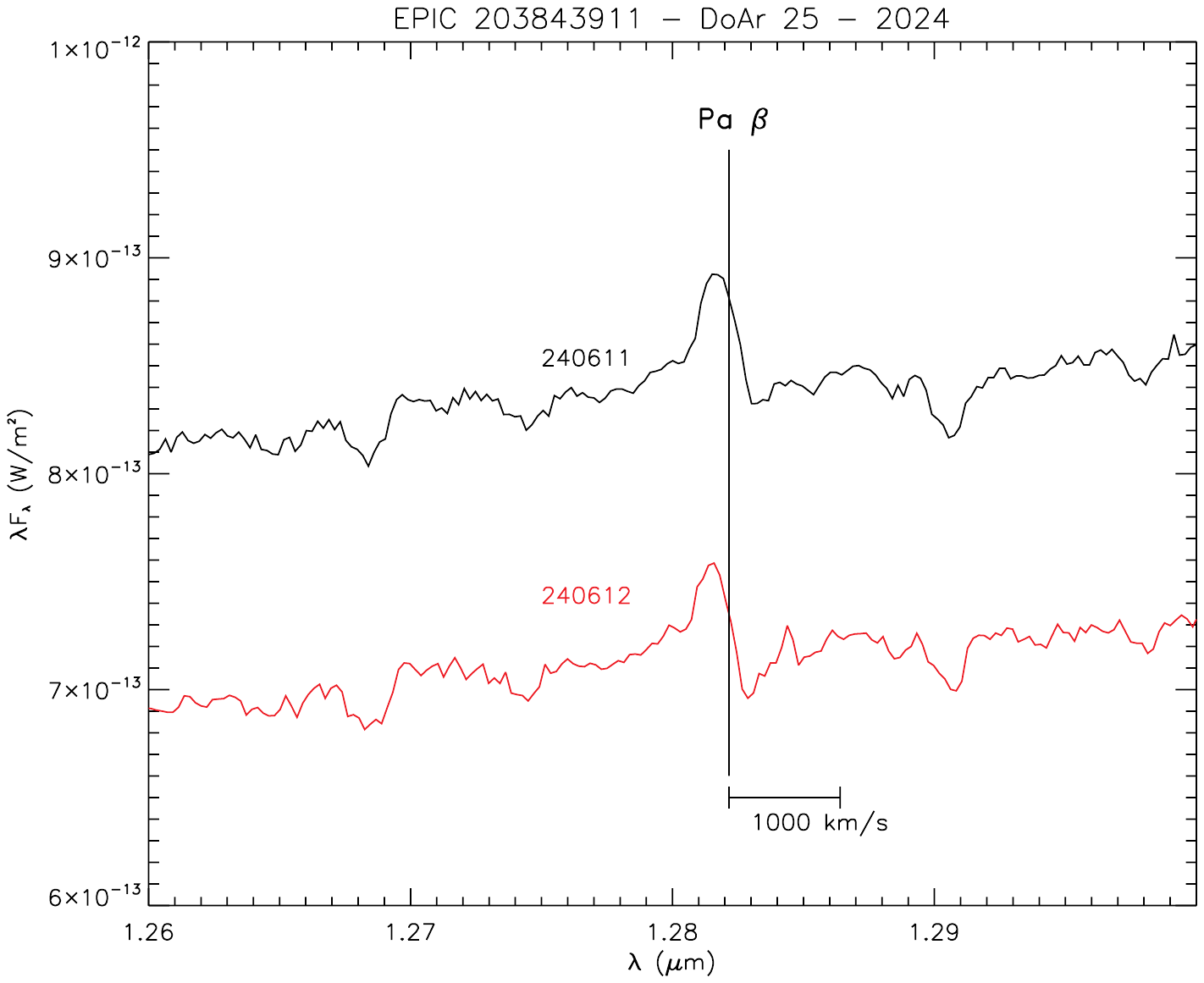}
\caption{The same as Figure 1, but for DoAr 25 - EPIC 203843911.\label{fig:f3}}
\end{figure}

\begin{figure}
\includegraphics[width=9.5cm, height=7cm]{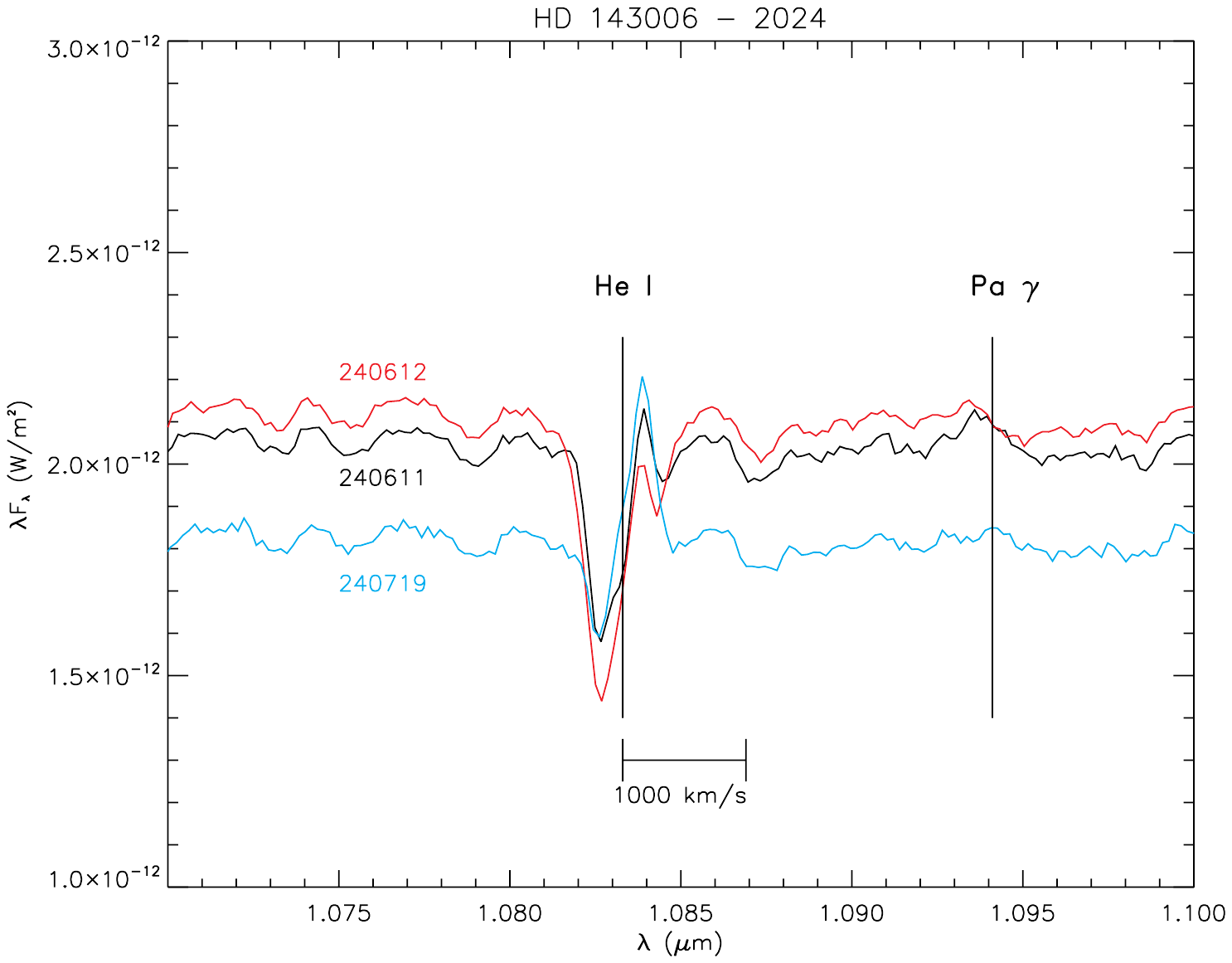}
\includegraphics[width=9.5cm, height=7cm]{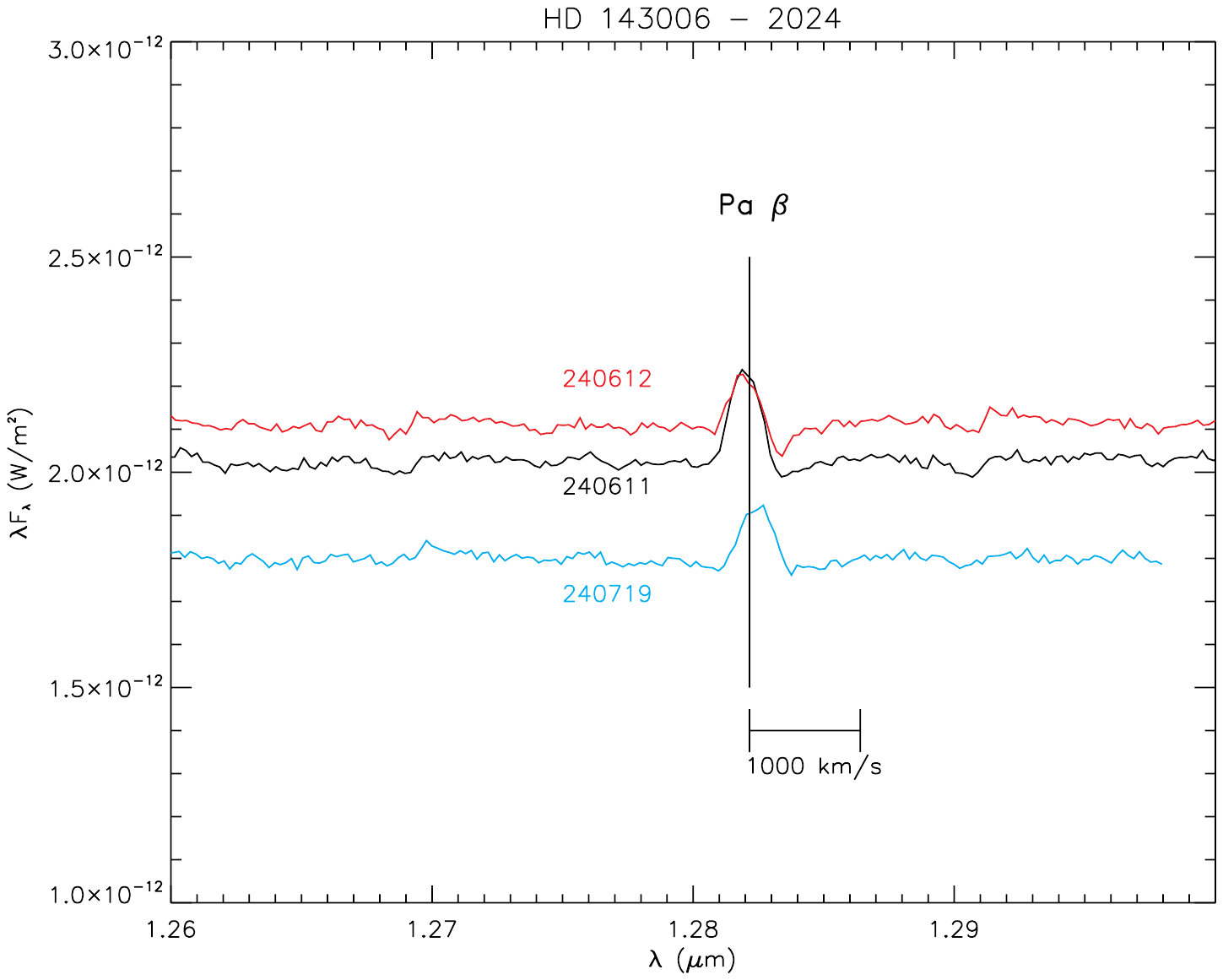}
\caption{The same as Figure 1, but for HD 143006. For HD 143006, the He I exhibits a P Cygni profile, indicating a disk wind.\label{fig:f4}}
\end{figure}

\begin{figure}
\includegraphics[width=9.5cm, height=7cm]{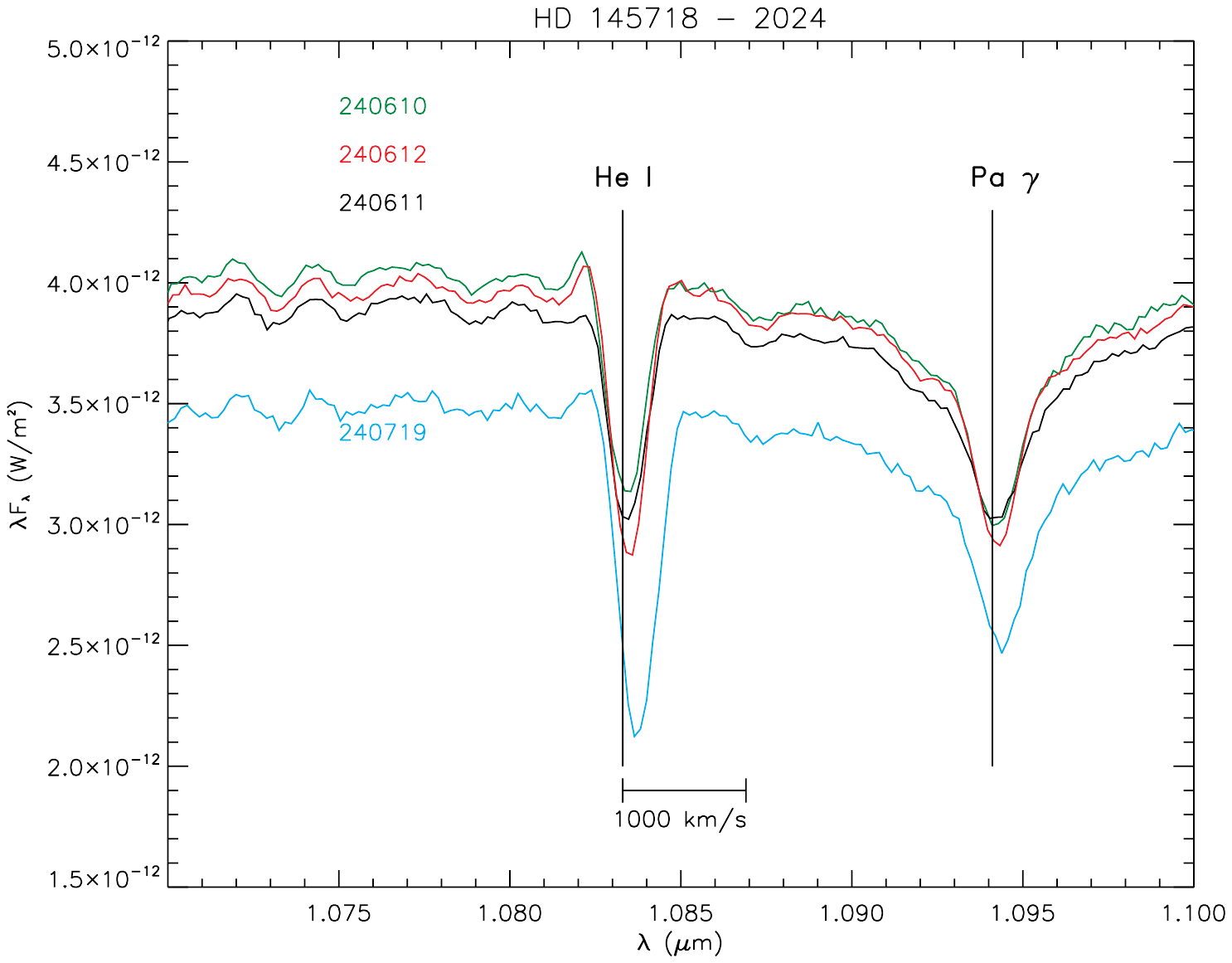}
\includegraphics[width=9.5cm, height=7cm]{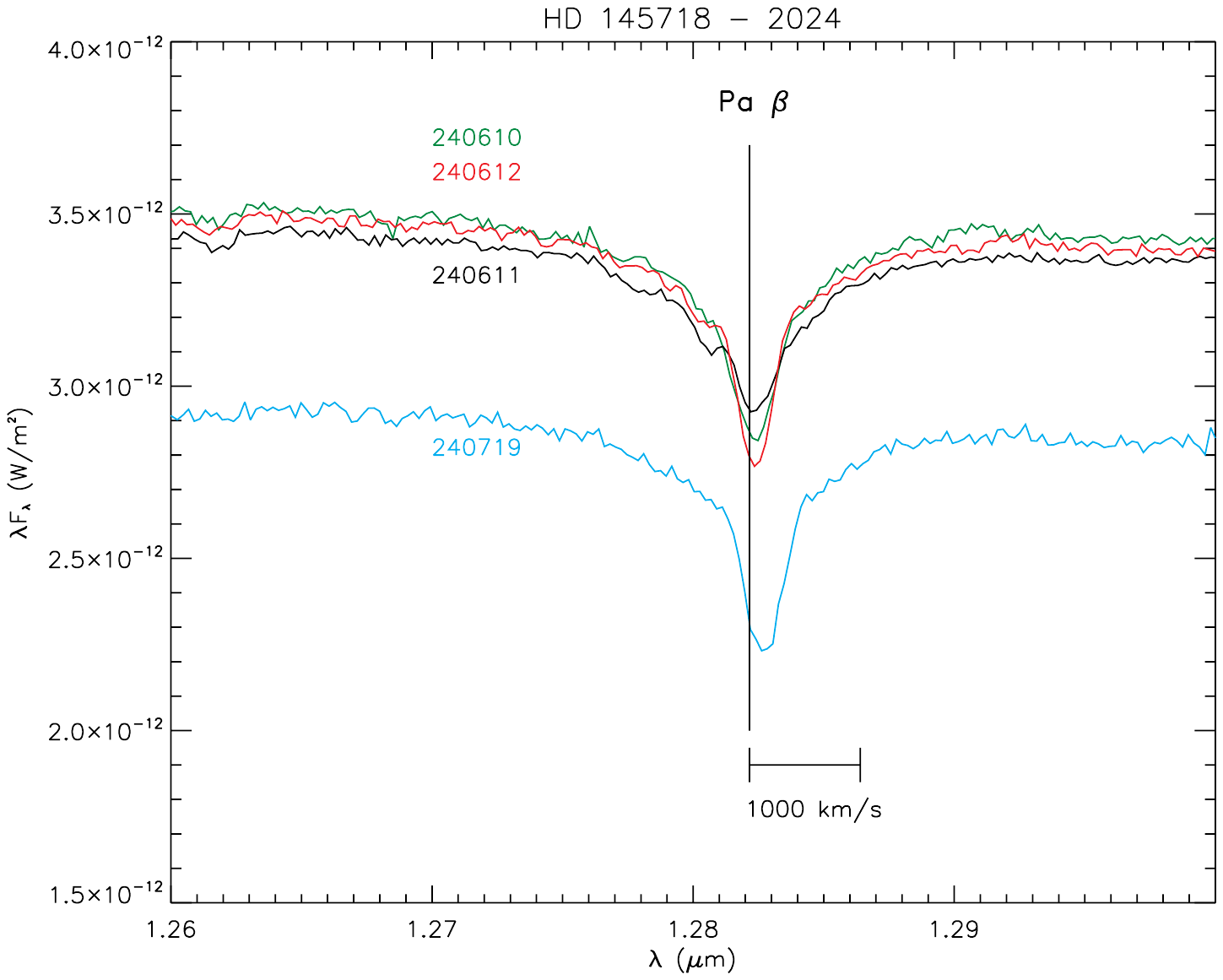}
\caption{The same as Figure 1, but for HD 145718.Here, the Pa $\gamma$ (and Pa $\beta$) absorption does not change in strength as the continuum decreases, but the He I becomes stronger. For all three lines, the line is shifted redward from the rest wavelength, indicating inflowing gas. The correlation between continuum flux and line absorption is indicates that the He I line is at least partially mixed with the dust. By contrast, the Paschen lines are less sensitive to the amount of dust in the line of sight. \label{fig:5}}
\end{figure}

\begin{figure}
\includegraphics[width=9.5cm, height=7cm]{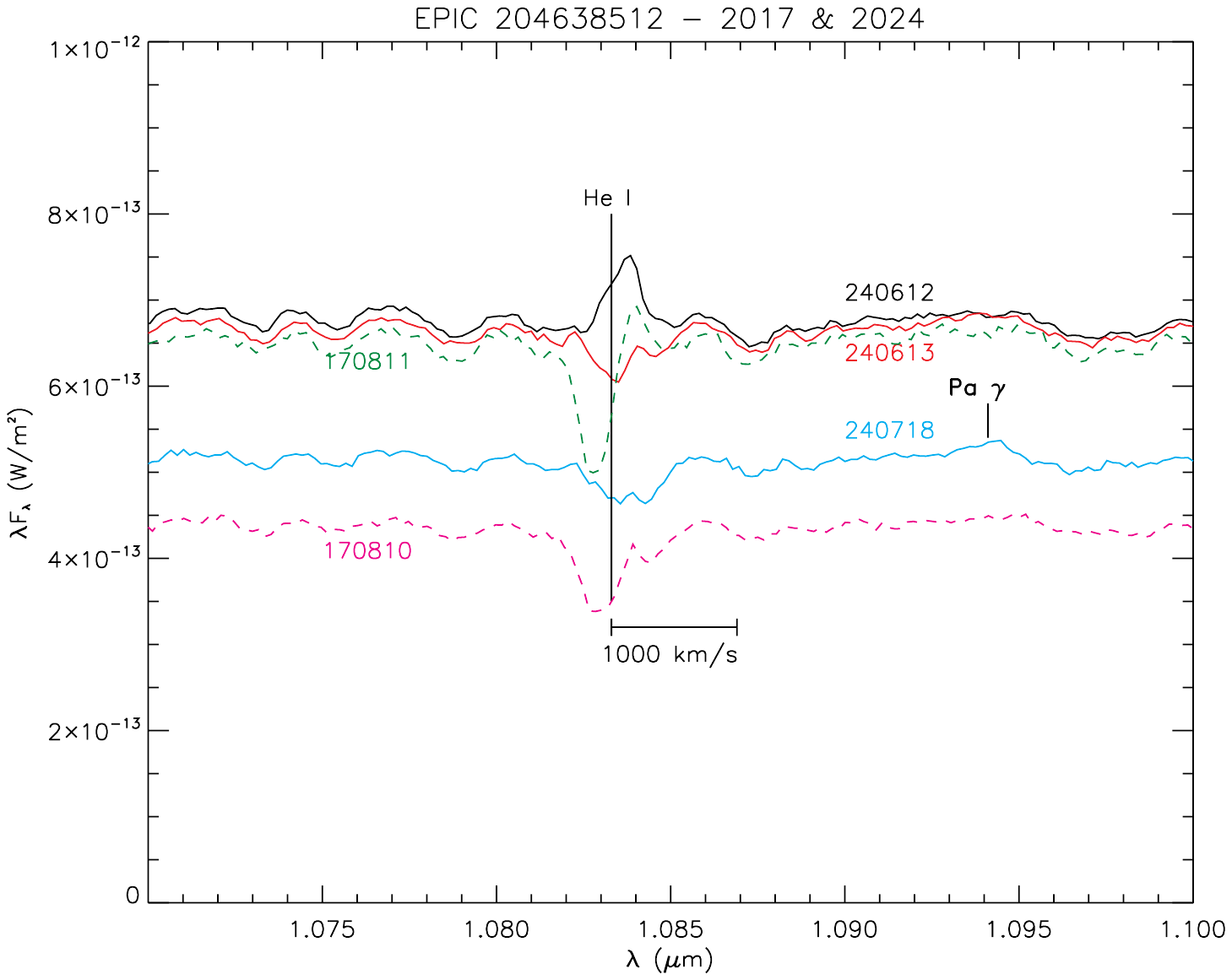}
\includegraphics[width=9.5cm, height=7cm]{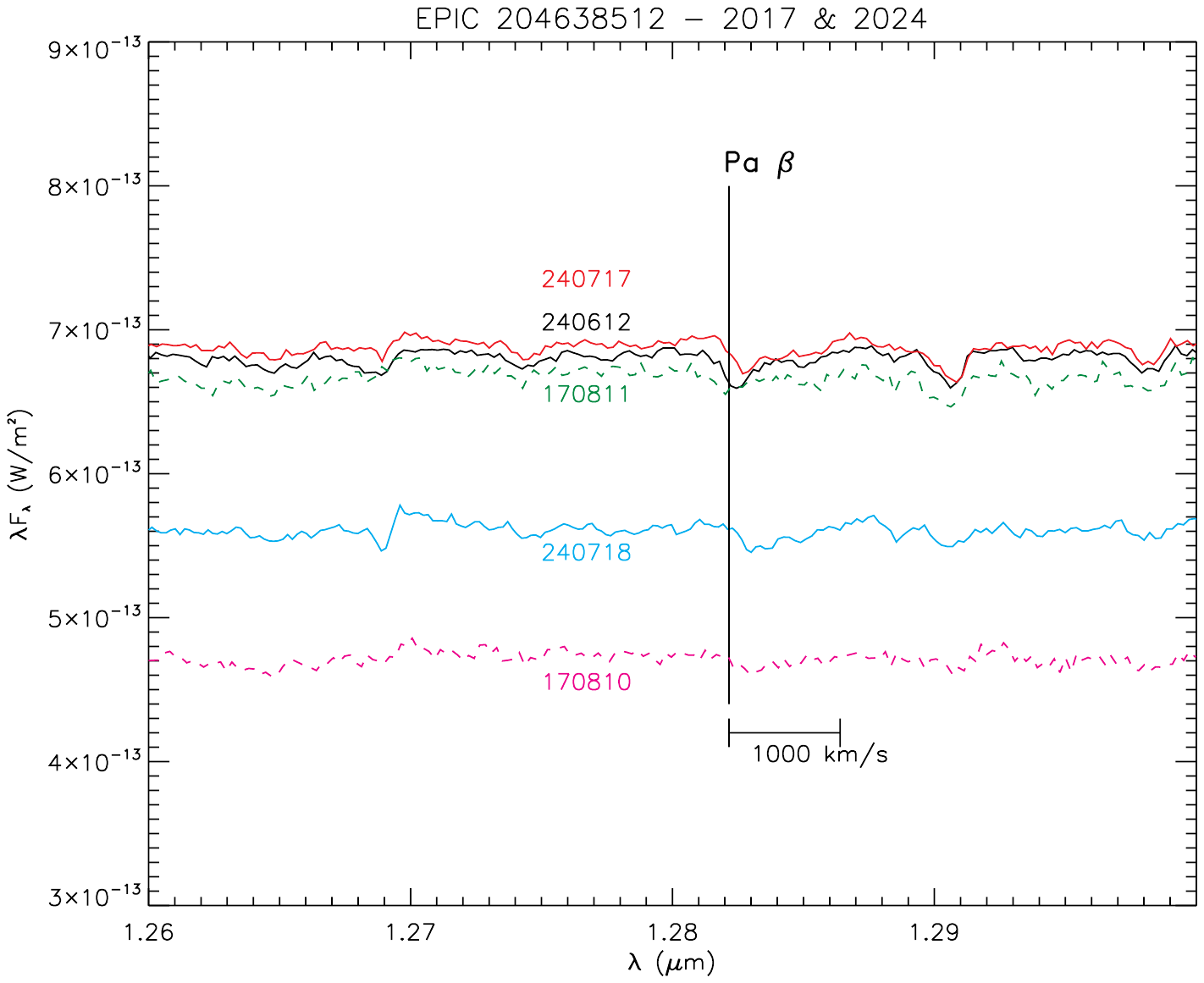}
\caption{The same as Figure 1, but for EPIC 204638512.\label{fig:6}}
\end{figure}

\begin{figure}
\includegraphics[width=9.5cm, height=7cm]{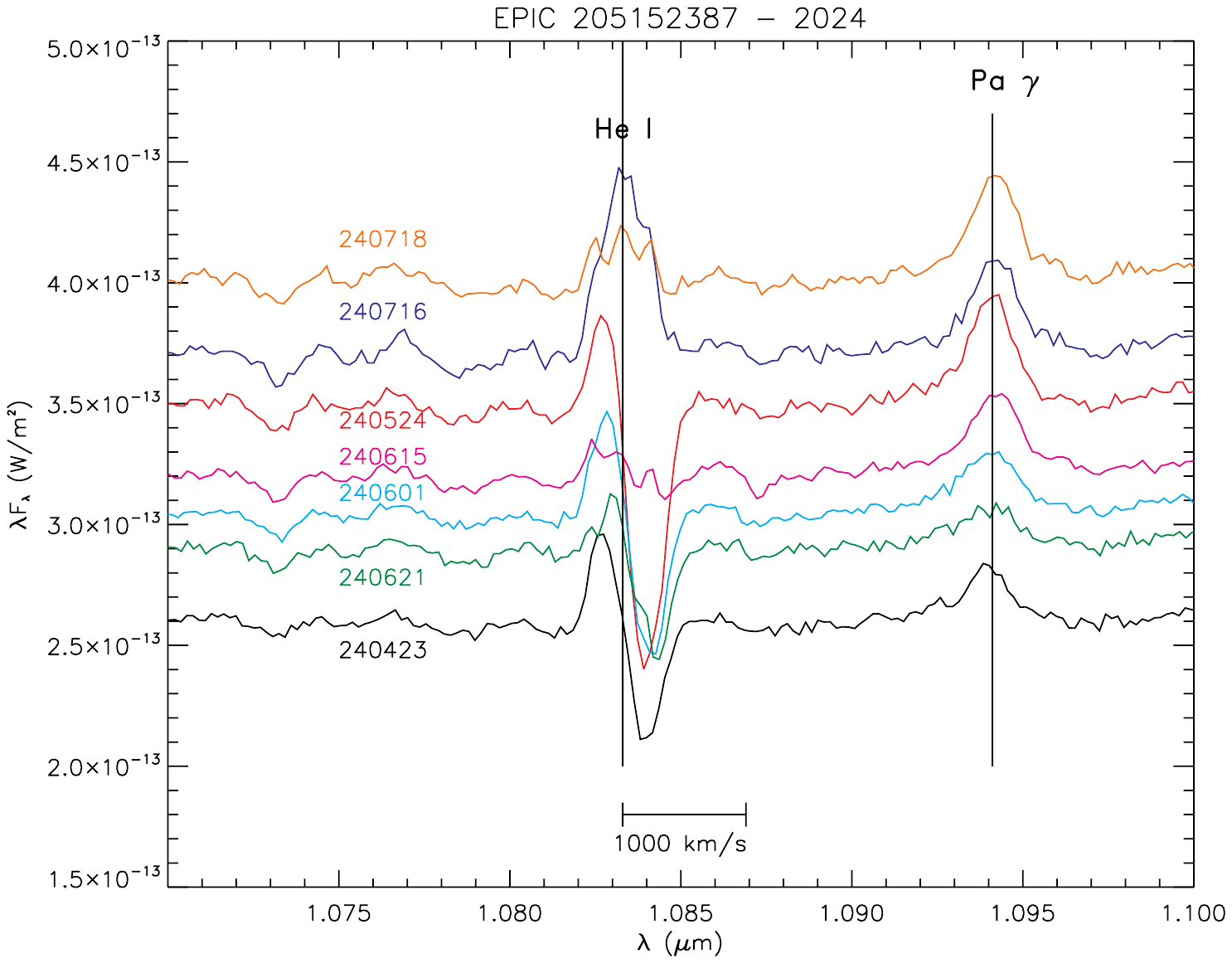}
\includegraphics[width=9.5cm, height=7cm]{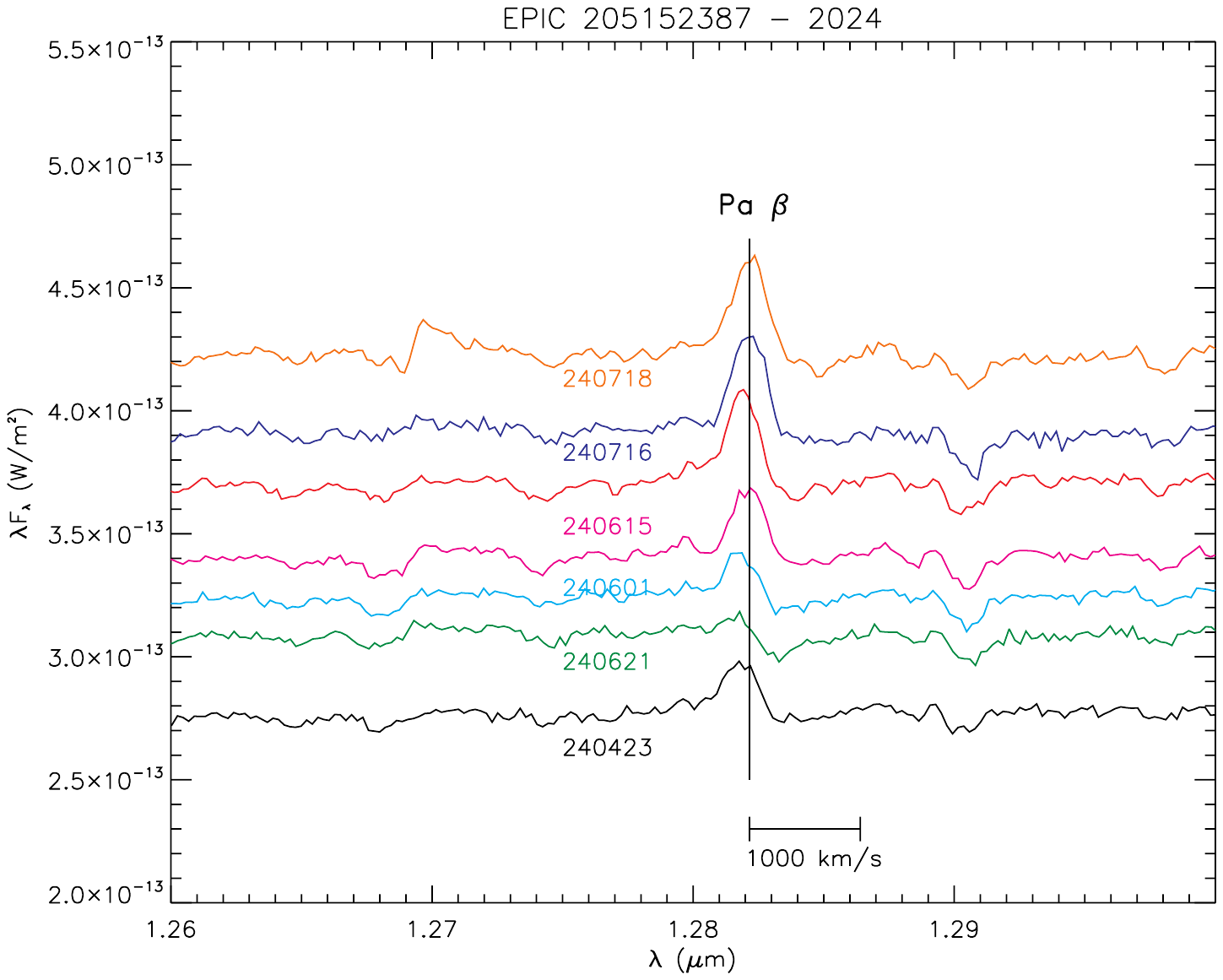}
\caption{The same as Figure 1, but for EPIC 205151387.\label{fig:7}}
\end{figure}

\begin{figure}
\includegraphics[width=9.5cm, height=11cm]{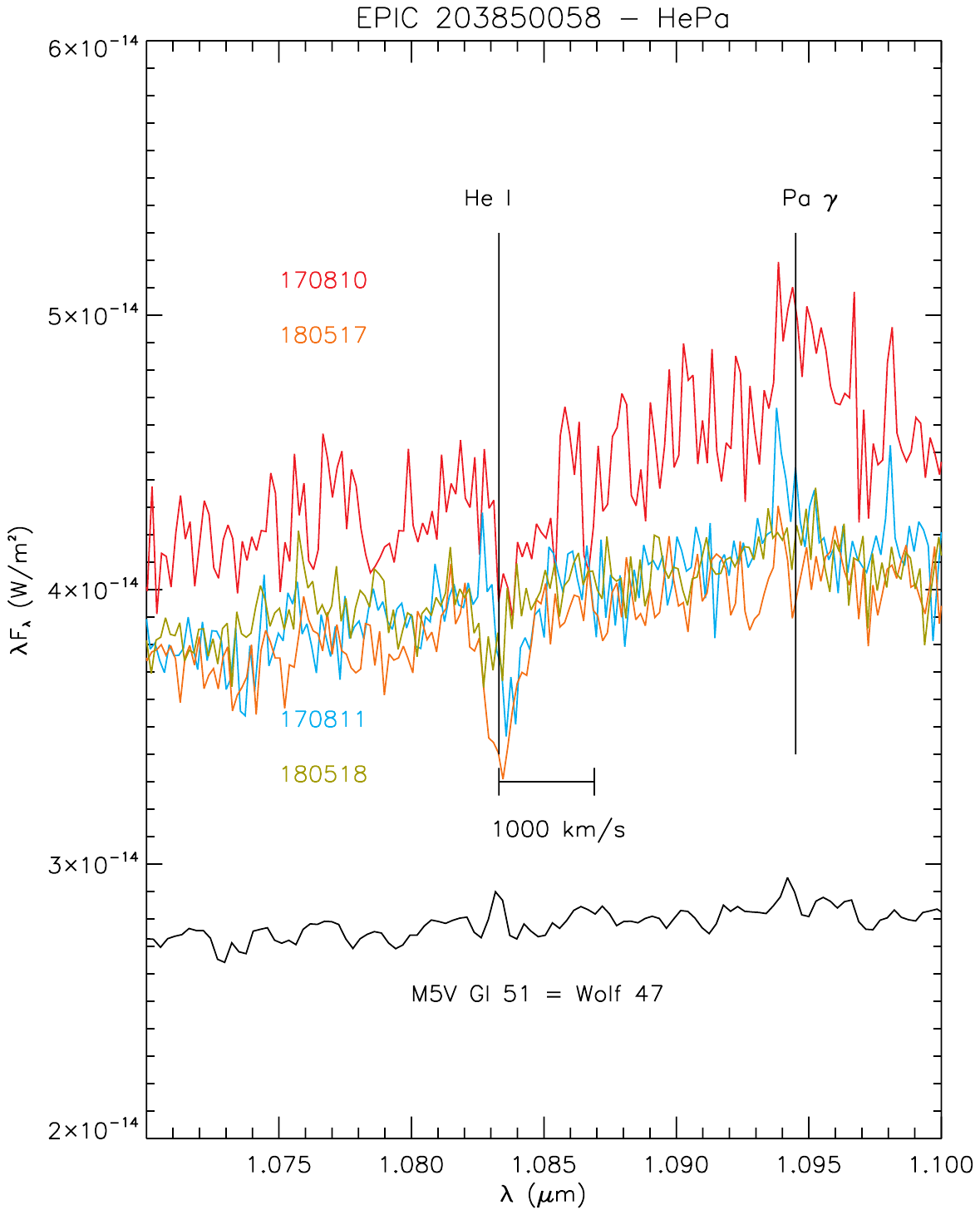}
\includegraphics[width=9.5cm, height=11cm]{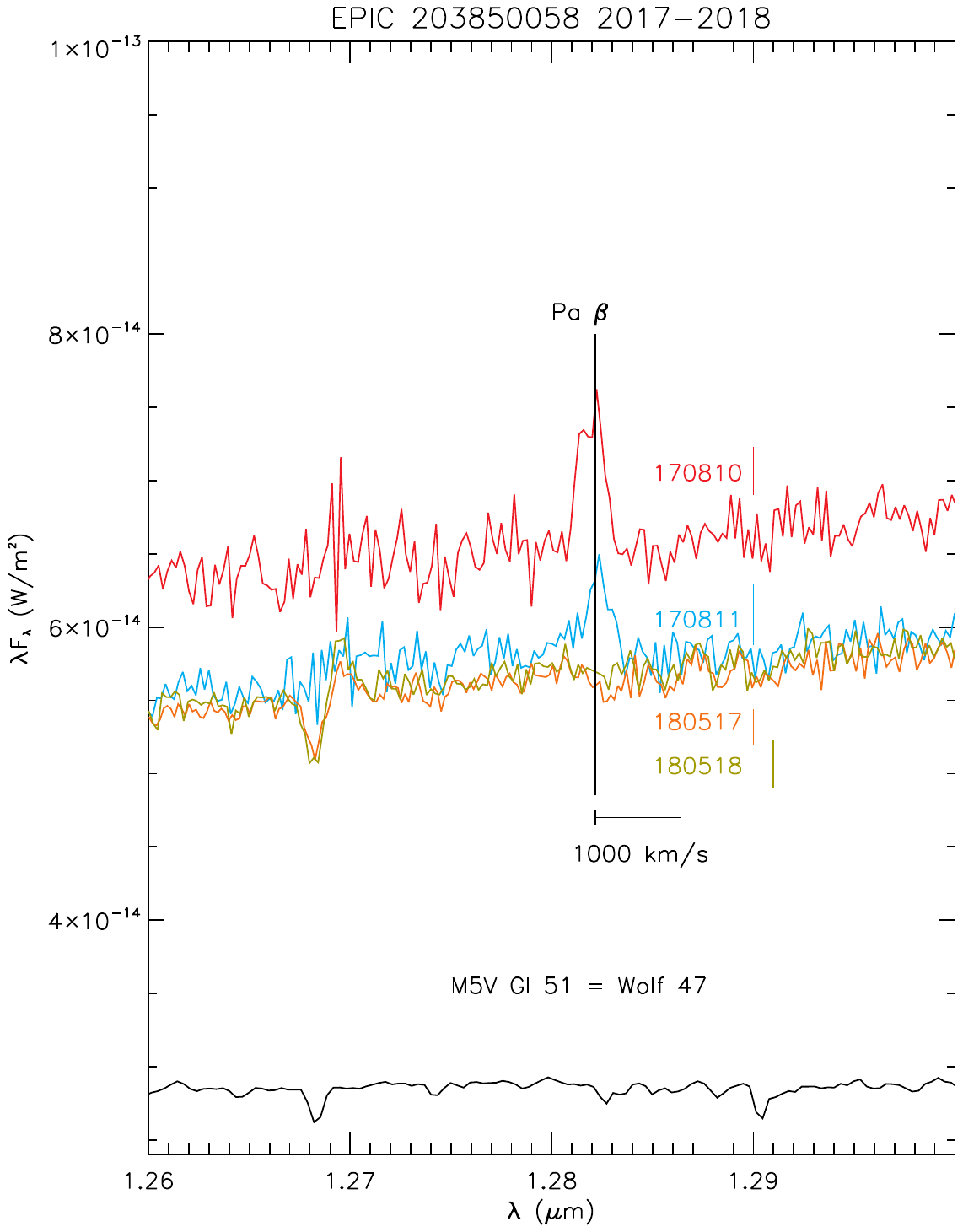}
\caption{The same as Figure 1, but for EPIC 203850058. For the Pa$\beta$ plot, the vertical bars next to the dates indicate the 1-$\sigma$ uncertainty in the photometry used to derive the absolute fluxes id the spectra.\label{fig:8}}
\end{figure}

Here, we employed the extraction methods used in \citet{sitko12} and \citet{pikhartova21} to derive the net line fluxes in Pa$\beta$ and Br$\gamma$.  These start with a general model of the continuum of the stellar photosphere with thermal dust emission added. - see Figure 9, in the Appendix. These are not \textit{physical} models, but rather just a tool to extract any line emission above the underlying continuum (both dust emission and the stellar photosphere) uniformly throughout the observed spectrum. From these, the line fluxes are extracted. A sample of the extracted Pa$\beta$ and Br$\gamma$ lines are presented in the Appendix in Figures 11-56.  \\

Finally, line luminosities were derived with the use of Gaia DR3 parallax data \citep{gaia23}. Using the mass and radius derived by plotting the locations of the stars on an HR diagram and with PMS evolutionary mass tracks and isochrones from \citet{tognelli11}, which are shown in Figure 9. From these we derived the mass accretion rates, using the calibration of both lines from \citet{alcala14}, which applies to stars with masses less than 1\(M_\odot\). For the A5 star, HD 145718,  and the F0 star HD 142666, the calibration of \citet{fairlamb17} was used. However, \citet{fairlamb17} showed that the calibrations for for Pa$\beta$ and Br$\gamma$ merge smoothly into the T Tauri star calibrations in the Alcala sample. This suggests that the Fairlamb calibration could be used for the T Tauri stars. It also implies that the uncertainties in the slope and intercept of these stars is likely smaller than those determined from the Herbig and T Tauri stars separately. \\

For determining the mass accretion rates in the T Tauri stars (HD 143006, V935 Sco, DoAr 25, EPIC 204638512, EPIC 205151387, and EPIC 203850058) we have included the correction factor described by \citet{pittman25a}, who found that the usually-assumed inner radius R$_{in}$=5R$_{*}$ is incorrect for the T Tauri stars. They found that R$_{in}$$\sim$2.8R$_{*}$ described the data in their survey. \citet{pittman25a} showed that mass accretion rates obtained with the older value of the co-rotation radius should be scaled upward by a factor of 1.25x. \\ 

For HD 142666, and HD 145718, the mass accretion rates were determined using the relation \.{M}=L$_{acc}$R$_{acc}$/GM, as is typically done for the Herbig Ae/Fe stars. \citet{sitko26} have shown that, while the Herbig stars tend to have magnetic field strengths that are weaker that those of the T Tauri stars, there is considerable overlap in the measured field strengths between the two classes of stars. Should the magnetic fields be strong enough to act as the T Tauri Stars do, the mass accretion rates should be multiplied by as much as 1.55x(R$_{in}$$\sim$2.8R$_{*}$). For these two stars, we only list the accretion rates with no added factor for a possible R$_{in}$. This is in keeping with the mass accretion rate determinations of \citet{db11} and \citet{fairlamb15}. These are likely lower limits for the Herbig stars, as discussed by \citet{sitko26}, however, and a more robust determination of R$_{in}$ is required for more precise determinations of the mass accretion rates. \\

\subsection{Fundamental Stellar Parameters}

\citet{kastner14}  has pointed out that the spectral type of TW Hya, the closest and best-studied T Tauri star, depends on what wavelength region is being used for the purpose of classification. Most classifications made at visible wavelengths place it between K6 and K8, but \citet{vacca11} found it to be M2.5V at near-IR wavelengths. \citet{vacca11}  suggested that this discrepancy is due to the two-component nature of the systems, with an accretion hot spot and inner disk contributing a $\sim$4000 K (K7) spectrum, plus a cooler M-type spectrum that dominated the flux at wavelengths longer than 1 $\mu$m.   \citet{hh13} found this to also be true for other stars in the TW Hydrae Association and Taurus Molecular Cloud, where the derived spectral classification depends on the size and shape of the spectrum of the accretionally-heated region. The presence of unresolved companions of later spectral types will tend to make the spectral type look cooler (V935 Sco is such an object, as discussed later in this paper). This will impact the effective temperature assumed for the stars in question.  This will influence both the mass and radius derived for the stars, particularly those cooler stars sill evolving along the PMS Hayashi tracks, where the tracks are spaced closely. \\

Hotter, higher mass stars suffer a different problem. Interferometric studies of such objects indicates that they are rotationally flattened, exhibiting a range of effective temperatures and surface gravity going from their rotational poles to their equators \citep{vanbelle06,mcalister05,aufdenberg06,peterson06a,peterson06b,monnier07,zhao09,che11}, and summarized by \citet{vbell12}. Because of this, the observed spectrum samples a range of both parameters.   \\

Our initial determination of the effective temperature luminosity, mass, and radius of each star begins with the spectral classification, brightness in the V photometric band, observed B-V color. For those stars where the effective temperature had been determined by modeling of existing spectra, these formed the basis of its actual value. In other cases, where merely the spectral type was known, a determination of the effective temperature was determined from the work of \citet{pm13} for PMS stars was used, which also provided the intrinsic B-V and Bolometric Correction. Distances came from the. The observed B-V and intrinsic B-V gave the reddening E(B-V), which was converted into total absorption in the V band, A${_V}$, assuming a standard interstellar extinction curve, from \citet{ccm89} and R${_V}$=3.1. From the distance, de-reddened V brightness, and bolometric correction, the total bolometric luminosity was determined. \\

Table 3 lists the adopted spectral classification and its associated effective temperature, intrinsic B-V, bolometric correction, and distance. Table 4, shows the parameters derived from the observed quantities and location of the stars in Figure 9 to get the stellar masses and ages. \\

\clearpage

\begin{figure}[!h]
\begin{center}
\includegraphics[width=15.0cm, height=18cm]{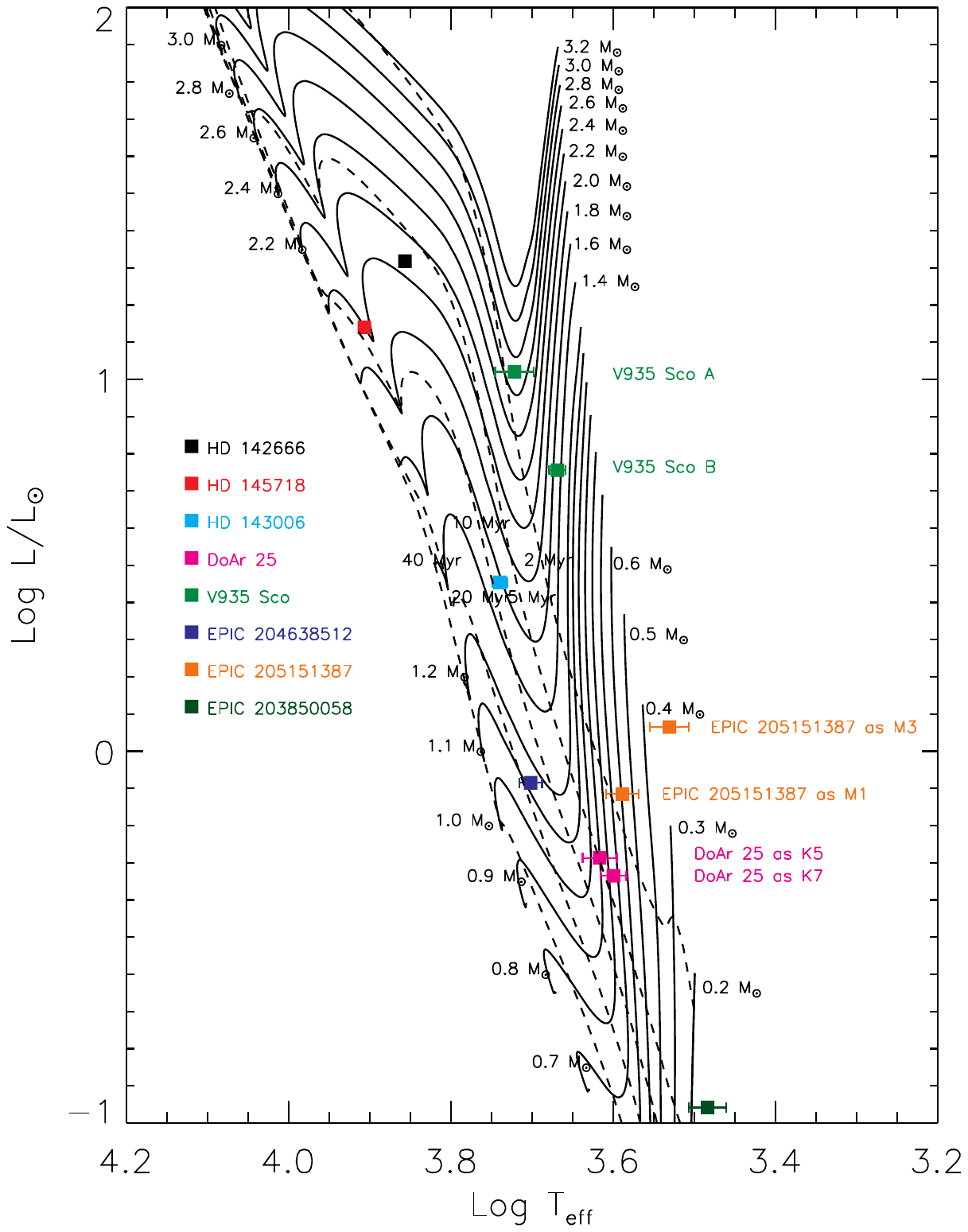}
\caption{Stellar luminosities versus effective temperatures. Overplotted are PMS evolution tracks and isochrones from \citet{tognelli11}, which include the revised cosmic abundances from \citet{asplund09}. For EPIC 205251387 and DoAr 25, we show 2 points, to include different spectral classifications found in the literature (DoAr 25 - K5 from \citet{ansdell20} and K7 (this paper); EPIC 205151387 - M1 from \citet{ansdell20} and M3 from this paper). The point for V935 Sco is that derived from the analysis in section 6.7.  \label{fig:9}}
\end{center}
\end{figure}

\clearpage

\begin{deluxetable}{lcccccc}
\tablecolumns{7}
\tablewidth{0pc}
%\rotate
\tabletypesize{\scriptsize}
\tablecaption{Star Spectral Classification}
\tablehead{
\colhead{Object} & \colhead{Sp.Type\tablenotemark{a}} & \colhead{Teff(uncert.)\tablenotemark{b}} & \colhead{Intrinsic (B-V)\tablenotemark{b}} & \colhead{B.C. at V\tablenotemark{b}} & \colhead{A$_{V}$\tablenotemark{c}} & \colhead{Distance (pc)} 
}
\startdata
HD 142666 & F0sh & 7200(45) & 0.053 & -0.01 & 1.54  & 142.6  \\  % Davies et al. 2018 CHARA chose A8
HD 143006 & G5IVe & 5500(120) & 0.70 & -0.17 & 0.25 & 167.3  \\ 
HD 145718 & A5Ve & 8080(135) & 0.10 & -0.03 & 1.00 & 154.7 \\ 
V935 Sco A & K7 & 4150(150) & 1.28 & -1.44 & 2.7  & 136.9 \\
DoAr 25 & K5 & 4140(200) & 1.24 & -0.80 & 0.5  & 138.2 \\ % RE-DO THIS LINE !!!!!!!!!!!!
EPIC 203850058 & M5.5 & 2880(300) & 1.65 & -3.21 & 2.2 & 141.6  \\ % Av from Manara 2015
EPIC 204638512 & K2 & 4760(165) & 0.93 & -0.46  & 2.1 & 145.3 \\
EPIC 205151387 & M1 & 3630(180) & 1.45 & -1.58 & 0.62 & 135.6 \\
EPIC 205151387 & M3 & 3360(180) & 1.47 & -2.03 & 2.2  & 135.6 \\ % DOUBLE CHECK Avs
\enddata
\tablenotetext{a}{Spectral types come from \citet{gray17}(HD 142666), \citet{pm16}(HD 143006), and  \citet{carmona10}(HD 145718). } 
\tablenotetext{b}{Values from \citet{pm13} for the given spectral type. Uncertainties are from changing the spectral classification by 1 subclass.}
\tablenotetext{c}{The values were derived using R$_{V}$=3.1}
\end{deluxetable}

\begin{deluxetable}{lcccc}[!ht]
\tablecolumns{5}
\tablewidth{0pc}
%\rotate
\tabletypesize{\scriptsize}
\tablecaption{Pre-Main Sequence Star Derived Stellar Parameters}
\tablehead{
\colhead{Object} &  \colhead{M$_{*}$/M$_{sun}$(uncert.)} &  \colhead{R$_{*}$/R$_{sun}$(uncert.)} &  \colhead{Age (yr) (uncert.)} &  \colhead{Log Lum.(W)(uncert.)}}
\startdata 
HD 142666 & 2.0(0.1) & 2.940(0.029) & 6(1) & 1.318(0.003)  \\
HD 143006 & 1.60(0.06) & 1.863(0.058) & 8(2) & 0.454(0.003)  \\ % check Log L for both values of Rv and HRD
HD 145718  & 2.0(0.1) & 1.9010.045) & 9.5(0.5) & 1.140(0.003)  \\
V935 Sco A & 1.5(0.1) & 3.7(0.2) & $<$2 & 0.76(0.03)  \\
DoAr 25 & 0.9(0.2) &1.40(0.10) & 9$^{+3.5}_{-2.5}$ & -0.287 (0.005) \\
EPIC 203850058 & 0.15(0.5) & 2.16(0.07) & 3.0(0.5) & -0.398(0.014) \\
EPIC 204638512 & 1.05(0.05) & 1.01(0.05) & 25(5) & -0.231(0.003)  \\
EPIC 205151387 (M1) & 0.6(0.1) & 1.90(0.15) & 2.0(0.2) & -0.135(0.003) \\
EPIC 205151387 (M3) & 0.6(0.1) & 3.03(0.19) & 2.0(0.2)  & 0.041(0.02) \\
\enddata
\end{deluxetable}

With the effective temperature and luminosity in hand, each object was plotted on a T-L plot, and PMS tracks and isochrones of \citet{tognelli11} were used. These included the revised ``cosmic'' abundances from \citet{asplund09} where the changes in the abundances of C and O (second only to H and He in the universe) were substantial. The abundance of O dropped by 40\% compared to the values a decade prior, and C dropped by 60\% over the same time span. As PMS evolution tracks derived prior to 2011 do not include these updated abundances, the masses and ages will differ from the earlier ones, for a given effective temperature and luminosity. \\

For each star, an uncertainty in effective temperature was used. For these we used the difference in effective temperature of one spectral subclass from \citet{pm13}. The PMS evolutionary tracks and isochrones, and the locations of each object are shown in Figure 9, while the adopted values are listed in Table 4. \\

The luminosity requires additional consideration. First, the objects are variable in brightness. Second, if the value of E(B-V) includes circumstellar (disk) extinction with a non-ISM extinction curve, the luminosity might be lower than that determined using the \citet{ccm89} extinction relation. For the very lowest-mass stars, this will not change the stellar mass significantly, as the PMS tracks are on the Hayashi convective portion of the PMS evolutionary tracks, and remain at nearly on the the same mass of the evolutionary track. \\

\subsection{Magnetic fields in A Stars} 

However, it is well-known that A and early F stars on the Main Sequence are rapidly rotating stars, with a range in T$_{eff}$ and log g, as determined using interferometric observations.  Vega (A0V), for example, is a rapidly-rotating star viewed nearly pole-on, exhibits a pole-to-equator difference of 2000 K \citep{peterson06a}, and Altair (A7Vn) behaves similarly \citep{peterson06b}. Its higher inclination leads to a broadening of its lines (the ``n'' in the classification being short for ``nebulous'' in spectral classifications).  In the past, A-type stars were typically assumed to surface layers dominated by radiative transport, as opposed to convective outer in stars of later spectral types, while the more massive members of the class will have convective cores, both of which can generate magnetic fields. \citet{hubrig09} have detected the presence of magnetic fields in some Herbig Ae stars. In surveys of Herbig Ae/Fe stars by \citet{ryspaeva23} and \citet{anilkumar24}, many exhibited  X-ray emission.  The latter two  studies conclude that the X-ray emission is \textit{not} coming from the accretion shock region, but from magnetically-driven coronal emission, such as magnetic reconnection events. \citet{anilkumar24} estimated magnetic field strengths to be from 10 G to 500 G. Thus, serious consideration must be given that even the A stars in our sample might host magnetic fields.  \\

In the calibration of the strengths of the Pa$\beta$ and Br$\gamma$ lines with the accretion luminosity by \citet{fairlamb17} and \citet{alcala14}, the correlation of the lines to the accretion luminosity transitions smoothly from the Herbig stars to the T Tauri stars, suggesting that the accretion physics may be similar in both classes of stars. This might be explained by the fact that, although the magnetic fields in the Herbig stars are generally weaker that those in the T Tauri stars, they often overlap the field strengths in those stars, to produce significant gas accretion using magnetically confined funnel flows. \citet{sitko26} have discussed this possibility, and, along with \citet{cauley15},  find an inner disk radius comparable to, and possibly smaller than, those of the T Tauri stars.\\

\subsubsection{Determination of the Mass Accretion Rates}

The mass accretion rate is defined as \\

\begin{equation}
\dot{M}= \frac{1}{\epsilon}  \frac{L_{acc}R_{*}}{GM}, \epsilon=1-\frac{R_{*}}{R_{in}}
\end{equation}

where R$_{in}$ is the innermost radius of the gaseous disk where the accreting gas flows from. \\

Thus, determining the mass accretion rate requires the determination of the mass and radius of the stars, which are fixed for any individual star. These will depend on the distance, brightness, and extinction toward the star, as well as the evolution tracks used. These will often be quite different among various sources in the literature, as well as those derived here. \\

Variations in the accretion rates may arise from changes in the inner star-disk regions in these systems. If the region near R$_{in}$ is clumpy, the availability to accrete material will vary with time. Variations in the value of R$_{in}$, as illustrated in Figure 7 of \citet{pittman25b} for the T Tauri stars CS Cha, will mean that $\epsilon$ will vary, and the accretion rate will vary. This is true for the T Tauri stars in this sample, and may also apply to the Herbig Ae/Fe  stars, which would increase the mass accretion rates that we have derived for the two Ae/Fe stars in our sample, HD 142666 and HD 145718.  \\

\subsection{Notes on Individual Objects}

For this study, we sought to sample these objects on a variety of time scales, from days to weeks. We were able to achieve this for all of the objects in this sample, except DoAr 25. A summary of the results for each star follows. In the Appendix, we show the  spectra after removal of the underlying ``background'' emission from the stellar photosphere and any other underlying continuum. \\

\subsubsection{HD 142666}

HD 142666, F0V, is a shell star \citep{gray17} that experiences  dips in brightness in its K2/C2 light curves of up to 40\%, but often has stretches of time where it does not exceed 10\% \citep{ansdell20}. We observed it twice per night on 240611 and 240612 (dates are \textit{yymmdd} UT), and once per night a month later, separated by 1-2 days, in an attempt to sample changes on time scales on short $\sim$few hours, days, and a month. Prior to subtraction of the spectrum of a field star of similar spectral type (HD 27397, F0IV), the He I line at 1.083$\mu$m exhibited an inverse P Cygni (hereafter IPC) profile  indicating gas flowing in toward the star (the opposite of that produced by a disk wind). Pa $\beta$ was in absorption, but at its brightest was accompanied by an emission peak at shorter wavelengths than the absorption core. After removal of the underlying photospheric spectrum, the situation becomes clearer. On 240611 and 240612, the He I line has a profile where the blue-shifted absorption core is many time stronger that the re-shifted emission component. For Pa$\beta$,  the net profile is also IPC, with a strong redshifted emission component and a very weak blue-shifted absorption core. The Br$\gamma$ line behaves in a similar manner. On 240717 the emission component was stronger, about half of the strength of the absorption component. Two days later (240719), the He I line was only weakly in absorption, surrounded by two weak emission components. At this point, it resembles the He I line sometimes seen in the Herbig Ae star HD 163296 \citep{sitko24}. \\
\begin{deluxetable}{ccccc}[ht]
\tablecolumns{5}
\tablewidth{0pc}
\tabletypesize{\scriptsize}
%\rotate
\tablecaption{Mass Accretion Rates for HD 142666 (M$\sim$2.05$\pm$0.05M$_{\sun}$, R=2.94$\pm$0.03R$_{\sun}$)}
\tablehead{
\colhead{Date (UT)} &  \colhead{\.{M}$_{acc,Pa\beta}$}  &  \colhead{\.{M}$_{acc,Br \gamma}$} &  \colhead{Difference (Br$\gamma$-Pa$\beta$)} &   \colhead{Significance}  \\ 
 \textit{yymmdd}  & (10$^{-10}$M$_{\sun}$y$^{-1}$) & (10$^{-10}$M$_{\sun}$y$^{-1}$)  & (10$^{-10}$M$_{\sun}$y$^{-1}$)  &  ($\sigma$) } 
\startdata
240611a & 77$\pm$10 & 119$\pm$16 & 42$\pm$19 &2.2 \\
240611b &86$\pm$11 & 140$\pm$19 & 54$\pm$20 &2.7  \\
240612a & 131$\pm$17 & 187$\pm$25  & 56$\pm$30 & 1.8 \\
240612b & 114$\pm$15 & 163$\pm$22  & 49$\pm$27 & 1.8\\
240716 & 73$\pm$19 & 102$\pm$14 & 29$\pm$24 & 1.2  \\
240717 & 73$\pm$9 & 102$\pm$14 & 29$\pm$17 & 1.7 \\
240719 & 129$\pm$17 & 169$\pm$22  & 40$\pm$28 & 1.4\\
210428 & 92$\pm$12 & 177$\pm$24 & 85$\pm$27 & 3.2 \\
\enddata
\end{deluxetable}

Overall, during the first two nights, as the continuum faded, the Pa$\beta$ IPC emission components did not change substantially. They were weaker, but detectable  on 240616 and 240617. By 240619, the P$\beta$ emission simply filled-in the photospheric core, and no red-shifted emission component was visible. The signal/noise on the data from these last three nights was lower than the ones obtained earlier. \\

 \citet{lazareff17}, using the PIONIER beam-combining instrument on the Very Large Telescope Interferometer (hereafter VLTI) found a resolved inner disk inner disk  with an inclination of $\sim$60$^\circ$. Unless the disk is unusually thick, variable disk extinctions would not be expected, but the presence of a puffed-up inner rim  for a thinner disk might suffice. \citet{sa20} suggest such a model for EPIC 204638512, discussed later in this section. \\
 
 For this object, the mass accretion rate determined from Pa$\beta$ was always lower, 8 out of 8 nights, than that determined from the Br$\gamma$ line.  Most were only at the $\sim$2$\sigma$ level, but collectively, they indicate convincing evidence of having the accretion rates derived from Pa$\beta$ being systematically lower than those derived from Br$\gamma$. On one night, 240719, was it a 3$\sigma$ level. This indicates that the Pa$\beta$ line was saturating faster than the Br$\gamma$ line, which would be expected if the line of sight passes through gas layers close to the disk surface. \\

\subsubsection{HD 143006} % PRE-TRANSITIONAL DISK!

HD 143006 is a T Tauri star (G5IVe) possessing an inner disk that is mis-inclined from the outer disk. As mentioned earlier, \citet{codron25} found that the inner disk has an inclination to the observer of $\sim$22$^{\circ}$$\pm$3$^{\circ}$, and that it is mis-inclined to the outer disk by $\sim$39$^{\circ}$$\pm$4$^{\circ}$
 \\

In the K2C2 light curves presented by \citet{ansdell20}, most of the periodic variations are 5\% or less, and with a period of $\sim$2.7d.  If these periodic variations are not due to dust along the line of sight, then star spots are a possible source, with magnetic fields linking the chromosphere to the disk at the co-rotation distance of R$\sim$2.8R$_{*}$.  \\ % for a stellar mass of 2.2 M$_{\sun}$.

\begin{deluxetable}{ccccc}[ht]
\tablecolumns{5}
\tablewidth{0pc}
\tabletypesize{\scriptsize}
%\rotate
\tablecaption{Mass Accretion Rates for HD 143006 (M=1.39$\pm$0.06M$_{\sun}$, R=1.86$\pm$0.06R$_{\sun}$)}
\tablehead{
\colhead{Date (UT)} &  \colhead{\.{M}$_{acc,Pa\beta}$}  &  \colhead{\.{M}$_{acc,Br \gamma}$} &  \colhead{Difference (Br$\gamma$-Pa$\beta$)} &   \colhead{Significance}  \\ 
 \textit{yymmdd}  & (10$^{-10}$M$_{\sun}$y$^{-1}$) & (10$^{-10}$M$_{\sun}$y$^{-1}$)  & (10$^{-10}$M$_{\sun}$y$^{-1}$)  &  ($\sigma$) } 
\startdata
240611 & 221$\pm$30 & 195$\pm$27 &  -26$\pm$48   & -0.6          \\
240612 & 145$\pm$20 & 167$\pm$23 &   22$\pm$30 &  0.7      \\
240719 & 144$\pm$19 & 139$\pm$19 &   -5$\pm$27 & -0.2          \\
\enddata
\end{deluxetable}

Our sample of 3 observations, 2 are separated by 1 day, and the third by 1 week, are likely under-sampled. The uncertainties in our derived mass accretion rates for the Pa$\beta$ and Br$\gamma$ lines are large enough to be consistent with no variations.  Stellar chromospheric activity might be the source of the observed K2/C2 light variations. Here, there is no systematic difference between the accretion rates determined by the two lines. The values of the accretion rates for Pa$\beta$  and Br$\gamma$ the same, to within 1$\sigma$ (see Table 5).  \\

\subsubsection{HD 145718 = V718 Sco}

\begin{deluxetable}{cccccc}[ht]
\tablecolumns{6}
\tablewidth{0pc}
\tabletypesize{\scriptsize}
%\rotate
\tablecaption{Mass Accretion Rates for HD 145718 (M=2.00$\pm$M$_{\sun}$, R=1.90$\pm$0.06R$_{\sun}$)}
\tablehead{
\colhead{Date (UT)} &  \colhead{\.{M}$_{acc,Pa\beta}$}  &  \colhead{\.{M}$_{acc,Br \gamma}$} &  \colhead{Difference (Br$\gamma$-Pa$\beta$)} &   \colhead{Significance} & \colhead{Notes} \\ 
 \textit{yymmdd}  & (10$^{-10}$M$_{\sun}$y$^{-1}$) & (10$^{-10}$M$_{\sun}$y$^{-1}$)  & (10$^{-10}$M$_{\sun}$y$^{-1}$)  &  ($\sigma$) } 
\startdata
240610 & 8 3$\sigma$ u.l.   & 12$\pm$2 & 4$\pm$2& 2? & Pa$\beta$ - weak. em. with broad absorption wings \\
240611 & 8$\pm$1               & 19$\pm$3 & 11$\pm$3.2 & 3.4 & \\
240712 & 12 3$\sigma$ u.l. &  29$\pm$4 &  17$\pm$4 & 4.2  & Pa$\beta$ - inv. P Cyg w. weak abs.\\
240719 & 16 3$\sigma$ u.l. & 17$\pm$2 &  & & Pa$\beta$ - inv. P Cyg w. equal em. \& abs.  \\
\enddata
\end{deluxetable}

HD 145718 is a Herbig Ae star - A5Ve. It often exhibits near-periodic photometric variations of $\sim$ 4 d, with amplitudes of 5-10\%, based on its K2/C2 photometry \citep{ansdell20}. Using PIONIER interferometry, \citet{lazareff17} determined the inclination of the inner disk to be 62-66$^{\circ}$ (using both ellipsoidal and ring models), while \citet{kluska20} derived a value of 48$\pm$3$^{\circ}$. Both derived the radial distance of the H-band flux to be $\sim$0.06 au, or 3.1R${_*}$, very close to the 2.8R${_*}$ favored by \citet{pittman25a} for T Tauri stars. \\

On all four nights for which we obtained data, the emission lines were weak - in three  cases only 3-$\sigma$ upper limits could be estimated. For Pa$\beta$, two of four nights (240611 and 240712) showed significant differences between the mass accretion rate determined from Pa$\beta$ and Br$\gamma$, with the former being smaller than the latter - a sign that the optical depth through the disk causes greater self-absorption in Pa$\beta$. For another night (240610), the upper limit on Pa$\beta$ was lower than significantly-detected Br$\gamma$ line. On the fourth night, (240719), there was no statistically significant difference between the accretion rates for Pa$\beta$ and Br$\gamma$. \\

The time series spectra, as shown in Figures A-12 through A-15, provide information on the time scales over which variation in the line shapes and strengths change.  On 10 June 2024 UT, the Pa$\beta$ was in emission. on 11 June, it was significantly greater in strength. On 12 June, a red-shifter absorption core developed. Weeks later, on 19 July, a distinctive IPC was present. 

\subsubsection{EPIC 204638512 = 2MASS J16042165-2130284}

EPIC 204638512 is a K2 star possessing photometric dips as large as 40\% in the K2/C2 light curves presented by \citet{ansdell20}. However, \citet{hedges18} demonstrated that the light curve often contains stretches of time where the brightness was stable to within a few \%. In the 4 epochs presented by \citet{sitko24} 3 were similar in brightness, while 1 other was 40\% lower. This indicates that 3 epochs of data were obtained during a relative ``standstill'', while the fourth was obtained during a major dip. The large dip event was likely caused by obscuration by dust whose grain sizes were larger than those seen in the general ISM, but comparable to those present in other circumstellar disks where grain growth has proceeded to a significant extent \citep{sitko23}. The strength of absorption in the He I line also grew as the dust extinction became larger, indicating that the He I gas was mixed in with the dust, so that the gas and dust largely sampled the same material - the gas and dust were inter-mixed. 

\begin{deluxetable}{ccccc}[ht]
\tablecolumns{5}
\tablewidth{0pc}
\tabletypesize{\scriptsize}
\tablecaption{Mass Accretion Rates for EPIC 204638512 \\   (M=1.05$\pm$0.10 M$_{\sun}$, R=1.01$\pm$0.05R$_{\sun}$)}
\tablehead{
\colhead{Date (UT)} &  \colhead{\.{M}$_{acc,Pa\beta}$}  &  \colhead{\.{M}$_{acc,Br \gamma}$} &  \colhead{Difference (Br$\gamma$-Pa$\beta$)} &   \colhead{Significance (Br$\gamma$-Pa$\beta$) }  \\ 
 \textit{yymmdd}  & (10$^{-10}$M$_{\sun}$y$^{-1}$) & (10$^{-10}$M$_{\sun}$y$^{-1}$)  & (10$^{-10}$M$_{\sun}$y$^{-1}$)  &  ($\sigma$) } 
\startdata
170810 & 3.9$\pm$ 0.6 & 5 3-$\sigma$ u.l. &   &  \\
170811 & 8.9$\pm$ 1.5 & 5 3-$\sigma$ u.l.& &  \\
180518 & 0.7$\pm$0.1 &  2.7$\pm$0.5 &  2.0$\pm$0.5  & 3.9 \\ 
180519 & 1.0$\pm$0.2 & 2.4$\pm$0.4 & 1.4$\pm$0.4 & 3.1 \\  % Pa beta estimated from plots
240612 &  6.4$\pm$ 1.0  & 7.8$\pm$1.3 & 1.4$\pm$1.6  & 0.8  \\
240717 &  8.3$\pm$1.4 & 13.6$\pm$2.3 & 5.5$\pm$2.7 & 2.0  \\
240718 &  4.3$\pm$0.7 & 6.6$\pm$1.1 &2.3$\pm$1.3 & 1.9 \\
\enddata
\end{deluxetable}

In 2024, we obtained 3 more epochs of data on EPIC 204638512. Figures 25-32 shows the continuum-subtracted spectrum for all 7 epochs.  Many of the line profiles of Pa$\beta$ are ether entirely in emission, while others exhibit a weak red-shifted absorption component indicative of inflowing gas. In all cases, the Br$\gamma$ line was the stronger of the two lines, except for the two nights in 2017, where only upper limits could be determined. For the nights where Br$\gamma$ was detected, two of those (180518 and 180519) had differences with a significance of 3-$\sigma$ or greater, while another two (240717 and 240718) were at the 2-$\sigma$ level. So, in EPIC204638512 the Pa$\beta$ line leads to mass accretion rates that are smaller than those derived from Br$\gamma$, as would be the case of Pa$\beta$ suffered more self-absorption than  Br$\gamma$. We note that the two nights with the most statistically significant difference were also the ones with the lowest mass accretion rates obtained using Pa$\beta$. Hence, the difference between the mass accretion rates from the two lines increased as the derived accretion rates dropped. Thus, as the self-absorption increased, it increased faster in Pa$\beta$ than in Br$\gamma$. \\

The high accretion rates on 240612 \& 240717 were accompanied by a strong emission line by He I. On many other nights, He I was in absorption, with weak emission wings on either side, suggesting a ring of orbiting gas close to thew star. On 170810, He I had an absorption core degraded to longer wavelengths, indicating gas inflow, with velocities reaching $\sim$ 800 km/s. \\

\subsubsection{EPIC 205151387}

EPIC 205151387 is a M1.0V star with 30\% photometric dips. As in the case of EPIC 204638512, the dips are consistent with dust grains larger than those seen in the ISM, and He I was stronger when the dip was deep \citep{sitko23}, again suggesting that the He I gas and the dust were likely inter-mixed. \\

\citet{sitko24} interpreted these using the radiative/hydrodynamic models of \citet{vinkovic21,vinkovic24}. In these models, gas observed above the surface of the dust disk are naturally flowing outward due to the radiation from the star \textit{and the disk}. going deeper into the disk, the situation changes, as accreting gas is flowing inward. These models would predict that as the dust extinction increases, the gas dynamics changes. On the 4 nights of spectra presented, when the dust extinction was least, no inflowing gas would be detected. As the extinction increased, inflowing gas would be detectable, and in the case of EPIC 205151387, an IPC would develop, as we have seen in the 2 epochs in 2018. Finally, when the dust extinction became much larger, the depth of the He I absorption would dominate, as the He I gas sampled would be well below the dust ``surface''. At the maximum depth of He I, the optical depth of the continuum was $\sim$0.3. 

\begin{deluxetable}{ccccc}[ht]
\tablecolumns{5}
\tablewidth{0pc}
\tabletypesize{\scriptsize}
%\rotate
\tablecaption{Mass Accretion Rates for EPIC 205151387 \\
(M=0.70$\pm$0.13 M$_{\sun}$, R=1.90$\pm$0.06R$_{\sun}$)}
\tablehead{
\colhead{Date (UT)} &  \colhead{\.{M}$_{acc,Pa\beta}$}  &  \colhead{\.{M}$_{acc,Br \gamma}$} &  \colhead{Difference (Br$\gamma$-Pa$\beta$)} &   \colhead{Significance}  \\ 
 \textit{yymmdd}  & (10$^{-10}$M$_{\sun}$y$^{-1}$) & (10$^{-10}$M$_{\sun}$y$^{-1}$)  & (10$^{-10}$M$_{\sun}$y$^{-1}$)  &  ($\sigma$) } 
\startdata
170810 & 10 3$\sigma$ u.l. & 6.1$\pm$1.2 & Pa$\beta$ has net absorption  \\
170811 & 4.6$\pm$0.9  & 3.3$\pm$0.7 & -1.3$\pm$1.1 & -1.1 \\
180518 & 5.4$\pm$1.1& 12.6$\pm$2.6 & 7.2$\pm$2.8 & 2.6 \\
180519 & 3.1$\pm$0.6 & 15.3$\pm$3.1  & 12.2$\pm$3.2 & 3.9 \\
240423 &  8.7$\pm$2.2  & 12.6$\pm$3.2 & 3.9$\pm$3.8 & 1.0  \\
240524 & 15.8$\pm$3.9  & 23.3$\pm$5.8 & 7.5$\pm$7.0 & 1.1  \\
240601 & 7.0$\pm$1.4 & 13.7$\pm$2.8 & 6.7$\pm$3.1 & 2.1  \\
240615 & 12.0$\pm$2.4 & 20.2$\pm$4.1 & 8.2$\pm$4.8 & 1.7 \\
240621 & 3.0$\pm$0.6 & 11.2$\pm$2.4 & 8.2$\pm$2.5 & 3.3 \\
240716& 18.2$\pm$3.7  & 21.2$\pm$4.3 &  3.0$\pm$5.8 & 0.5 \\
240818 & 17.1$\pm$3.5  & 20.6$\pm$4.2 & 3.5$\pm$5.5 & 0.6  \\
\enddata
\end{deluxetable}

It was the suggestive behavior of the lines in 2017 \& 2018  by see \citet{sitko24} that prompted a more extensive campaign in 2024, where an additional 7 epochs of data were obtained. In the raw spectra (Figure 7)  it is apparent that as the red-shifter absorption wing of He I increased, the Pa$\beta$ line became more asymmetric (Figures 33-44). \\

As shown in Table 8, the mass accretion rates determined using Pa$\beta$ were generally smaller than those using Br$\gamma$. On three of the eleven nights, the difference was at the 2.6-$\sigma$ level or greater. On one night (170810), Pa$\beta$ had a net absorption, not emission. On only one night was the accretion rate determined from the Pa$\beta$ not smaller that that derived from Br$\gamma$, and that was not a statistically significant difference. In all but one night (170811) the measured Br$\gamma$ resulted in an accretion rate less than that of Pa$\beta$, while in the other ten nights, the reverse was the case. So, like EPIC 204638512, the evidence shows that the Pa$\beta$ line shows evidence of greater self-absorption than Br$\gamma$. \\

On 180518, 180519, 240423, 240424, and 240601, He I possessed an IPC, and on many of these nights, Pa$\beta$ had a weak IPC absorption core.  On 240613, the very strong He I absorption, like that on 170810, had developed.  But unlike the earlier date, Pa$\beta$ had an IPC profile, but retained a significant emission component. Unfortunately, that night was not photometric, so no net mass accretion rate was determined. Two night later, on 240615, the He I line was in net emission, and the accretion rate was higher than on most nights. On 240621, the IPC profile of He I had returned, with Pa$\beta$ developing a weak IPC absorption component. \\

On 240716, He I was strongly in emission, and the accretion rate derived from the Pa$\beta$ line was the highest of the  observations in our sample. Two nights later, on 240718, the He I emission was weaker, but the mass accretion rate was comparable. \\

It is apparent that the He I line can change dramatically over one (170810 \& 170811, ) or two nights (240716 \& 240717). Looking at Figure 7, the net absorption in the He I profile has nearly the same depth on 4 of the 5 occasions, all with lower continuum emission. On one night, 240615, however, the net absorption was not present, and in the continuum-subtracted line, it is weakly in emission. The same was true for 240718, while two days prior, on 240716, it was strongly in emission.\\

Thus, while the He I line absorption component tends to  get stronger when the extinction gets larger, there is no one-to-one correlation between the two. As was mentioned in Sec. I,  \citet{zhu24} and \citet{zhang24} have looked at the star-disk interaction through the connecting magnetic field, and have suggested that near the co-rotation radius, the magnetic field can set up vortices in the gas (see their Figure 24). These magnetic fields can thread the disk in a complex manner. Should dust sublimation occur in these regions, some pockets of gas could develop in regions of lower dust concentrations than others. A study at higher spectral resolution with accurate photometry on the same night will likely be required to provide greater insight as to nature of these complex interactions. \\

\subsubsection{EPIC 203850058}

EPIC 203850058 is an M5 star, and the faintest observed in the sample of 3 dippers of \citet{sitko23,sitko24}. The light K2 light curves presented by \citet{hedges18} and \citet{ansdell20} indicate a periodicity with a time scale of $\sim$2.5d, with amplitudes of $\sim$15\%. If the variations are due to star spots connected to the disk, then the co-rotation distance would be  $\sim$2.7 R$_{*}$. \\

Unfortunately, EPIC 203850058 was not observed in our 2024 observations, so all of the discussion here is derived from the data obtained in 2017 \& 2018. In 2017, the He I line was in absorption, with a profile that degraded to the red, indicating inflowing gas (incorrectly labeled as \textit{outflowing} gas in \citet{sitko24}), with a maximum velocity of $\sim$550 km/s. \\

In 2017, the Pa$\beta$ line is clearly in emission in the 2 nights in 2017, but was missing in the 2 nights in 2018. Most notably, the continuum on 170811 was the same as in 2018, (within the photometric uncertainties), but the Pa$\beta$ line flux was not. This suggests that the line flux changes are not due to changes in line-of-sight dust, but more closely associated with changes in the observed mass accretion rate. Whether this is due to the observed location of the accretion column, or that the accretion is simply sputtering out on those nights. Its location with respect to the isochrones in Figure 9 suggests that the star must still accrete more mass before reaching the MS, but the is noting that forbids the accretion being sporadic. \\ 

In 2018, unsurprisingly, only upper limits to the mass accretion rates could be derived for Pa$\beta$ and Br$\gamma$. In 2017, the accretion was detectable in Pa$\beta$ but at the 2.1- and 2.8-$\sigma$ level for 170810 and 170811, respectively. Much of the uncertainty is due to the relatively low precision of the photometry used to scale the spectra. On the latter night, Br$\gamma$ was detected only at the 2.4-$\sigma$ level. As seen in Figure 8 and 47\&48, when the Pa$\beta$ line was not in emission, the Na I line at 1.2682 $\mu$m developed into a P Cygni profile with weak absorption (blended with the photospheric line) and emission components, indicating a disk wind was present. \\

\begin{deluxetable}{ccccc}[ht]
\tablecolumns{5}
\tablewidth{0pc}
\tabletypesize{\scriptsize}
%\rotate
\tablecaption{Mass Accretion Rates for EPIC 203850058 (M=0.15$\pm$0.05M$_{\sun}$, R$\sim$1.53$\pm$0.24R$_{\sun}$)}
\tablehead{
\colhead{Date (UT)} &  \colhead{\.{M}$_{acc,Pa\beta}$}  &  \colhead{\.{M}$_{acc,Br \gamma}$} &  \colhead{Difference (Br$\gamma$-Pa$\beta$)} &   \colhead{Significance (Br$\gamma$-Pa$\beta$) }  \\ 
 \textit{yymmdd}  & (10$^{-10}$M$_{\sun}$y$^{-1}$) & (10$^{-10}$M$_{\sun}$y$^{-1}$)  & (10$^{-10}$M$_{\sun}$y$^{-1}$)  &  ($\sigma$) }   
\startdata
20170810 & 15$\pm$7 & 13 3$\sigma$ u.l &   \\
20170811 & 11$\pm$4 & 12$\pm$5 & Difference (Br$\gamma$-Pa$\beta$) = 3.5$\pm$5.5 & 0.6 \\
20180517& 13 3$\sigma$ u.l& 13 3$\sigma$ u.l & No detection above photospheric spectrum \\
20180518 & 13 3$\sigma$ u.l & 13 3$\sigma$ u.l  & No detection above photospheric spectrum  \\
\enddata
\end{deluxetable}

\subsubsection{V935 Sco = EPIC 204176565}
We were unable to fit the spectrum of V935 Sco with the spectrum of a single star. \citet{davies19} list V935 Sco as a binary, with a separation of 0.02 sec ($\sim$3 AU) and a $\Delta$K of 0.42 mag. The primary is classified as K5$\pm$2. The separation was determined using Keck II sparse aperture masking, and appears in \citet{rodriguez16} as WSB 12 in their Table 7. \\

Because of this, we examined several possible models with multiple components: a binary star with and without additional non-photospheric contributions, such as the inner-disk NIR excesses or accretion UV-optical excesses. Our best solution was one with two stellar components and NIR excesses: a primary star with T$_{1}$=5278$^{+270}_{-322}$ K, and a secondary star with T$_{2}$=4663$^{+164}_{-81}$ K and the inner disk modeled as a blackbody with T=1200K. The spectral types of the stellar components are K0 and K4, respectively (Pecaut and Mamajek, 2013). For these stellar components, we used the BOSZ/Phoenix spectral library (REF). The best solution was found using the Asexual Genetic Algorithm \citep{canto09,adame24}  \\

Some close-binary T Tauri stars exist with only one component possessing a disk. RW Aur is one well-documented example, where one component (A) has a massive disk with sporadic X-ray flaring, likely due to the destruction of an asteroidal-mass body, while the other component  (B) has no significant disk emission \citep{lisse22,lisse24}. V982 Tau is also a binary with 2 stars and a substellar companion, where the disk orbits the substellar object \citep{vandam20}. \\

V935 Sco is classified as a rotationally-modulated object by \citet{hedges18}, but the K2 light curve in \citet{ansdell20} exhibits more random fluctuations likely due to dust. The rotationally-modulated variations have a period  of $\sim$ 4d, which for a mass of 0.6 M$_{\sun}$ and radius of 0.86R$_{\sun}$, would make the co-rotational radius at $\sim$7R$_{*}$, larger than the assumed  inner radius R$_{in}$$\sim$2.8R$_{*}$.  \\

\begin{deluxetable}{cccccc}[ht]
\tablecolumns{6}
\tablewidth{0pc}
\tabletypesize{\scriptsize}
%\rotate
\tablecaption{Mass Accretion Rates for V935 Sco (M=0.87$\pm$0.03M$_{\sun}$, R=0.84$\pm$0.04R$_{\sun}$)}
\tablehead{
\colhead{Date (UT)} &  \colhead{\.{M}$_{acc,Pa\beta}$}  &  \colhead{\.{M}$_{acc,Br \gamma}$} &  \colhead{Difference (Br$\gamma$-Pa$\beta$)} &   \colhead{Significance (Br$\gamma$-Pa$\beta$) } & \colhead{Notes} \\ 
 \textit{yymmdd}  & (10$^{-10}$M$_{\sun}$y$^{-1}$) & (10$^{-10}$M$_{\sun}$y$^{-1}$)  & (10$^{-10}$M$_{\sun}$y$^{-1}$)  &  ($\sigma$) & }   
\startdata
240610 & 230$\pm$32   & 223$\pm$32  & -7$\pm$45   &  -0.2 & \\
240611 & 241$\pm$33    & 231$\pm$33  & -10$\pm$46 & -0.2 & \\
240612 & 23.1$\pm$3.2 &  79$\pm$11 & 56$\pm$11 & 4.9 & inv. P Cyg w. weak em. \& abs.   \\
240616 & 103$\pm$14 & 130$\pm$18 &  27$\pm$22 & 1.2 & inv. P Cyg w. weak em. \& abs.   \\ 
240717 & 70 3-$\sigma$ u.l & 42$\pm$5 &  & & Pa$\beta$ inv. P Cyg w. equal em. \& abs.   \\
240718& 70 3$\sigma$ u.l Pa$\beta$ & 70 3$\sigma$ u.l   Br$\gamma$ &  &  &  \& inv. P Cyg  abs. dominates em. \\
\enddata
\end{deluxetable}

% INCLUDES REVISED Rin correction of 1.25x

The behavior of the Pa$\beta$ line of V935 Sco, seen in Figure 2, is even more extreme than in HD 205151387. A better comparison will be seen in the figures containing both He I and Pa$\gamma$ (Figures 2 and A-40 through A-45). The He I line goes from a modest IPC profile at high continuum (low dust absorption) to a strong IPC profile as the continuum faded. Over that same time scale, the continuum flux decreased, and the Pa$\gamma$ line went from an emission line to a moderately strong IPC. Putting it in its simplest terms, when there was more dust in the line-of-sight, there was more He I and H I gas absorption along the line of sight. The IPC profile indicates inflowing gas, and if the models of Vinkovi\'{c} \& \v{C}emelji\'{c} are applicable, This is what would be expected if one looks through deeper and deeper layers of the accretion disk.  \\

Of the 6 nights of observation, the derived mass accretion rates for Pa$\beta$ versus Br$\gamma$ resulted in only upper limits for one (270717) or both (240718) of the lines. On another night (240616) Pa$\beta$ had a smaller derived accretion rate than Br$\gamma$, but only at the  $\sim$1-$\sigma$ level. On yet another night (240612), Pa$\beta$ produced a derived mass accretion rate that was significantly ($\sim$5-$\sigma$)  smaller than that derived from Br$\gamma$, but the lines were weak and had complex structures indicative of self-absorption in these lines. On the first two nights, 240610 \& 240611, the derived accretion rates from the emission line strengths were not compromised by the addition of absorption components, resulting in the same accretion rates being derived using both lines. These likely indicate the true accretion rates, unaffected by self-absorption in the lines. \\

\subsubsection{EPIC 203843911 = DoAr 25}

DoAr 25 is a K5 star, and is another rotationally-modulated dipper \citep{hedges18} with a period of $~\sim$2.9d, and the co-rotation distance would be at $\sim$7R$_{*}$. Here we assumed that the inner disk radius R$_{in}$ is at 2.8 R$_{*}$.

\begin{deluxetable}{ccccc}[ht]
\tablecolumns{5}
\tablewidth{0pc}
\tabletypesize{\scriptsize}
%\rotate
\tablecaption{Mass Accretion Rates for DoAr 25 (M=1.12$\pm$0.10M$_{\sun}$, R=1.33$\pm$0.08R$_{\sun}$)}
\tablehead{
\colhead{Date (UT)} &  \colhead{\.{M}$_{acc,Pa\beta}$}  &  \colhead{\.{M}$_{acc,Br \gamma}$} &  \colhead{Difference (Br$\gamma$-Pa$\beta$)} &   \colhead{Significance (Br$\gamma$-Pa$\beta$) }  \\ 
 \textit{yymmdd}  & (10$^{-10}$M$_{\sun}$y$^{-1}$) & (10$^{-10}$M$_{\sun}$y$^{-1}$)  & (10$^{-10}$M$_{\sun}$y$^{-1}$)  &  ($\sigma$) }  
\startdata
240611 & 337$\pm$107 & 559 $\pm$178 &222$\pm$208 & 1.1   \\
240612 & 206$\pm$65 & 551 $\pm$163 & 345$\pm$175 & 2.0 \\
\enddata
\end{deluxetable}

In DoAr 25, the two observations suggest that some self-absorption might be present Figures A-46 \& A-47), but more observations are needed to make any confident conclusions regarding self-absorption. 

\subsubsection{The Overall Sample} 

Within our sample of observations, the accretion rate derived from Pa$\beta$ is smaller than that derived from Br$\gamma$ for HD 142666, HD 145718, EPIC 204638512, and EPIC 205251387. In the case of HD 143006, there is no evidence of any difference between the two lines, and  the Kepler light curve in \citet{ansdell20} shows only sporadic dimming of $\sim$5\%. For EPIC 203850058, the data were of insufficient quality to determine any difference at a statistically meaningful level. In the case of V935 Sco, the one night where a meaningful difference cold be detected was also the one with the weakest Pa$\beta$ line. This indicates that the line-of-sight is predominantly not significantly attenuated by the innermost disk. With only 2 measurements, the data suggests a lower accretion rate derived from Pa$\beta$  than Br$\gamma$, but more data is required to become more confident in the difference. So, for 4 of the 8 objects in our sample, greater self-absorption in Pa$\beta$ than in Br$\gamma$ is secure. In those where it is not definitive, the attenuation is likely too small most of the time (HD 143006), the fluxes are too weak (EPIC 203850058),  or the results were suggestive, but not conclusive (V935 Sco and DoAr25). \\

For the He I line seen in Figures A-2 through A-47 , the following was found:  \\

HD 142666: The line was either in absorption or exhibited an IPC profile. \\

HD 143006: two nights had simple absorption, while the other was a PC profile.\\

HD 145718: all 4 nights showed strong absorption, with a very weak emission at shorter wavelengths. \\

EPIC 204638512: emission, absorption, and IPC lines were all present. The emission ``spikes'' surrounding the He I lines suggest that, if the spectral classification is correct, then the photospheric lines  are being weakened. This might be due to ``veiling'' of the spectrum due to the accretion shock. \\

EPIC 205151387: emission or IPC. \\

EPIC 203850058: weak absorption, sometimes degraded the red. \\

V935 Sco:  emission or IPC. \\

DoAr 25: both nights showed an IPC profile. \\

Because the lower state in the hydrogen line transitions are quickly depopulated by downward transitions from the lower energy state, their absorption components are suppressed. In a few cases, however, it is clearly visible: HD 145718 on 240719, EPIC 205151387 on 170810, and V935 Sco on 240612, 240717, and 240718.  On these nights, the He I line was in deep absorption. In all cases, when the Pa$\beta$ exhibited a significant absorption component, the He I did as well.  On numerous nights, the profile of the Pa$\beta$ emission line showed a sharper decline in intensity on the longer wavelength side, indicating a very weak absorption component was present. \\

\subsubsection{Short Time Scales for Variability in He I}

HD 142666: as short as 1 day (240716 \& 240717) \\

HD 143006: 1 month, but under-sampled on daily time scales \\

HD 145718: 1 month, and in the three measured at daily intervals - no change \\

EPIC 204638512: 1 day (180518 \& 180519) \\

EPIC 205151387: 1 day (170810 \& 170811), 2-day (240716 \& 240718) \\

EPIC 203850058:  unknown, based on this data set \\

V935 Sco: 1 day (240611 \& 240612) \\

DoAr 25:  unknown, based on this data set \\

Within the sampling constraints of the observations, HD 142666, EPIC 204638512, and EPIC 205151387 all show changes in the structure of the He I line profiles with time scales as short as 1 day. However, only 1 pair for EPIC 204638512 exhibited this (in 2018), and only 2 for EPIC 205151387 (2017 \& 2024). So, fortuitous sampling may dominate the success of daily changes. \\

HD 142666 is an interesting case. Stars of spectral class A and early F are known to be rapid rotators. For example, Altair (A7IV-V) has a rotational period of $\sim$2 days \citep{peterson06b}. This indicates that the He I gas extends close to the surface of the star. \\

Table 13 contains a summary of these results. 

\begin{deluxetable}{lcc}[ht]
\tablecolumns{3}
\tablewidth{0pc}
\tabletypesize{\scriptsize}
\tablecaption{Smallest Time Scales for Variability in Days}
\tablehead{
\colhead{Object} &  \colhead{He I}  &  \colhead{TESS and/or Kepler (K2)}
}    
\startdata
HD 142666 & 1 & 1.7 (K2) \\
HD 143006 & -  & 2.0 (TESS), 2.3 (K2), 2.5 (K2 but more sporadic) \\
HD 145718 & - &  1.3 (TESS but mostly sporadic), 5.2 (K2) \\
EPIC 204638512 & 1 & 1? (TESS, mostly sporadic), 5.2 (K2) \\
EPIC 205151387 & 1, 2 & sporadic (TESS \& K2) \\
EPIC 203850058  & - & 2.8 (K2) \\
V935 Sco & 1 & sporadic (TESS), $\sim$2.4 (K2) \\
DoAr 25 & - & 1.1 (TESS), 1.06 \& 4.28=4x1.07 (K2) \\
\enddata
\end{deluxetable}

We have also examined images of the TESS and Kepler light curves from the MAST data base in order to find the shortest time scales for variability. The results ae also listed in Table 13. Many are sporadic, but often contain segments that are at least quasi-periodic. The Kepler data on DoAr 25 exhibited two time scales, with one being four times longer than the other. \\

\section{Discussion}

It is apparent that the gas and dust usually change together. The most clear cases are those of EPIC 205151387 and V935 Sco. \\

In EPIC 205151387, the IPC profile in He I is what would be expected for emission from the disk surface flow (or possibly  the accretion column). For this star, the dust sublimation radius (assumed to be where T=1500 K) would be between 3.3R$_{*}$ (M1)  and 2.6R$_{*}$ (M3), comparable to the revised gas inner radius of $\sim$ 2.8 R$_{*}$  found by \citet{pittman25a} for T Tauri stars in general. \\

Is it possible for the He I line to change with the dust extinction, if the line of sight is near the surface of the disk? \citet{vinkovic21} and \citet{vinkovic24}  have explored the trajectories of dust grains affected by a combination of gas drag (entrainment) and radiation pressure from both the star \textit{and the disk}. In their models, a slight change in the line of sight through the ``surface''\footnote{In \citet{vinkovic21} this is where the outward radiation pressure is approximately equal to the inward flowing drag on the dust grains, which occurs roughly where the disk becomes optically thick.}, of the disk can produce a difference in dust and gas trajectories. Consequently, going deeper into a disk can change the direction in which the gas moves. In this scenario, blue-shifted emission would also be seen coming from the disk surface on the far side of the star. Detailed modeling using spectra with a much higher resolving power will be needed to decide this issue. \\

V935 Sco not only shows substantial changes in the line profile of He I, but in the H I Paschen lines as well. As the continuum drops, both Pa$\gamma$ and Pa$\beta$ transition from being in emission to an IPC profile with a deep red-shifted absorption core and a weaker blue-shifted emission wing. When the Paschen lines are in emission, they are symmetric and located as zero wavelength shift. The simplest explanation for the IPC at lower continuum flux levels would be the addition of a redshifted absorption core as the continuum flux level dropped. \citet{tb23}, using spectra at a much higher resolving power, found that in the low-accreting T Tauri stars, a significant stationary emission component was present, which they identified as stellar chromospheric emission. Because of our much lower resolving power, the FWHM velocity width is only $\sim$350 km/s, comparable to the observed emission component. In our opinion, the simplest explanation for the behavior of the Paschen lines is a component produced by the stellar chromosphere, with the addition of varying amounts of infalling gas mixed with dust. This is precisely what would be expected if the line of sight were along the surface of the disk, but included some vertical irregularities, or ``roughness''. \\

The faintest star in our sample, EPIC 203850058, shows no discernible emission in the He I line. Rather, it exhibits absorption only (at our resolving power), perhaps with a red-shifted wing indicating mass infall. The Pa$\beta$ line also behaves in a telling manner. In 2017, it was in emission. In 2018, it vanished. The net emission in 2017 is identified as the net stellar chromospheric emission line. In 2018, sufficient absorption in Pa$\beta$ due to an increase in disk gas along the line of sight effectively absorbed the chromospheric emission. As shown in Figure 8, an absorption line at 1.2685 $\mu$m also appeared in 2018. This is likely the 1.26826 $\mu$m line of Na I, which is also present in the photospheric spectra of M5V stars \citep{rayner09}. \\

The presence of mis-inclined disks begs the question: ``From which disk is the accreting gas coming from? The inner disk, the outer disk, both?" \citet{sitko24} found that for the data in 2017 \& 2018, both EPIC 204638512 and EPIC 205252387, the absorption component in the He I line increased when the surrounding continuum flux decreased, indicating that much of the He I gas was mixed in with the dust producing the dipper phenomenon. The observations of HD 145718 in 2024 show the same effect. The Pa$\beta$ line was not similarly affected in all three cases, indicating that these two species are not coincident. Due to the higher ionization potential of He I compared to H I, the He I line likely traces distances closer to the star than H I. For V935 Sco, on the other hand, the changes in the He I line, Pa$\gamma$ and Pa$\beta$ are much more dramatic than the change in continuum over the same span of time. \\

If there is no radial gap between the inner disk and the outer disk - with different inclinations - these disks must cross, and would interact. What spectral signatures would be imposed on the data is far beyond the scope of this investigation. Regardless, the fate of the accreting gas from the outer disk as it passes the radial extent of the inner disk needs to be determined. For EPIC 204638512, \citet{sa20} envisioned the mis-inclined inner and outer disks as not intersecting (their Figure 11). By contrast, for the non-dipper star SAO 206462 (=HD 135344B), \citet{muller11} suggested that the gas from the outer disk might make it past the mis-inclined inner disk (their Figure 1). \\

\section{Conclusions}

Multi-epoch spectroscopic observations of the dipper stars have provided a great deal of information on the structure and behavior of these objects. \\

The accretion rates can vary substantially with time. Using the Br$\gamma$ line, the derived mass accretion rate - based on the line flux integrated over the line profile, of HD 142666 changes by up to a factor of 2x (3-$\sigma$). For V935 Sco, it reaches 5x. While actual changes in the true accretion rate may be present, they are heavily affected by the presence of absorption components. \\

The derived accretion rates using the H I  lines of Pa$\beta$ and Br$\gamma$ are often different. Generally the rate derived from Pa$\beta$ are smaller than those of Br$\gamma$. If the column density of gas along the line of sight is large, as would be expected when viewing along the disk surface, then Pa$\beta$ would suffer more self-absorption that Br$\gamma$ would, as the lower quantum level will become more over-populated for Pa$\beta$ than for Br$\gamma$. This effect is not typically present in non-dipper PMS stars (see \citet{sitko26} for example). The drawback of using Br$\gamma$ in low-accretors like the stars in our sample, is that the line can be very weak, and would require much better signal-to-noise observations than some included here to achieve a more robust determination of the mass accretion rates. \\

As \citet{pittman25a} indicated, decreased line emission is correlated with decreased continuum brightness in WTTs (more dust extinction), so that the derived mass accretion rates are biased toward lower values. Our sample of dippers confirms this. The ``true'' mass accretion rates are closer to those derived when the dust extinction is at a minimum. But even these values may be biased if significant dust extinction is still present when the lines are brightest, and the star sill suffers from dust extinction. \\

The most challenging objects, from the standpoint of understanding the highly variable line profiles, and associated gas dynamics, are EPIC 205251387 and V935 Sco. Much higher resolving powers, like those of \citet{tb22,tb23} and \citet{pittman25a} are required  to provide the needed detailed modeling of the line profiles. Additional epochs of data on EPIC 203850058 would provide a better understanding of this object. \\

Overall, we conclude that as the line of sight passes through more material in the disk, the \textit{derived} mass accretion rates become less that the \textit{true} mass accretion rates, due to the self-saturation in the lines as well as the addition of absorption components in the line profiles. This occurs first in the Pa$\beta$ line and then the Br$\gamma$ line. When the line of sight is close to the surface of the disk, even minor changes in disk height above the mid-plane will alter the extinction in both the dust continuum and the atomic lines. The derived mass accretion rates will often be smaller than what the star is actually accreting. \\

The authors would like to thank Nuria Calvet for her help and advice on the subject throughout the years on the subject of accretion onto PMS stars. This paper was based on observations taken with the IRTF/SpeX 0.8-5.5 Micron Medium-Resolution Spectrograph and Imager, funded by the National Science Foundation and NASA and operated by he NASA Infrared Telescope Facility.The authors would like to thank the tireless support of of the staff of NASA's Infrared Telescope Facility.  We wish to emphasize the pivotal cultural role and reverence that the summit of Maunakea has always had within the indigenous Hawaiian community. We are most fortunate to have the privilege to conduct scientific observations from this mountain. This work was supported in part by NASA XRP grants NNX17AF88G and NNX16AJ75G.

{\it Facilities:} \facility{IRTF (SpeX)}
\software{Spextool(Vacca et al. 2003, PASP, 115, 389; Cushing et al. 2004, PASP, 116, 362)}

\clearpage

\appendix

\renewcommand{\thefigure}{A-\arabic{figure}} % redefine the command that creates the figure number; the new numbering scheme is now A-1, A-2, ... 
\setcounter{figure}{0}  % reset counter

\section{Line Extractions}
% Figure 5 shows the spectra of EPIC 205151387 obtained on 11 August. Pa$\beta$ exhibits an IPC, indicative of inflowing gas. The Br$\gamma$ line is too weak to obtain a net line flux, so we list this as an upper limit in Table 2. Note that the Ca II emission is weak to nonexistent, and the  O I 8446 $\mu$m light exhibits an IPC (as did Pa$\beta$).

\begin{figure}
\includegraphics[width=9.5cm, height=11cm]{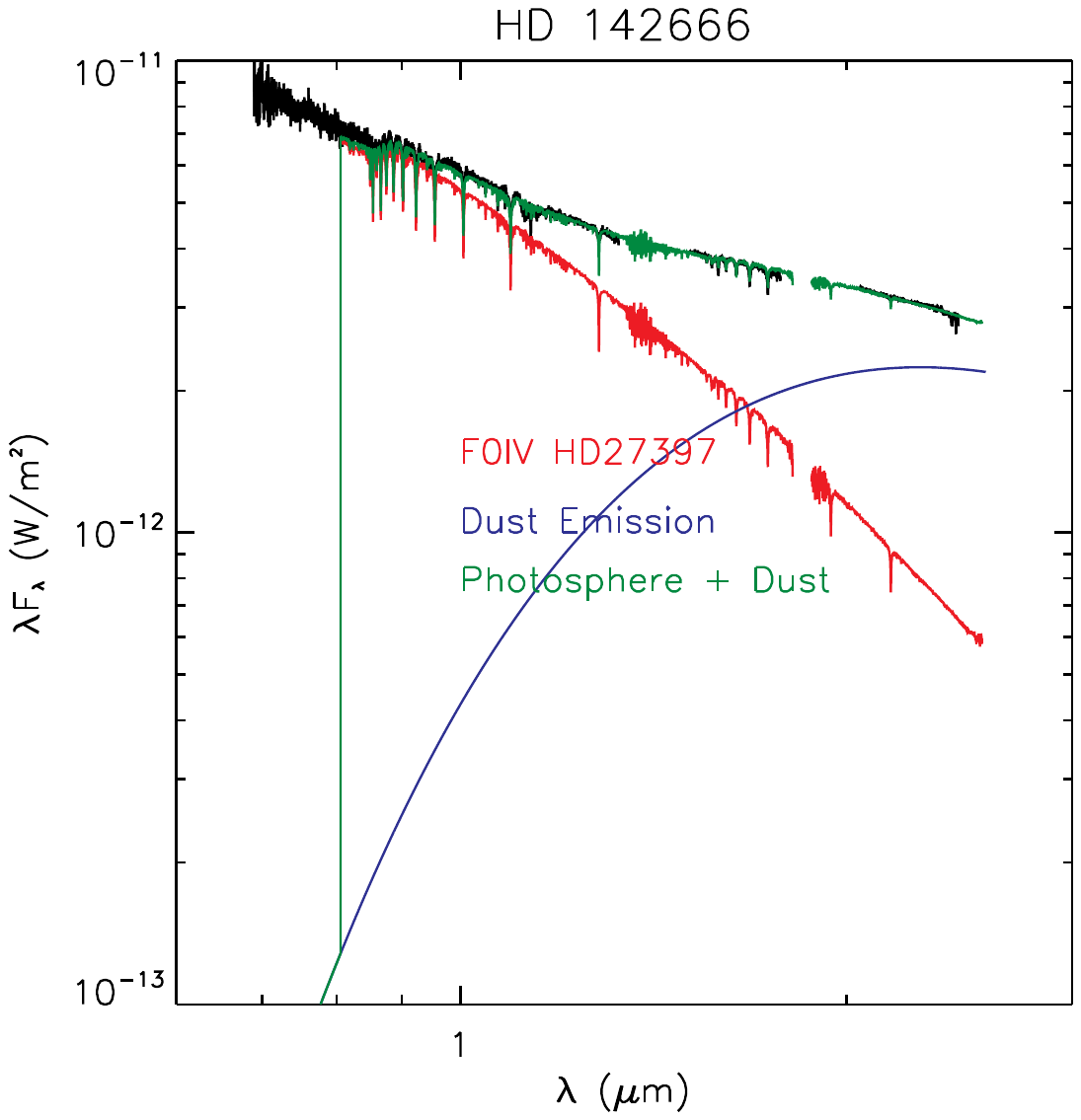}
\includegraphics[width=9.5cm, height=11cm]{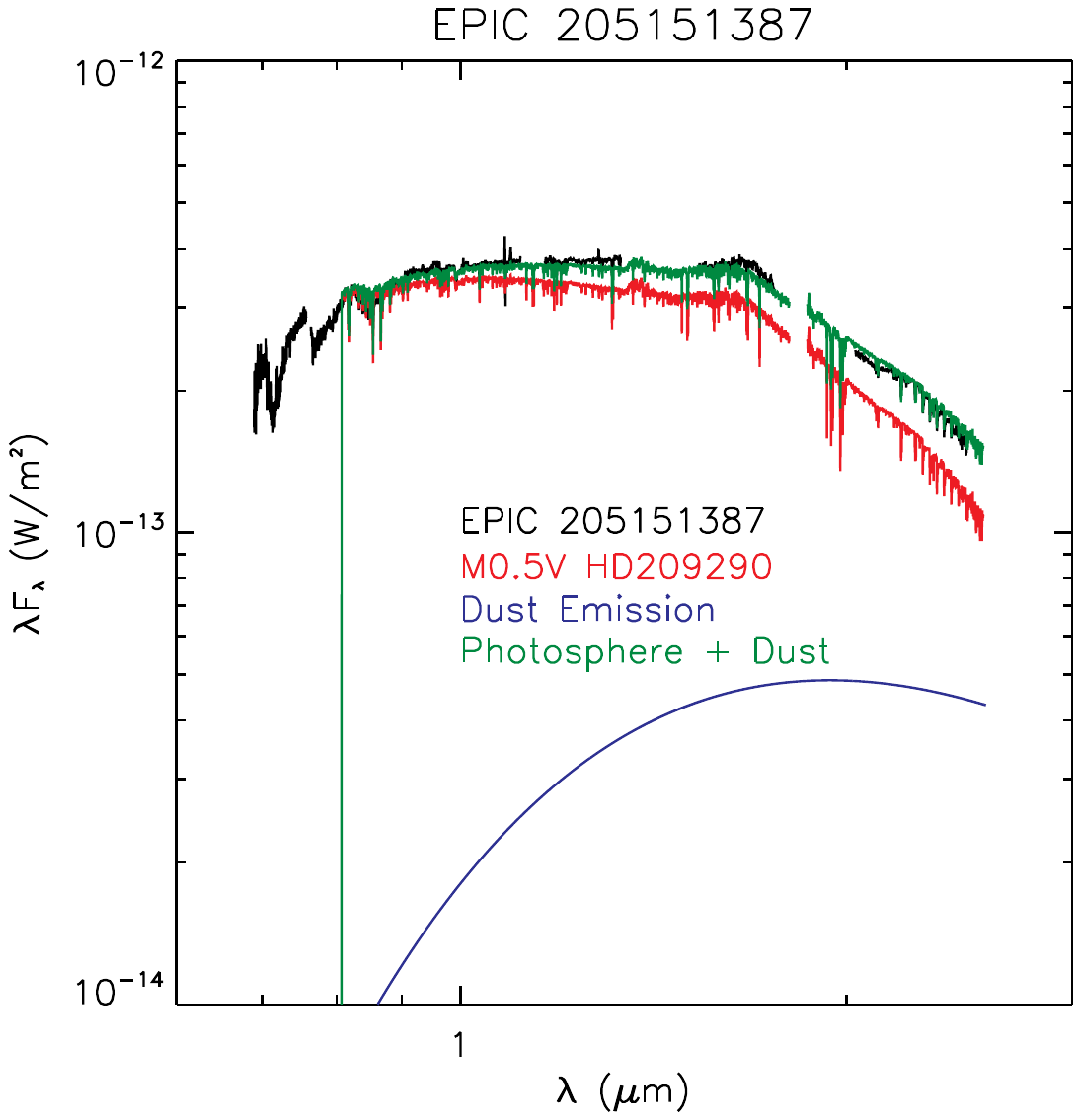}
\caption{Left: SED Model for HD 142666 (first spectrum). Black: the first observed spectrum of HD 142666 on 240611 UT. Red: the photospheric spectrum of the F0IV star HD 27397 (suitably scaled to the flux of HD 142666 at the shortest wavelengths). Blue: the thermal emission by the dust in the disk. Green: The sum of the photospheric and dust emission, to simulate all of the observed flux not containing the emission lines in the system. Right: the same for EPIC 205151387 on 240601 UT. In cases where the model does not fall on top of the observed spectrum, a localized additional adjustment in the vicinity of the line has been added to improve the fit and line flux extraction. \label{fig:A-1}}
\end{figure}

\begin{figure}
\includegraphics[width=6.0cm, height=6.0cm]{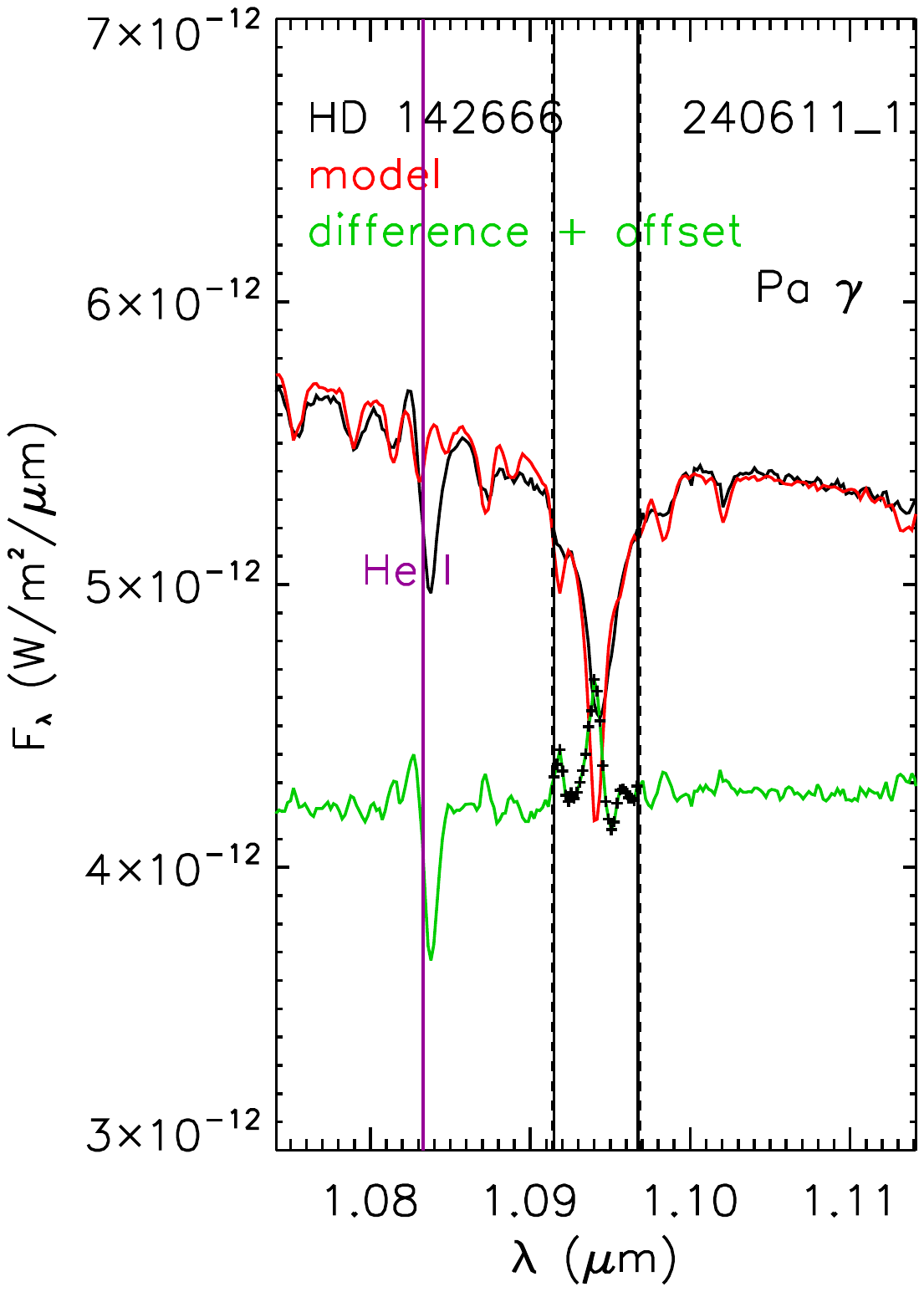}
\includegraphics[width=6.0cm, height=6.0cm]{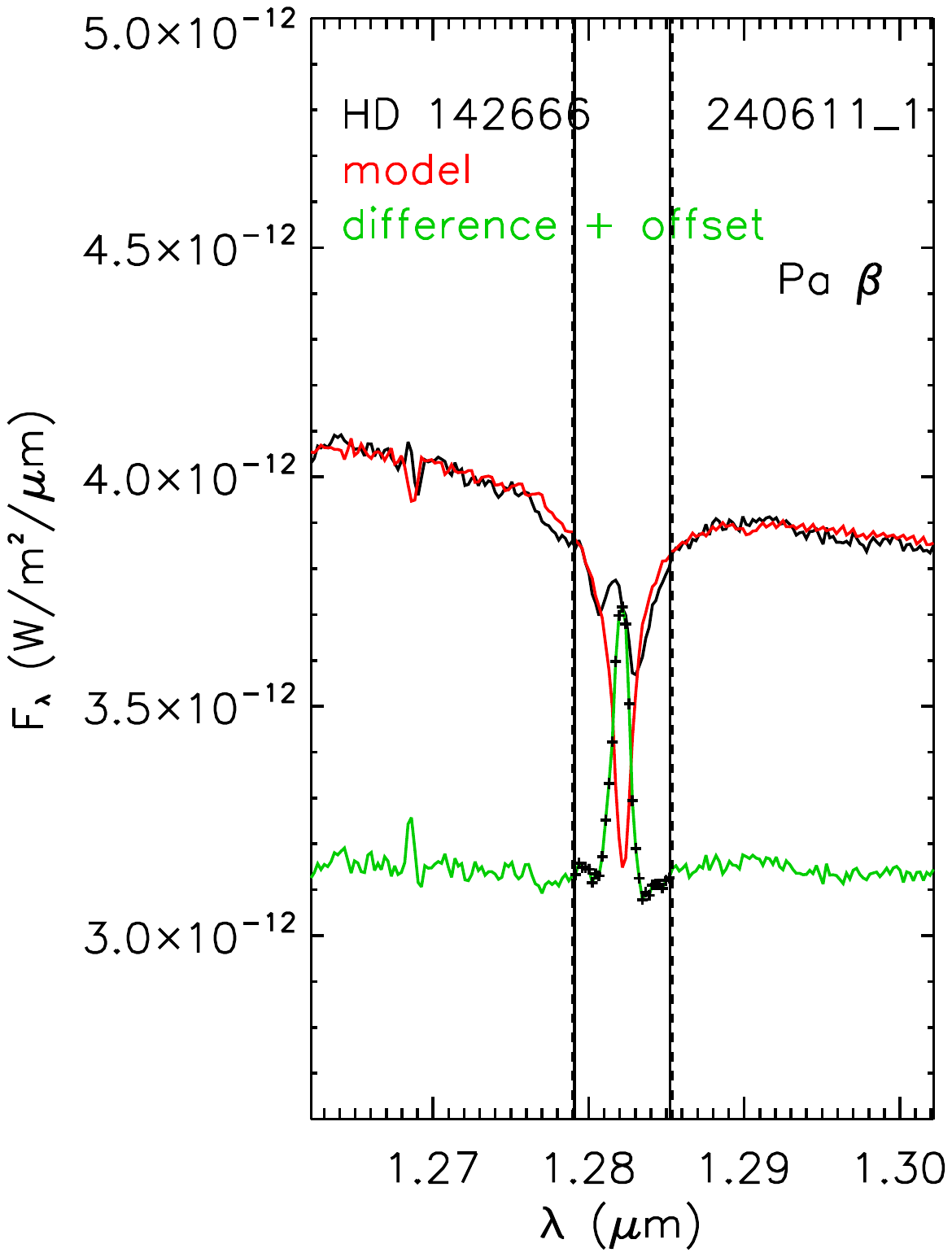}
\includegraphics[width=6.0cm, height=6.0cm]{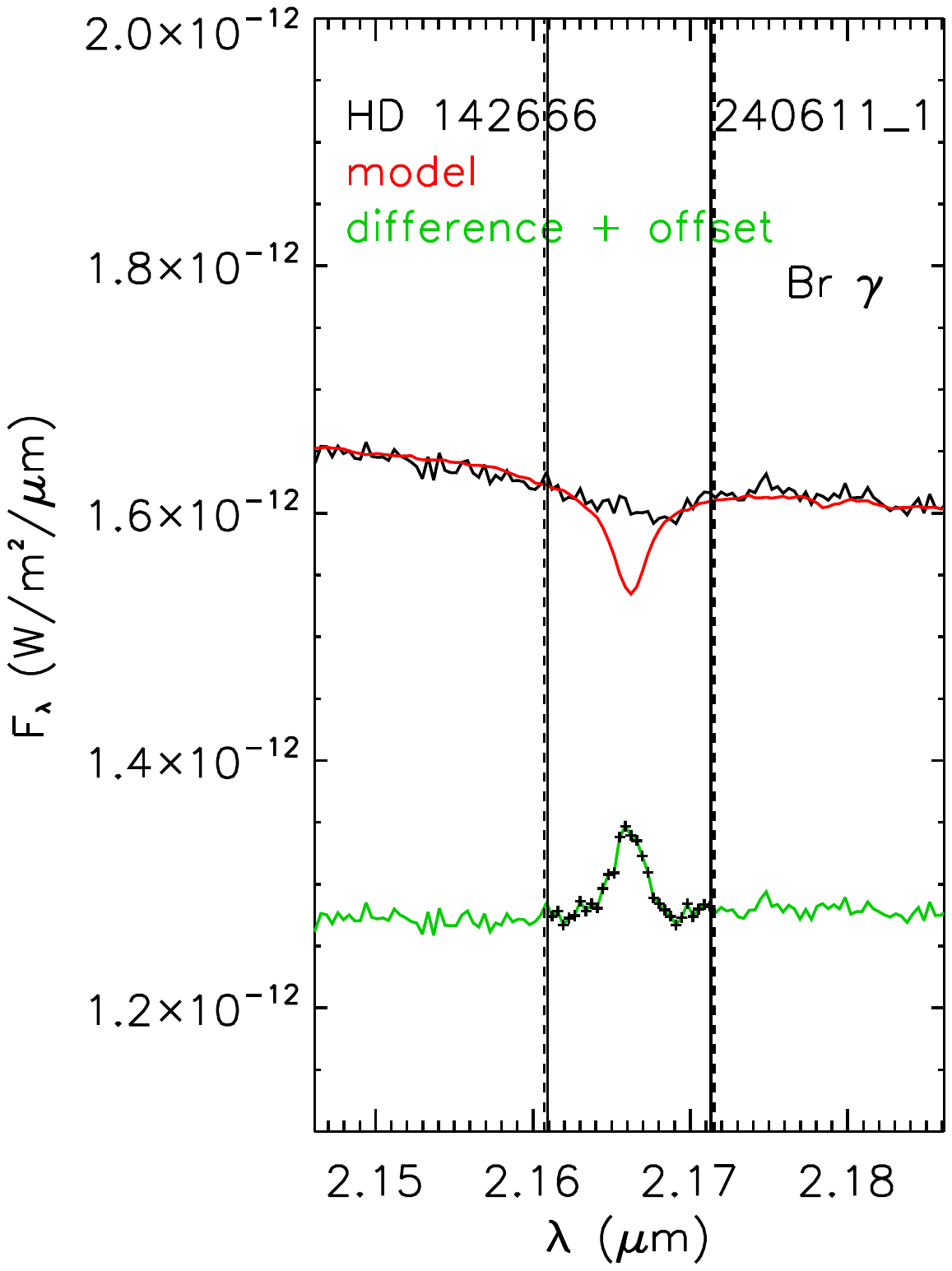}

\caption{Pa$\gamma$ \& and He I, Pa $\beta$, and Br $\gamma$ for HD 142666 on 240611 (first measurement). The observed spectrum is shown in black, the model (photospheric plus dust emission) is in red, and the difference of the two, vertically shifted above zero flux (to be included in the figure) is shown in green. The vertical lines indicate the wavelengths over which the net line flux is extracted, with the solid lines indicating that the flux was to be measured using individual histograms for each point, while the dashed lines used a trapazoidal fit between points. Due to the similarity of the two, we used only the former to extract the line fluxes used to determine mass accretion rates via the two lines. The profile of the extracted profile for Pa$\beta$ exhibits an IPC, indicating inflowing gas, while the redshifted absorption component is weak, indicating that only a small column of gas was observed along the line of sight to the stellar surface.This component might be present in Br$\gamma$ but weakened due to its lower energy state being above that of Pa$\beta$, and less likely to be populated. In both lines, the extracted flux can be measured with high confidence. Note that the local peak in the core of Pa$\beta$ is located at a wavelength than the peak in the ``difference'' spectrum, where the photospheric profile has been removed. \label{fig:A-2}}
\end{figure}

\begin{figure}
\includegraphics[width=6.0cm, height=6.0cm]{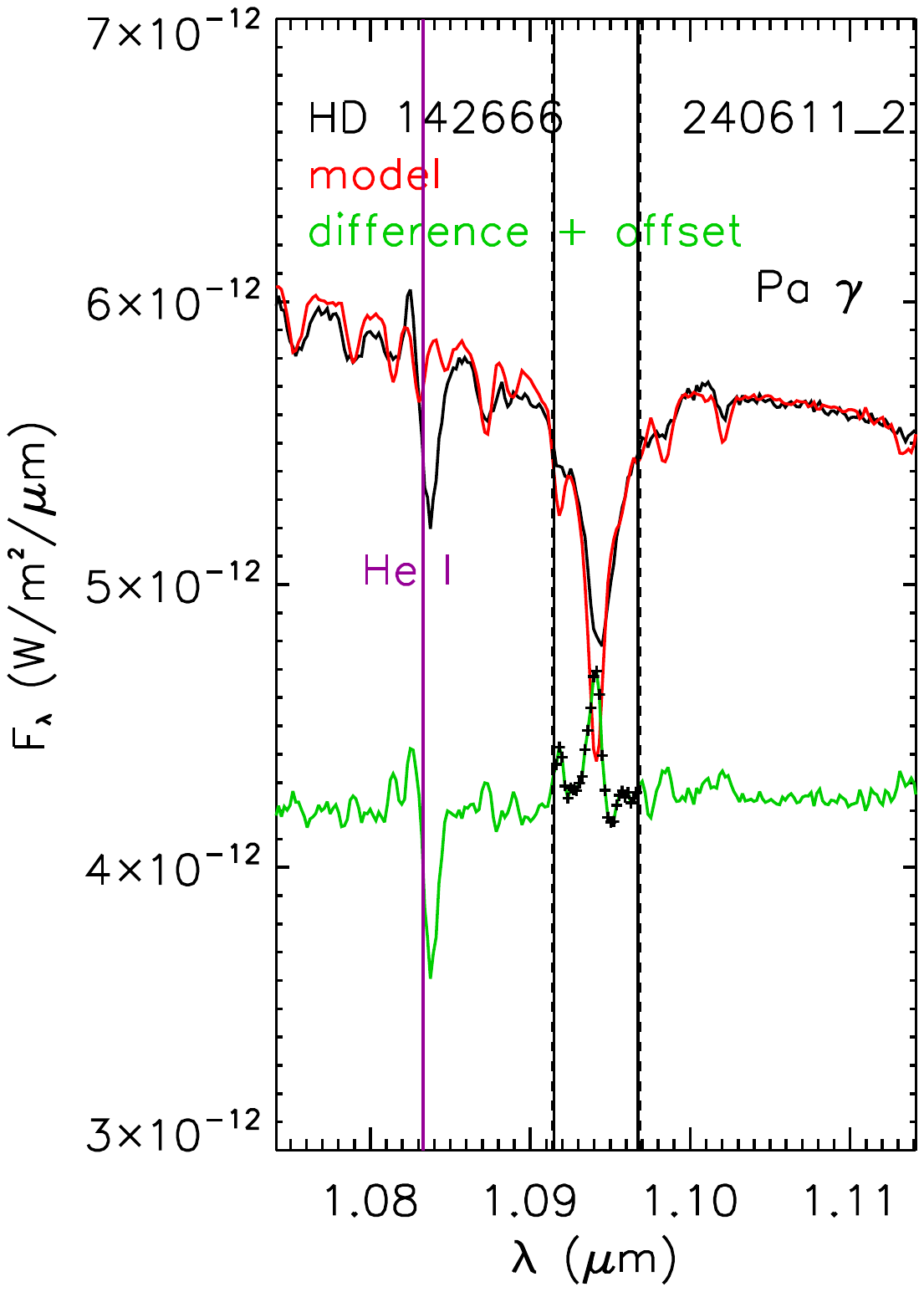}
\includegraphics[width=6.0cm, height=6.0cm]{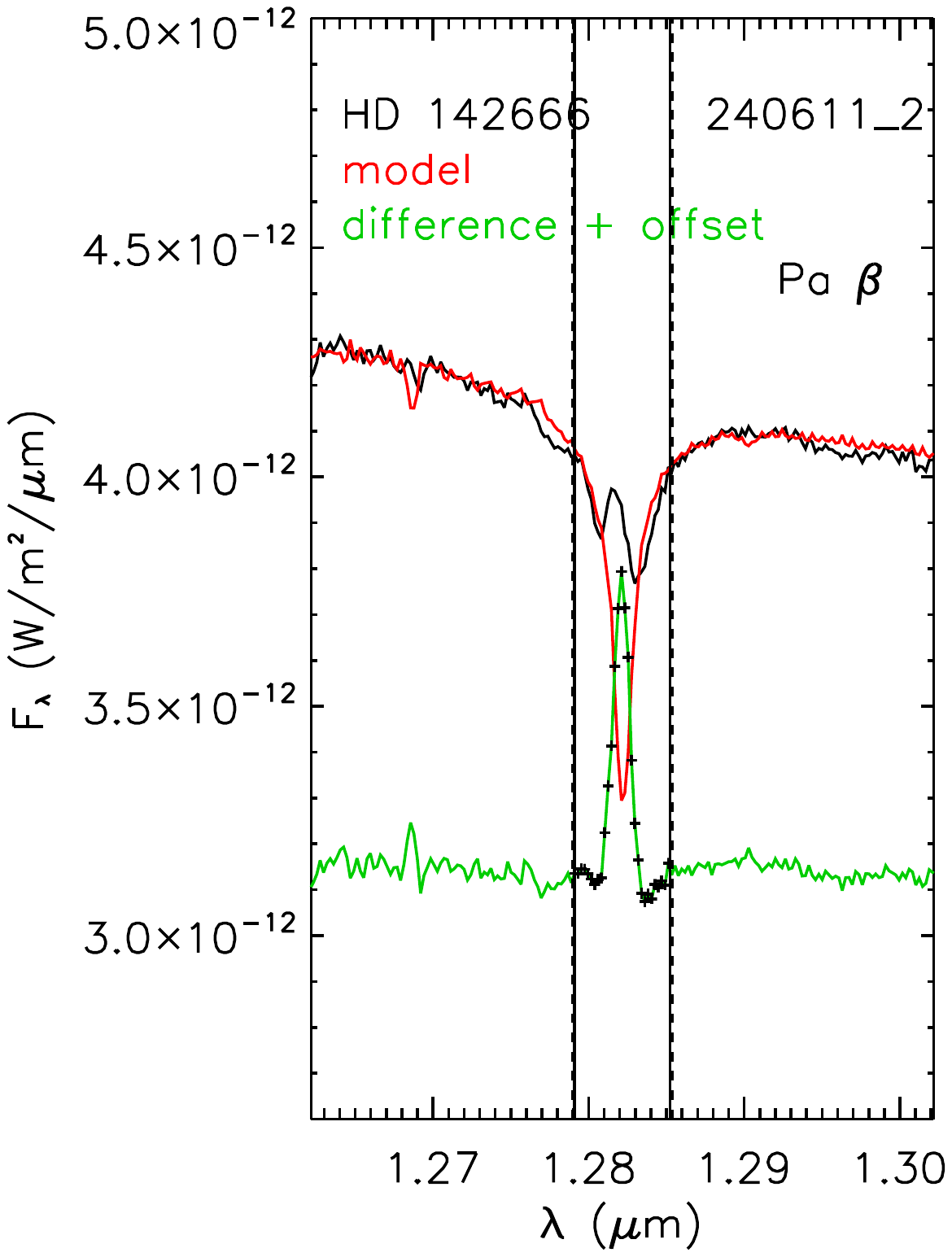}
\includegraphics[width=6.0cm, height=6.0cm]{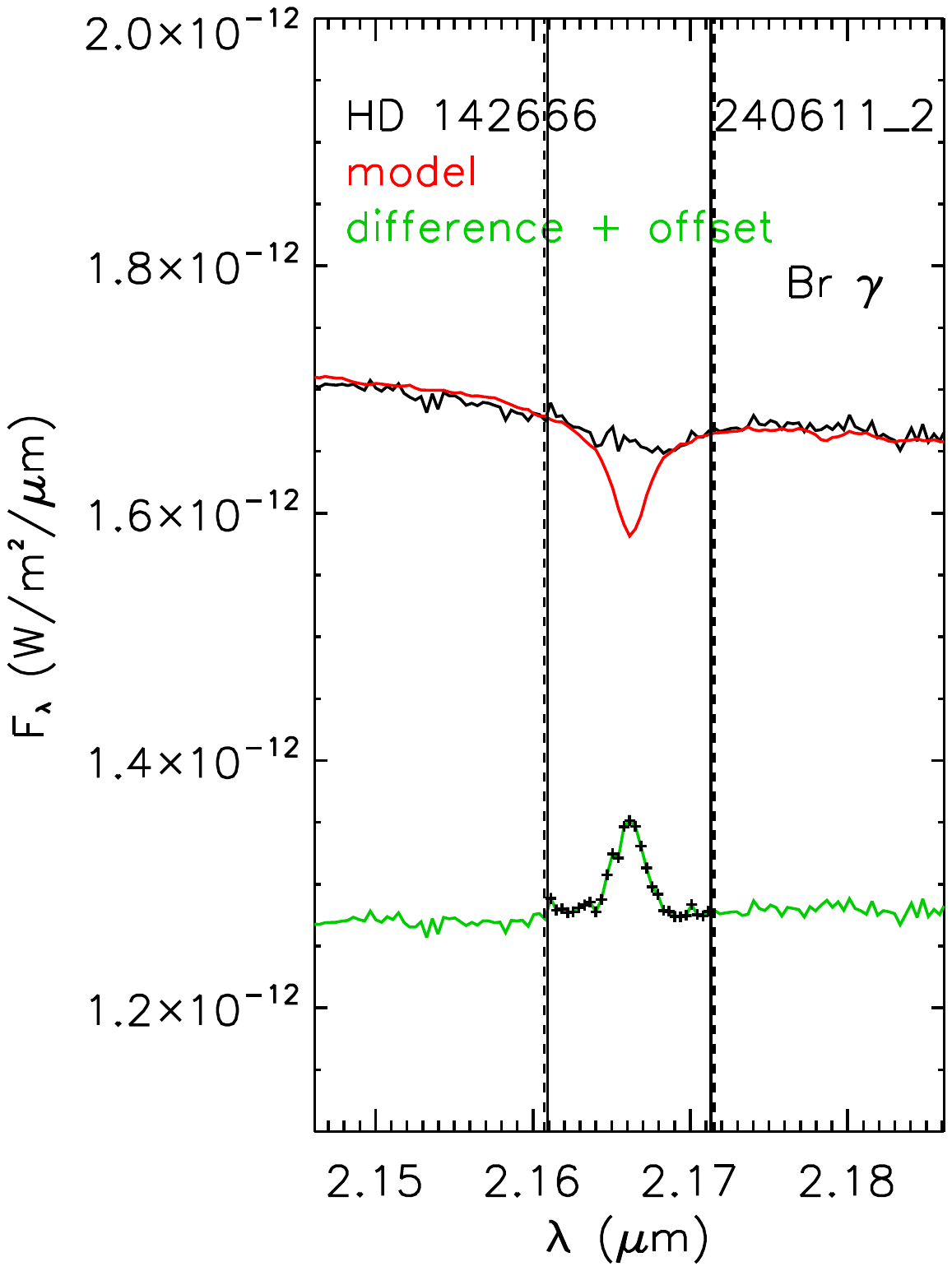}
\caption{The same as Figure A-2, except for the second measurement of HD 142666 on 240611. \label{fig:A-3}}
\end{figure}

\clearpage

\begin{figure}
\includegraphics[width=6.0cm, height=6.0cm]{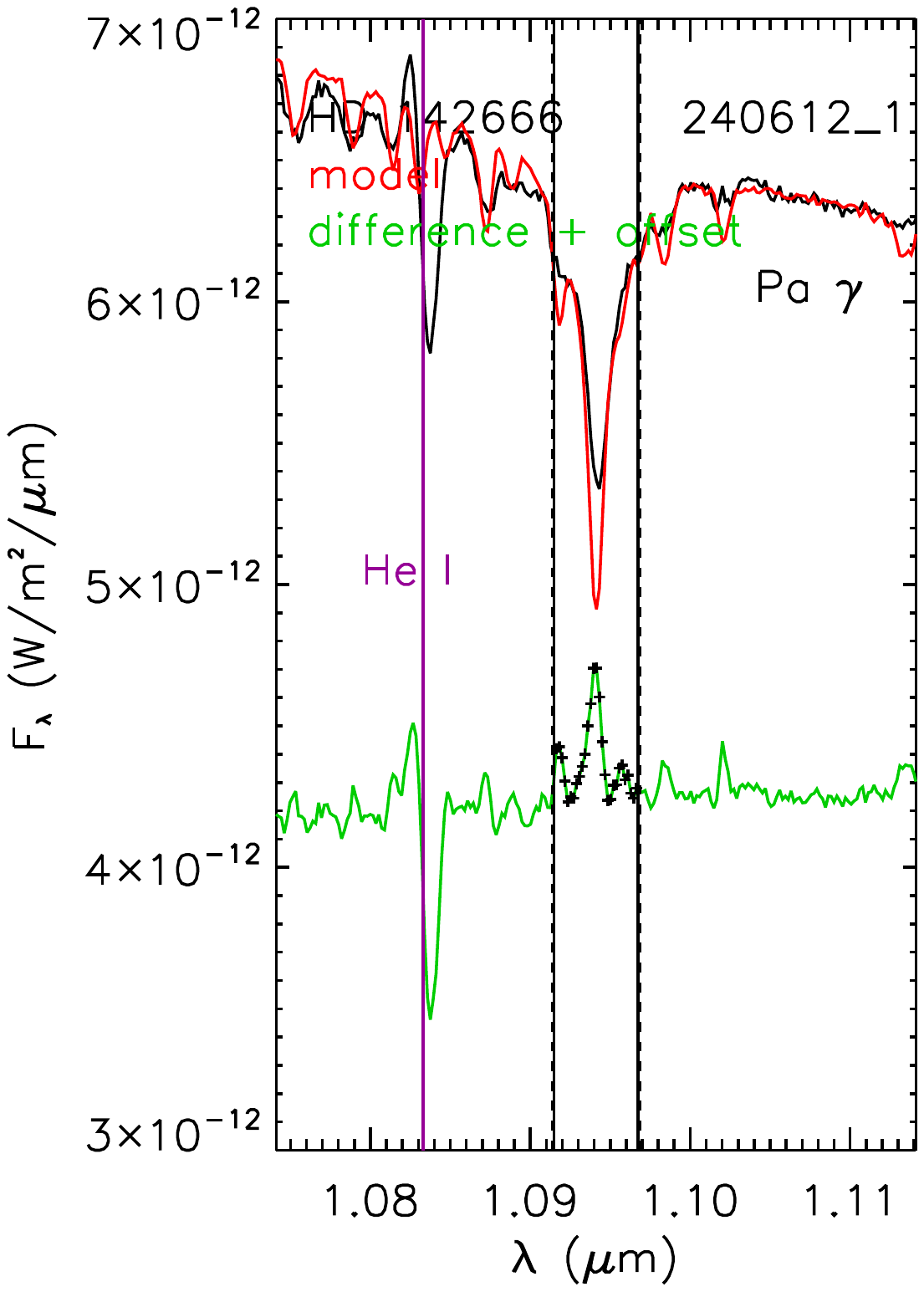}
\includegraphics[width=6.0cm, height=6.0cm]{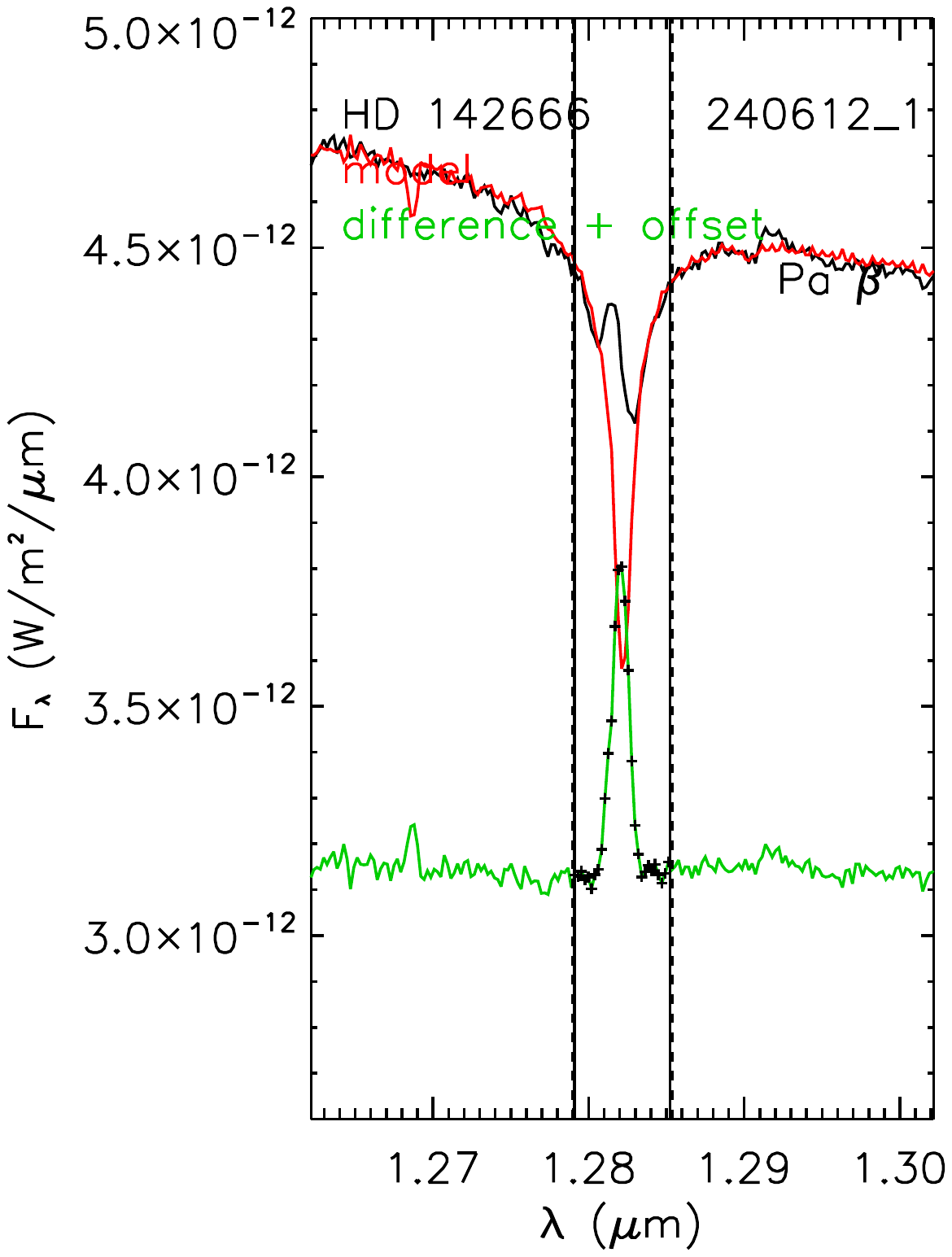}
\includegraphics[width=6.0cm, height=6.0cm]{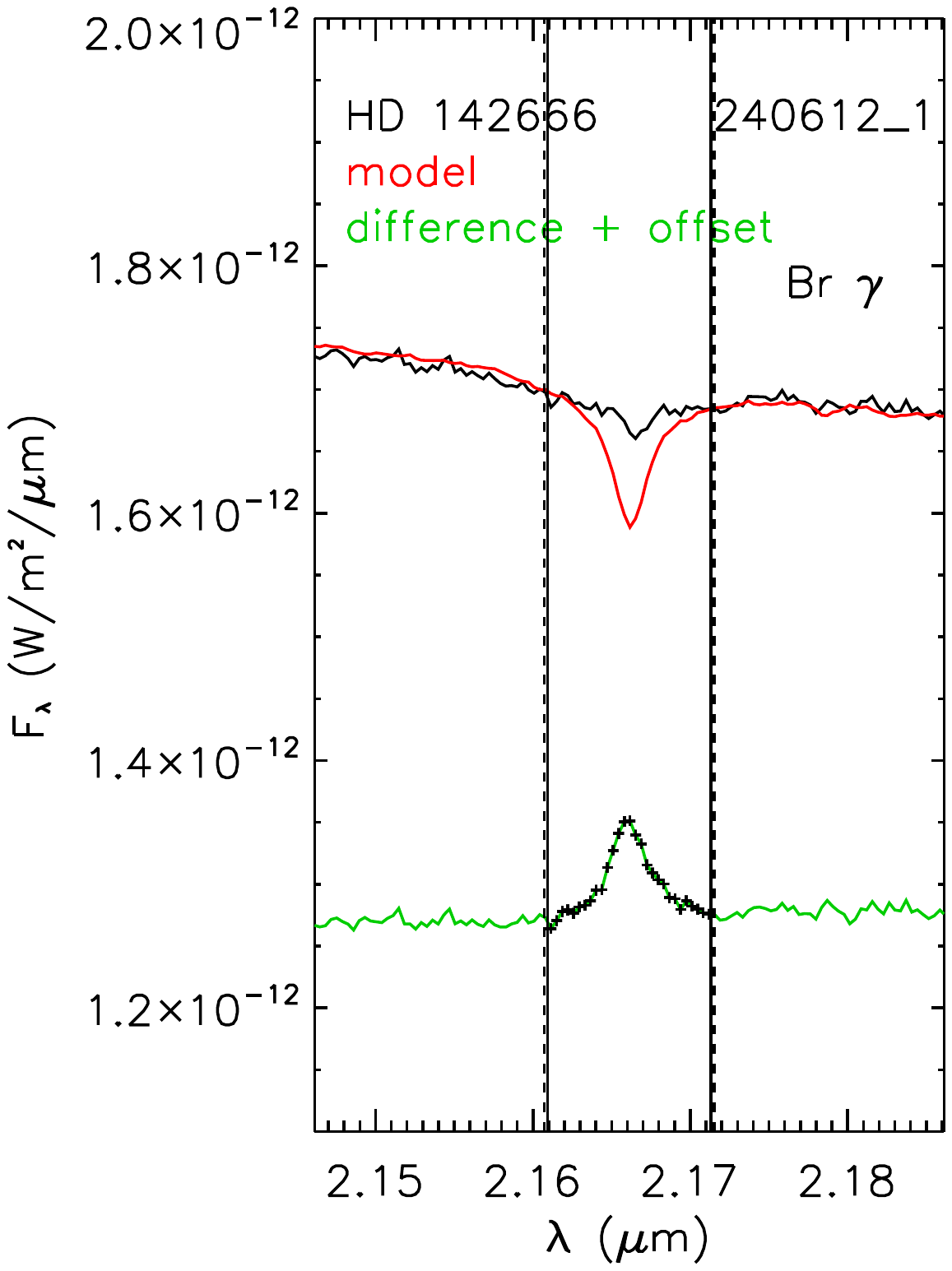}
\caption{The same as Figure A-2, except for the first measurement of HD 142666 on 240612. \label{fig:A-4}}
\end{figure}

\begin{figure}
\includegraphics[width=6.0cm, height=6.0cm]{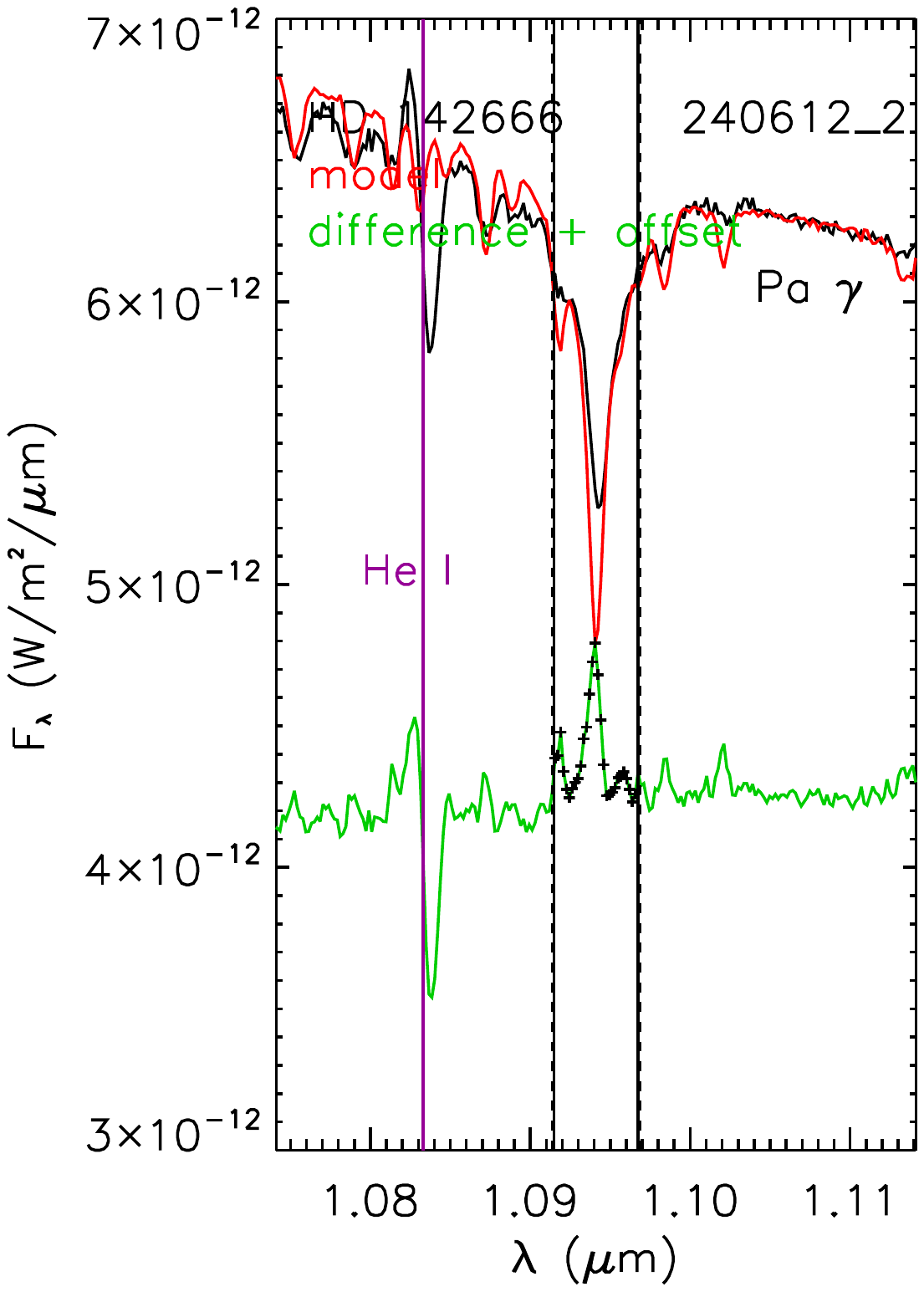}
\includegraphics[width=6.0cm, height=6.0cm]{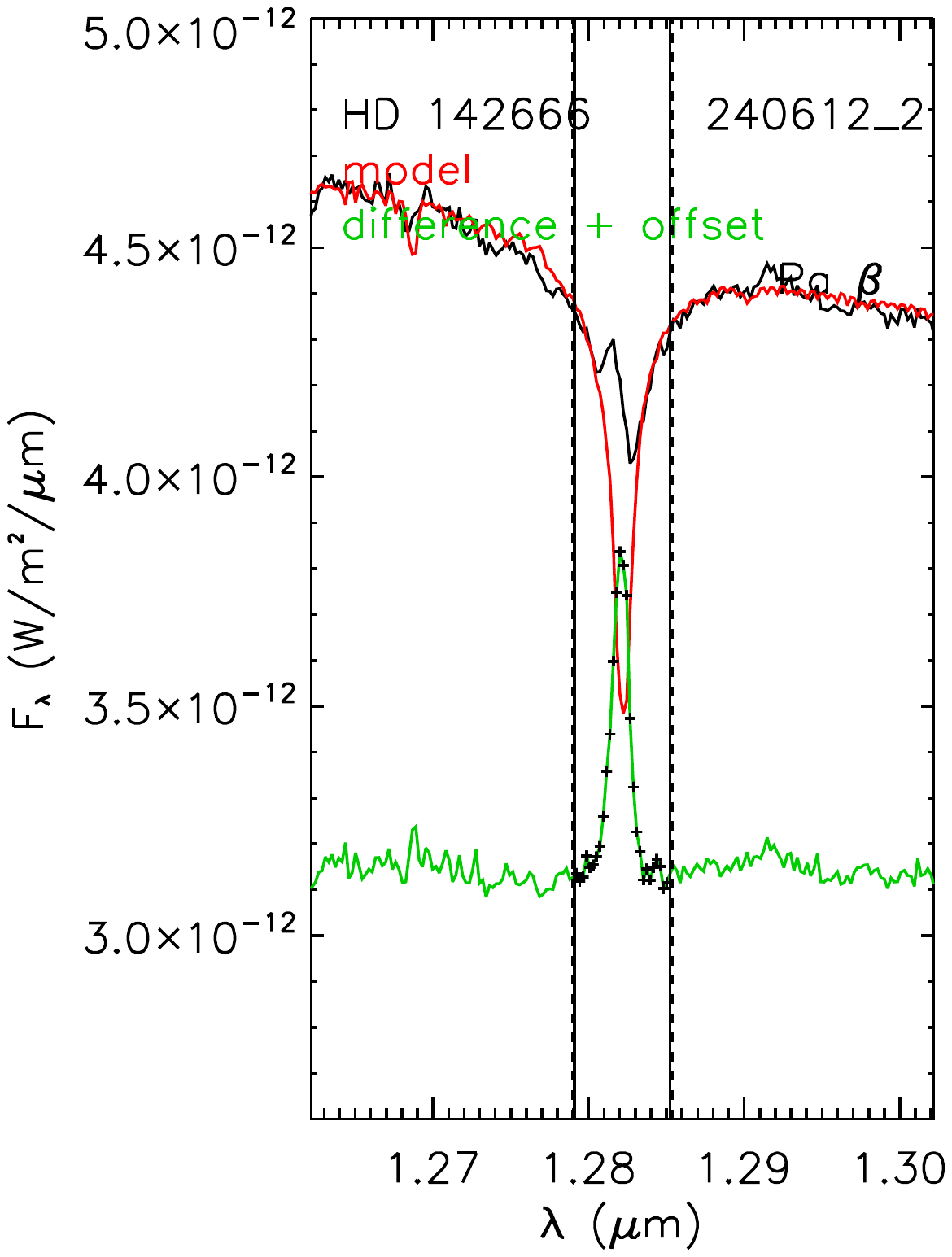}
\includegraphics[width=6.0cm, height=6.0cm]{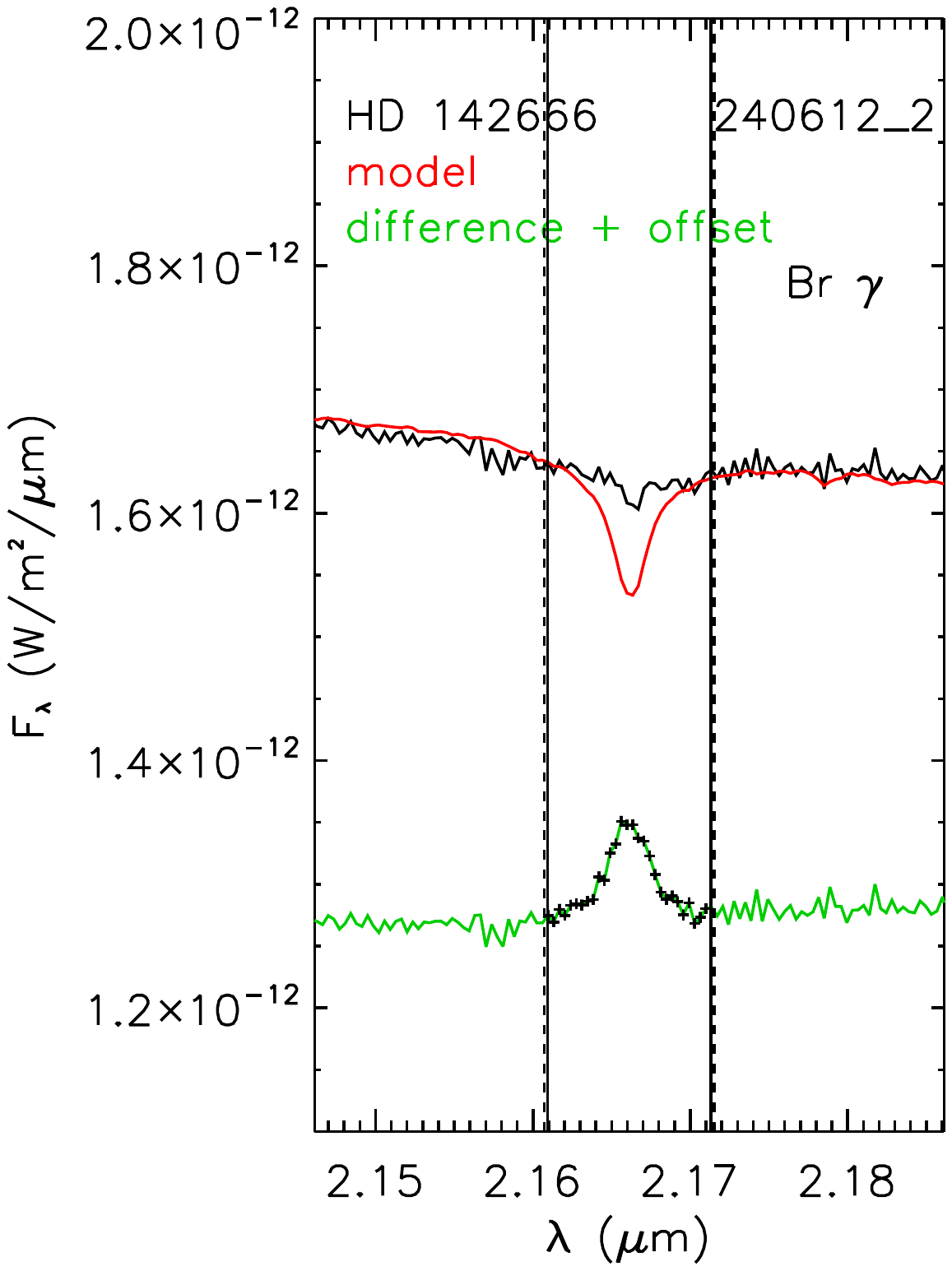}
\caption{The same as Figure A-2, except for the second measurement of HD 142666 on 240612. This spectrum was obtained $\sim$1.5 hours after the previous observation, and is virtually identical. \label{fig:A-5}}
\end{figure}

\begin{figure}
\includegraphics[width=6.0cm, height=6.0cm]{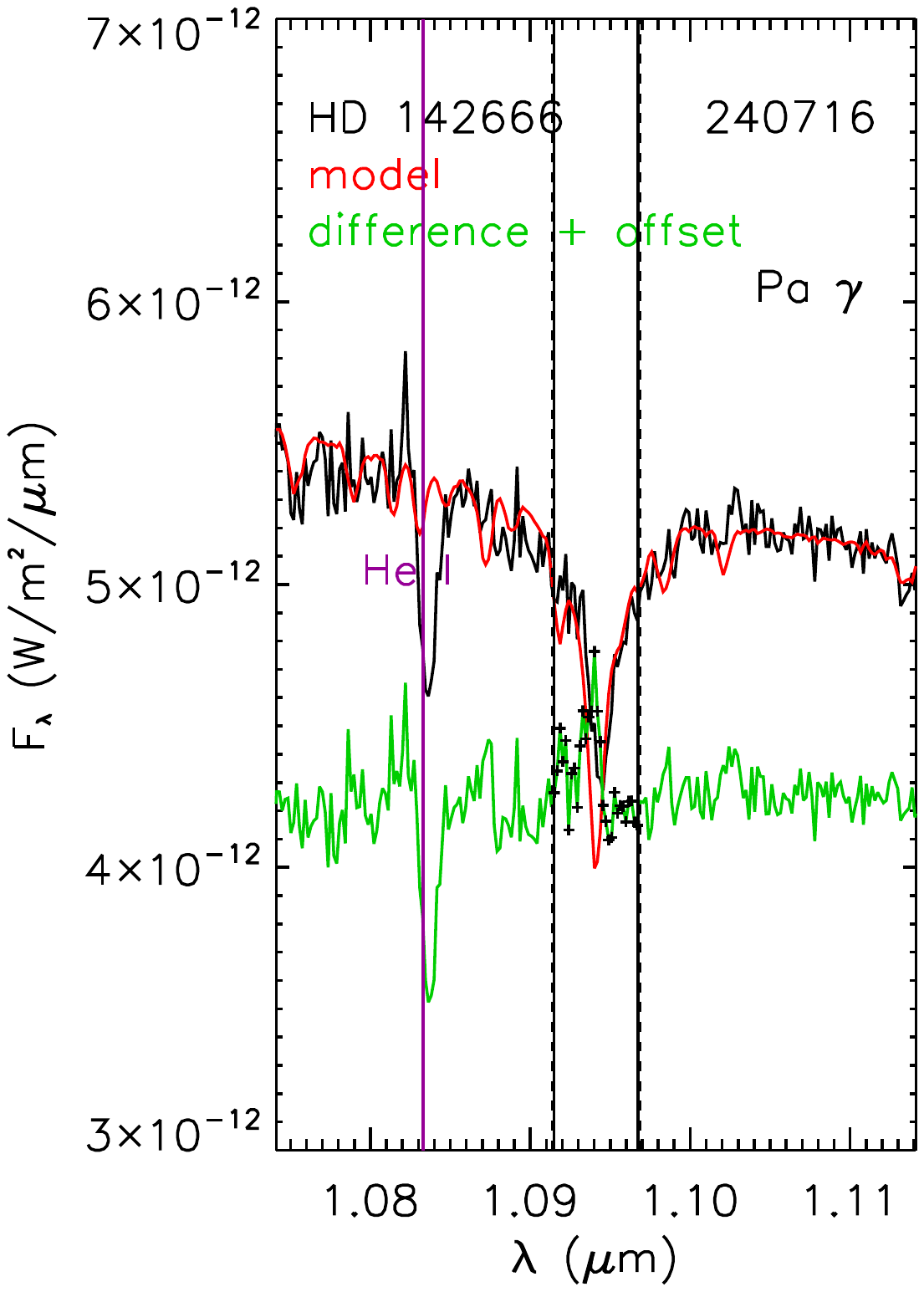}
\includegraphics[width=6.0cm, height=6.0cm]{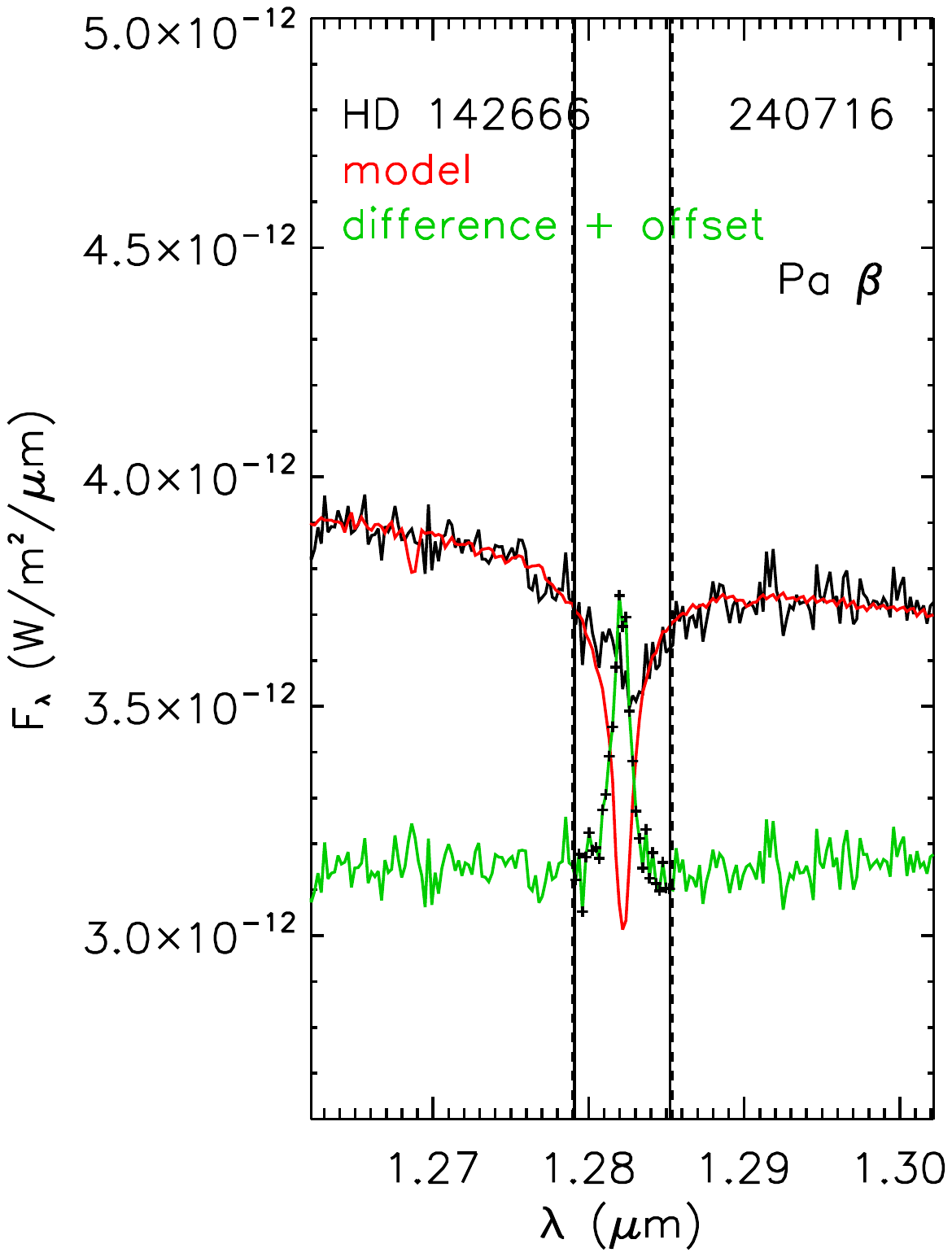}
\includegraphics[width=6.0cm, height=6.0cm]{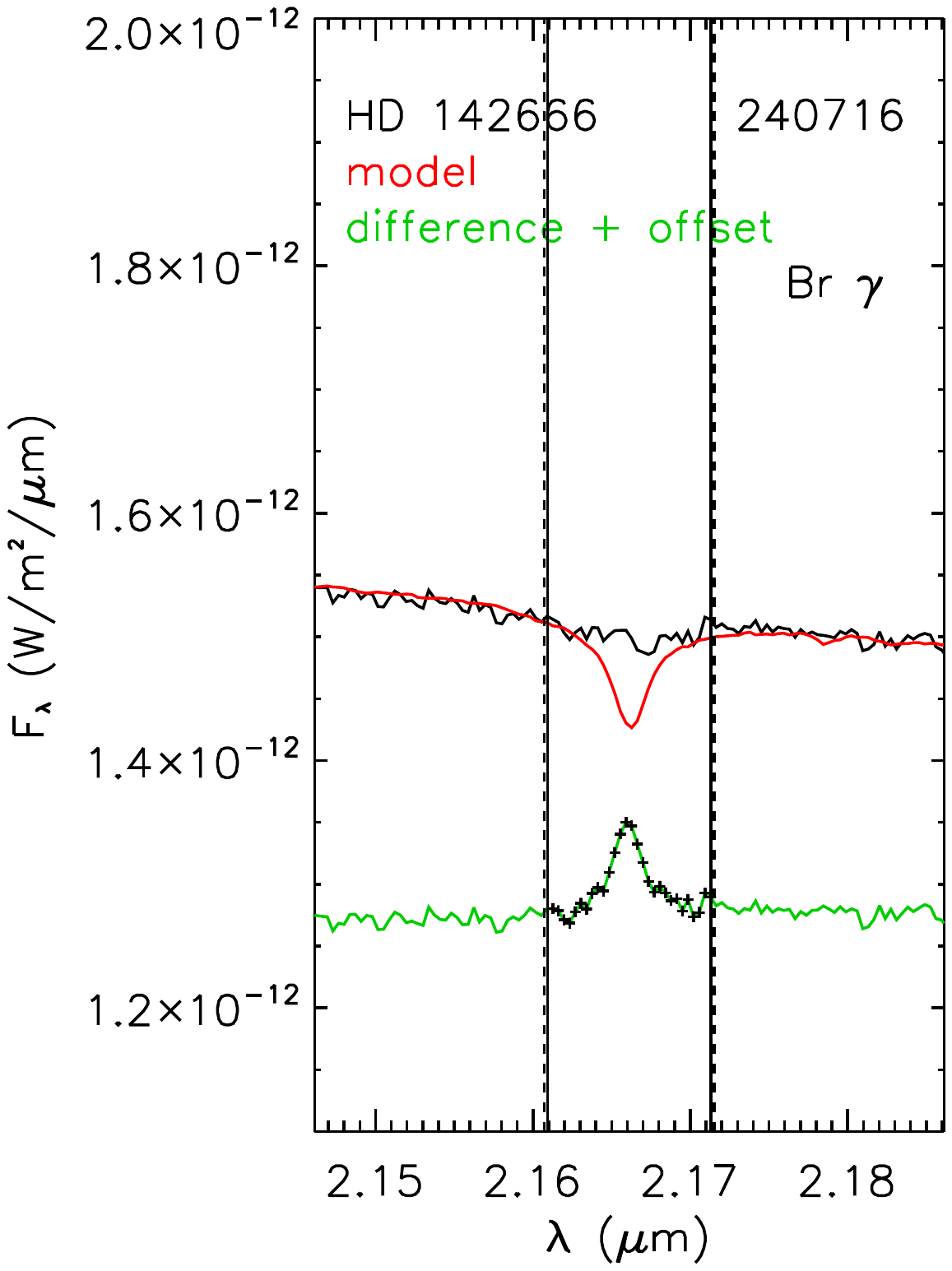}
\caption{The same as Figure A-2, except for HD 142666 on 240716. \label{fig:A-6}}
\end{figure}

%\clearpage

\begin{figure}
\includegraphics[width=6.0cm, height=6.0cm]{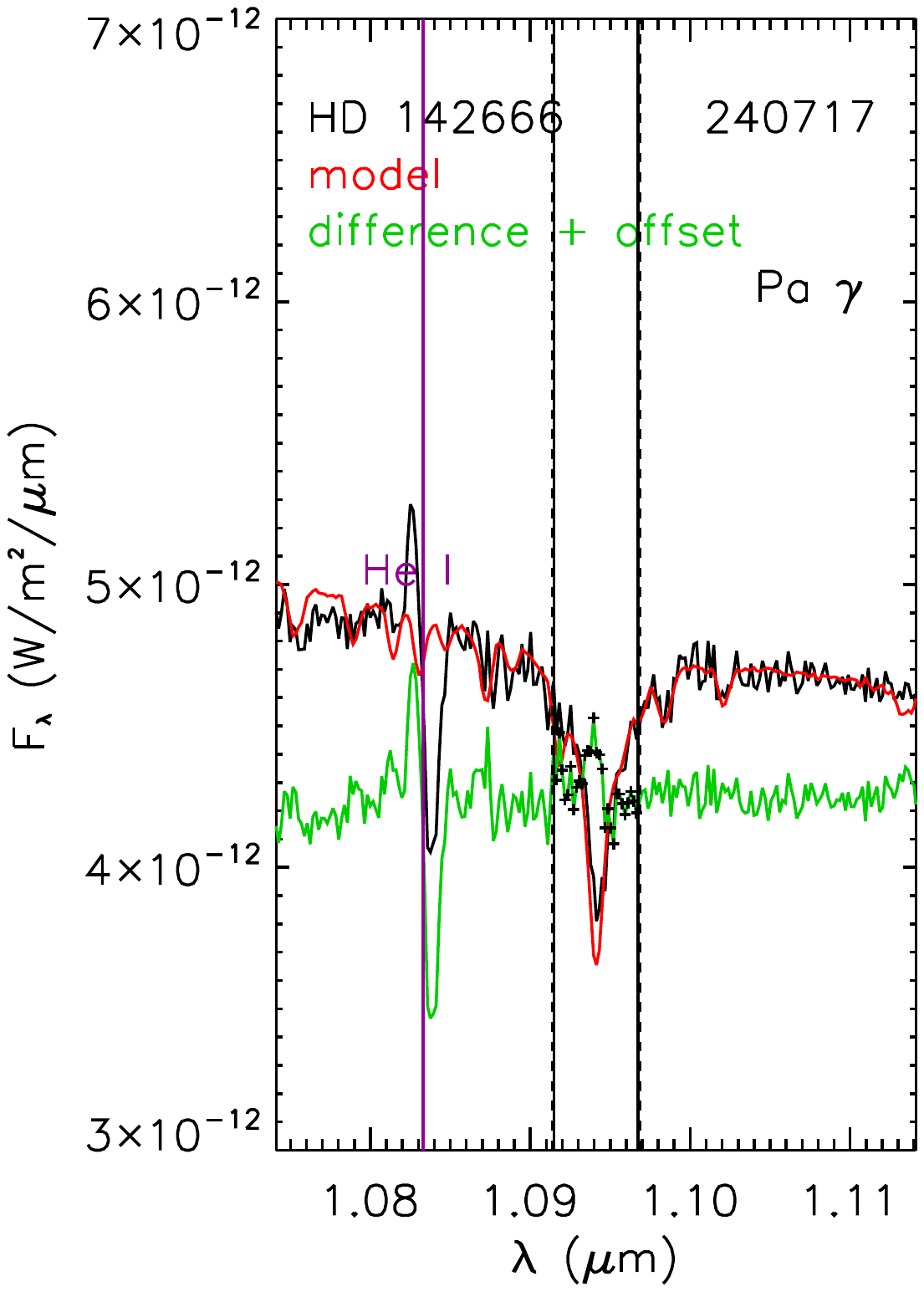}
\includegraphics[width=6.0cm, height=6.0cm]{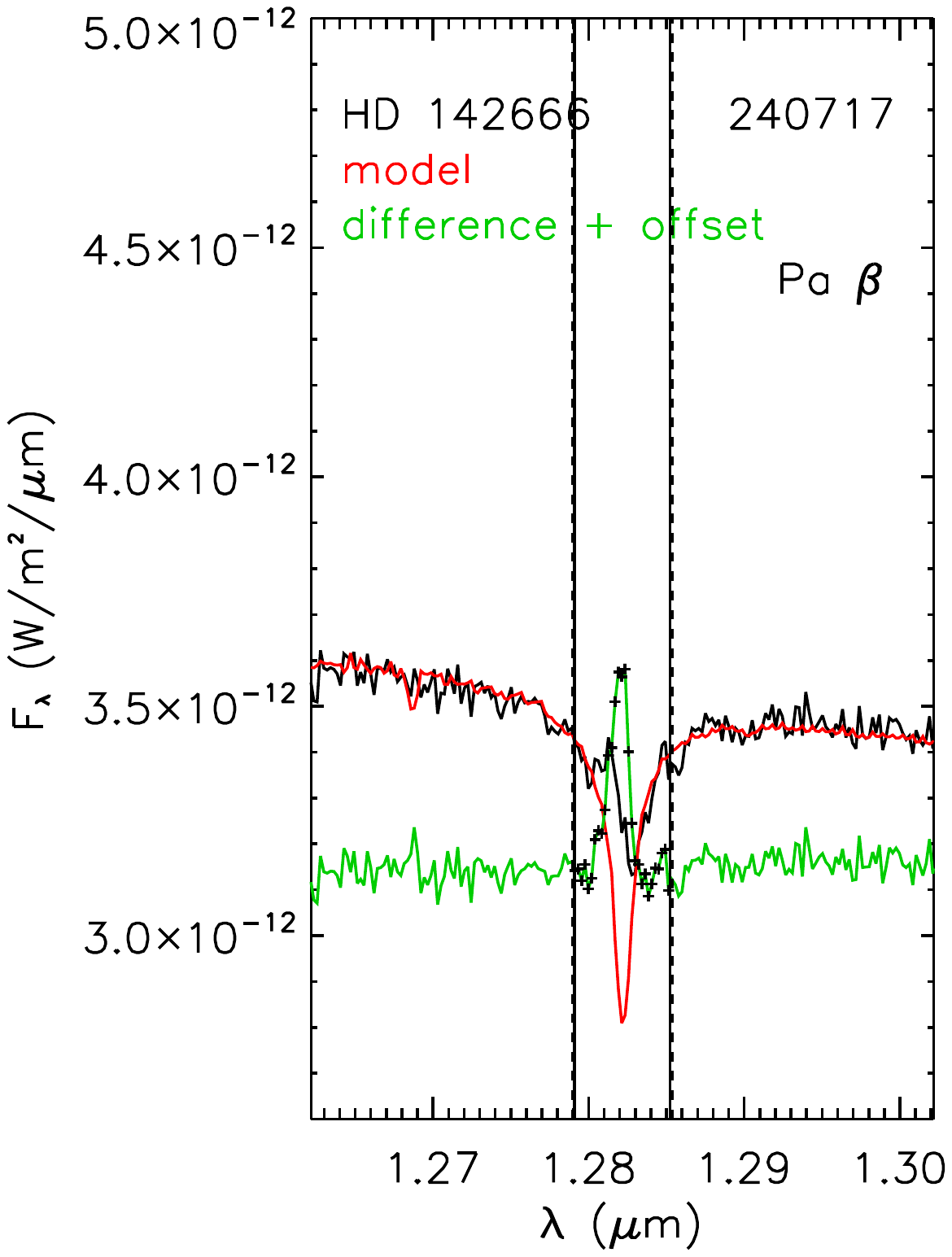}
\includegraphics[width=6.0cm, height=6.0cm]{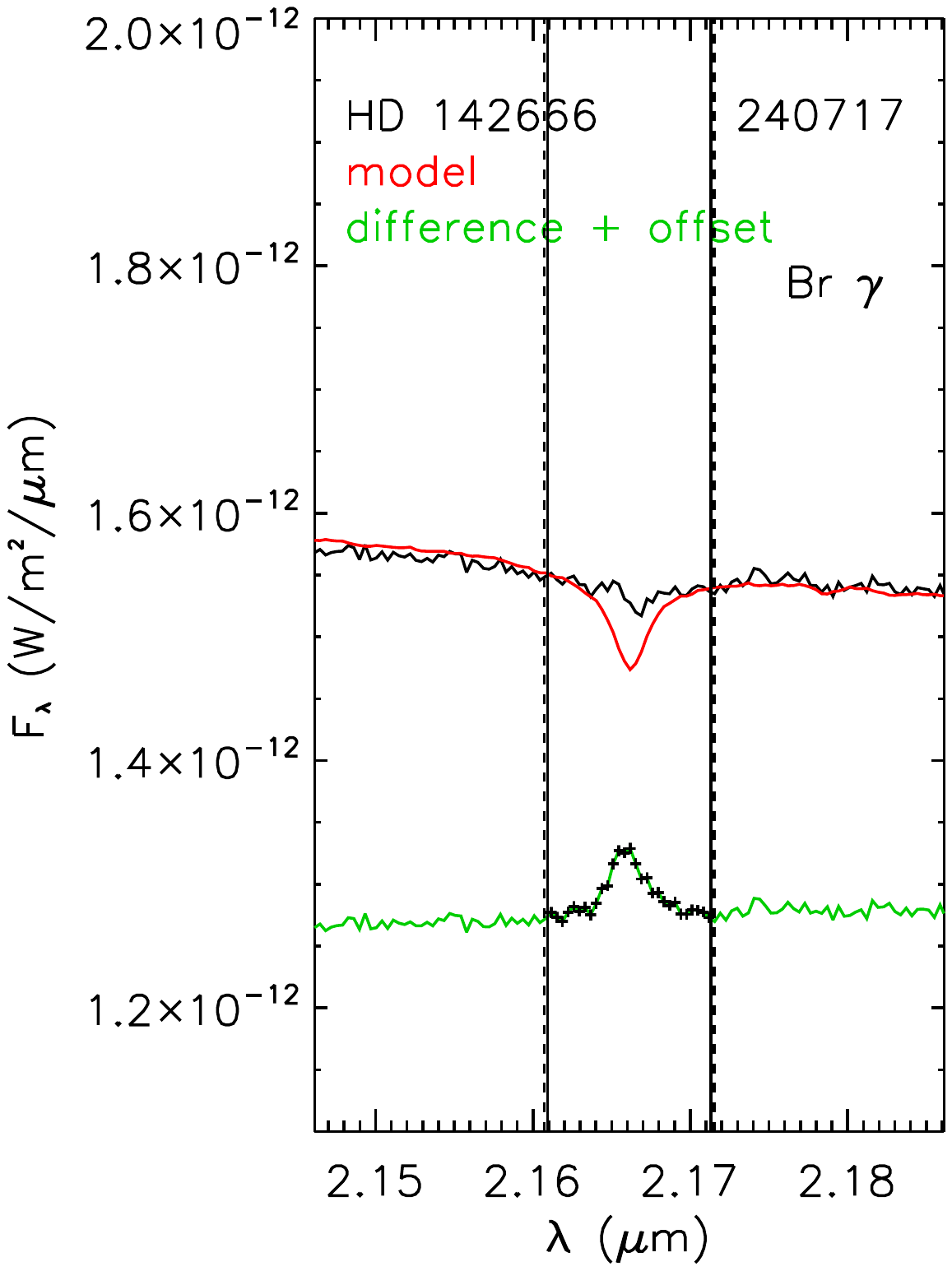}
\caption{The same as Figure A-2, except for HD 142666 on 240717. \label{fig:A-7}}
\end{figure}

\begin{figure}
\includegraphics[width=6.0cm, height=6.0cm]{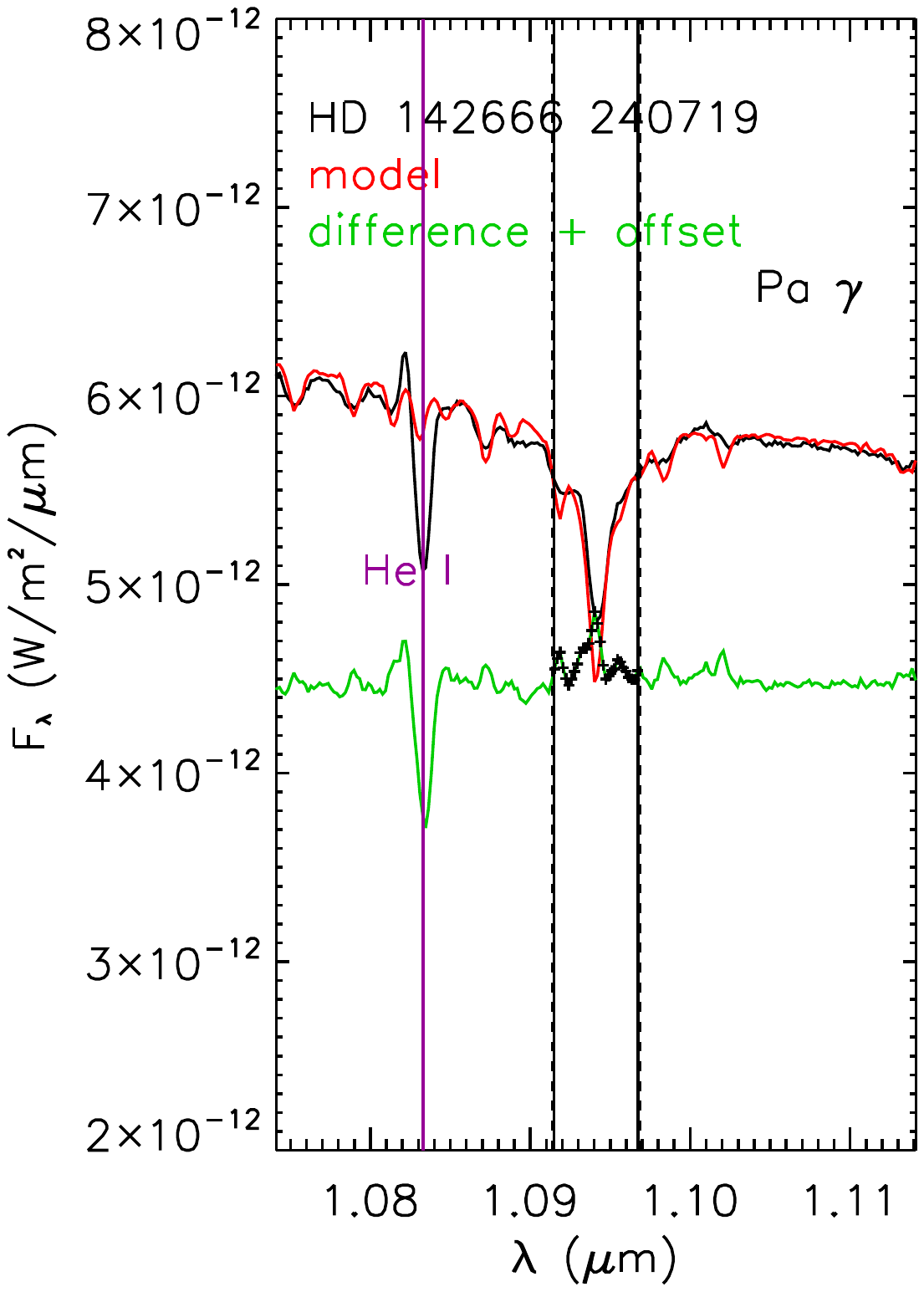}
\includegraphics[width=6.0cm, height=6.0cm]{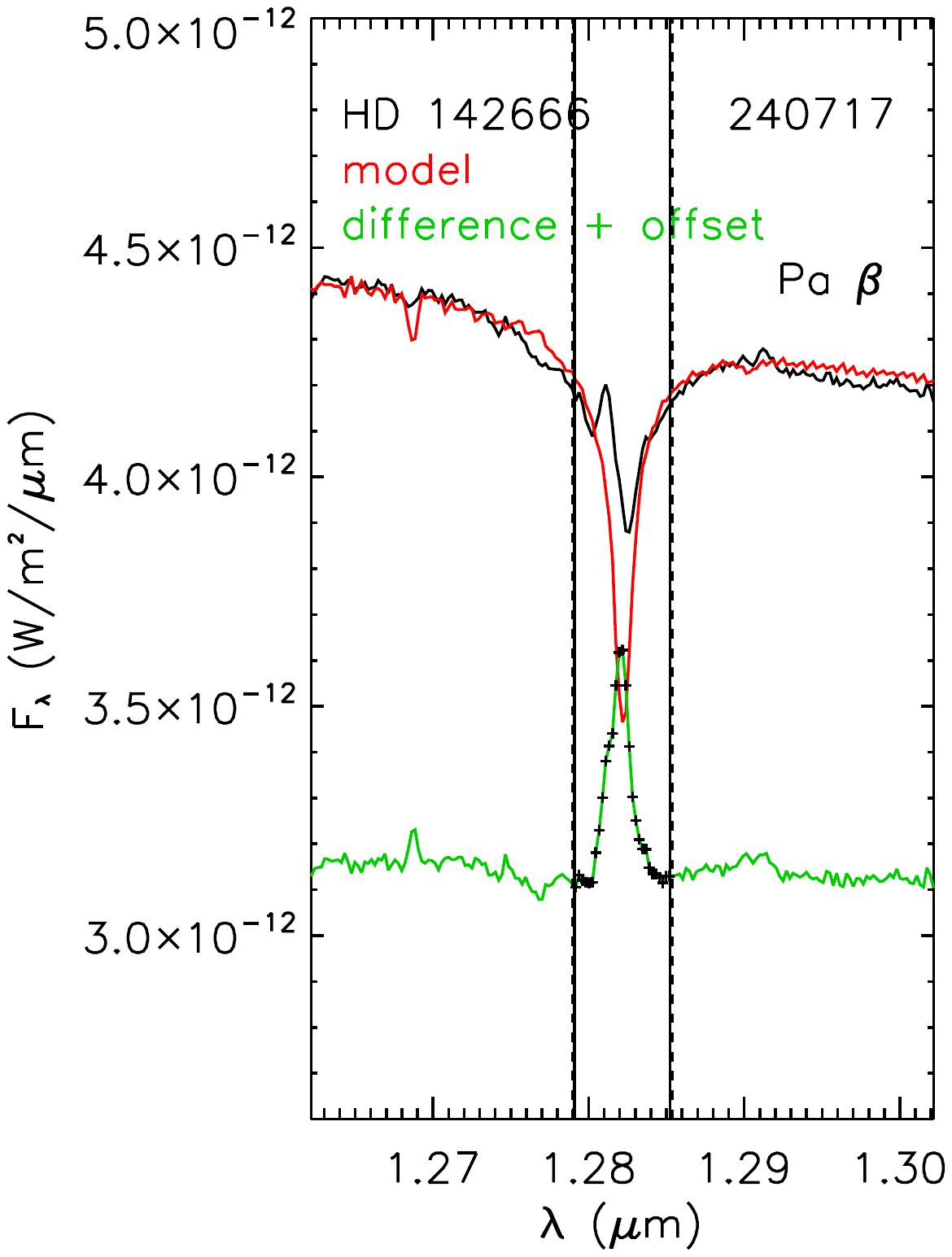}
\includegraphics[width=6.0cm, height=6.0cm]{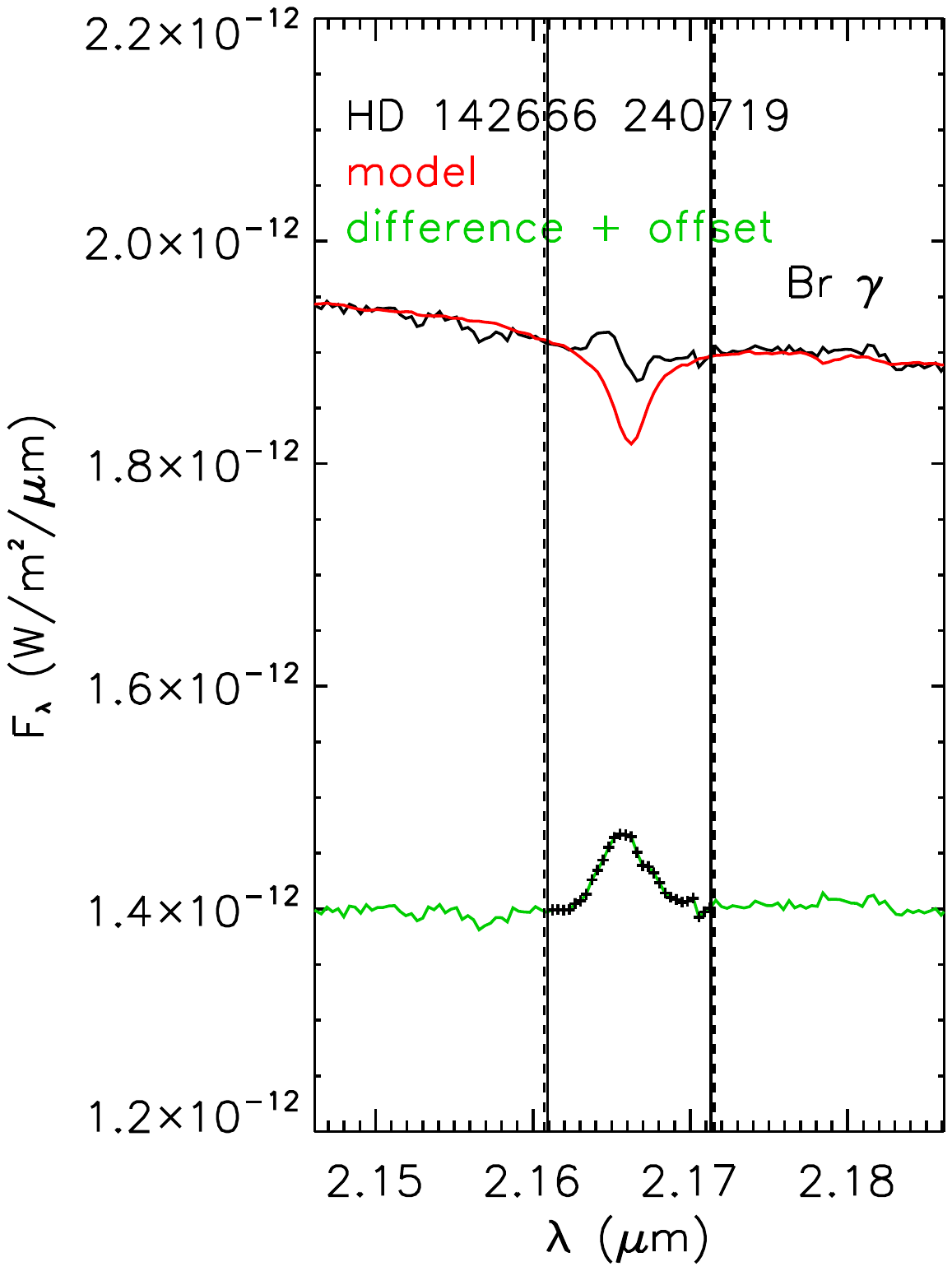}
\caption{The same as Figure A-2, except for HD 142666 on 240719. \label{fig:A-8}}
\end{figure}

\begin{figure}
\includegraphics[width=6.0cm, height=6.0cm]{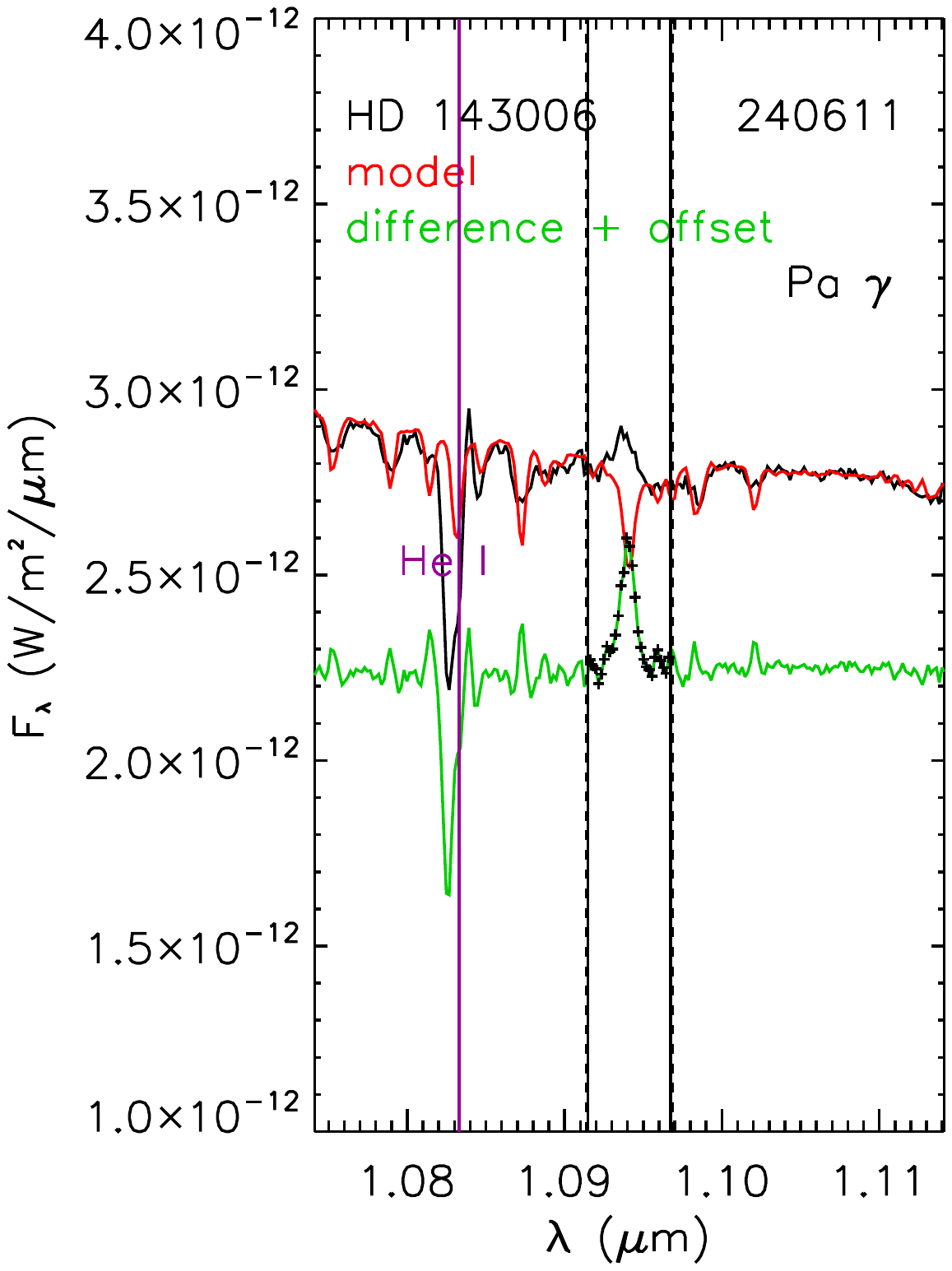}
\includegraphics[width=6.0cm, height=6.0cm]{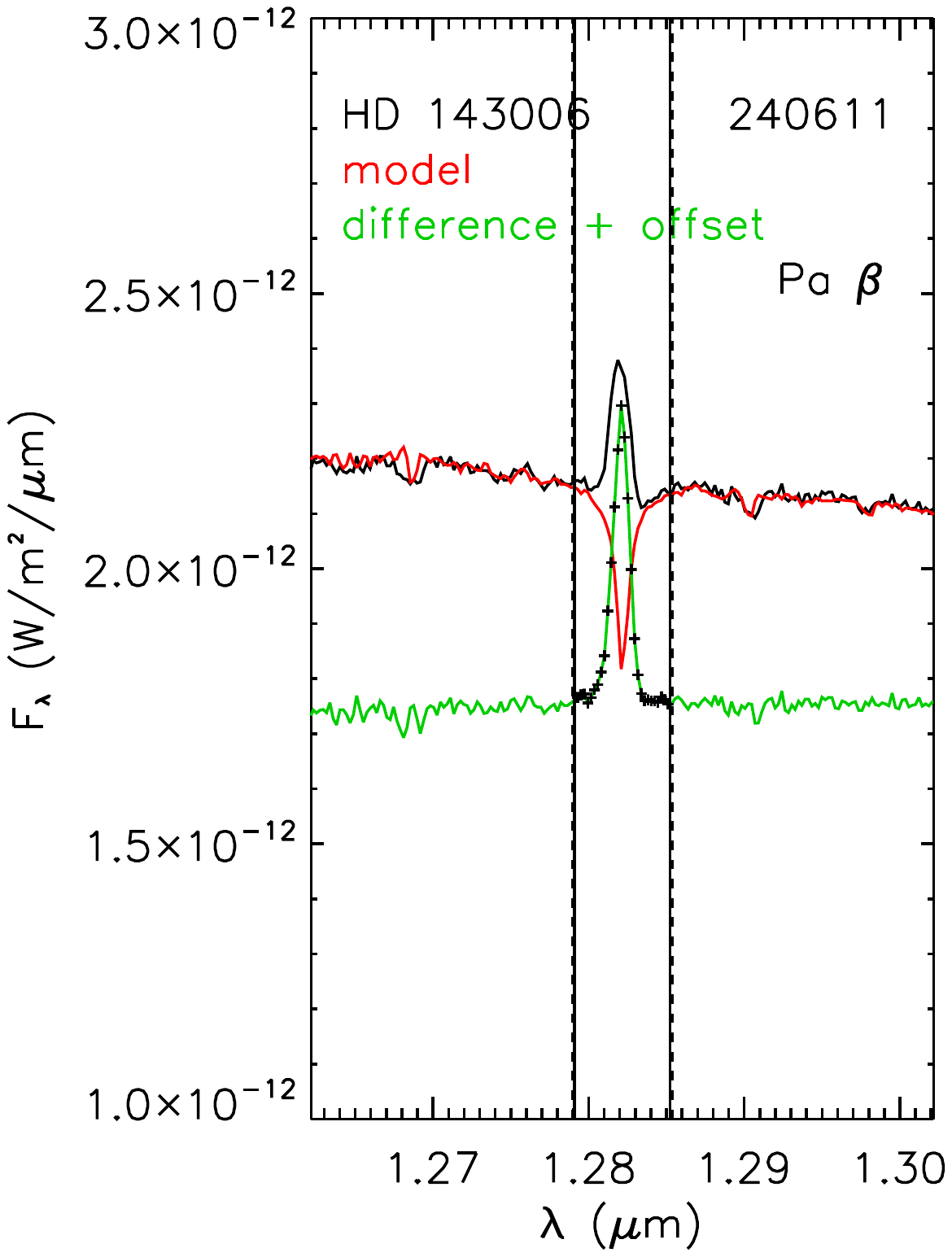}
\includegraphics[width=6.0cm, height=6.0cm]{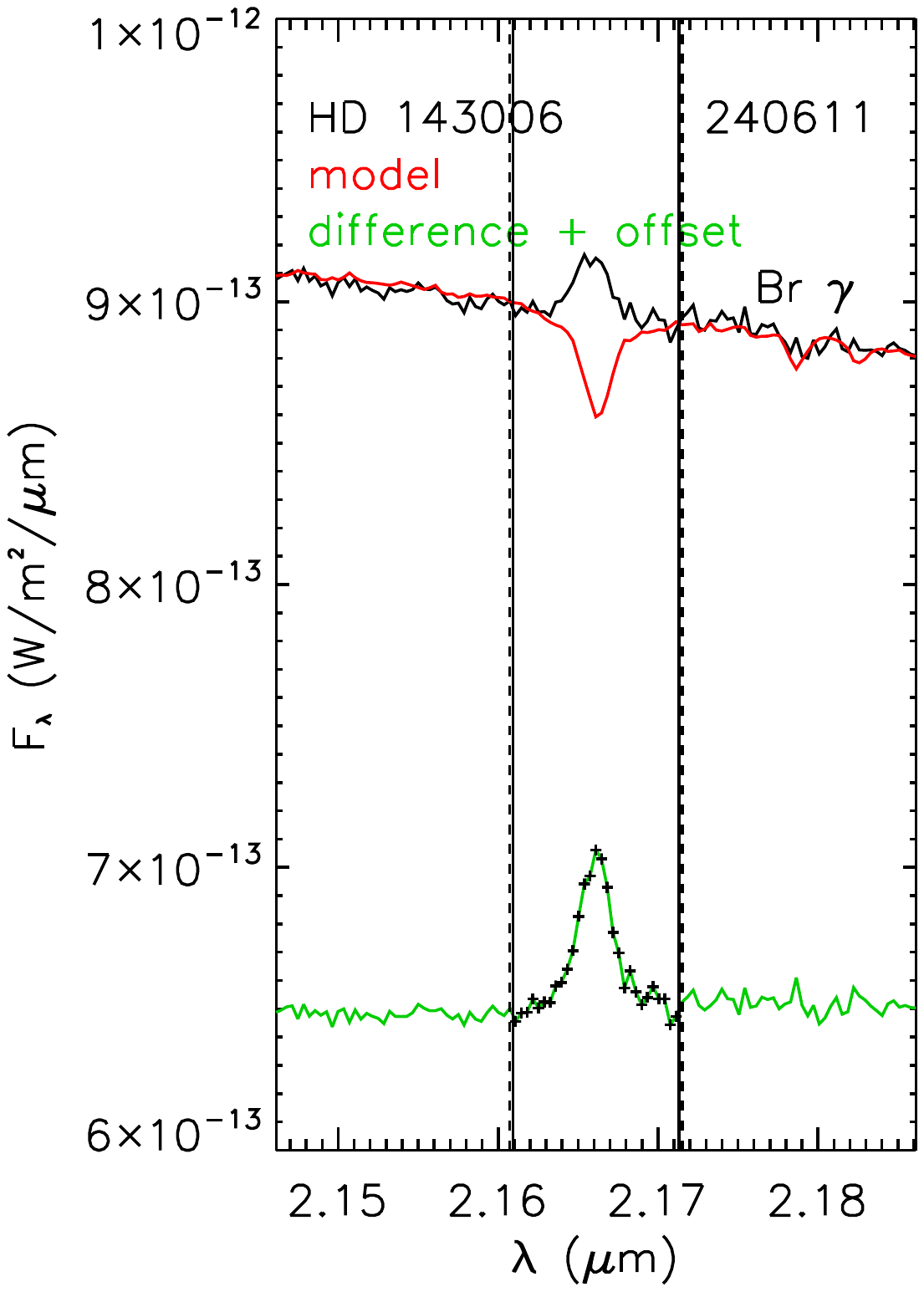}
\caption{The same as Figure A-2, except for HD 143006 on 240611. \label{fig:A-9}}
\end{figure}

\begin{figure}
\includegraphics[width=6.0cm, height=6.0cm]{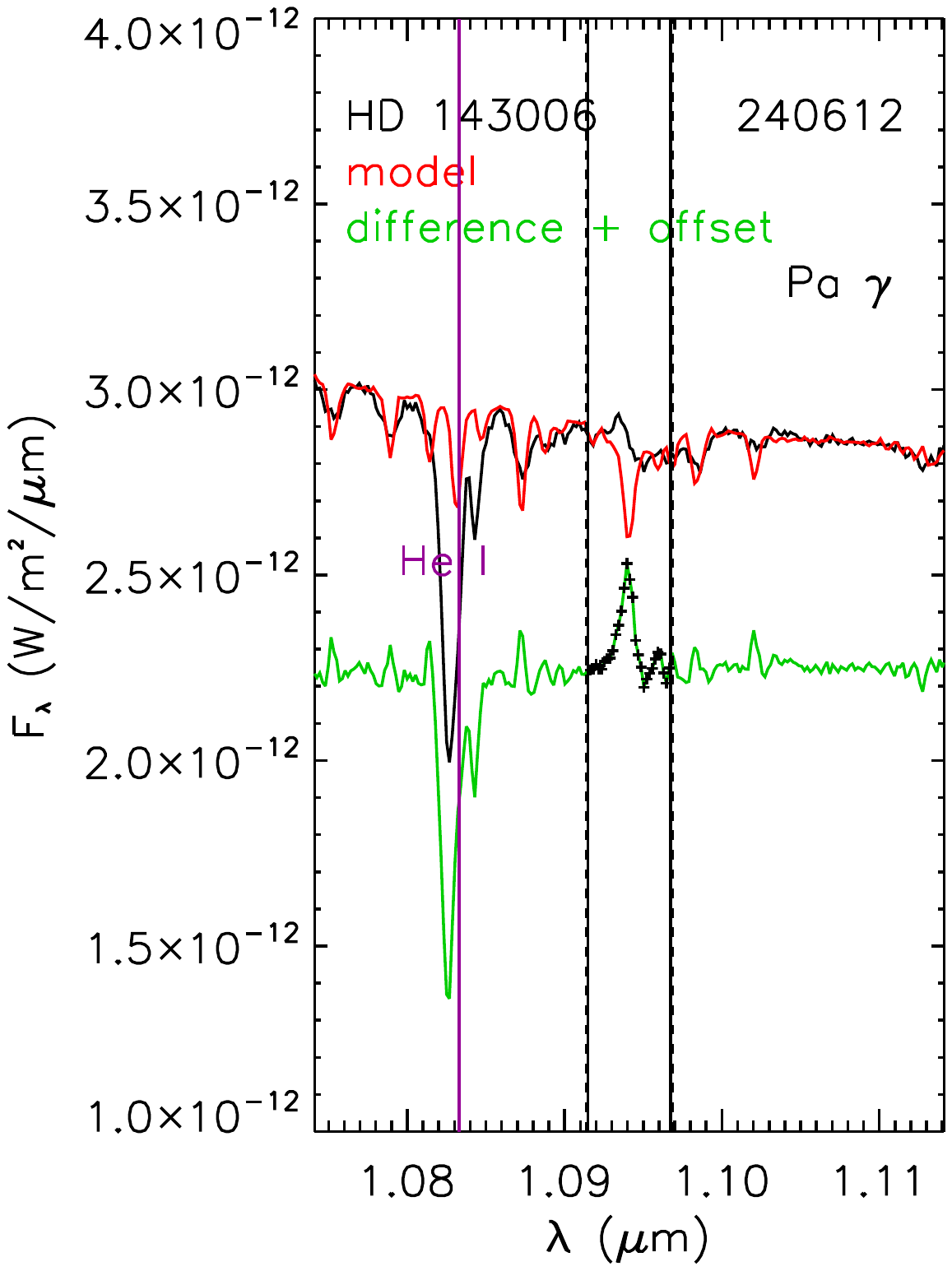}
\includegraphics[width=6.0cm, height=6.0cm]{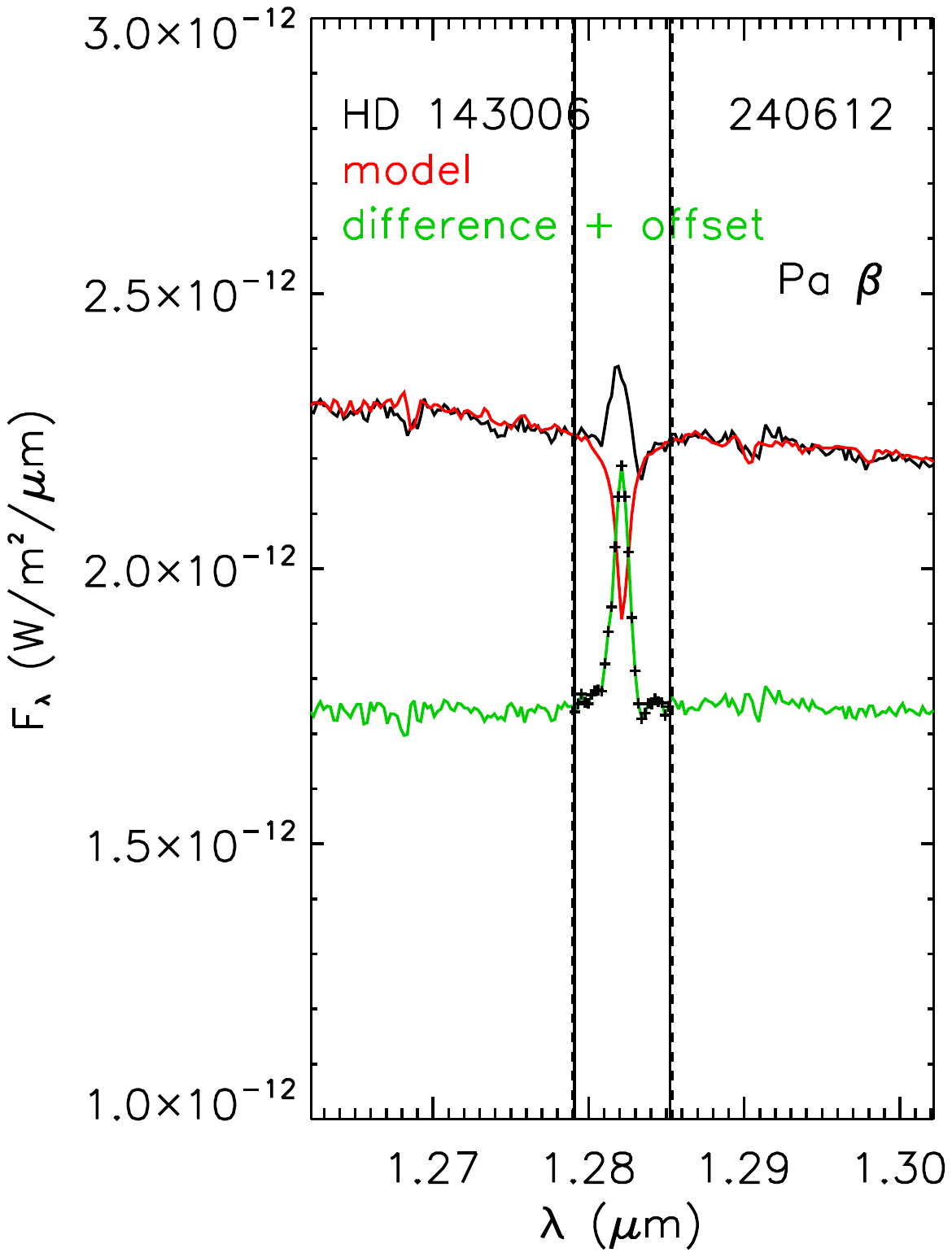}
\includegraphics[width=6.0cm, height=6.0cm]{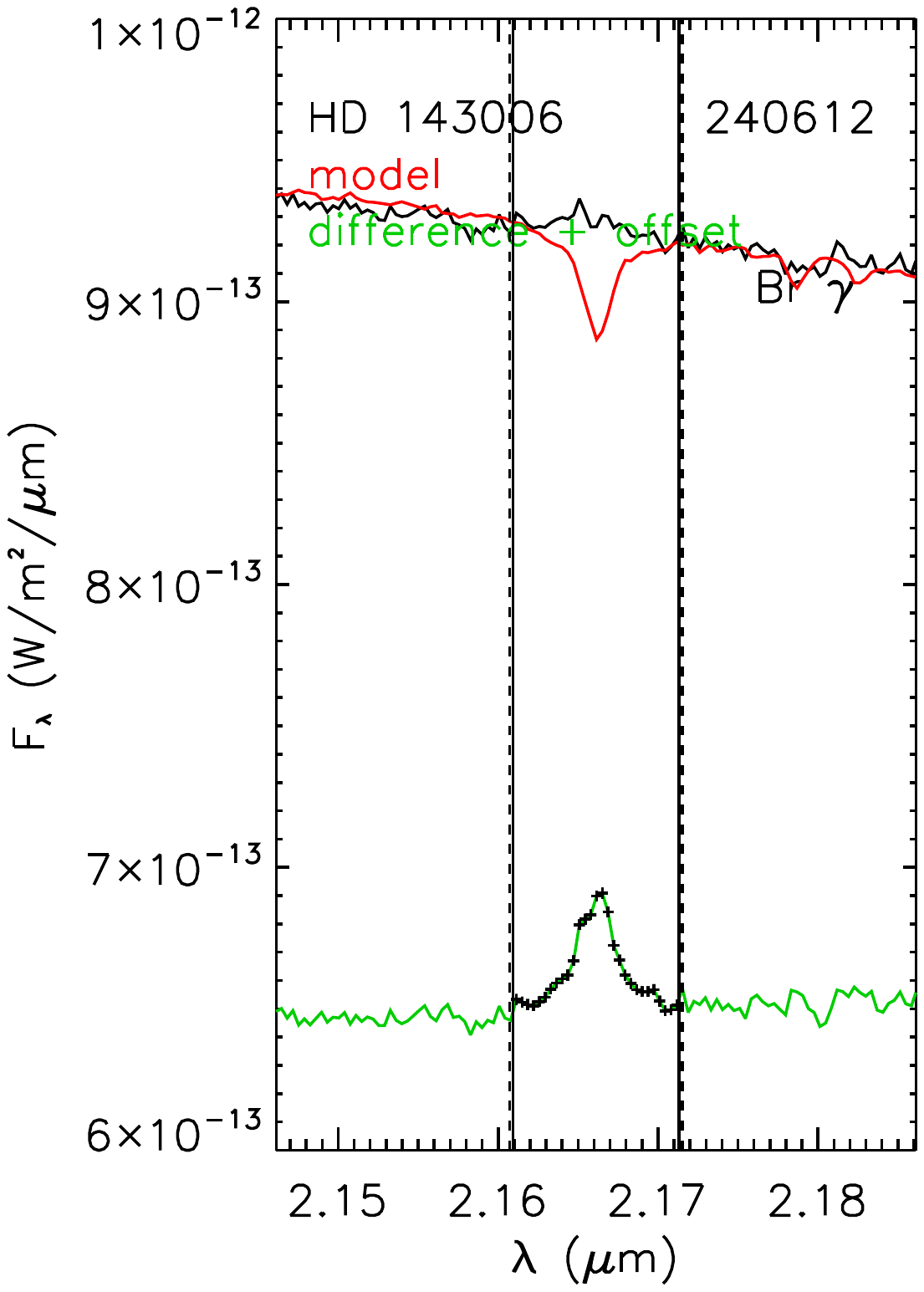}
\caption{The same as Figure A-2, except for HD 143006 on 240612. \label{fig:A-10}}
\end{figure}

\begin{figure}
\includegraphics[width=6.0cm, height=6.0cm]{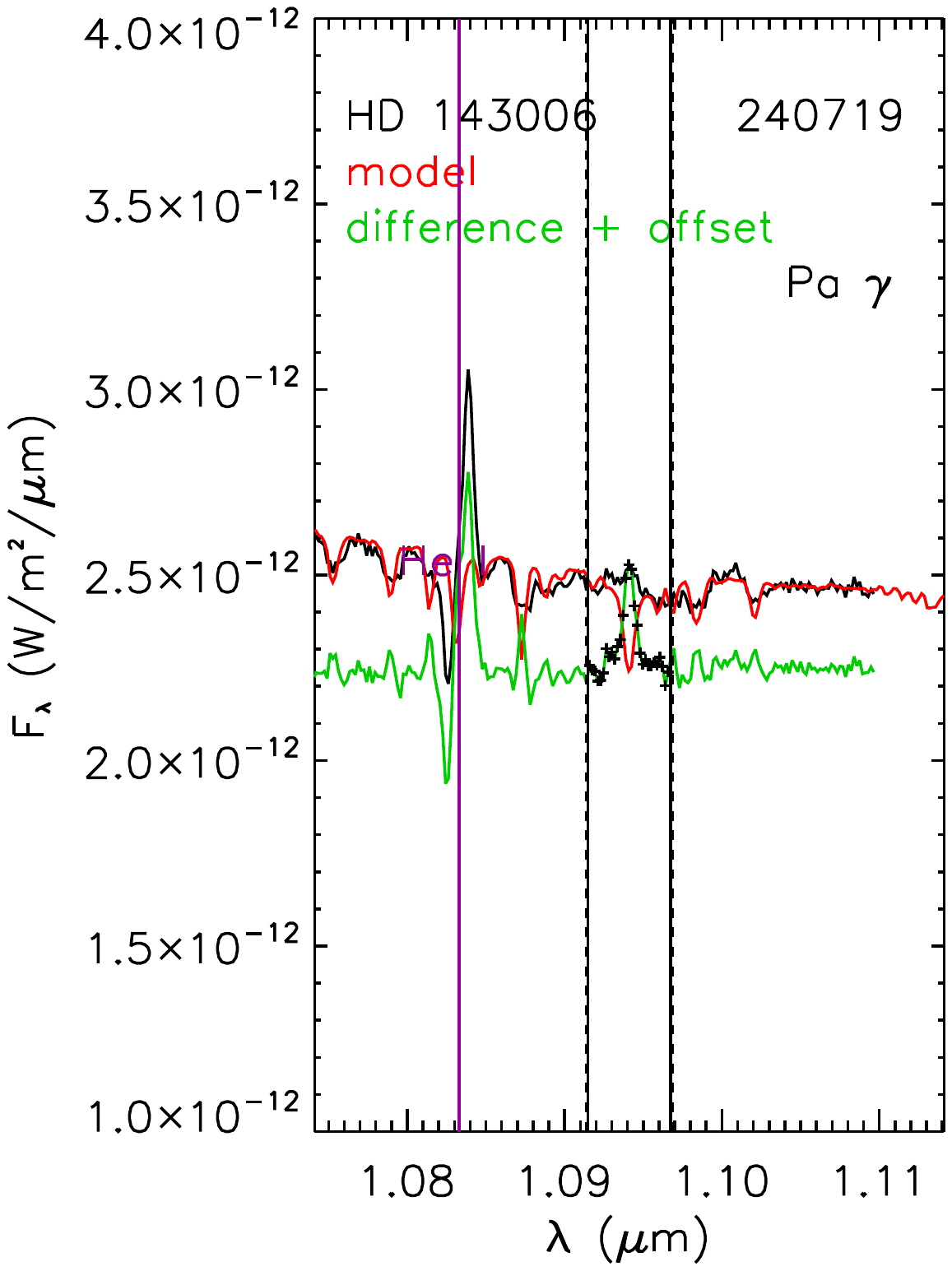}
\includegraphics[width=6.0cm, height=6.0cm]{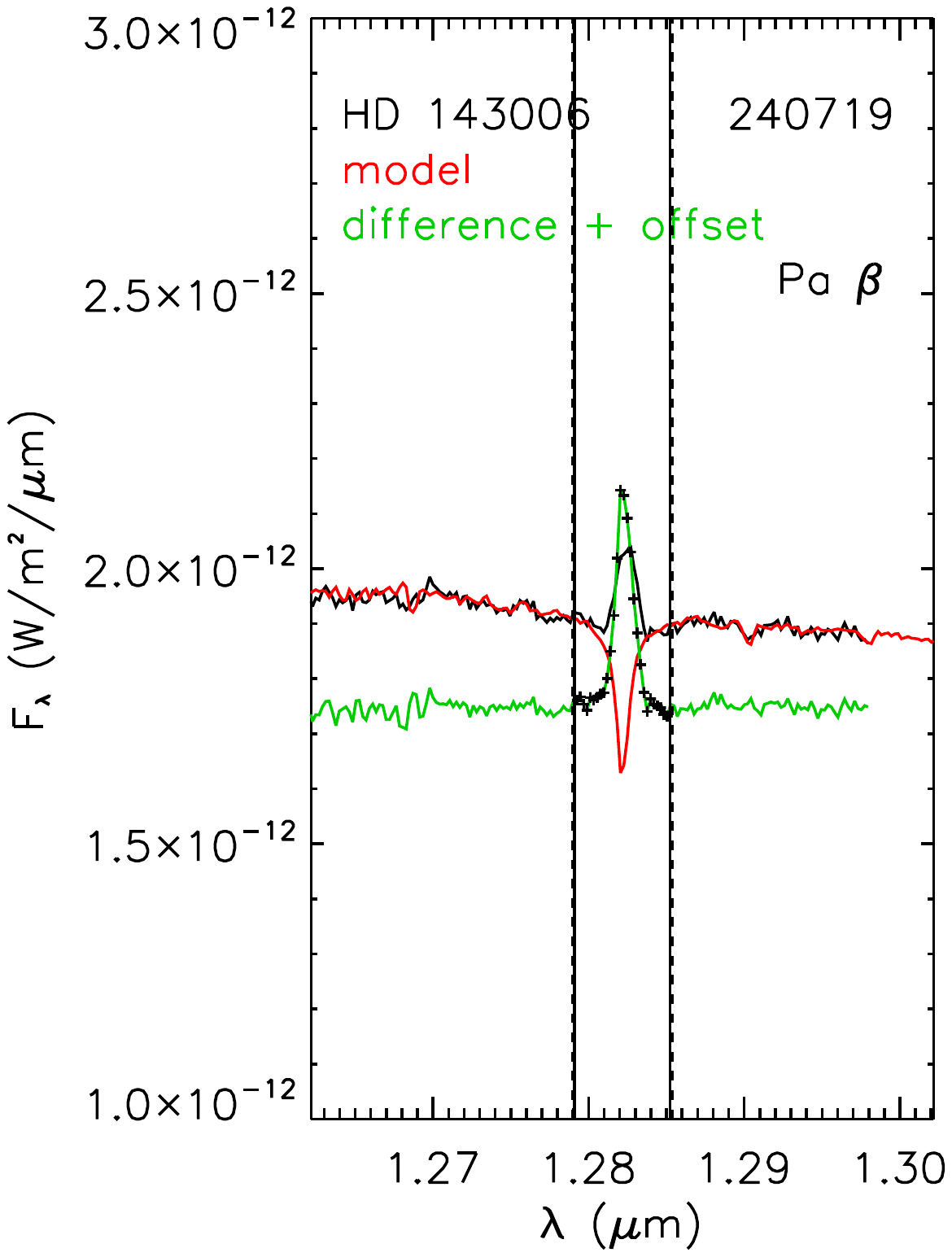}
\includegraphics[width=6.0cm, height=6.0cm]{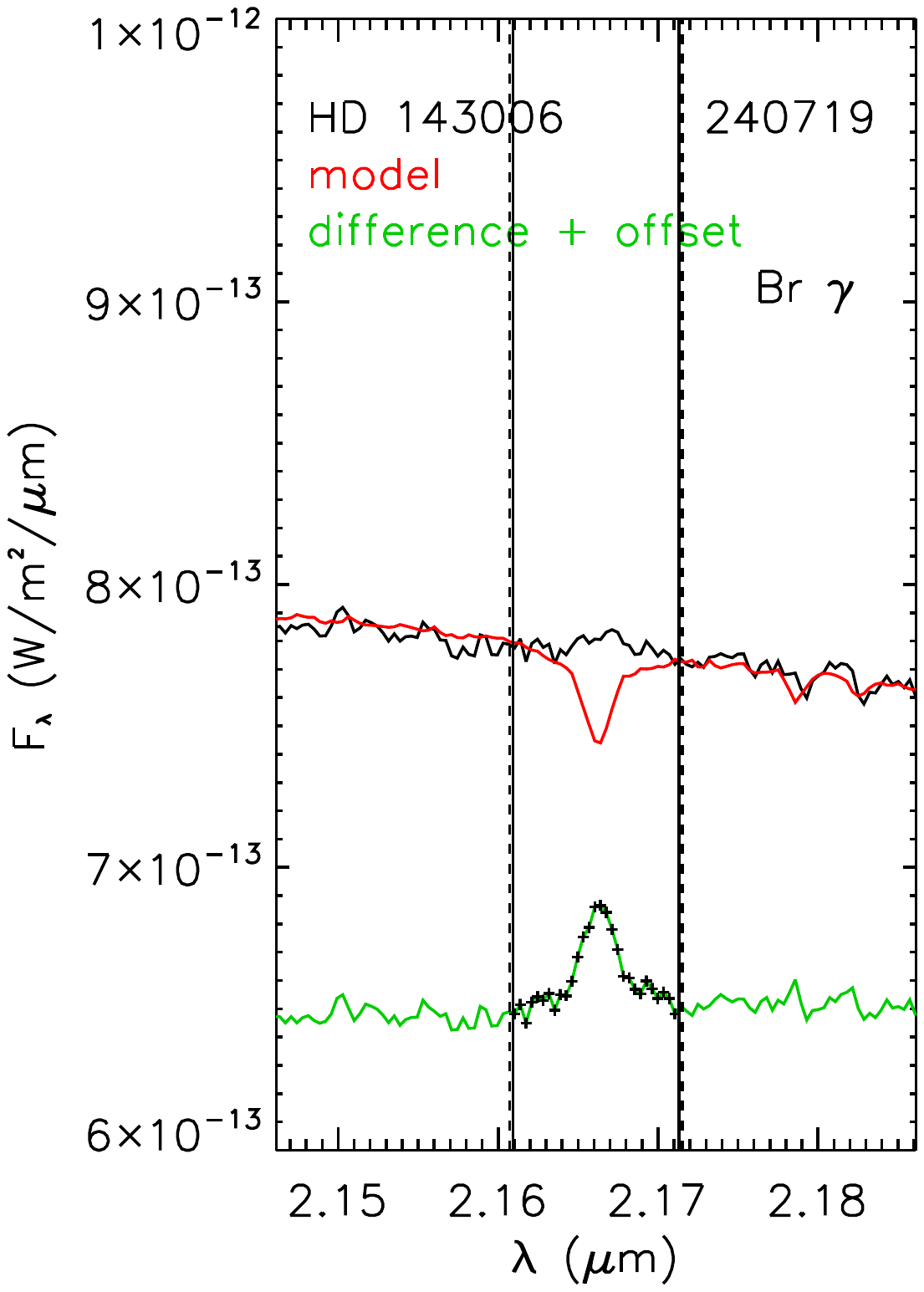}
\caption{The same as Figure A-2, except for HD 143006 on 240719. \label{fig:A-11}}
\end{figure}

\begin{figure}
\includegraphics[width=6.0cm, height=6.0cm]{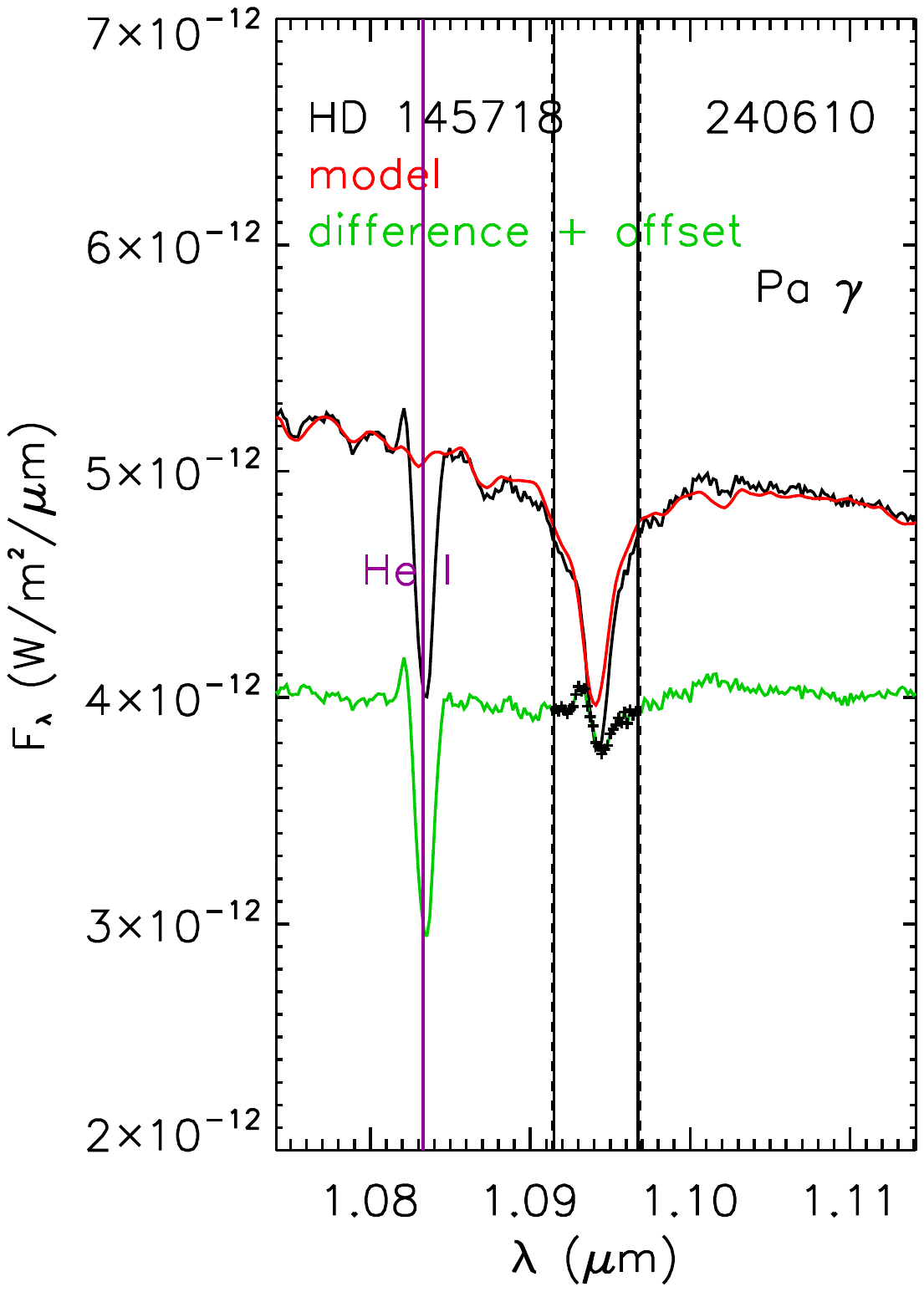}
\includegraphics[width=6.0cm, height=6.0cm]{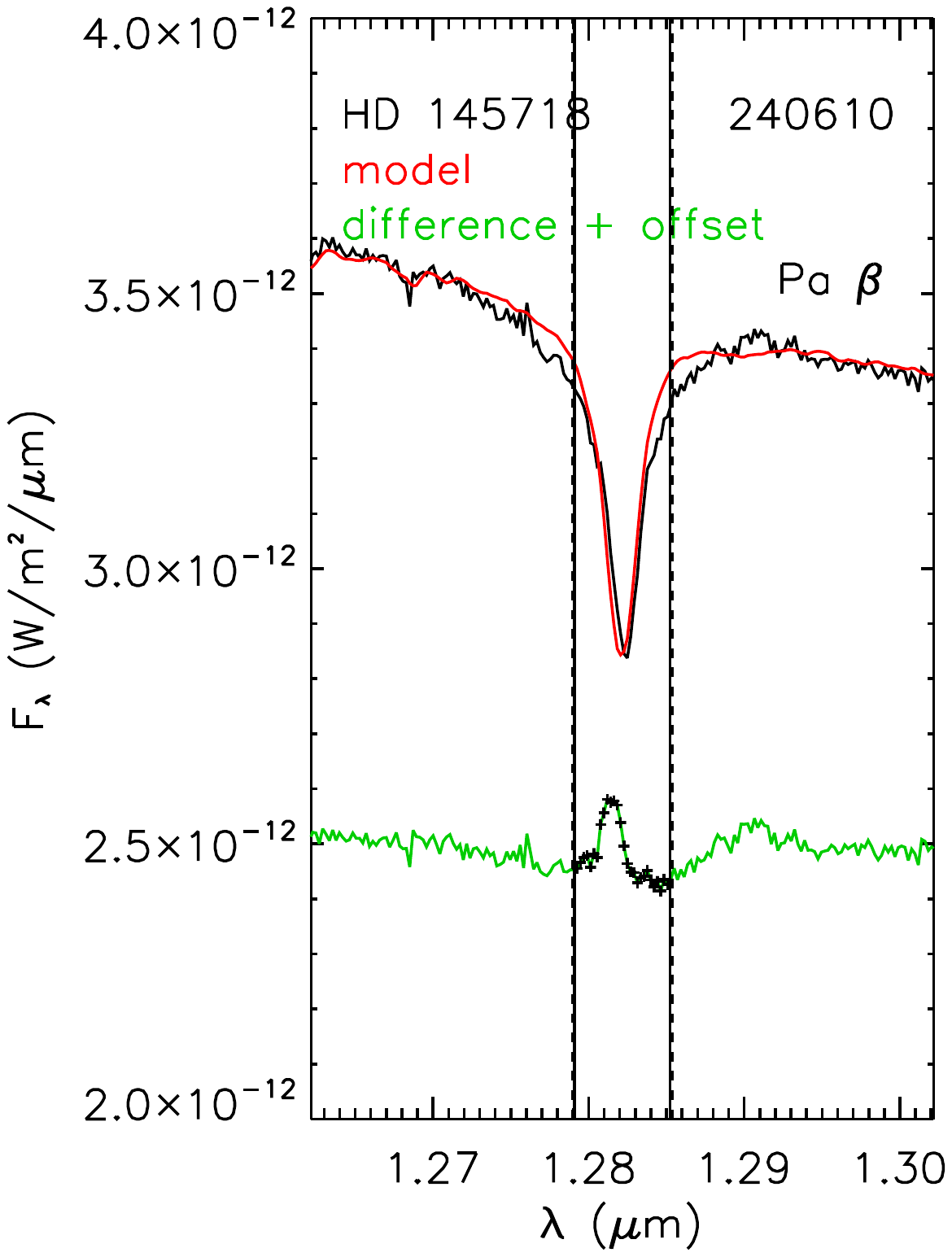}
\includegraphics[width=6.0cm, height=6.0cm]{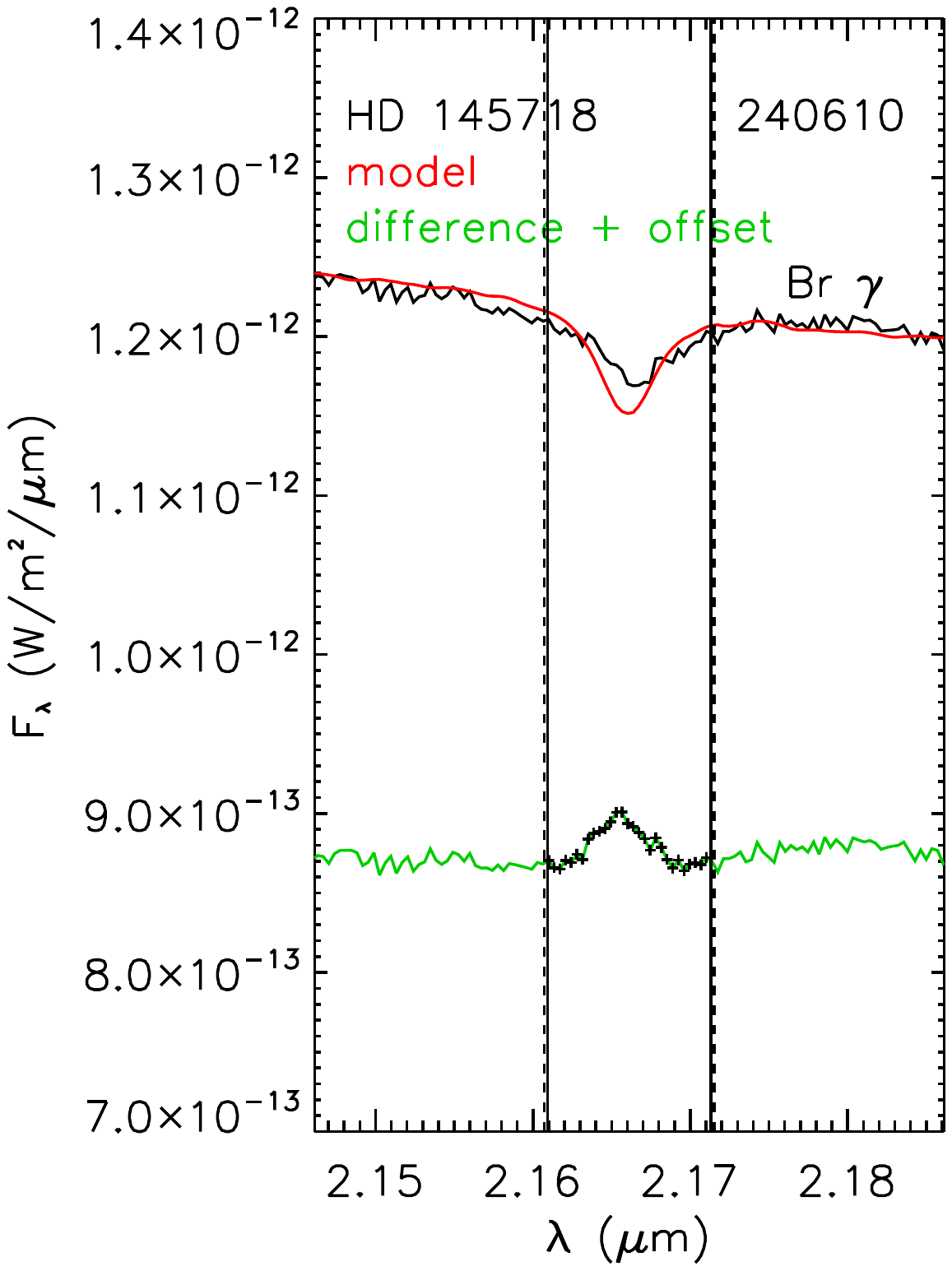}
\caption{The same as Figure A-2, except for HD 145718 on 240610. \label{fig:A-12}}
\end{figure}

\clearpage

\begin{figure}
\includegraphics[width=6.0cm, height=6.0cm]{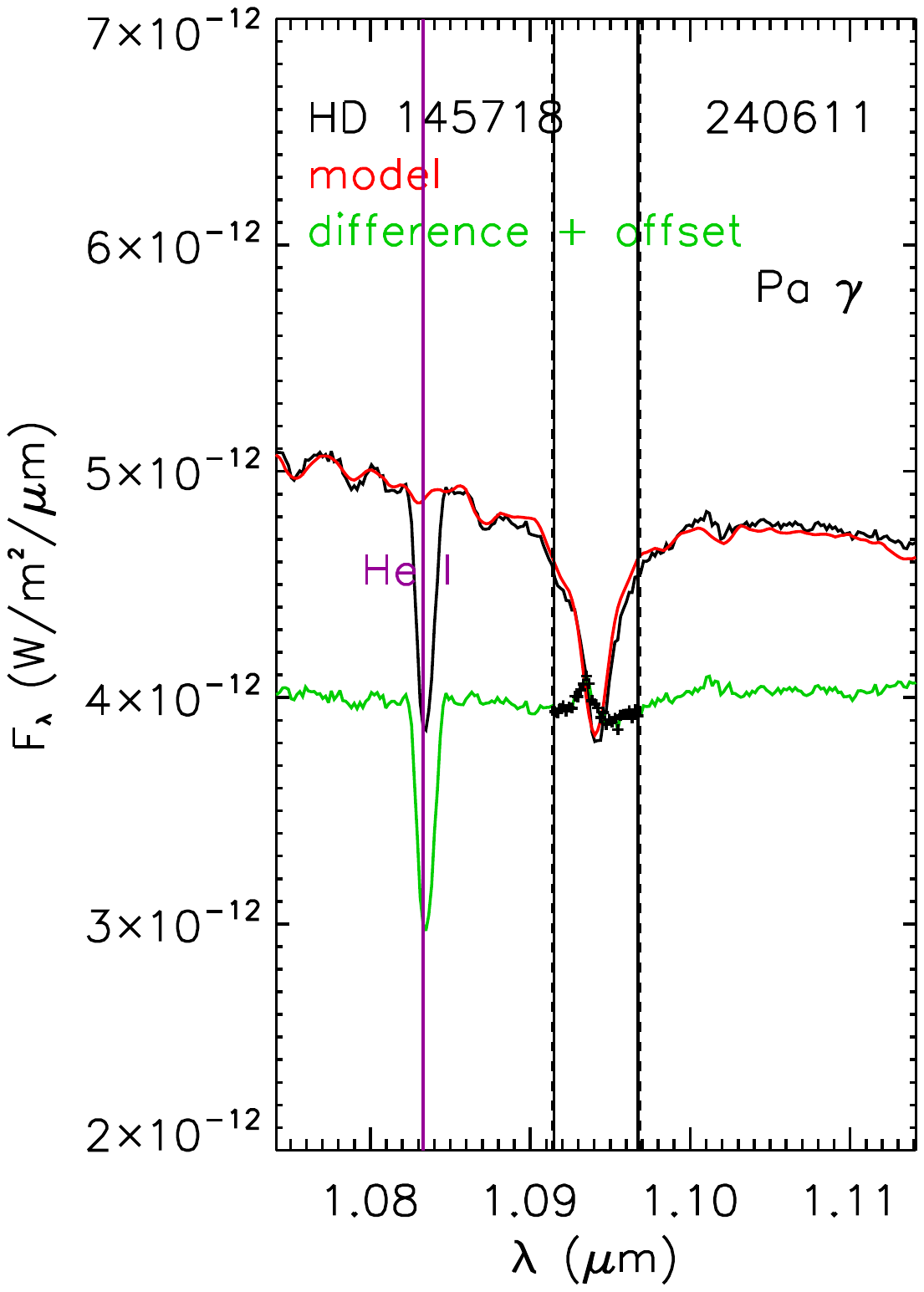}
\includegraphics[width=6.0cm, height=6.0cm]{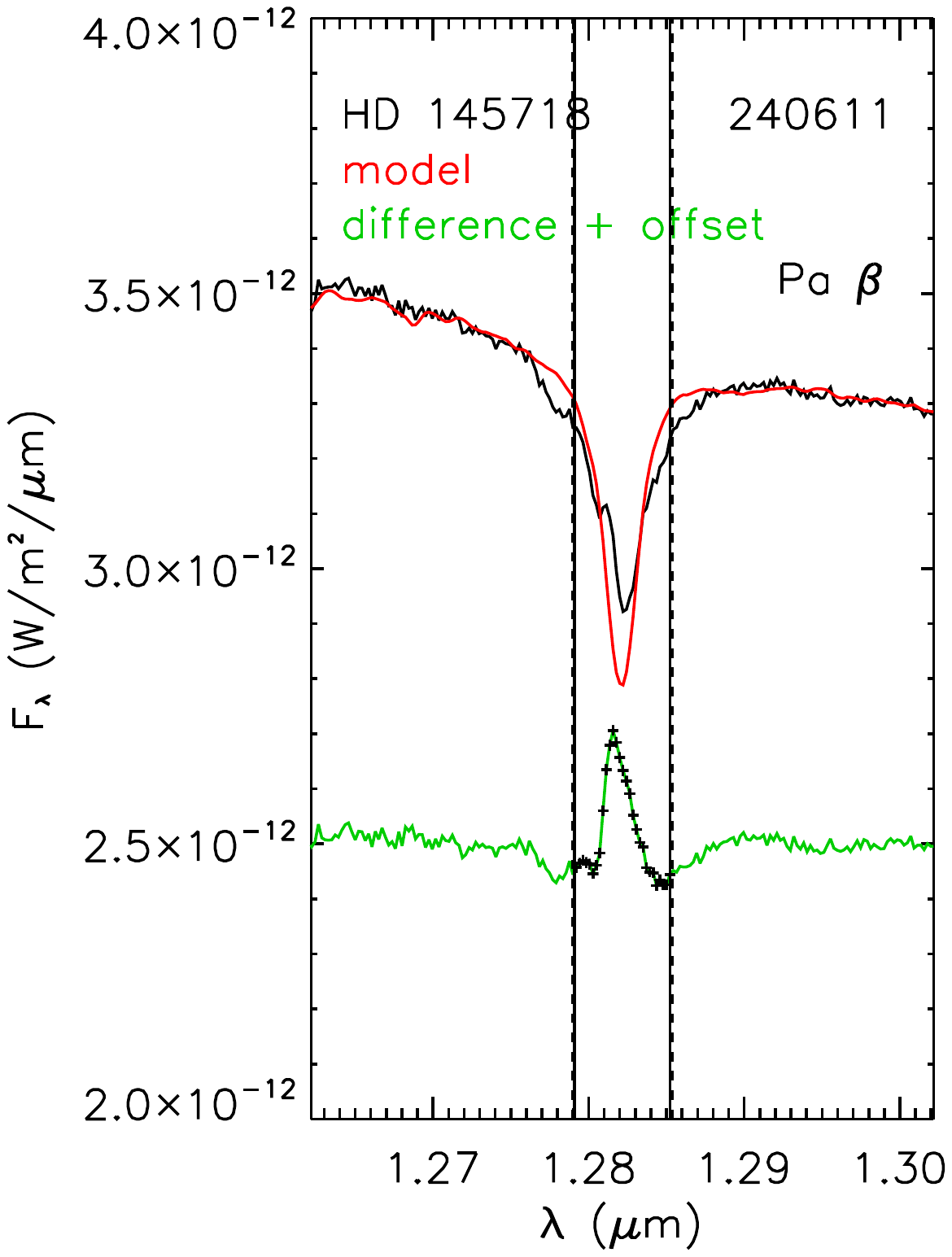}
\includegraphics[width=6.0cm, height=6.0cm]{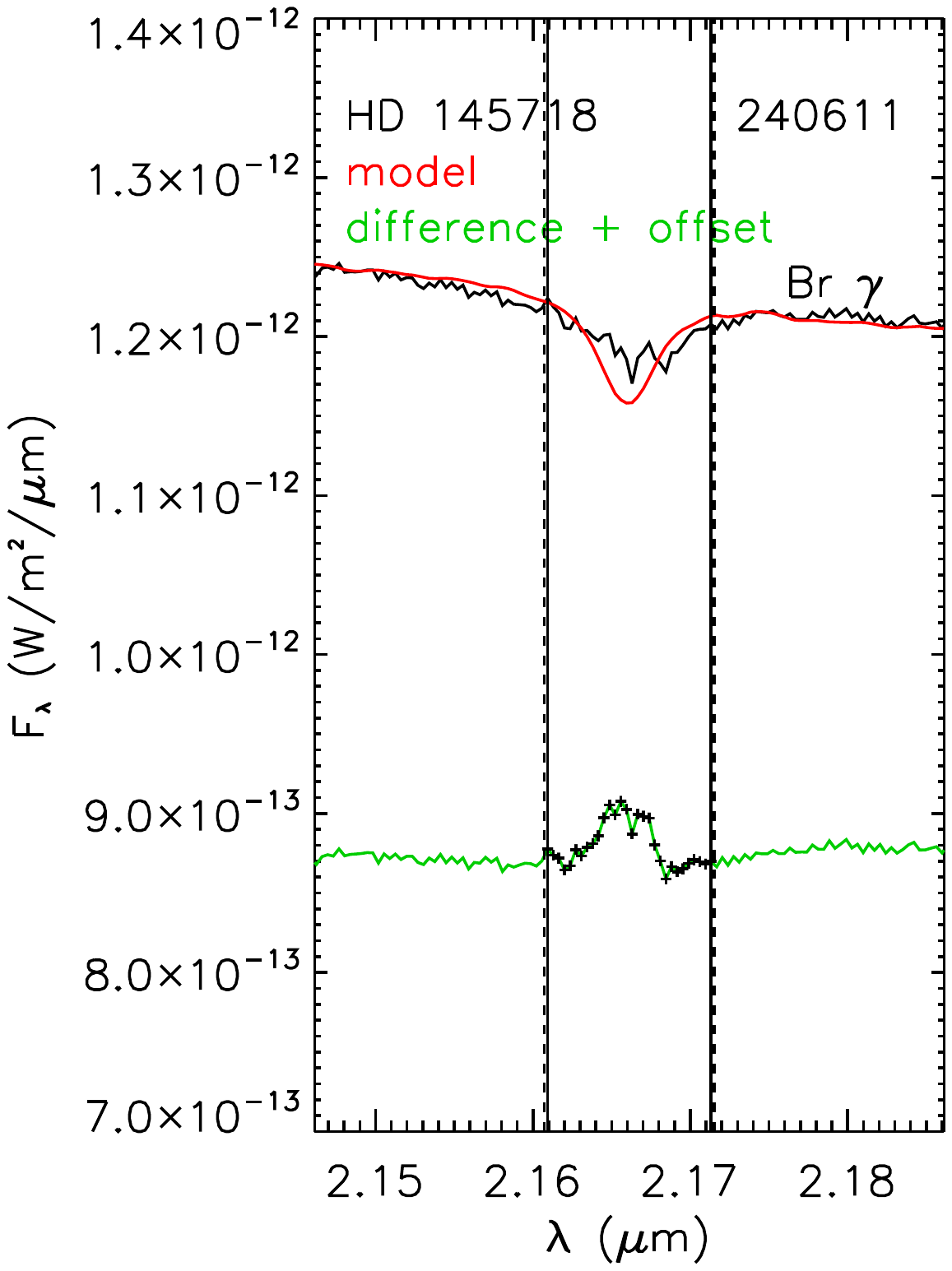}
\caption{The same as Figure A-2, except for HD 145718 on 240611. \label{fig:A-13}}
\end{figure}

\begin{figure}
\includegraphics[width=6.0cm, height=6.0cm]{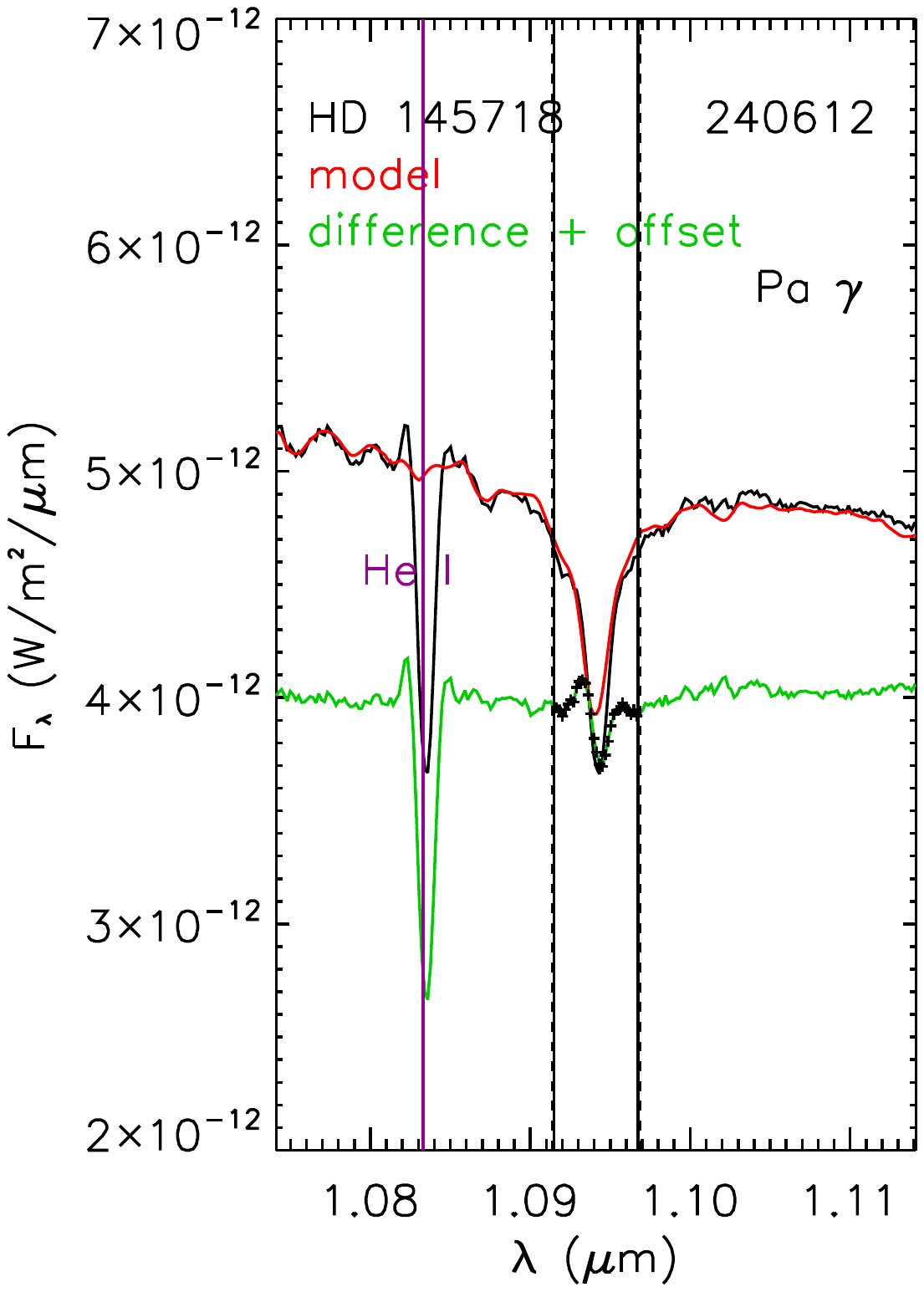}
\includegraphics[width=6.0cm, height=6.0cm]{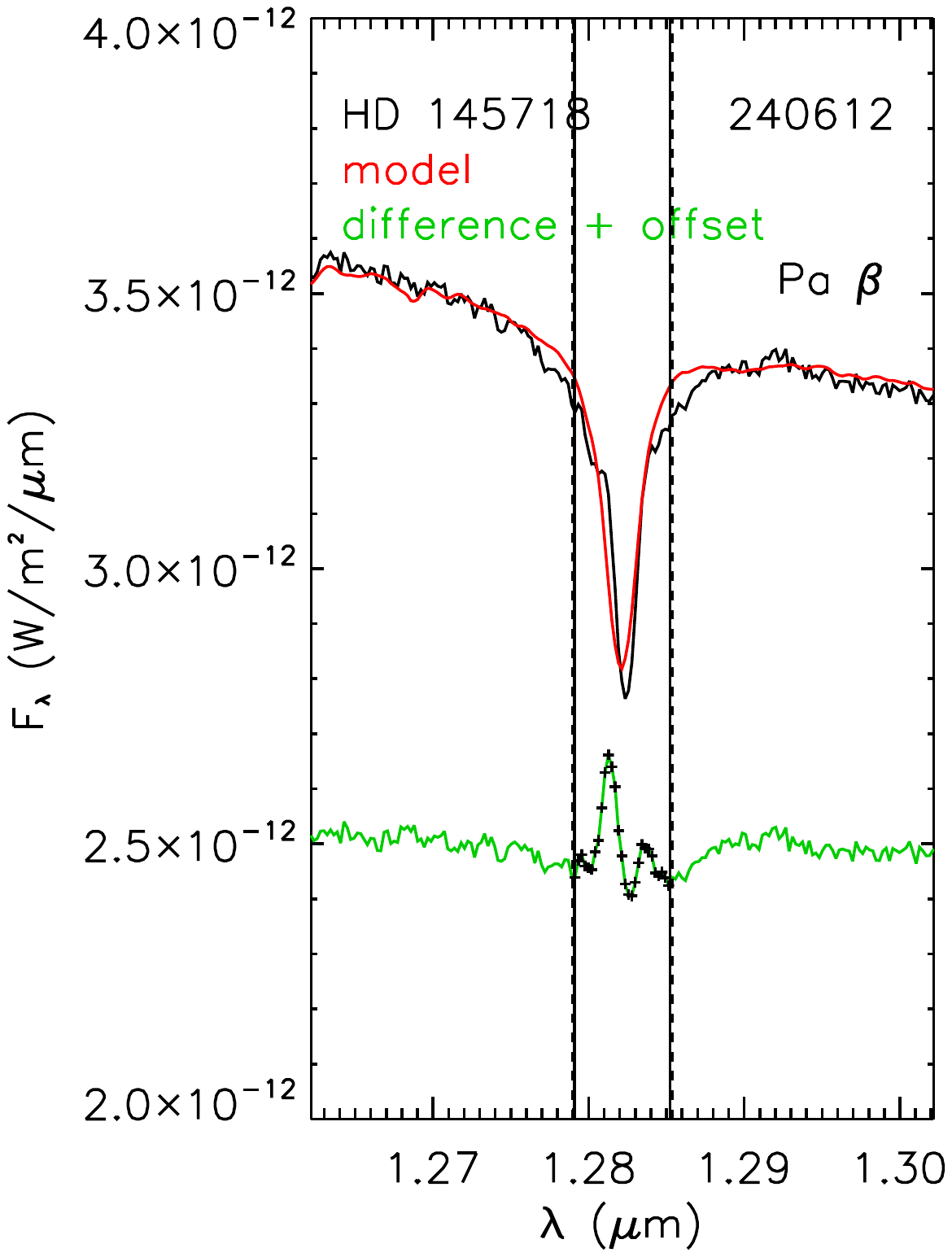}
\includegraphics[width=6.0cm, height=6.0cm]{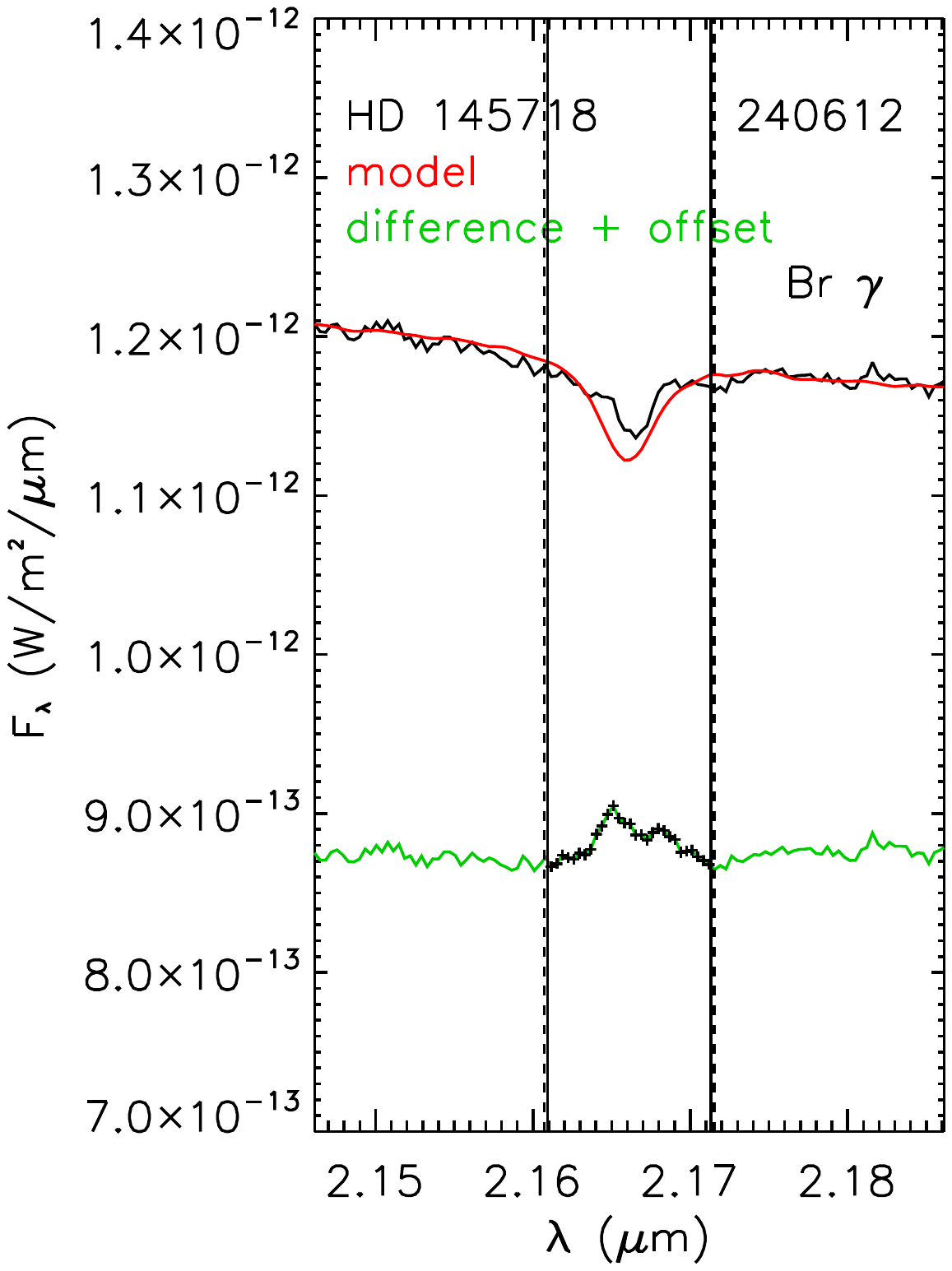}
\caption{The same as Figure A-2, except for HD 145718 on 240612. \label{fig:A-14}}
\end{figure}

%\clearpage

\begin{figure}
\includegraphics[width=6.0cm, height=6.0cm]{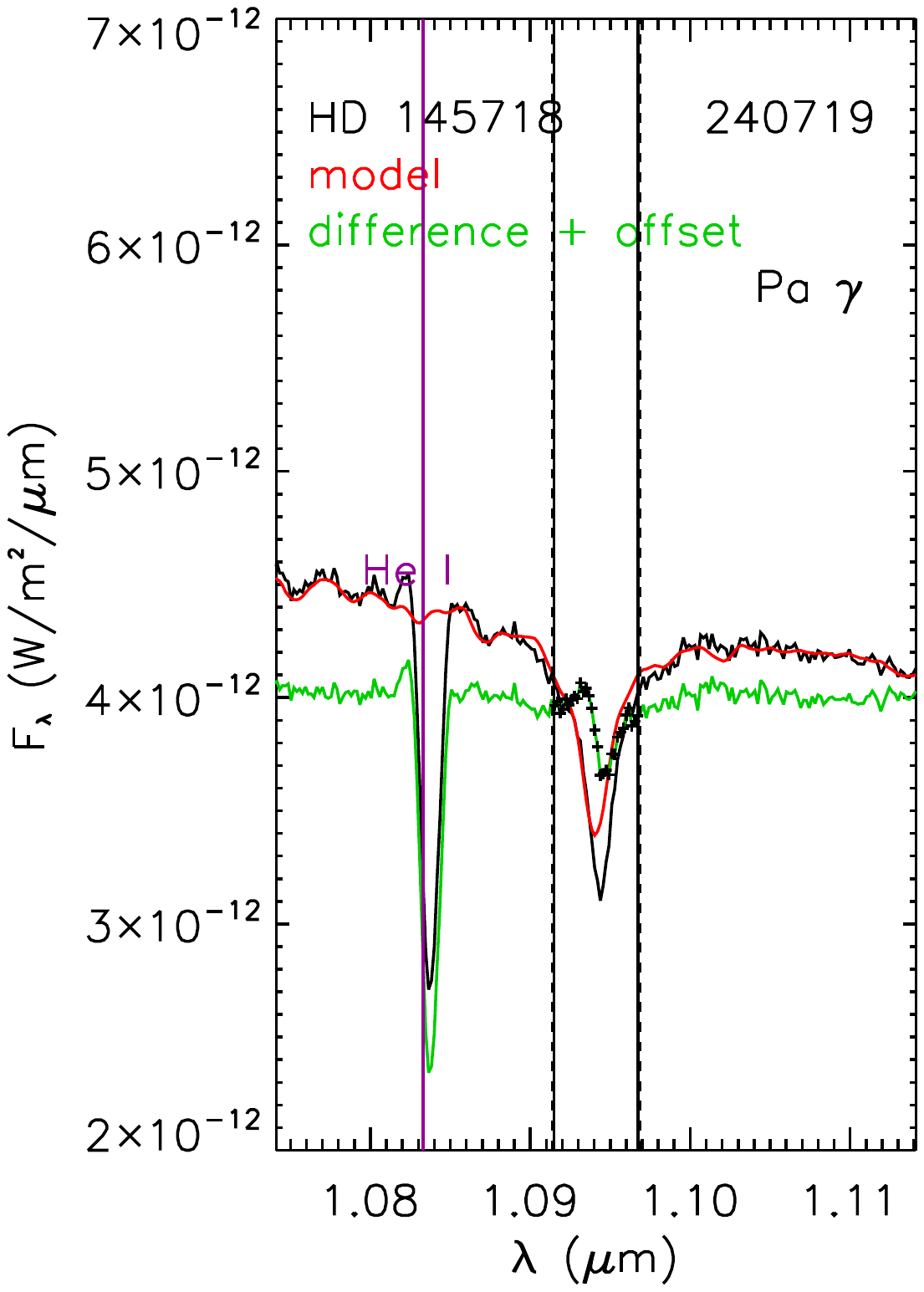}
\includegraphics[width=6.0cm, height=6.0cm]{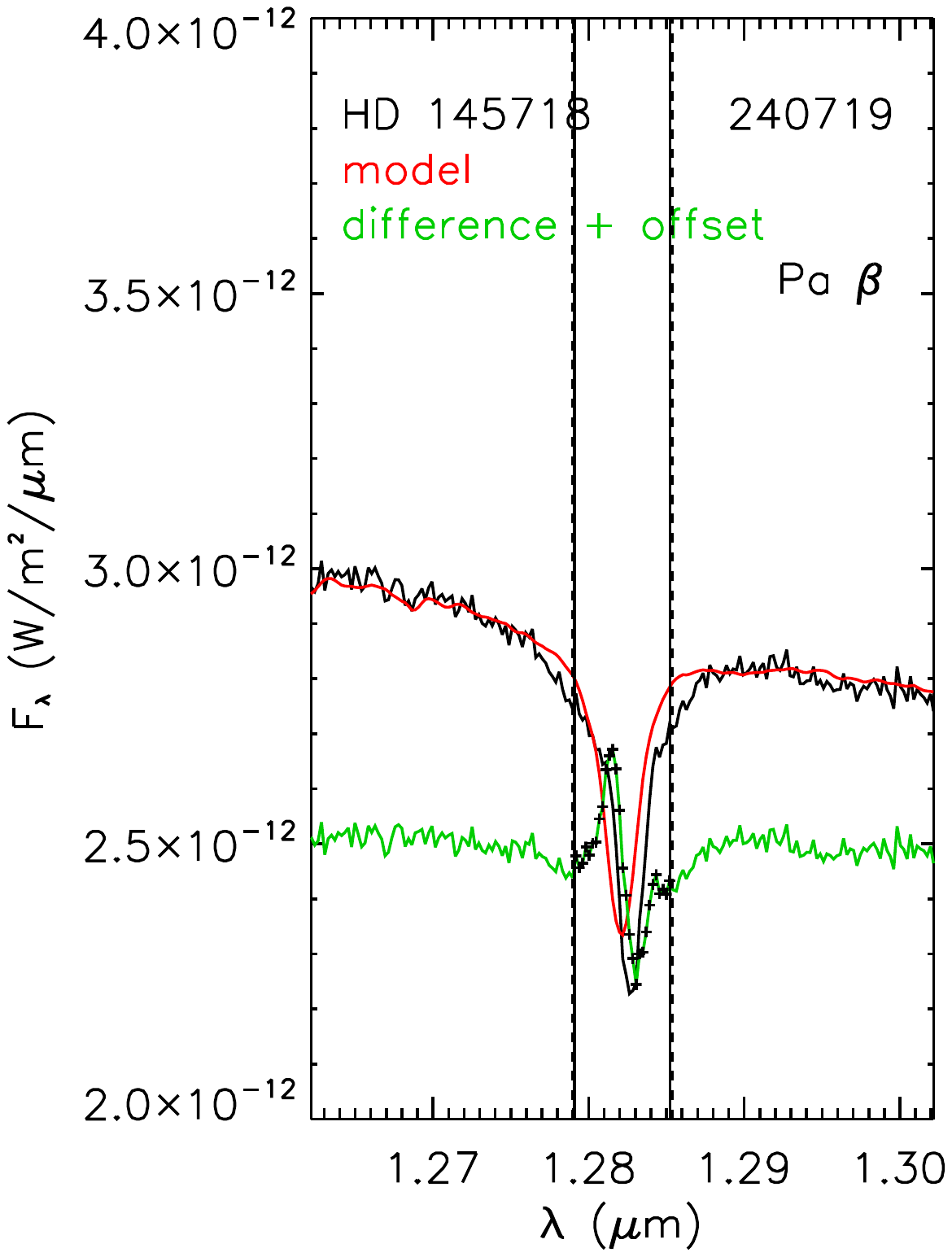}
\includegraphics[width=6.0cm, height=6.0cm]{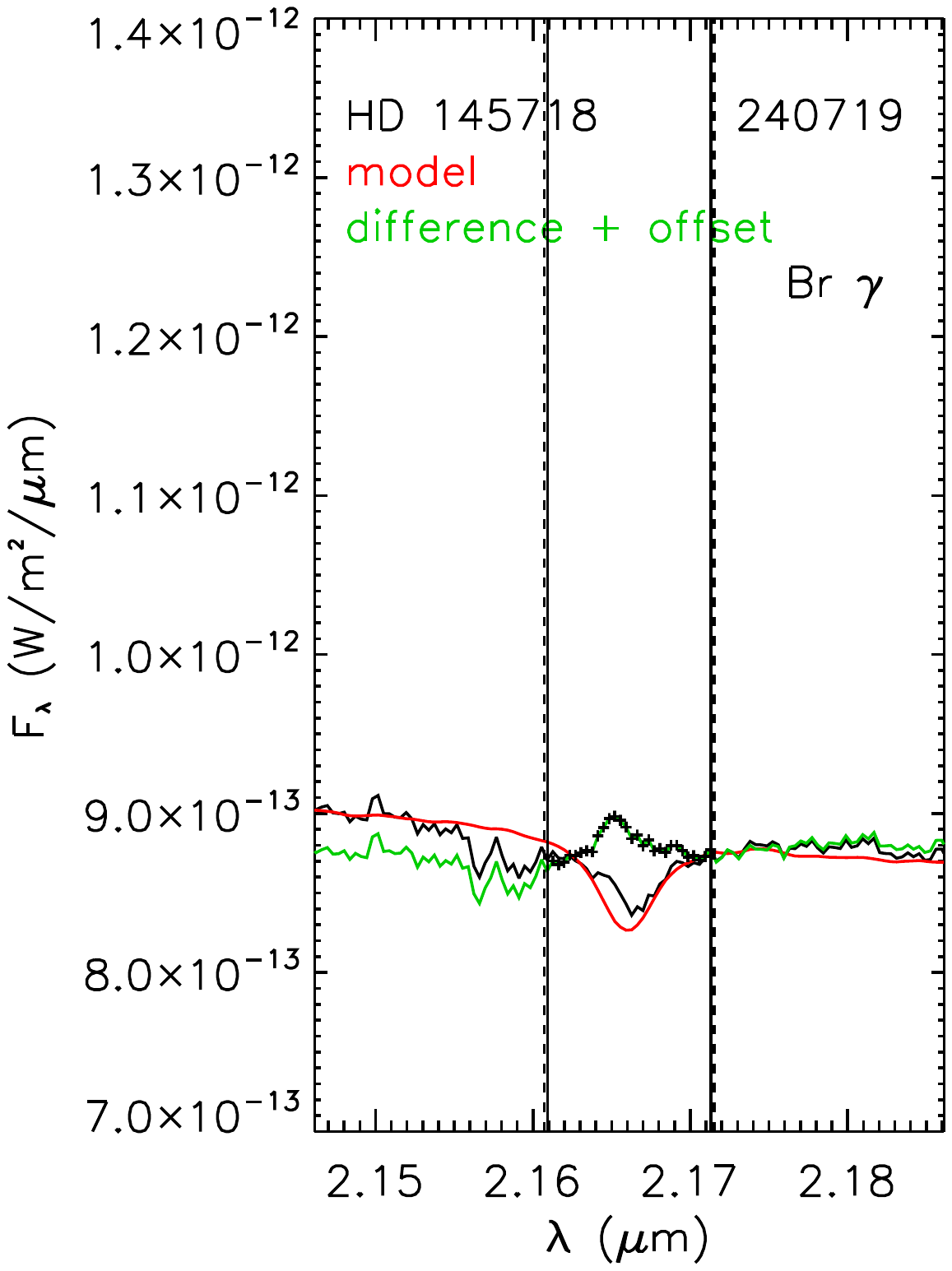}
\caption{The same as Figure A-2, except for HD 145718 on 240719. \label{fig:A-15}}
\end{figure}

\begin{figure}
\includegraphics[width=6.0cm, height=6.0cm]{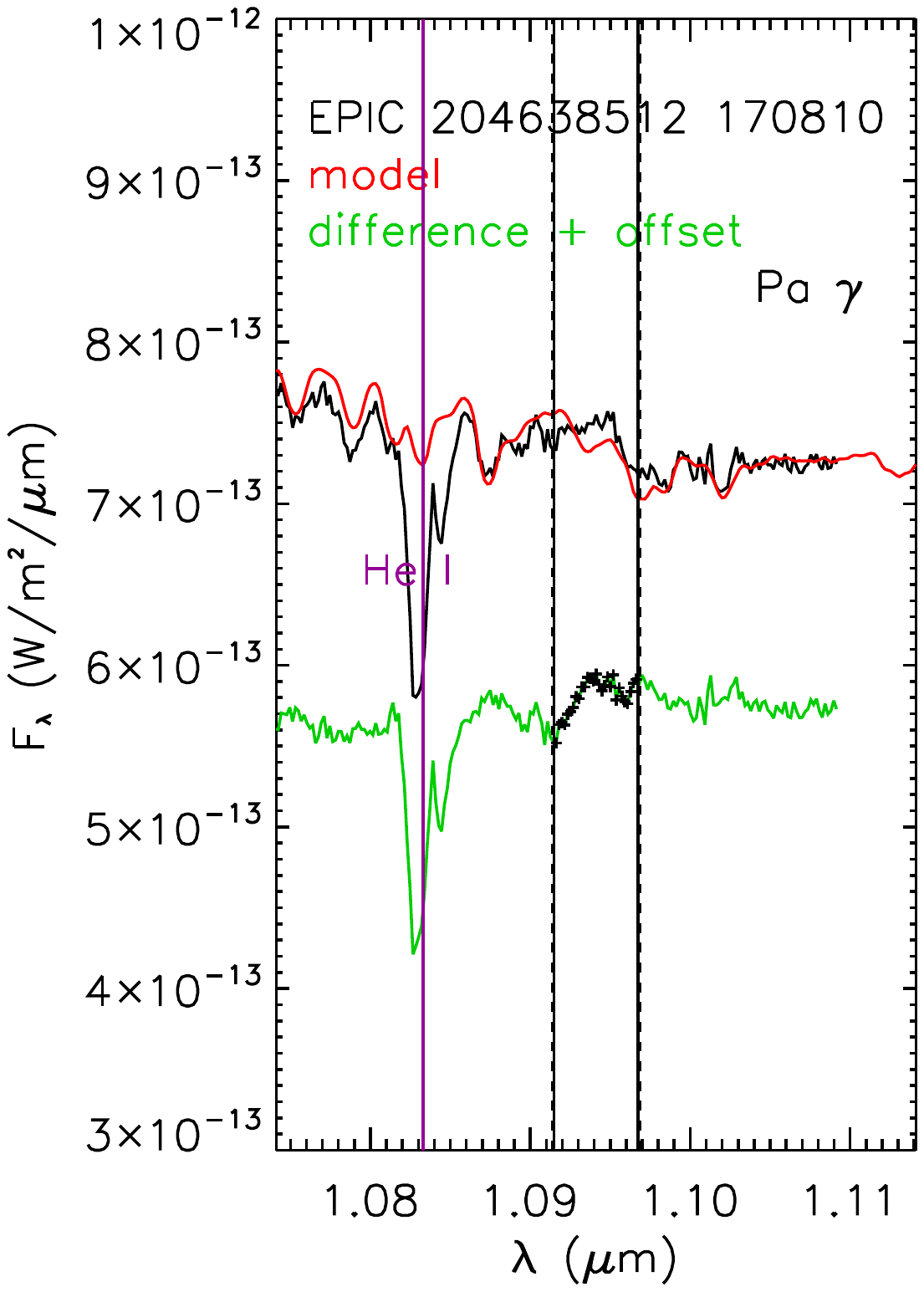}
\includegraphics[width=6.0cm, height=6.0cm]{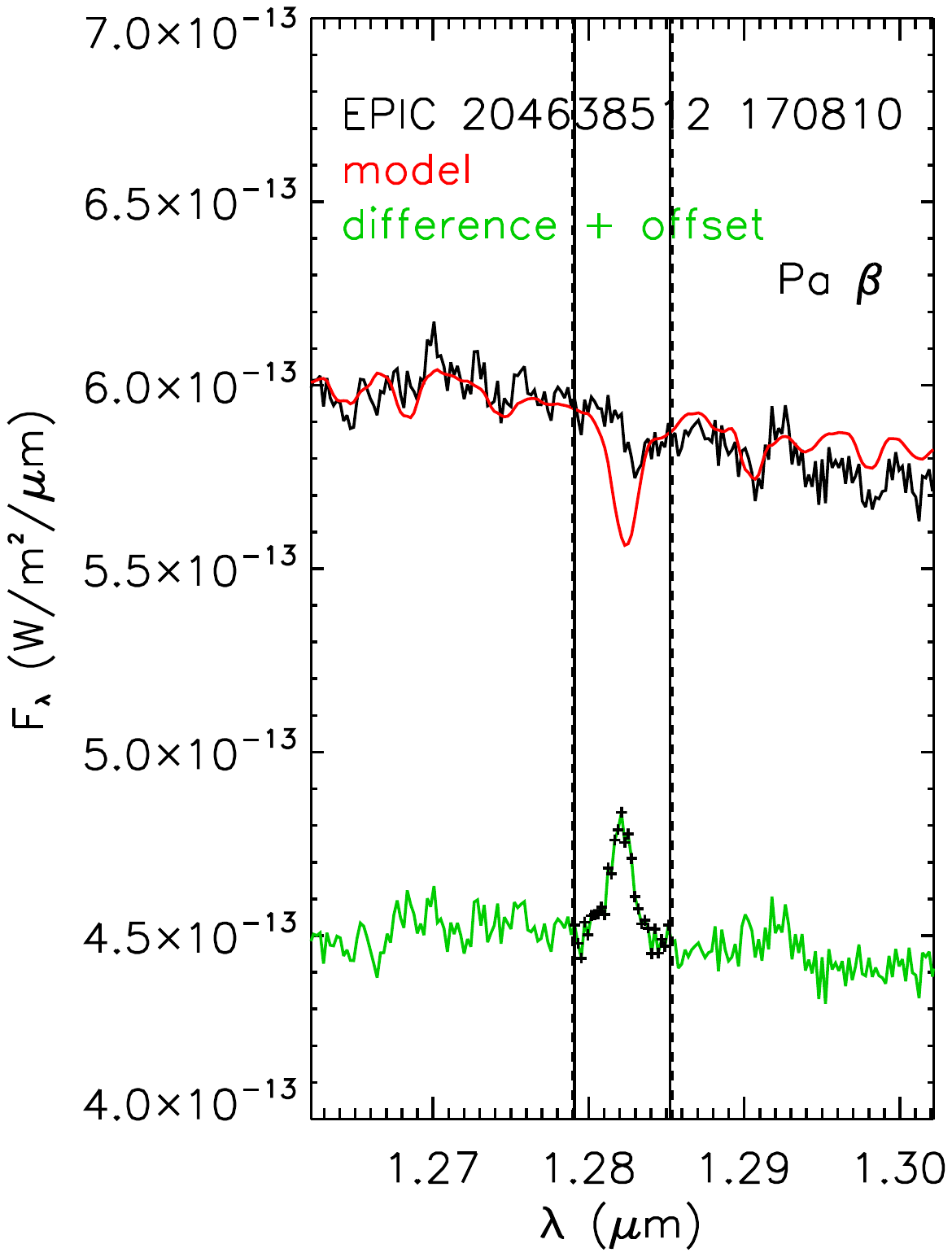}
\includegraphics[width=6.0cm, height=6.0cm]{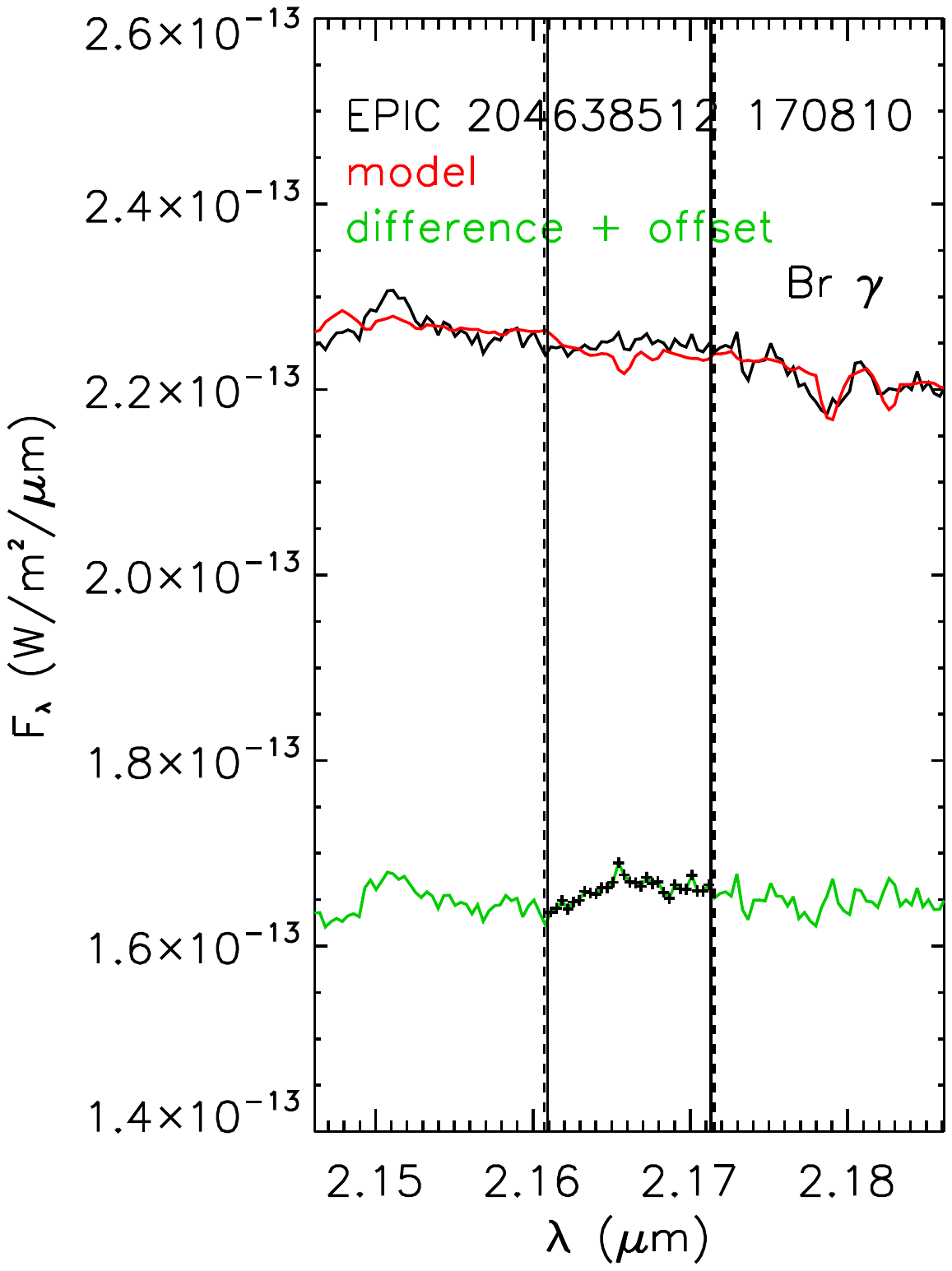}
\caption{The same as Figure A-2, except  for EPIC 204638512 on 170810. \label{fig:A-16 }}
\end{figure}

%\clearpage

\begin{figure}
\includegraphics[width=6.0cm, height=6.0cm]{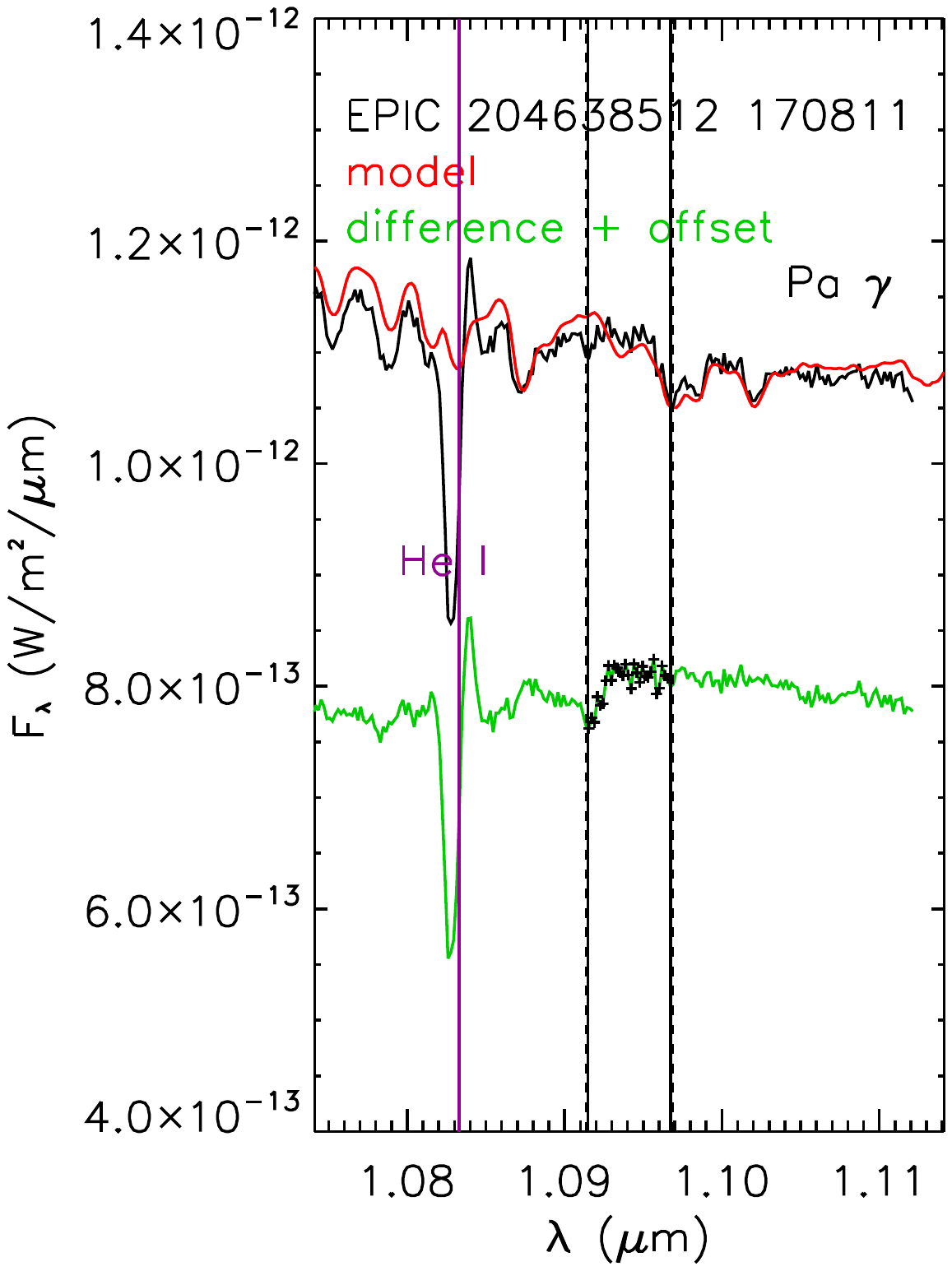}
\includegraphics[width=6.0cm, height=6.0cm]{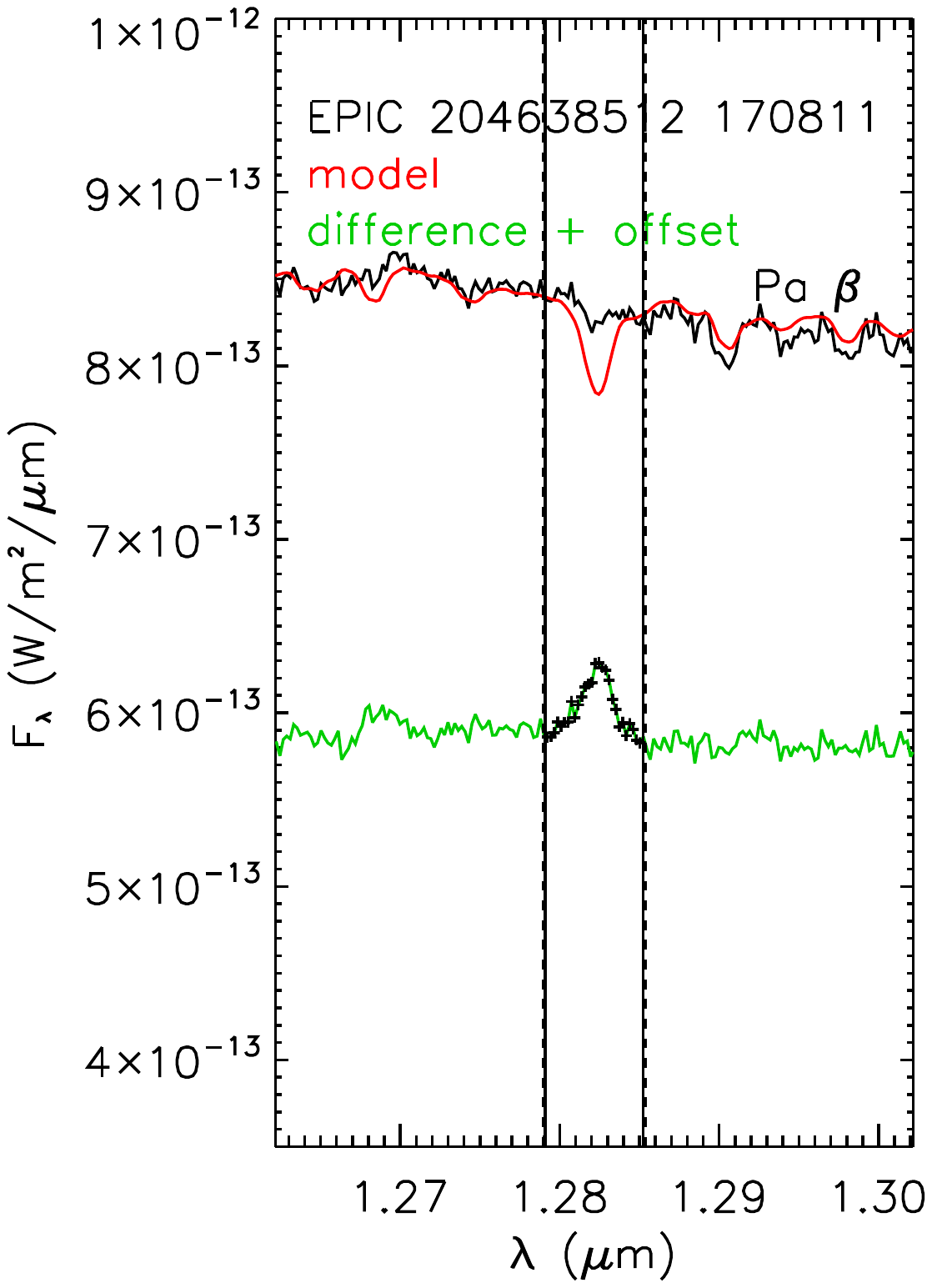}
\includegraphics[width=6.0cm, height=6.0cm]{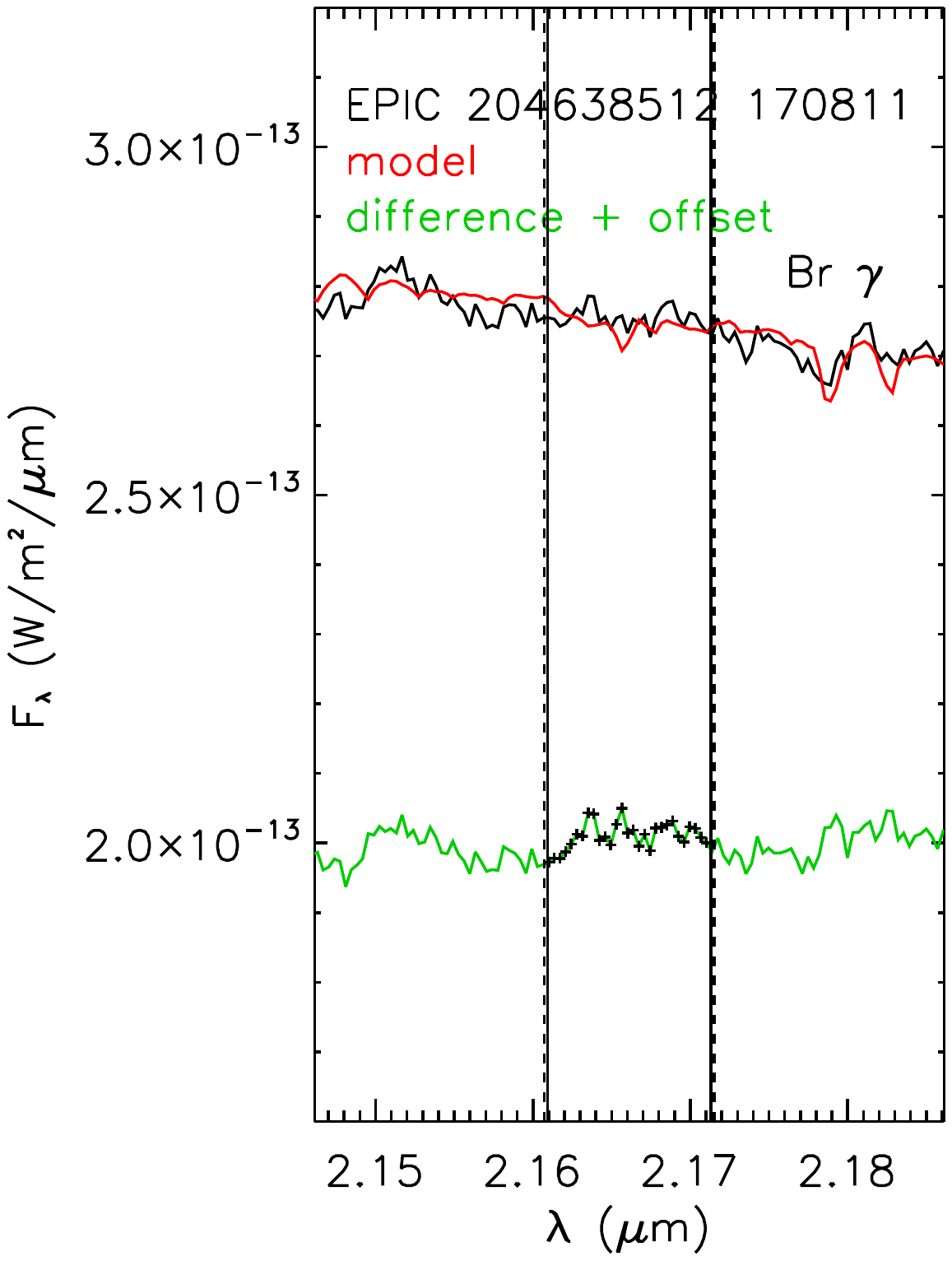}
\caption{The same as Figure A-2, except  for EPIC 204638512 on 170811. \label{fig:A-17}}
\end{figure}

\begin{figure}
\includegraphics[width=6.0cm, height=6.0cm]{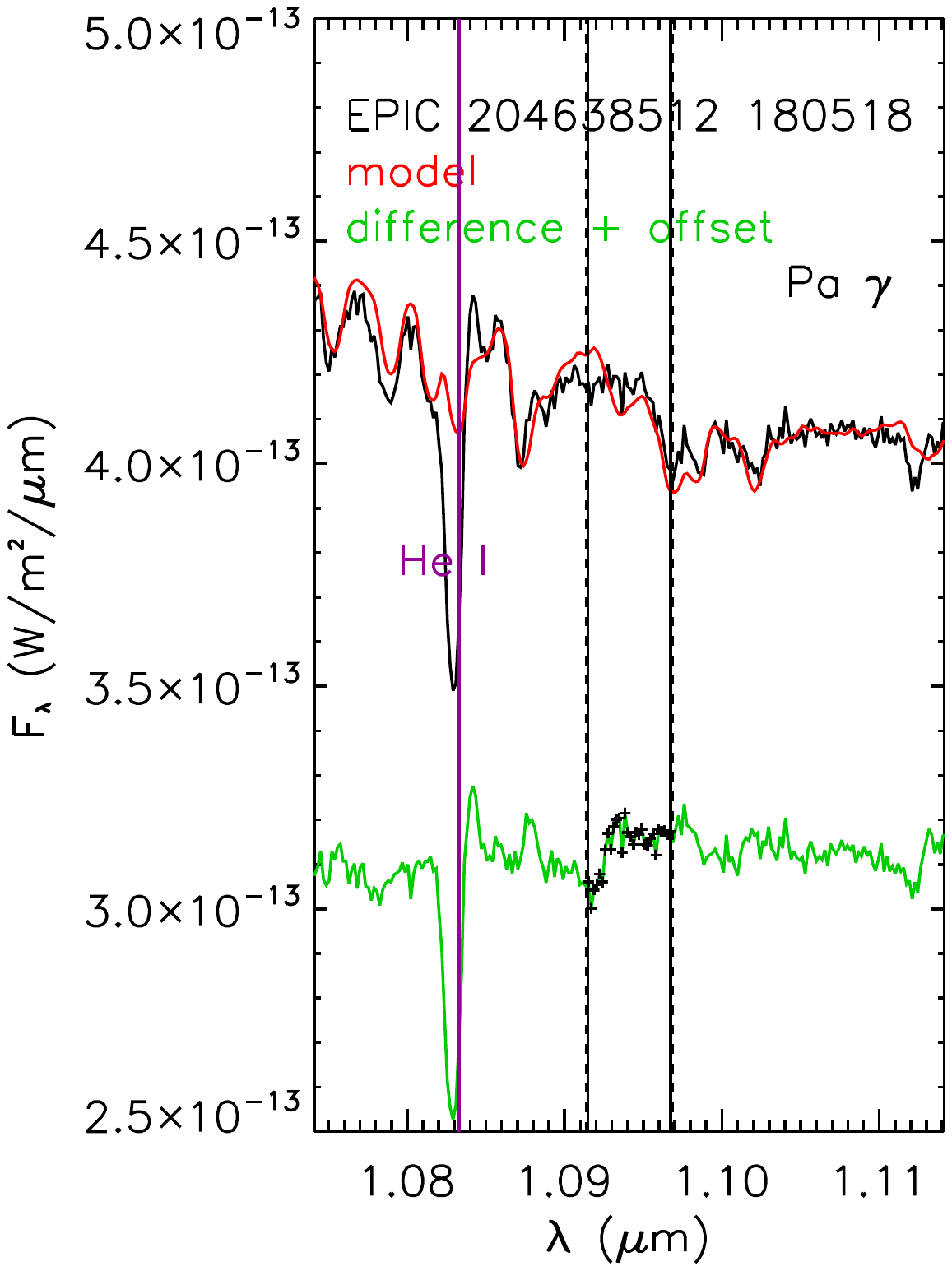}
\includegraphics[width=6.0cm, height=6.0cm]{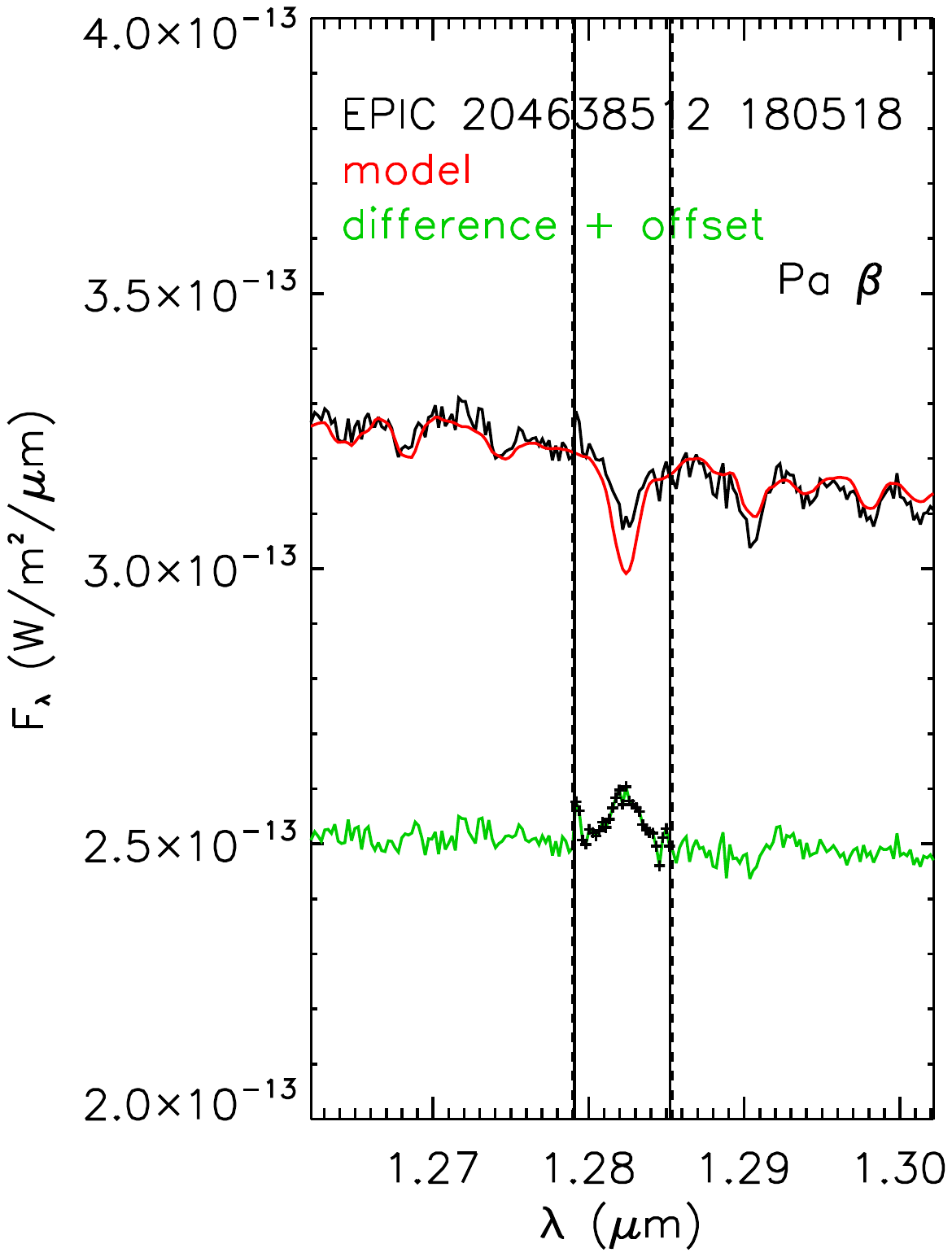}
\includegraphics[width=6.0cm, height=6.0cm]{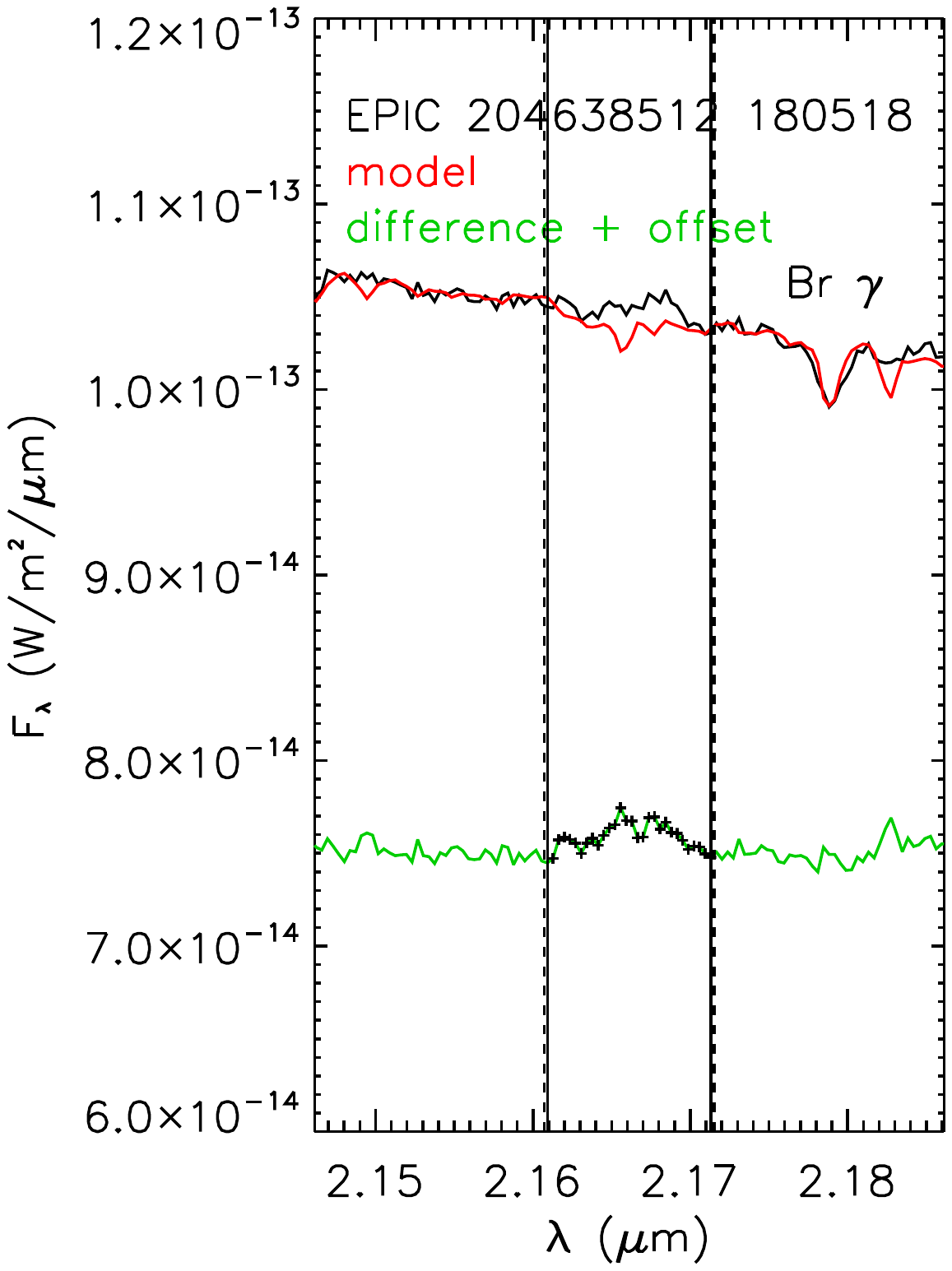}
\caption{The same as Figure A-2, except for EPIC 204638512 on 180518. \label{fig:18}}
\end{figure}

%\clearpage

\begin{figure}
\includegraphics[width=6.0cm, height=6.0cm]{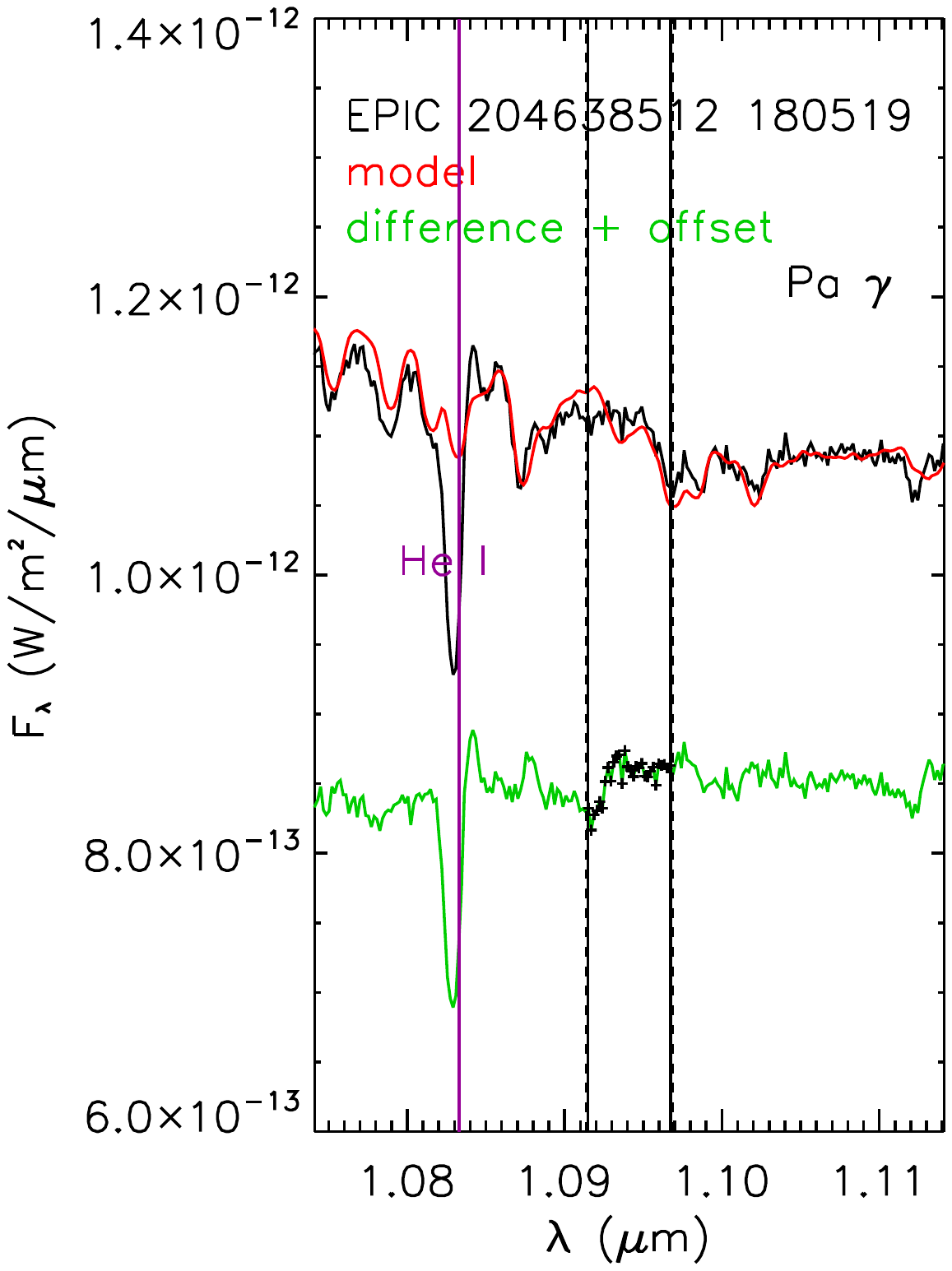}
\includegraphics[width=6.0cm, height=6.0cm]{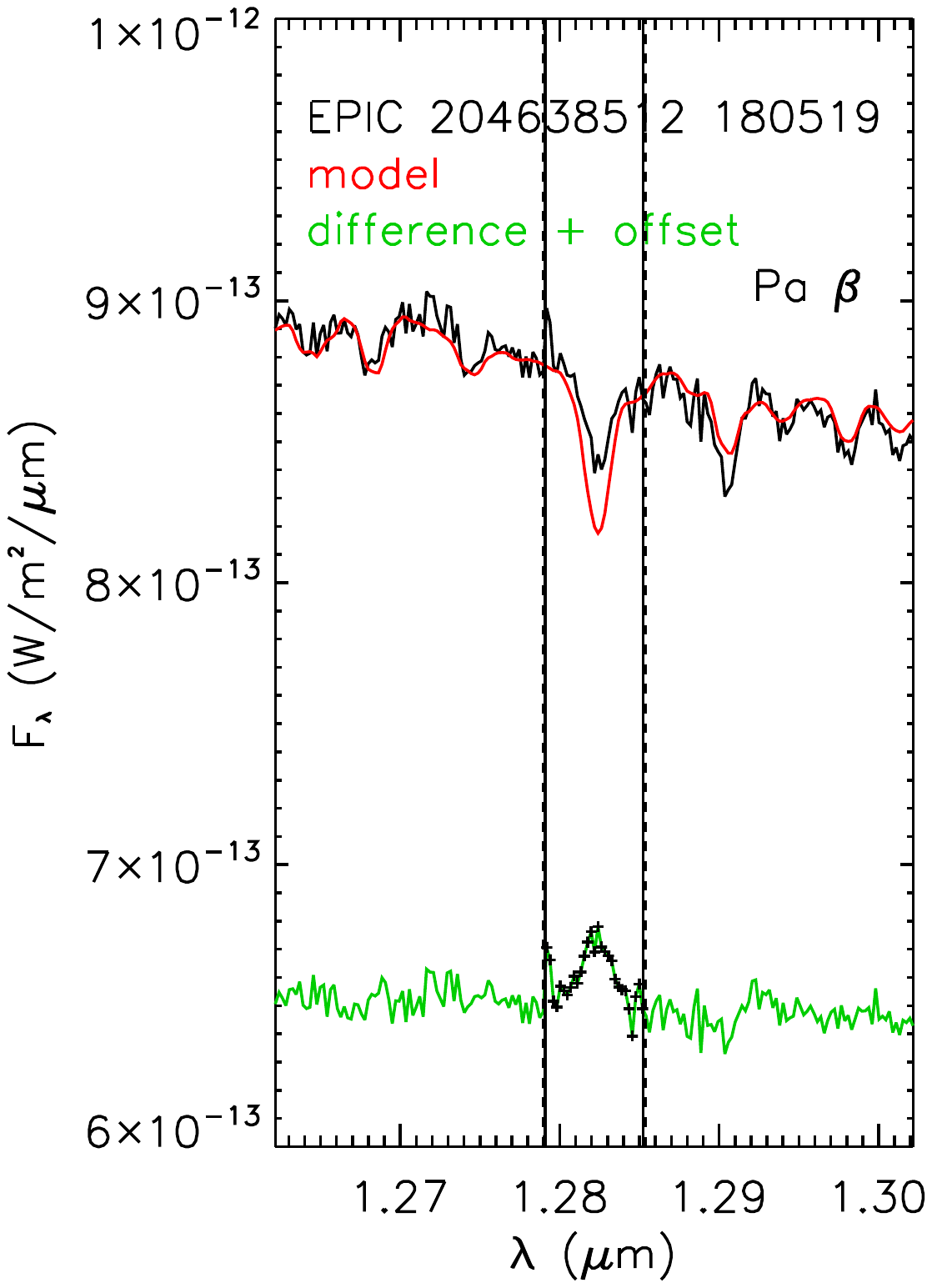}
\includegraphics[width=6.0cm, height=6.0cm]{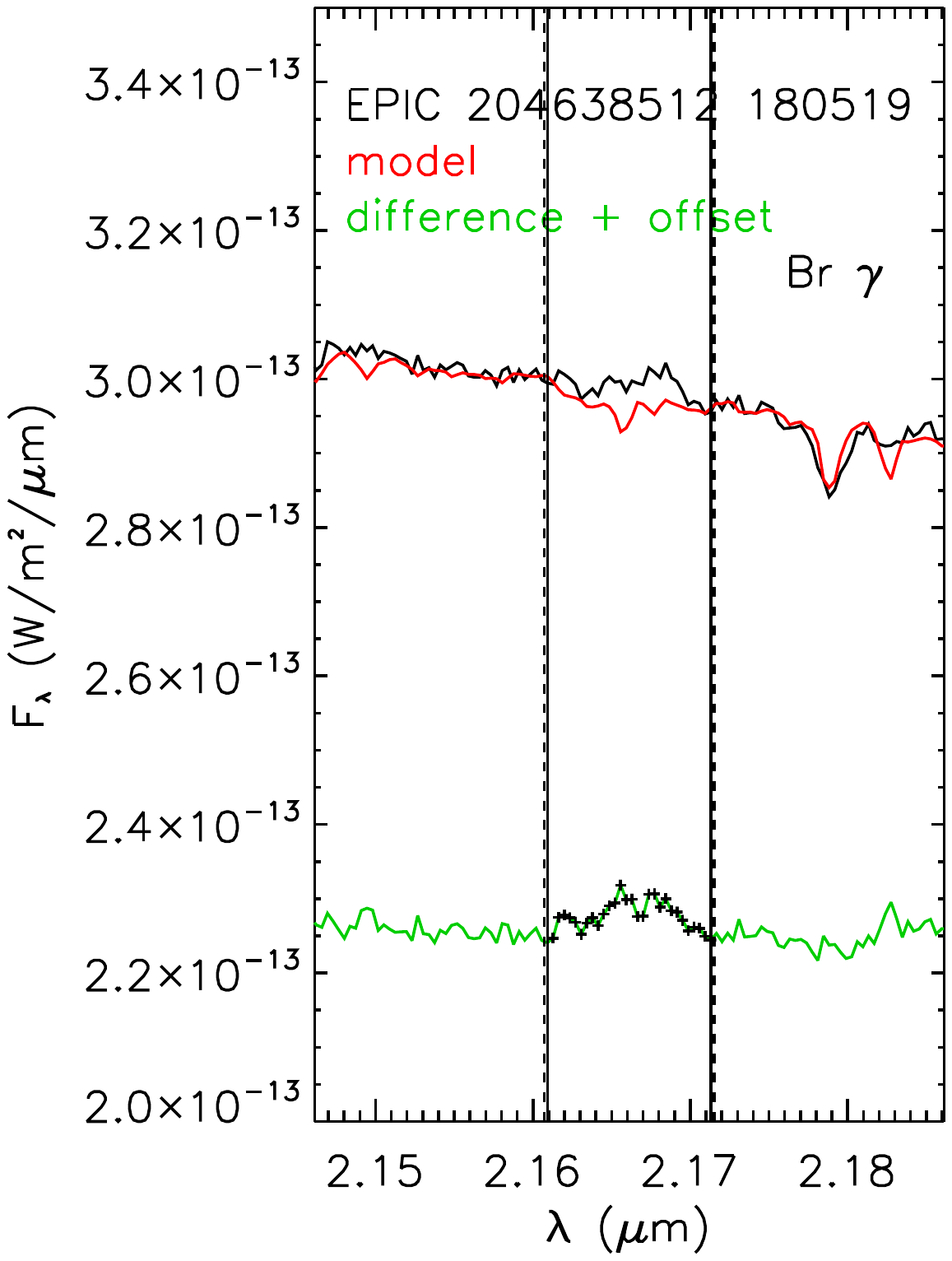}
\caption{The same as Figure A-2, except for EPIC 204638512 on 180519. \label{fig:A-19}}
\end{figure}

\begin{figure}
\includegraphics[width=6.0cm, height=6.0cm]{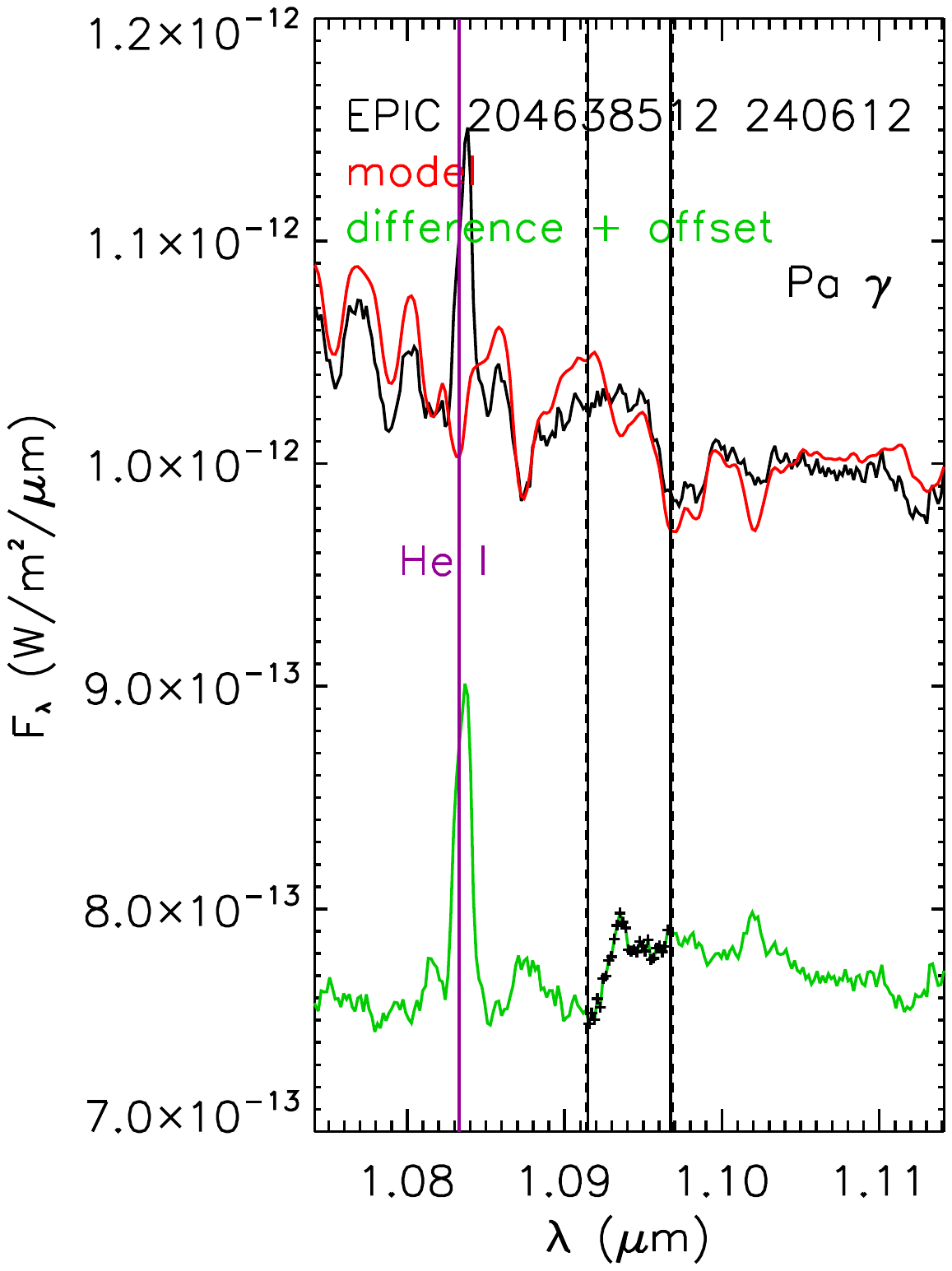}
\includegraphics[width=6.0cm, height=6.0cm]{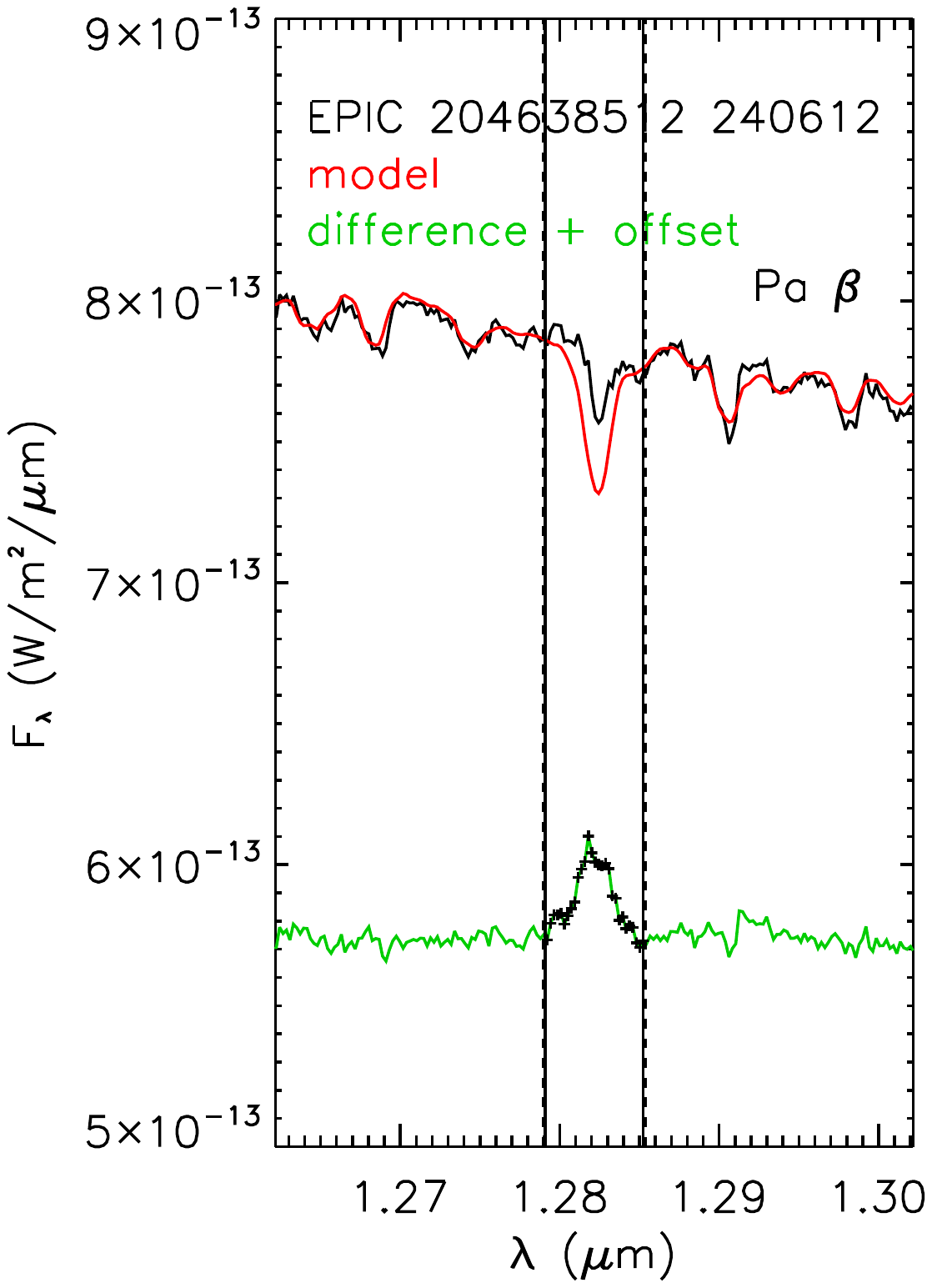}
\includegraphics[width=6.0cm, height=6.0cm]{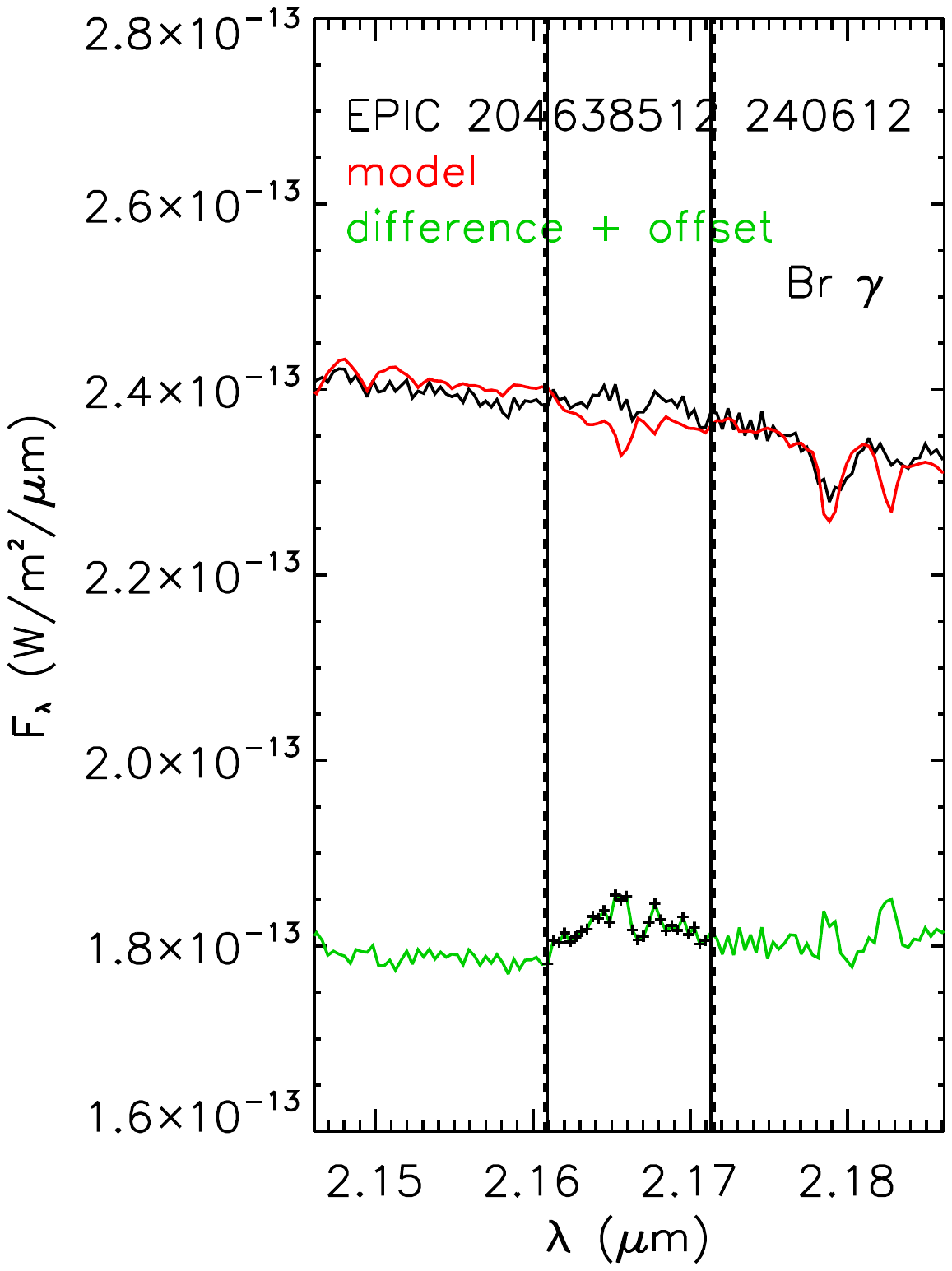}
\caption{The same as Figure A-2, except for EPIC 204638512 on 240612. \label{fig:A-20}}
\end{figure}

%\clearpage

\begin{figure}
\includegraphics[width=6.0cm, height=6.0cm]{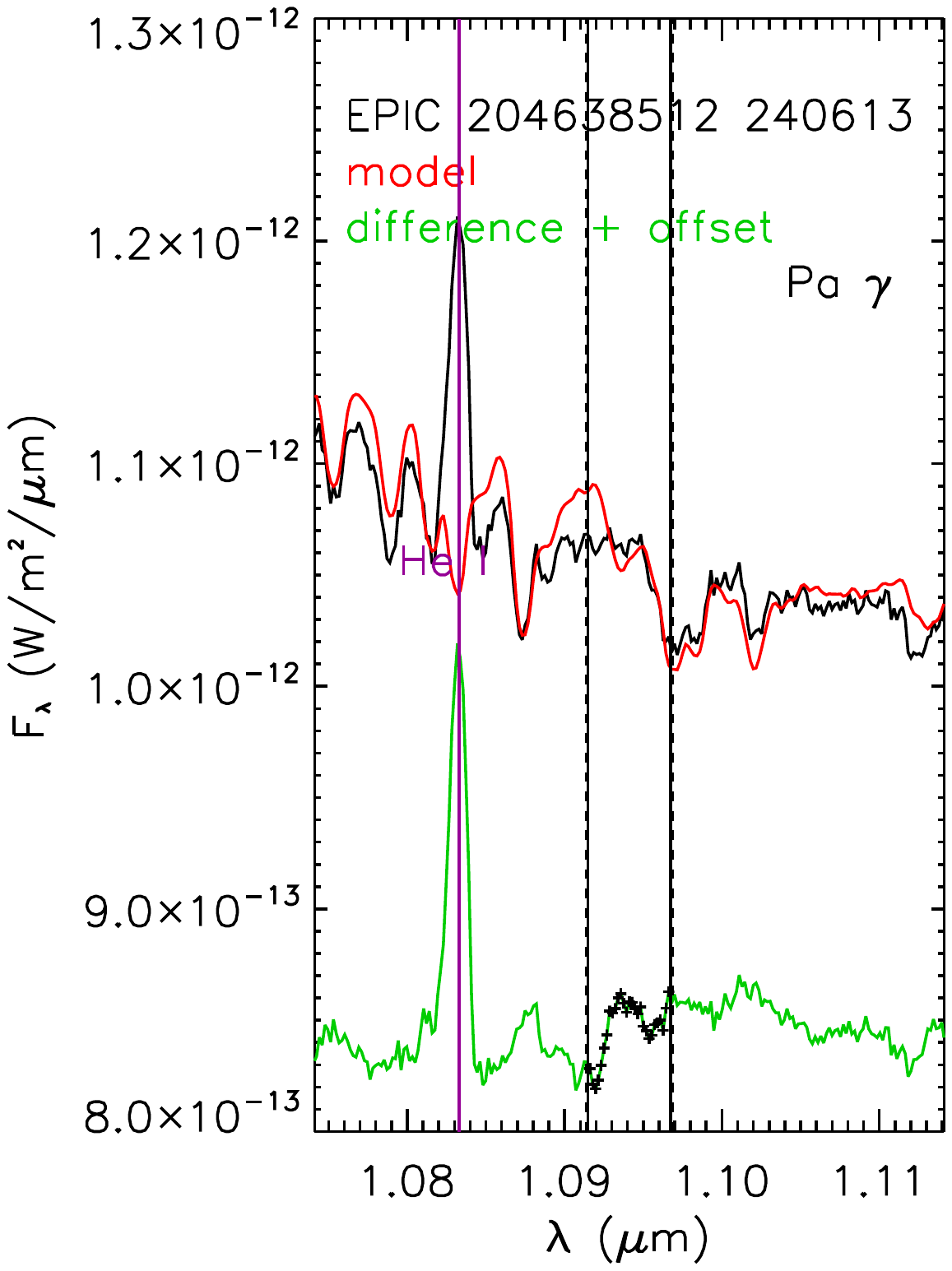}
\includegraphics[width=6.0cm, height=6.0cm]{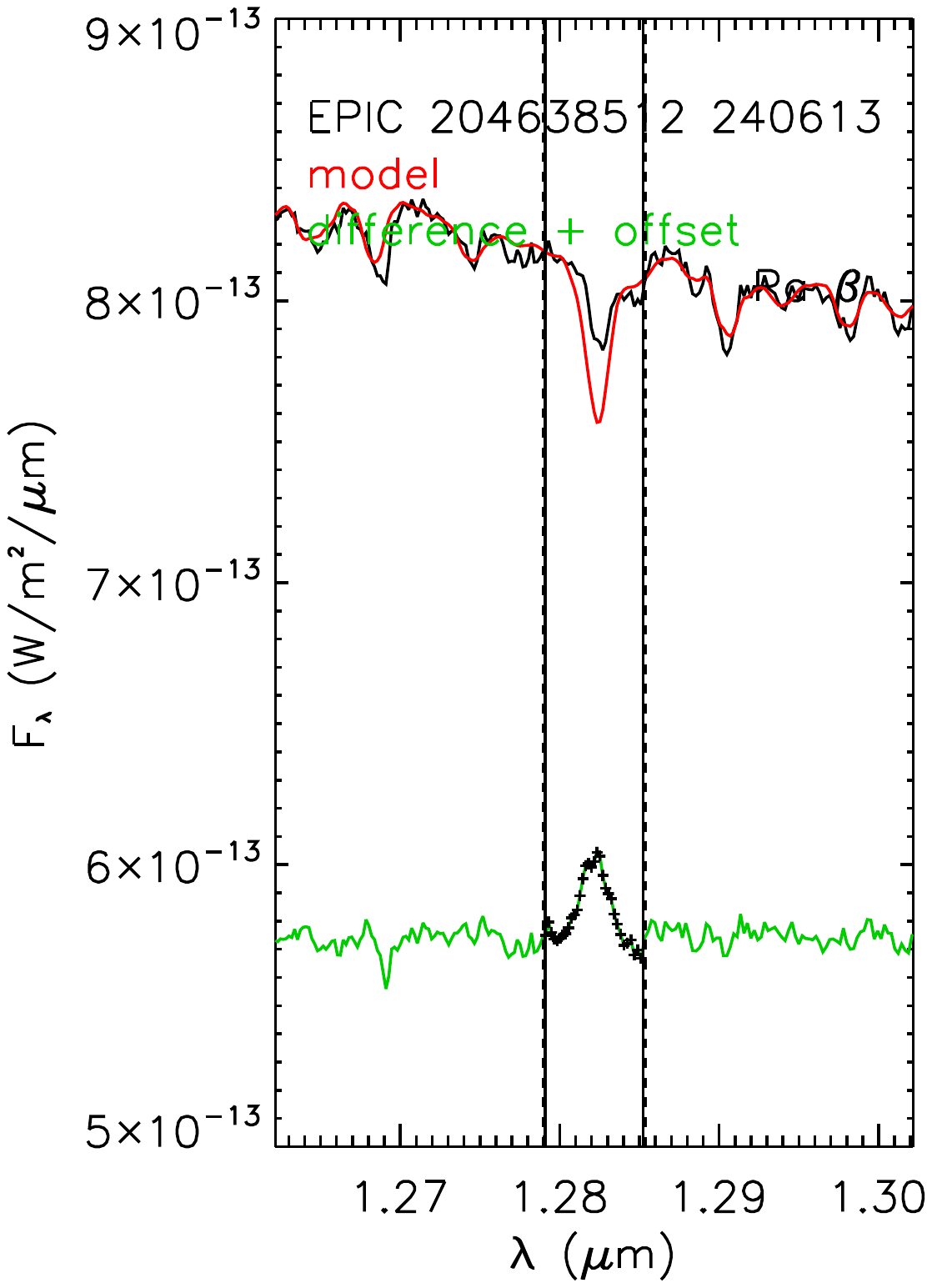}
\includegraphics[width=6.0cm, height=6.0cm]{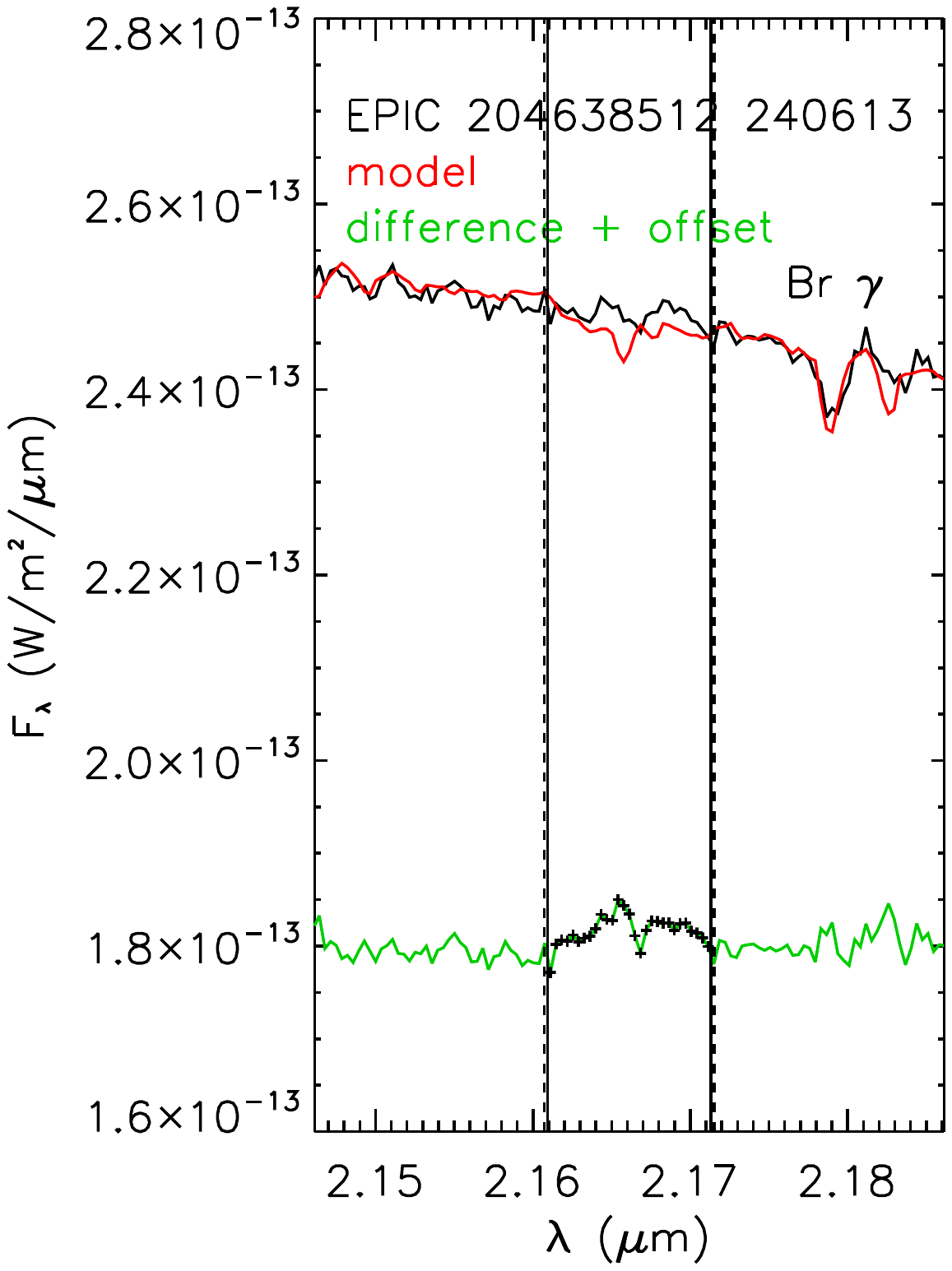}
\caption{The same as Figure A-2, except for EPIC 204638512 on 240613. \label{fig:A-21}}
\end{figure}

\begin{figure}
\includegraphics[width=6.0cm, height=6.0cm]{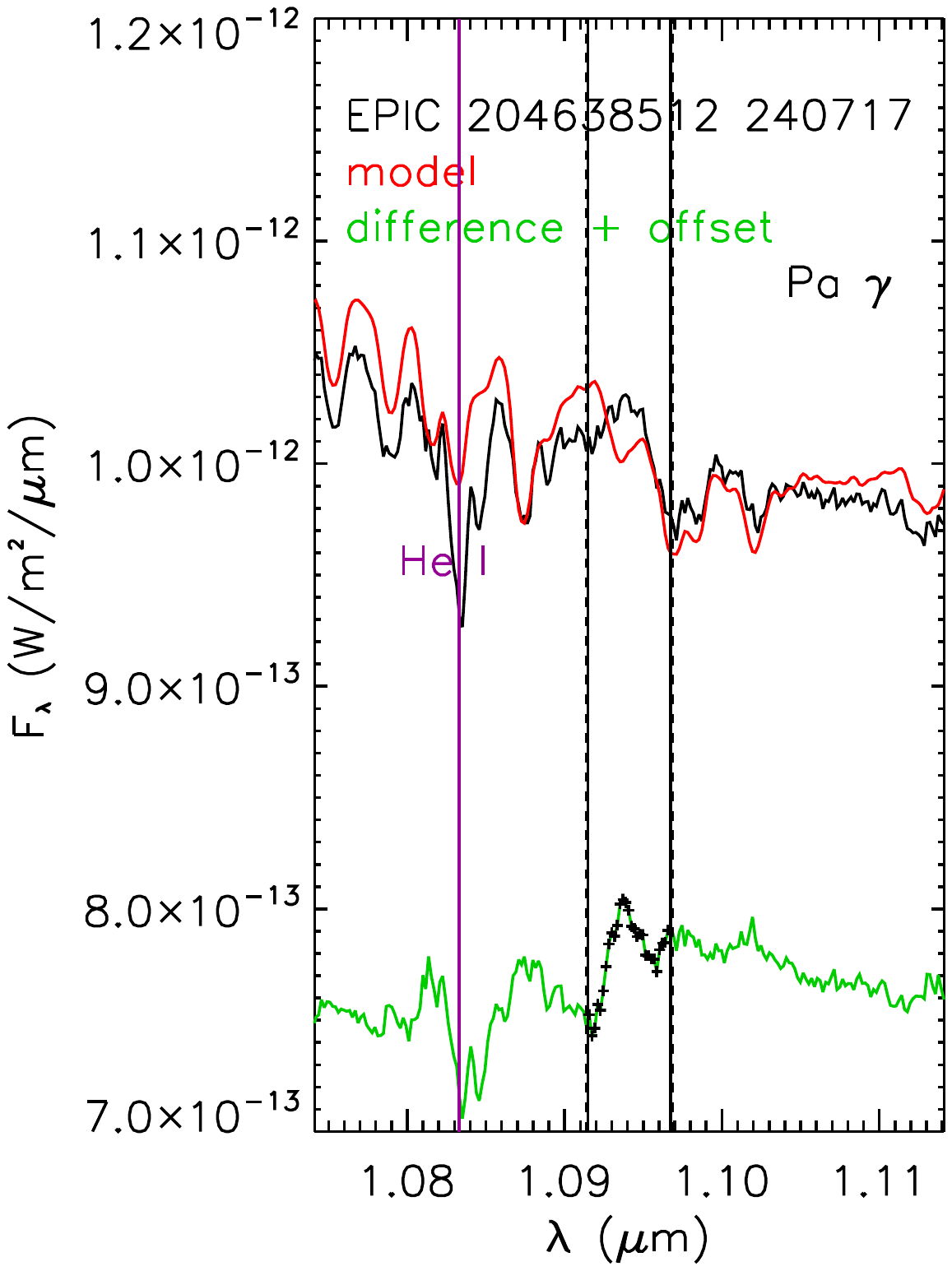}
\includegraphics[width=6.0cm, height=6.0cm]{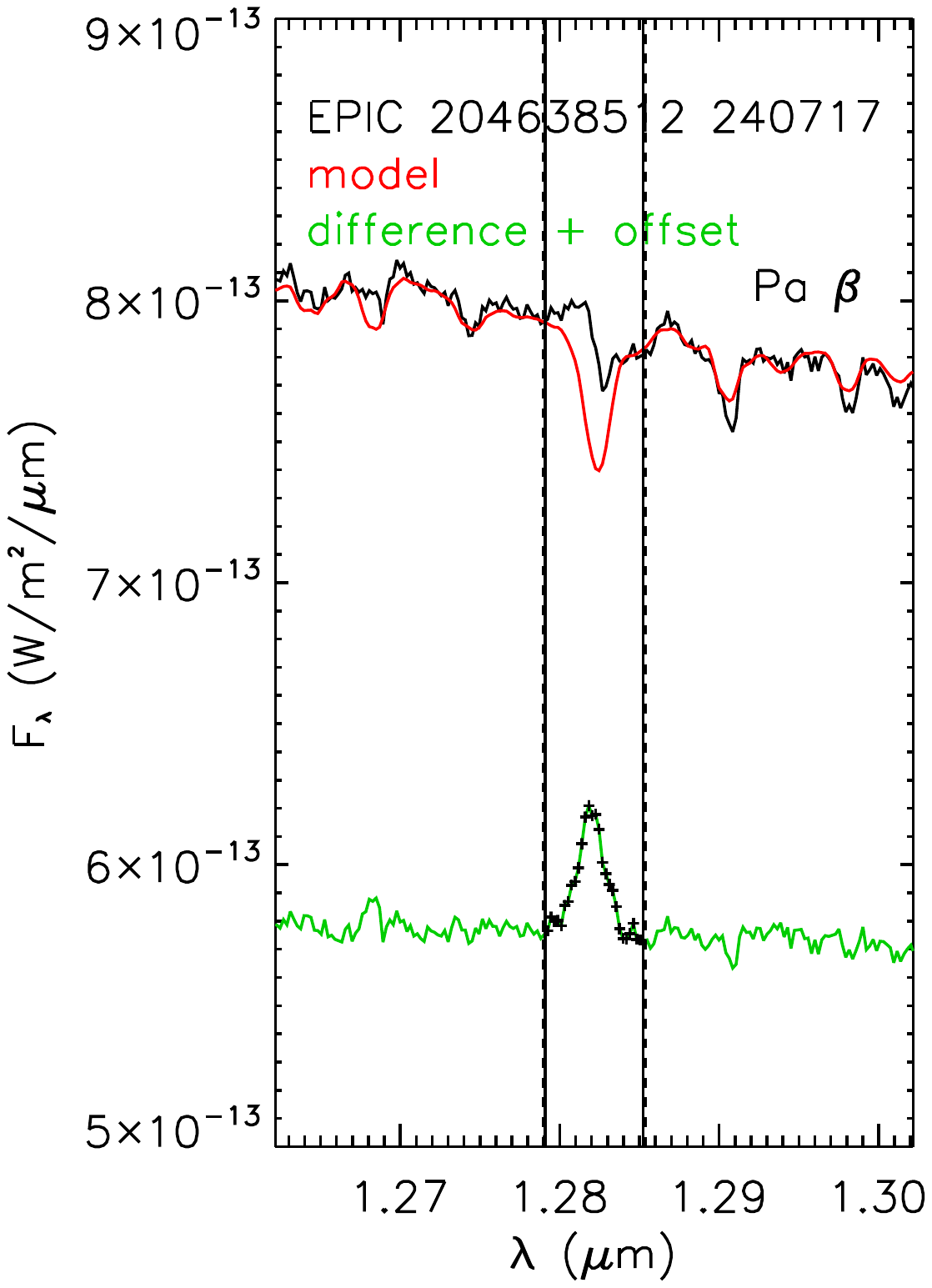}
\includegraphics[width=6.0cm, height=6.0cm]{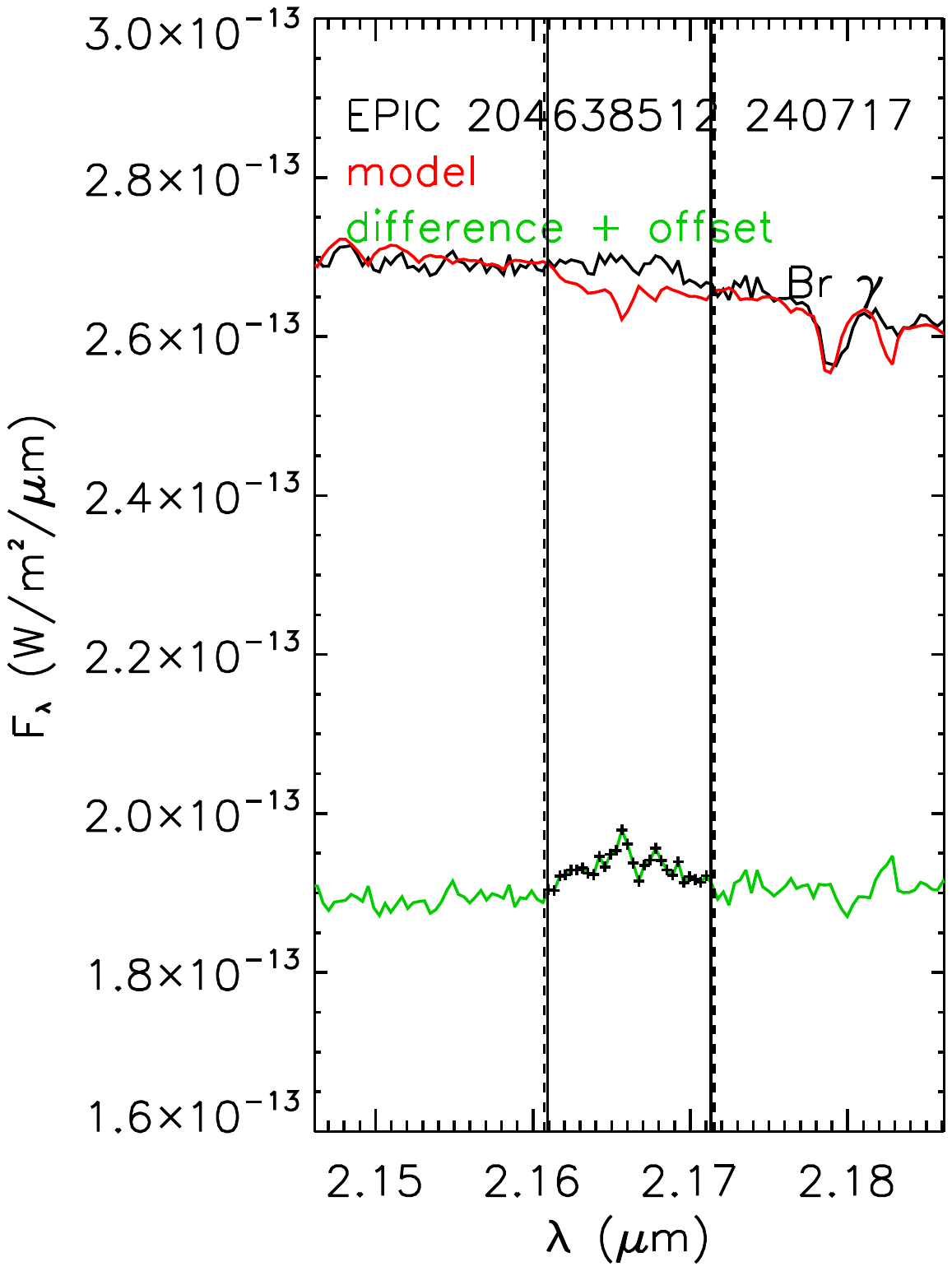}
\caption{The same as Figure A-2, except  for EPIC 204638512 on 240717. \label{fig:A-22 }}
\end{figure}

%\clearpage

\begin{figure}
\includegraphics[width=6.0cm, height=6.0cm]{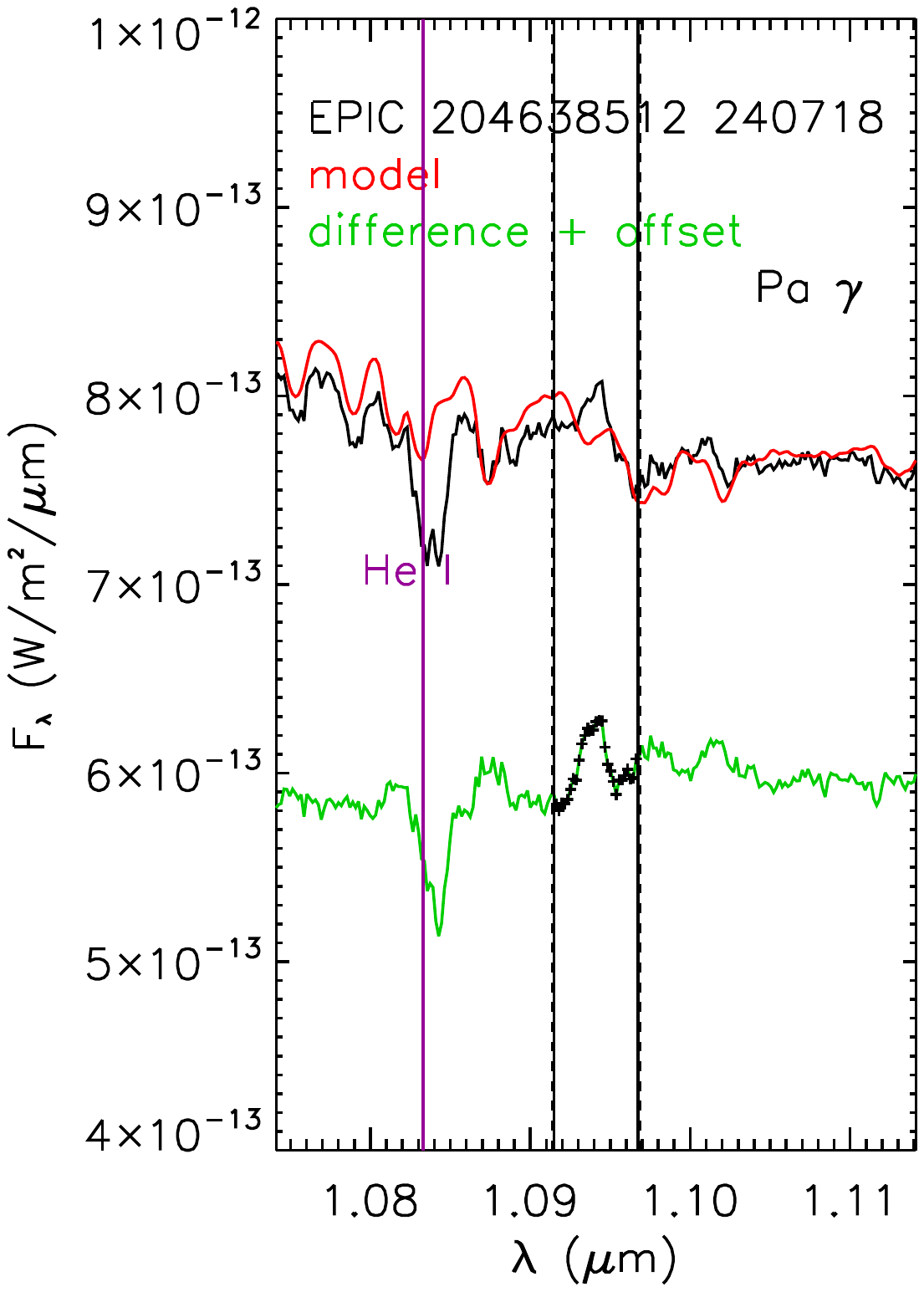}
\includegraphics[width=6.0cm, height=6.0cm]{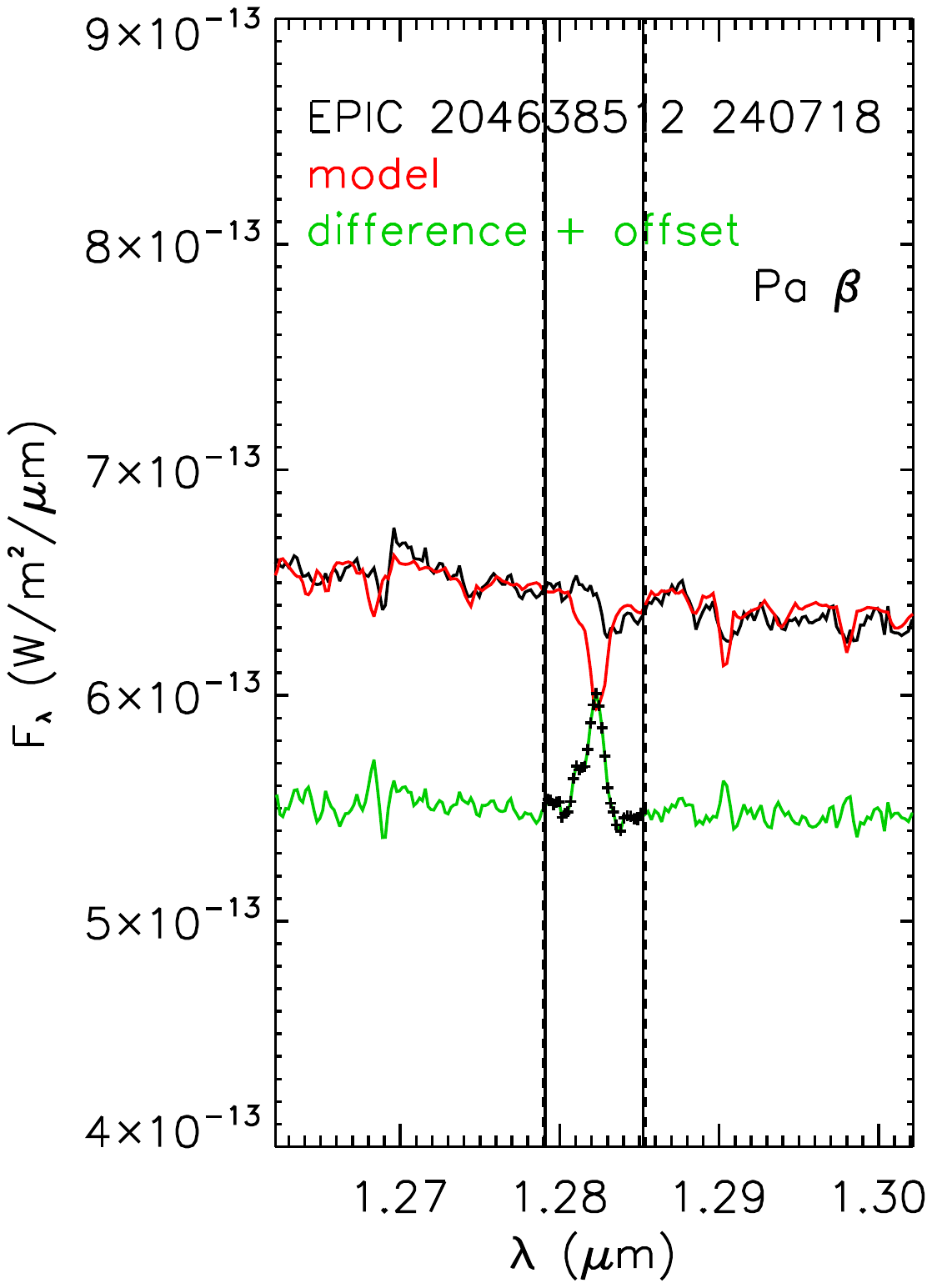}
\includegraphics[width=6.0cm, height=6.0cm]{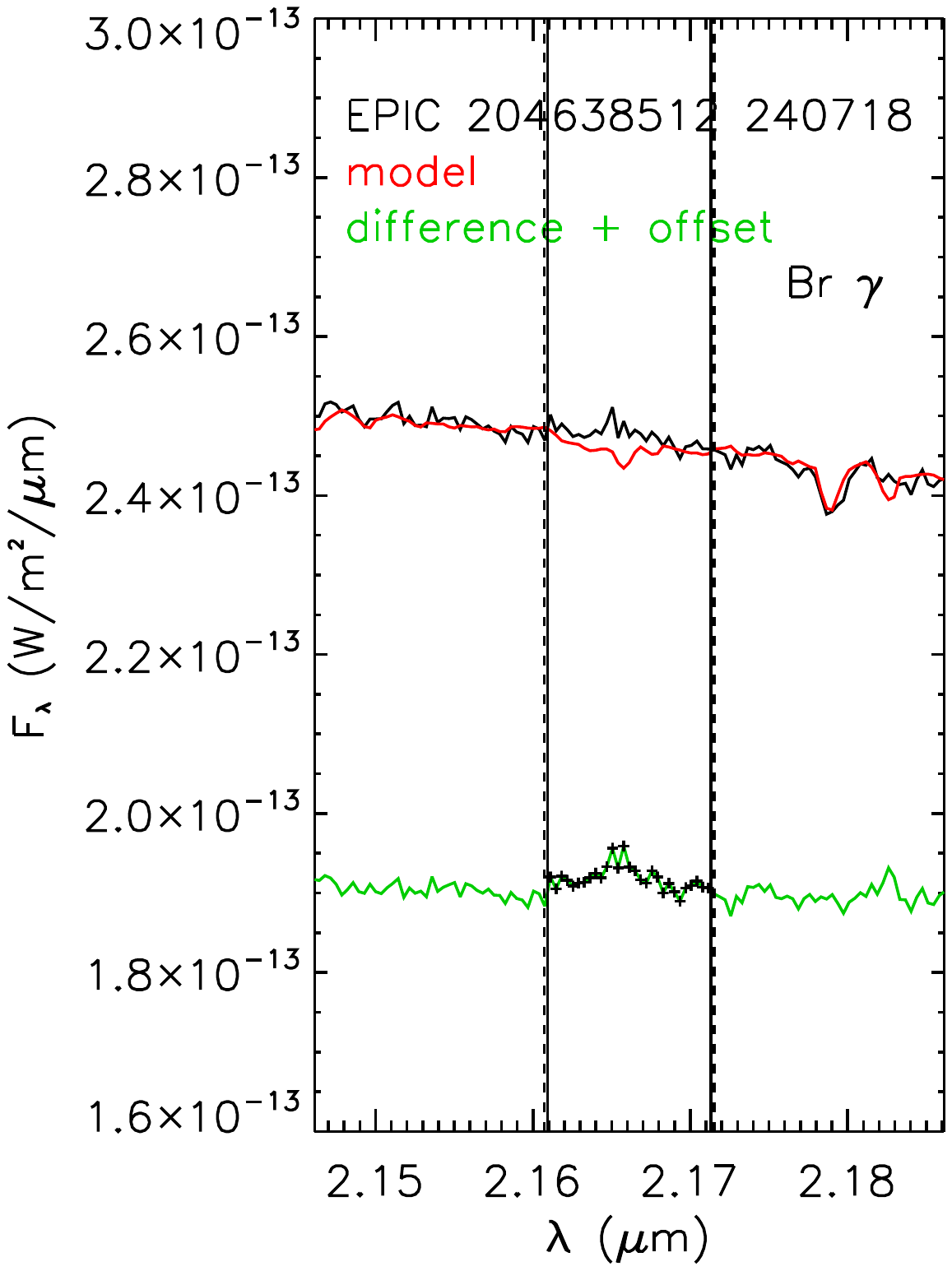}
\caption{The same as Figure A-2, except for EPIC 204638512 on 240718 UT. \label{fig:A-23 }}
\end{figure}

\begin{figure}
\includegraphics[width=6.0cm, height=6.0cm]{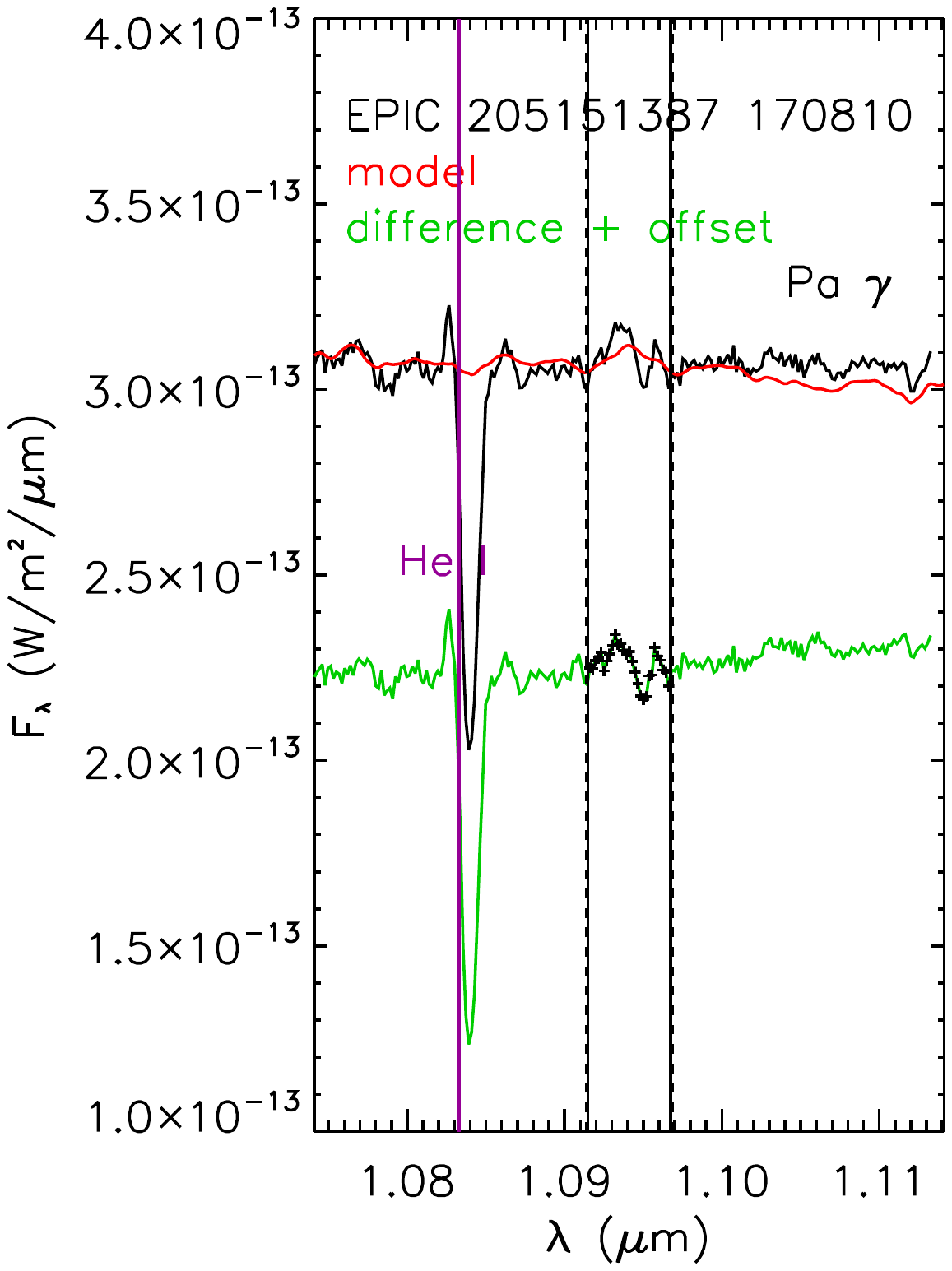}
\includegraphics[width=6.0cm, height=6.0cm]{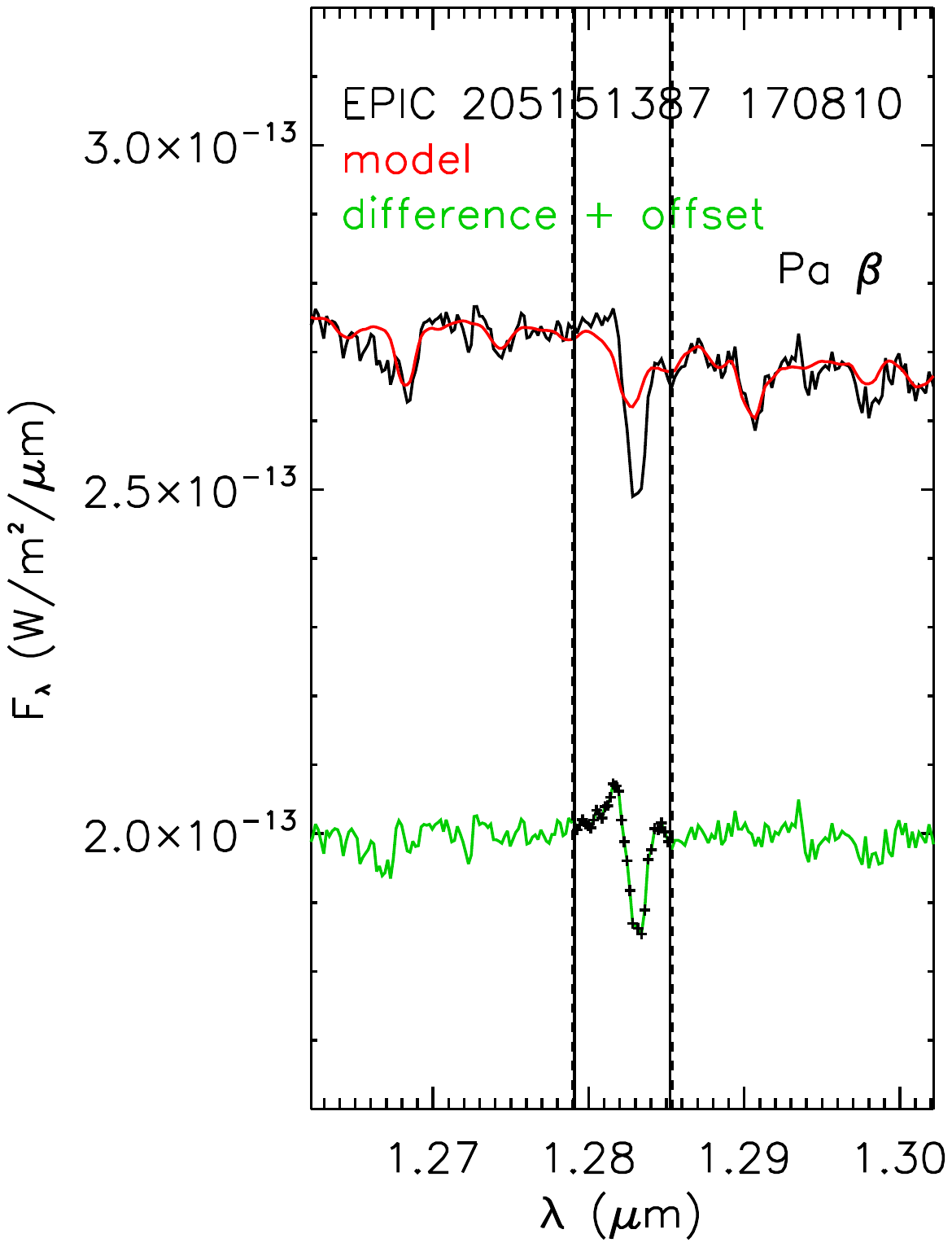}
\includegraphics[width=6.0cm, height=6.0cm]{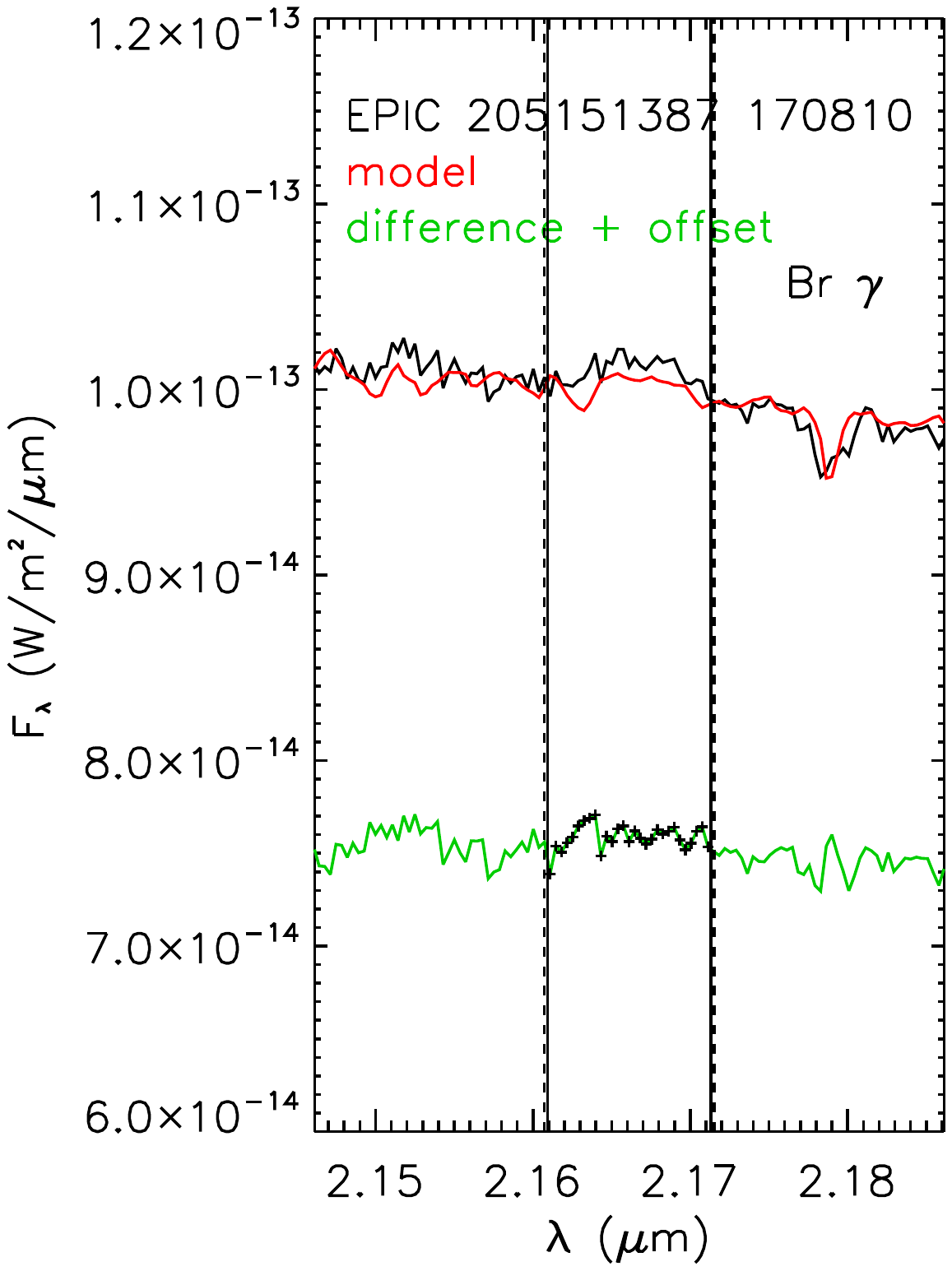}
\caption{The same as Figure A-2, except for EPIC 205151387 on 170810 UT. \label{fig:A-24 }}
\end{figure}

\begin{figure}
\includegraphics[width=6.0cm, height=6.0cm]{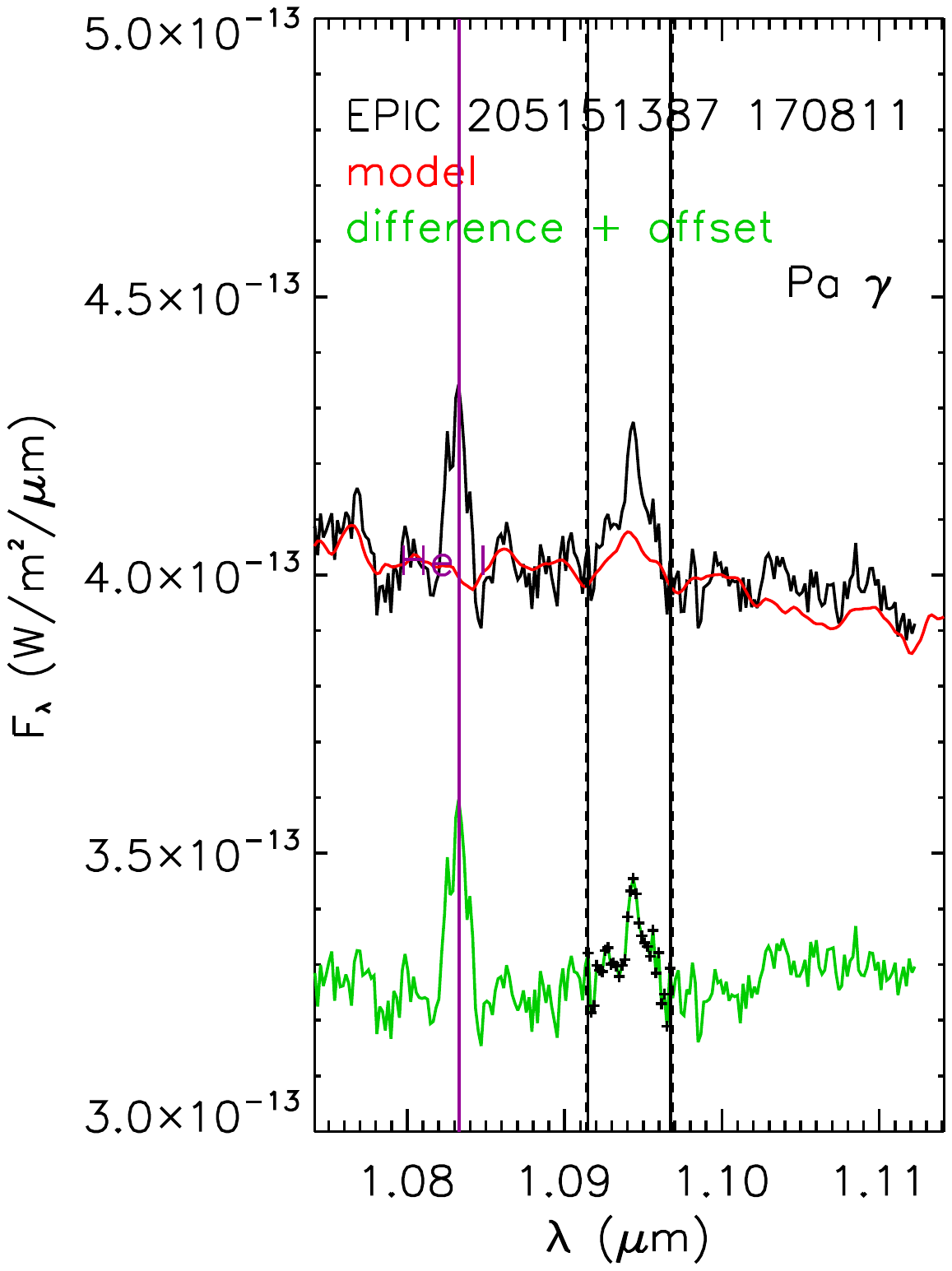}
\includegraphics[width=6.0cm, height=6.0cm]{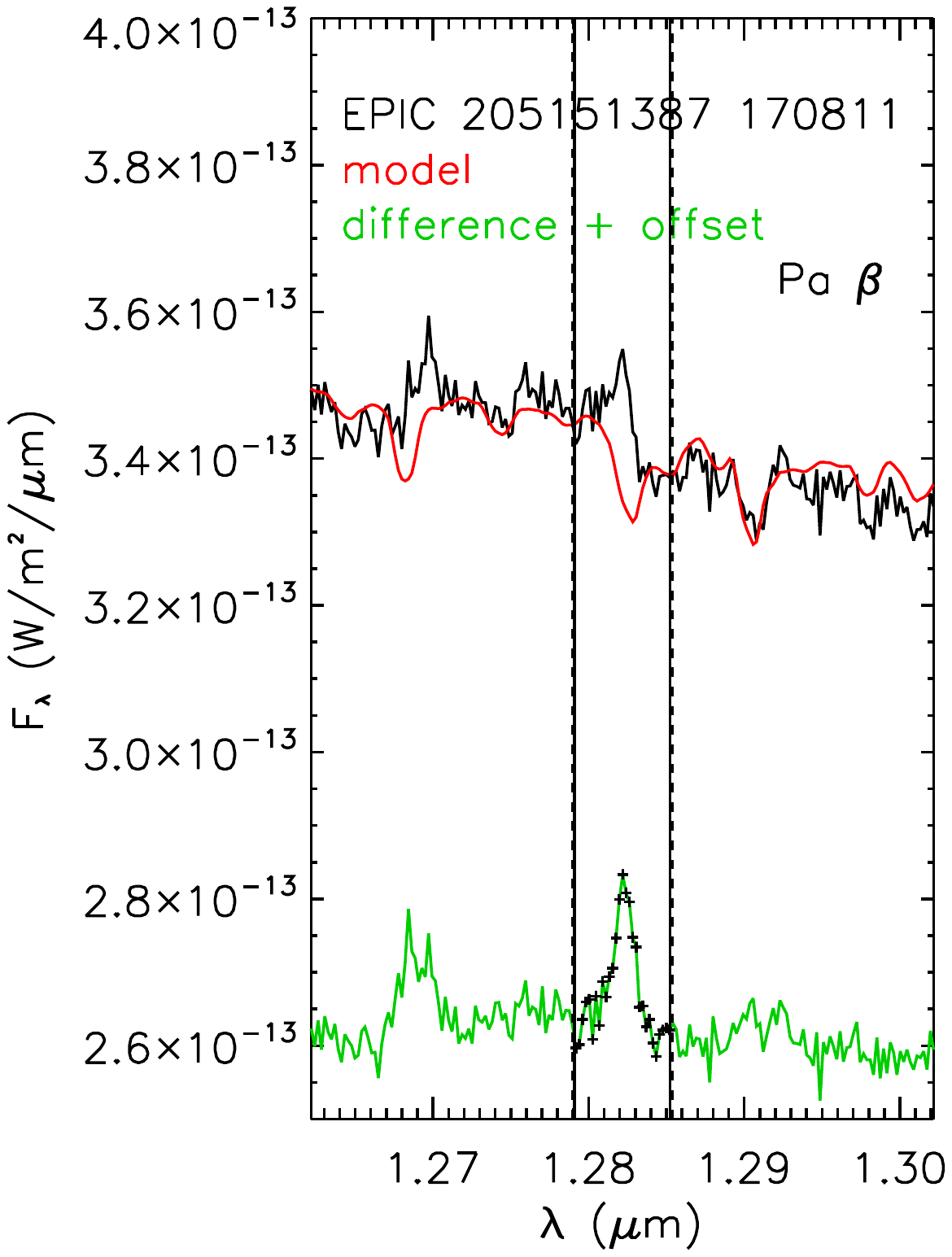}
\includegraphics[width=6.0cm, height=6.0cm]{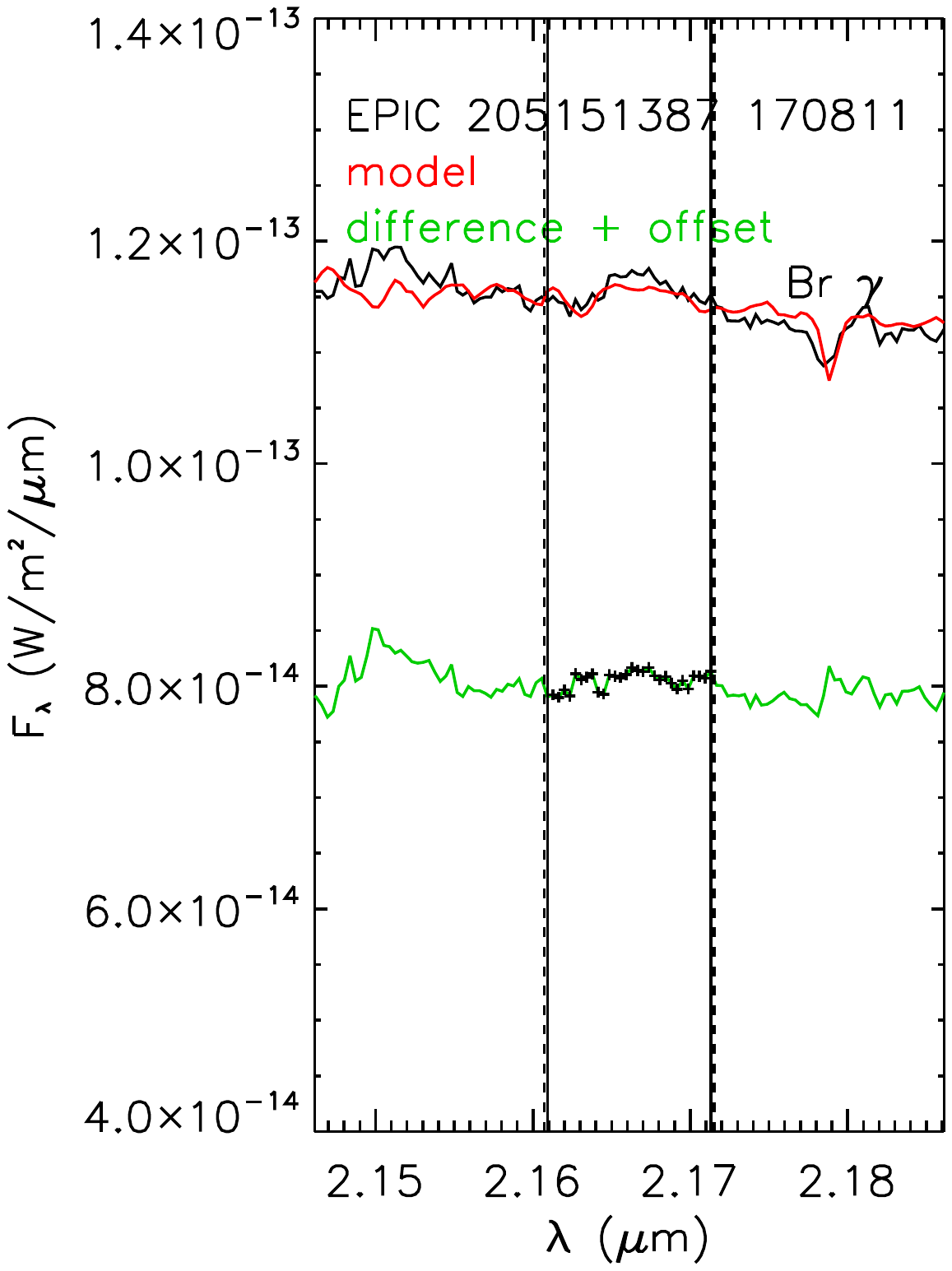}
\caption{The same as Figure A-2, except for EPIC 205151387 on 170811 UT. Obtained only 1 day after the previous spectrum, both the He I and Pa$\beta$ lines are in net emission, indicating that they can change on daily time scales. \label{fig:A-25}}
\end{figure}

\begin{figure}
\includegraphics[width=6.0cm, height=6.0cm]{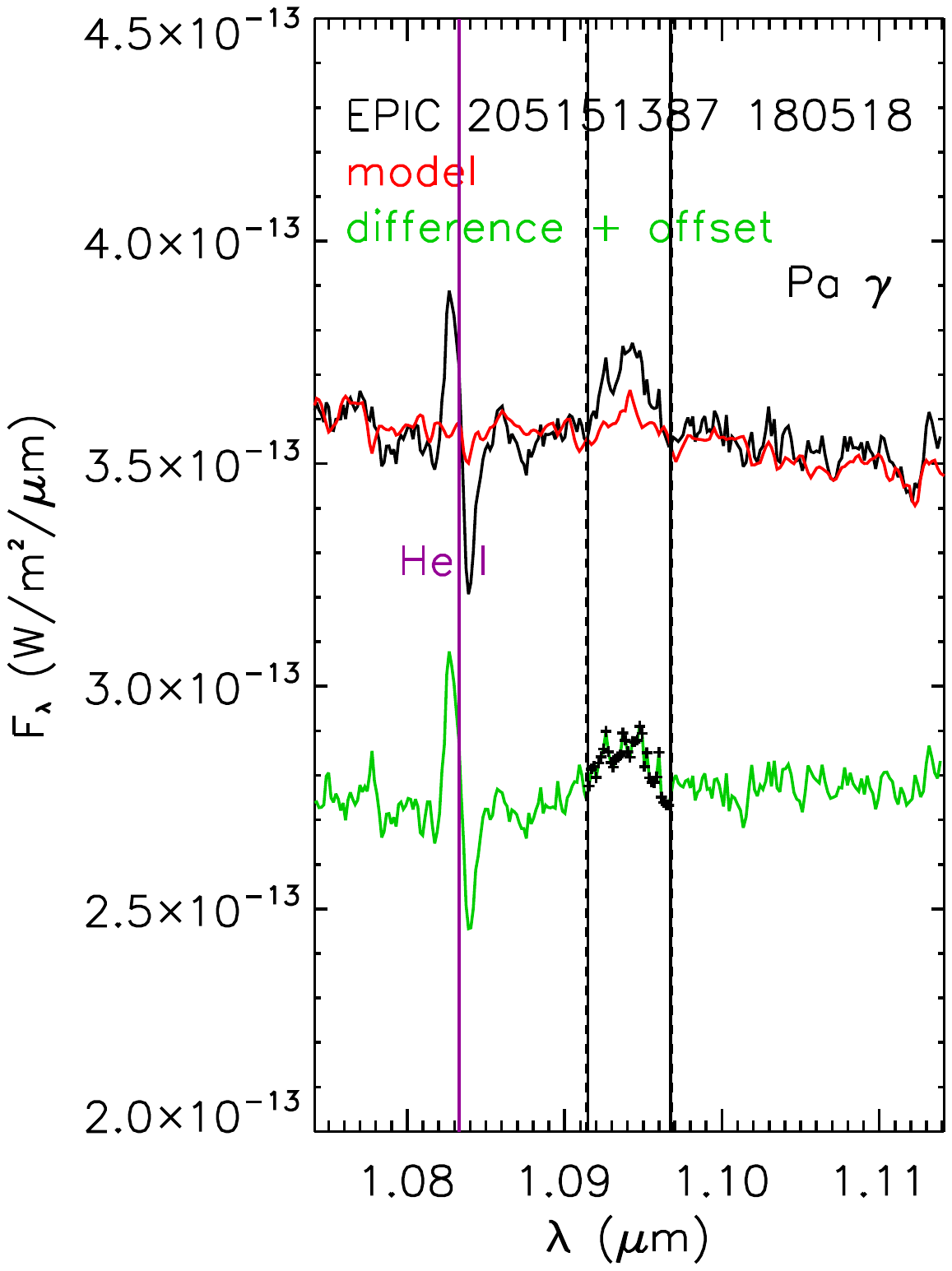}
\includegraphics[width=6.0cm, height=6.0cm]{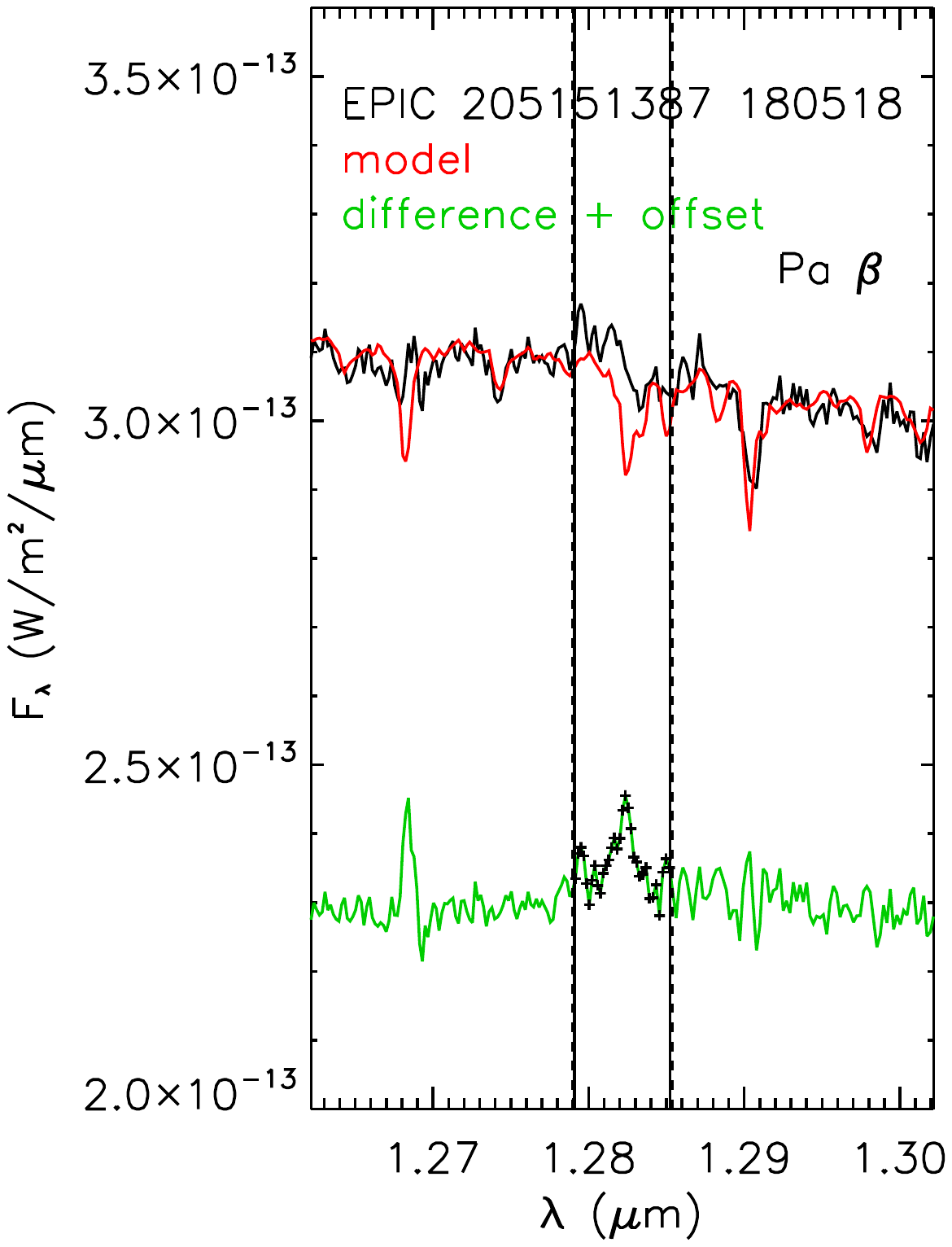}
\includegraphics[width=6.0cm, height=6.0cm]{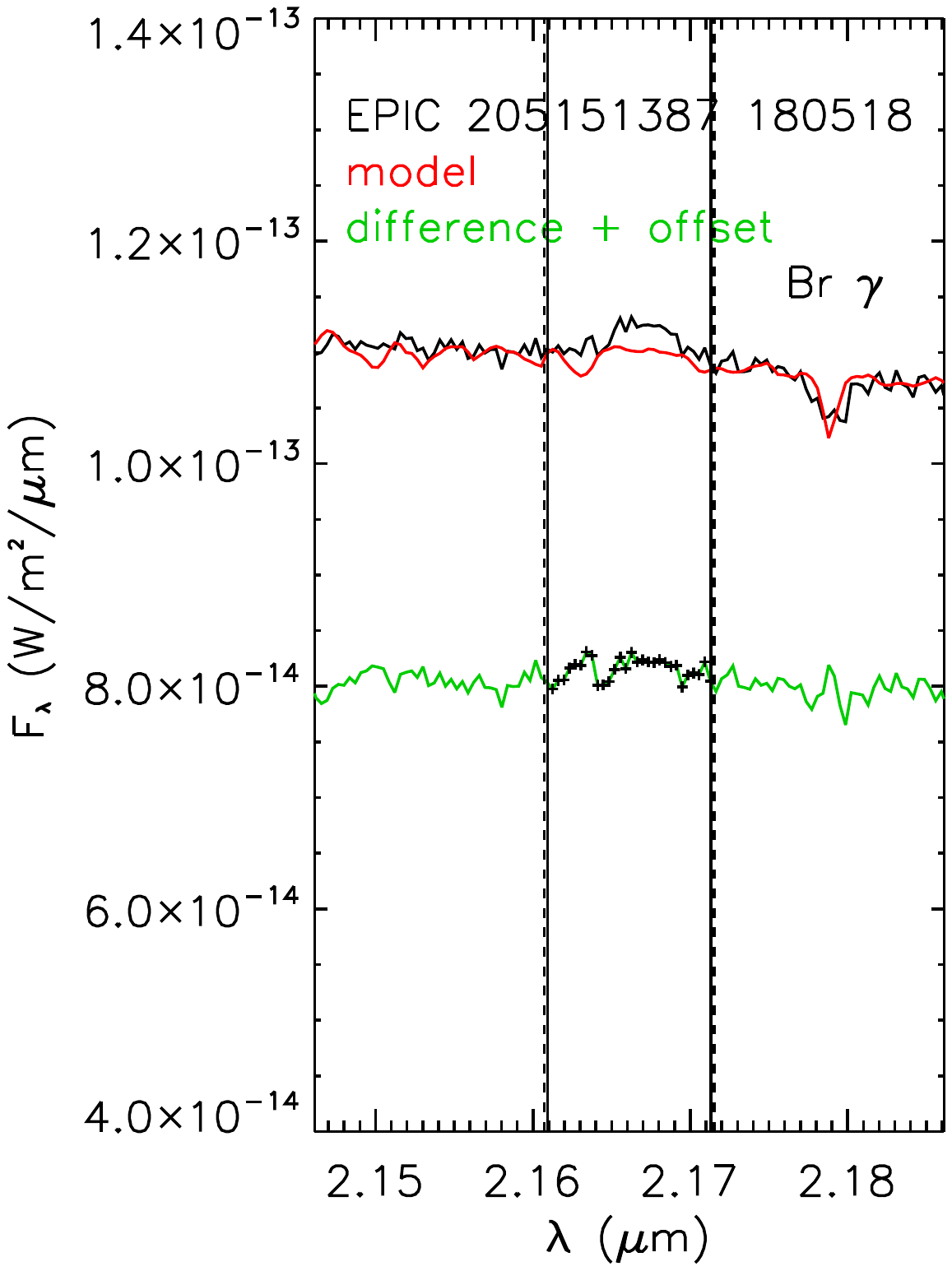}
\caption{The same as Figure A-2, except for EPIC 205151387 on 180518 UT. \label{fig:A-26}}
\end{figure}

\begin{figure}
\includegraphics[width=6.0cm, height=6.0cm]{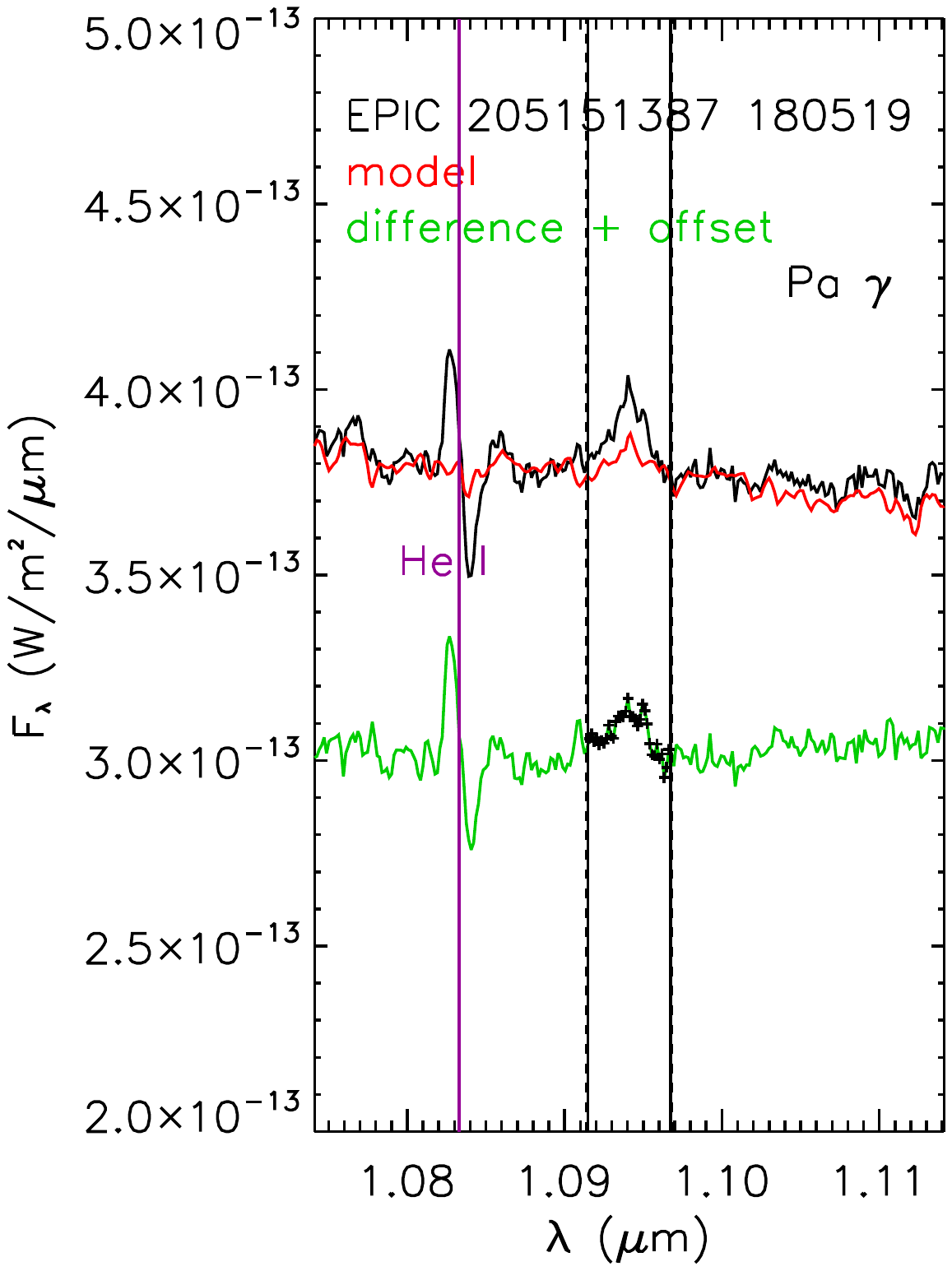}
\includegraphics[width=6.0cm, height=6.0cm]{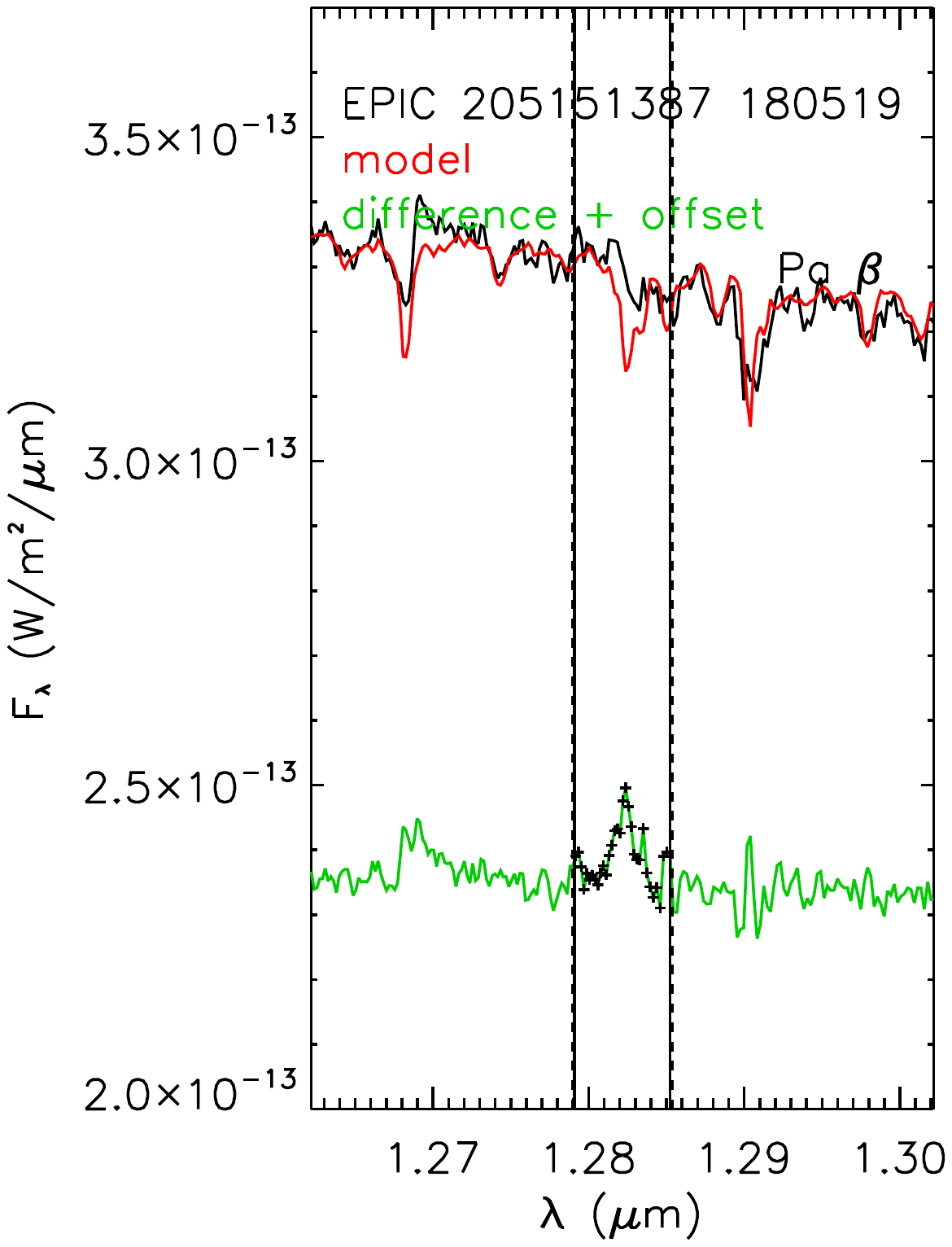}
\includegraphics[width=6.0cm, height=6.0cm]{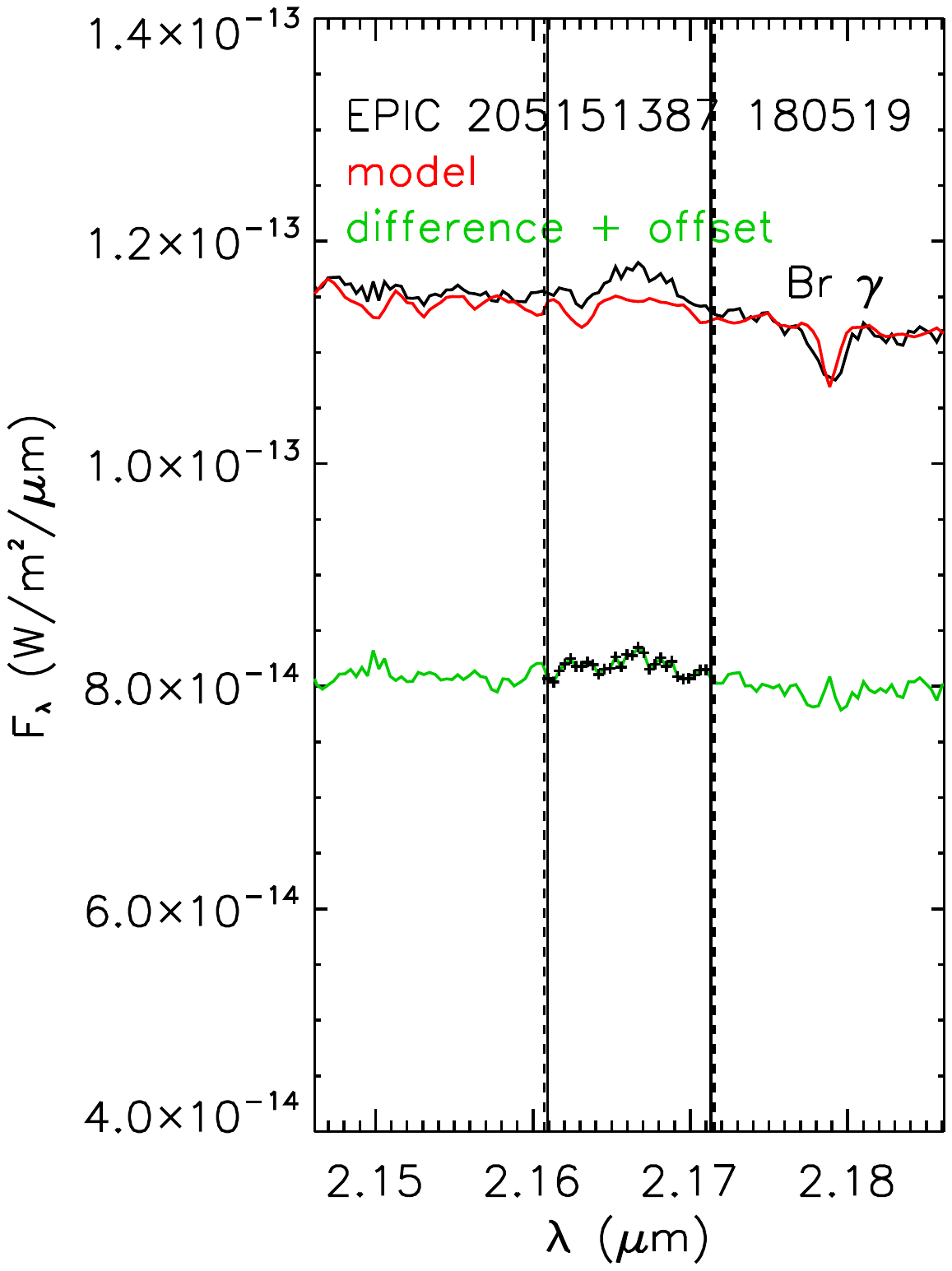}
\caption{The same as Figure A-2, except for EPIC 205151387 on 180519 UT. \label{fig:A-27}}
\end{figure}

\begin{figure}
\includegraphics[width=6.0cm, height=6.0cm]{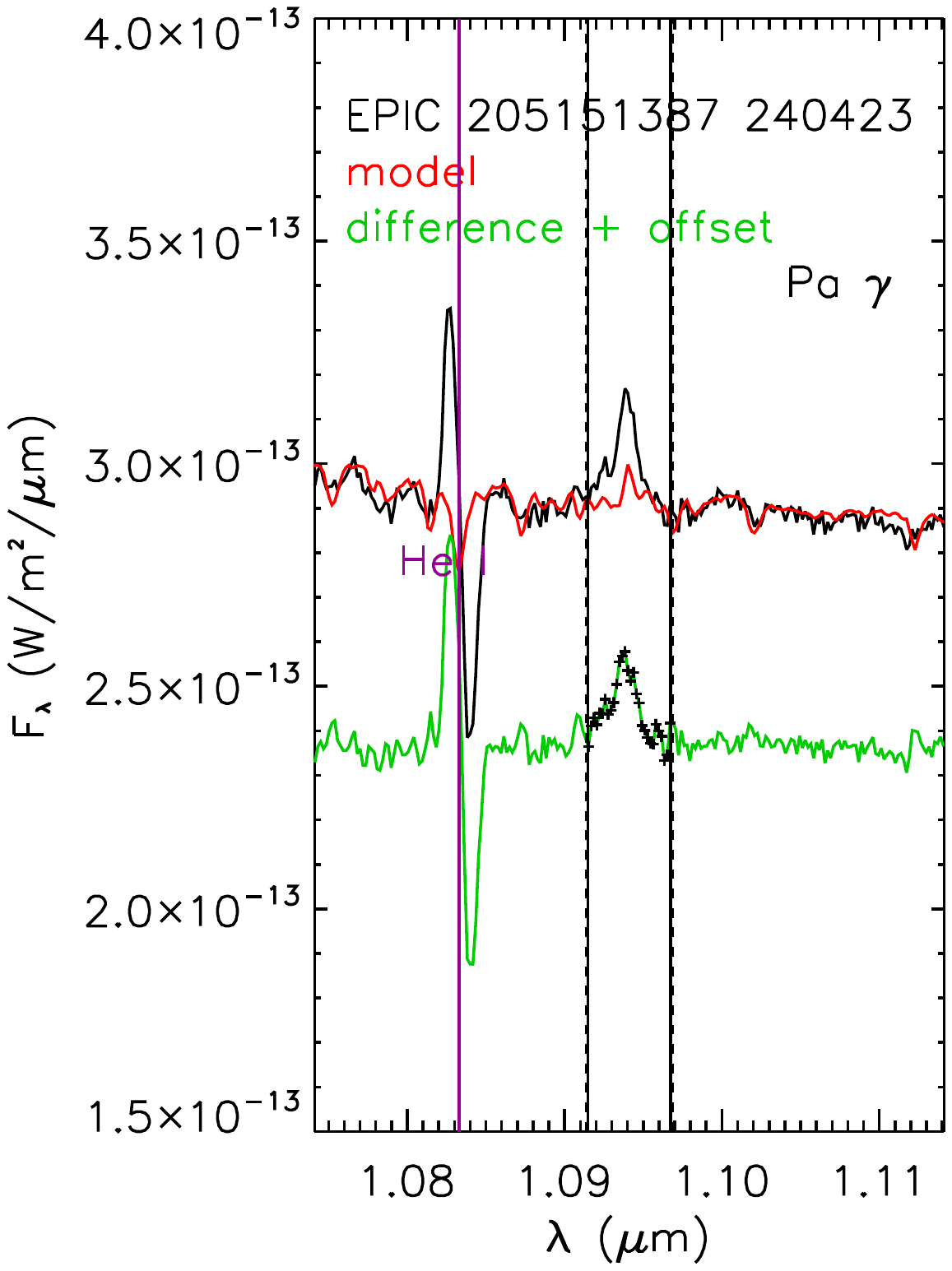}
\includegraphics[width=6.0cm, height=6.0cm]{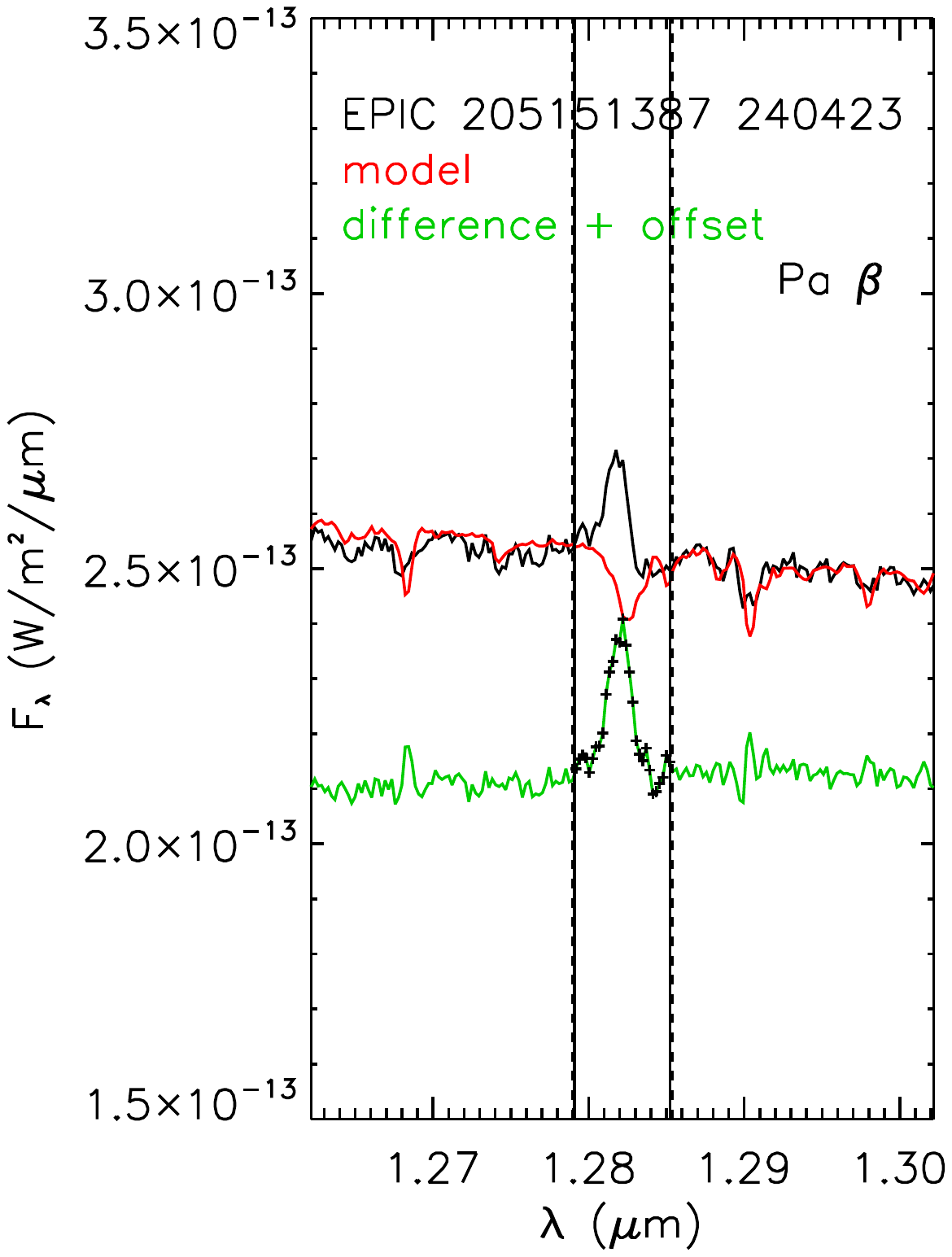}
\includegraphics[width=6.0cm, height=6.0cm]{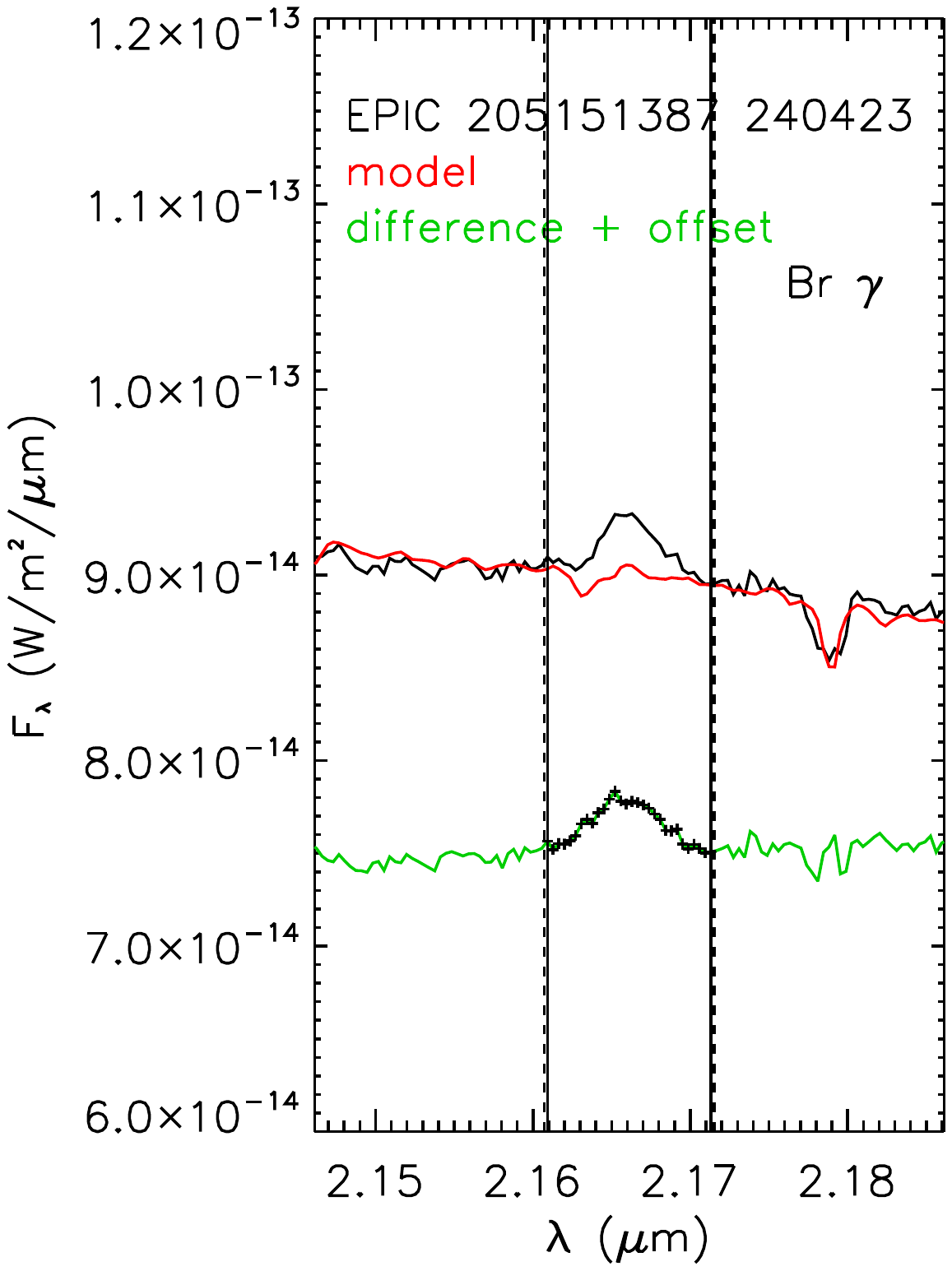}
\caption{The same as Figure A-2, except for EPIC 205151387 on 240423 UT. \label{fig:A-28}}
\end{figure}

%\clearpage

\begin{figure}
\includegraphics[width=6.0cm, height=6.0cm]{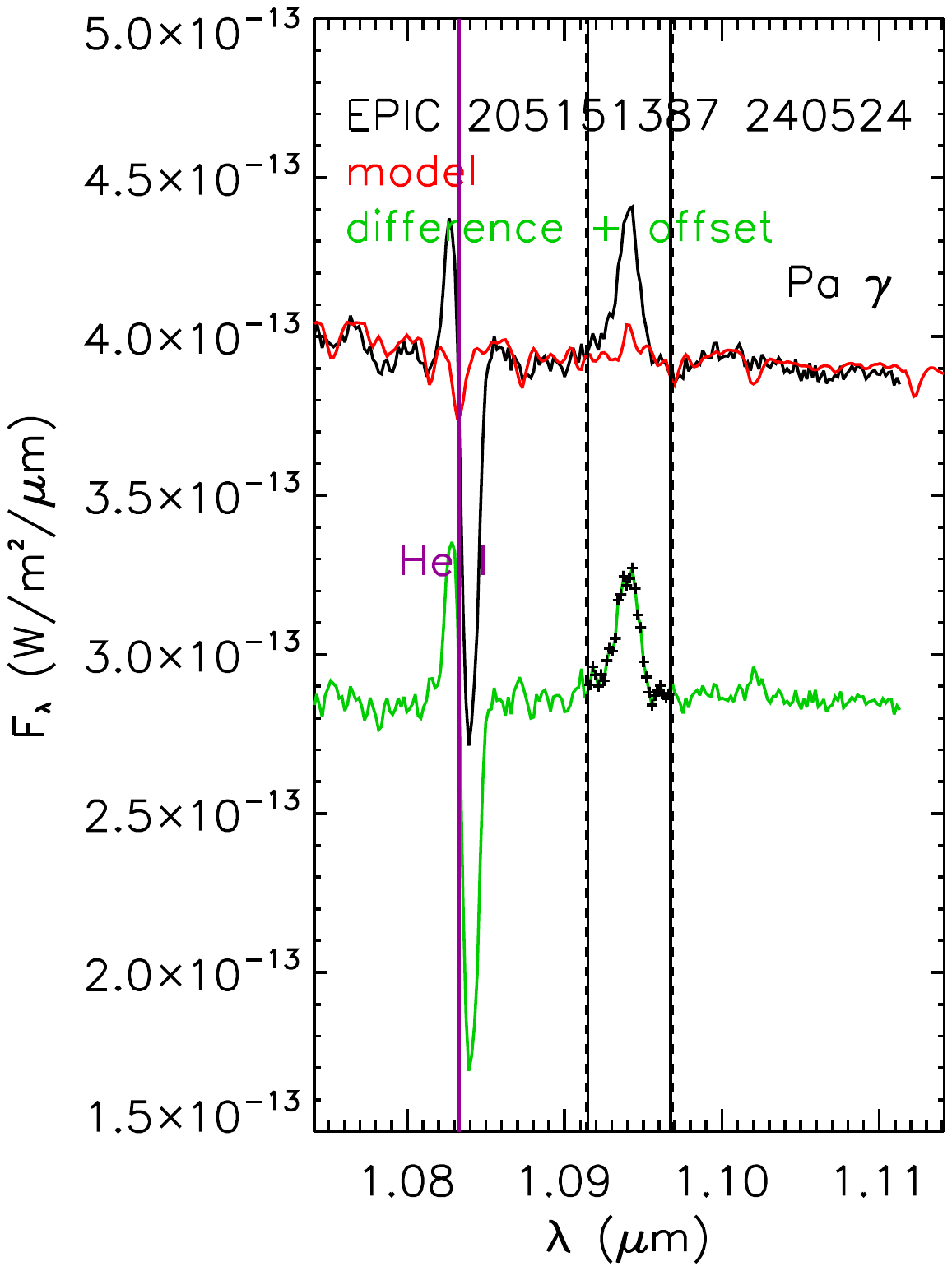}
\includegraphics[width=6.0cm, height=6.0cm]{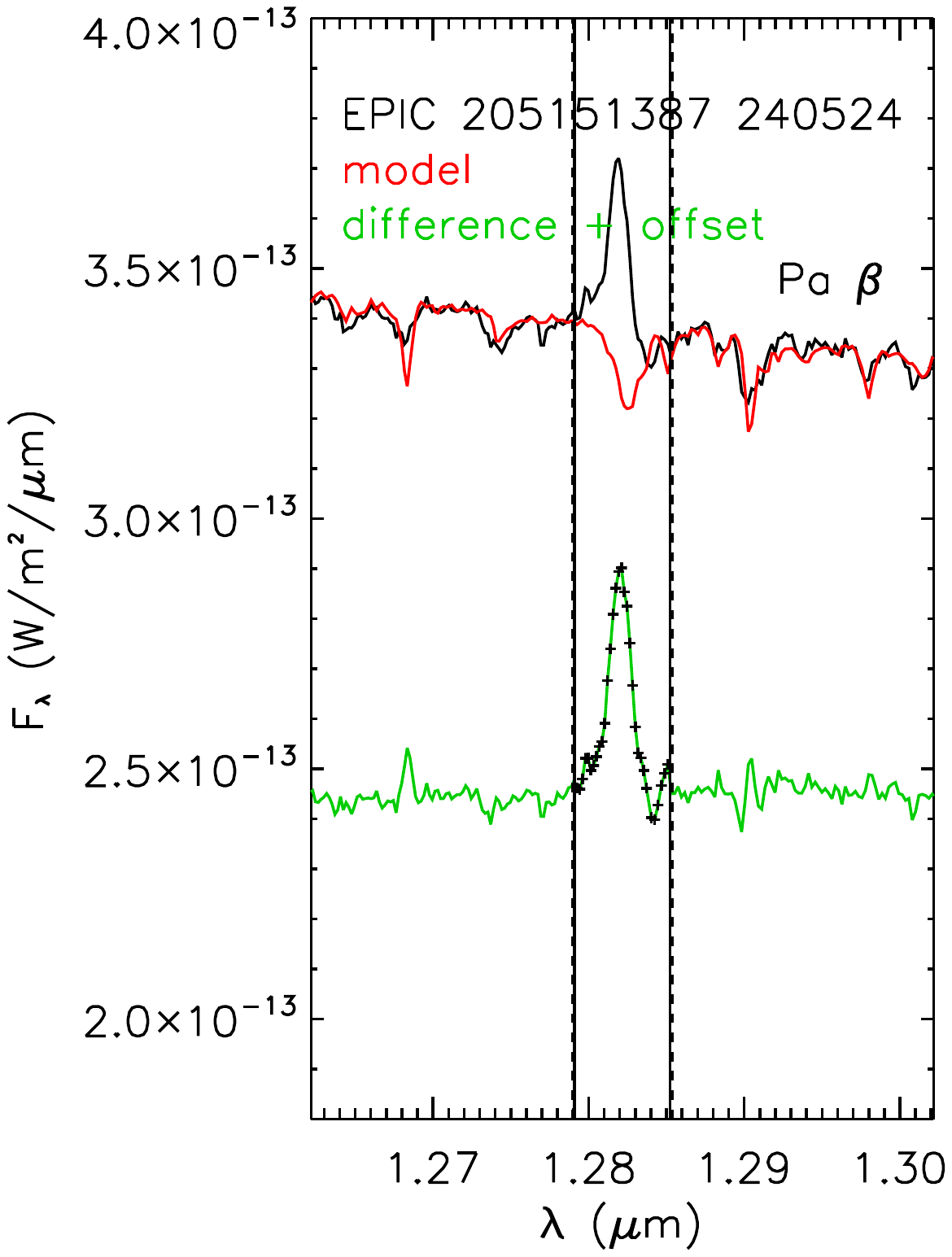}
\includegraphics[width=6.0cm, height=6.0cm]{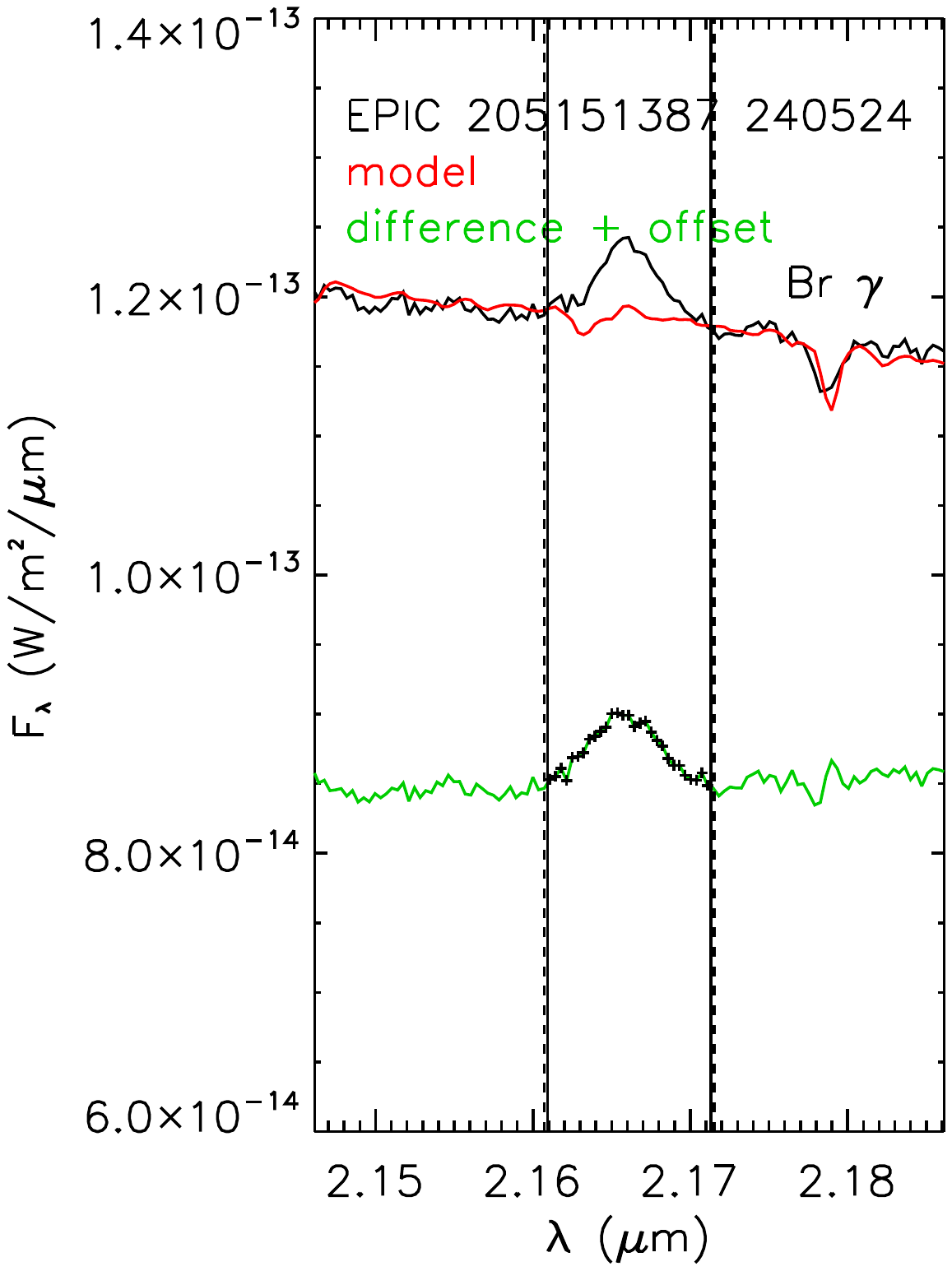}
\caption{The same as Figure A-2, except for EPIC 205151387 on 240524 UT. \label{fig:A-29}}
\end{figure}

\begin{figure}
\includegraphics[width=6.0cm, height=6.0cm]{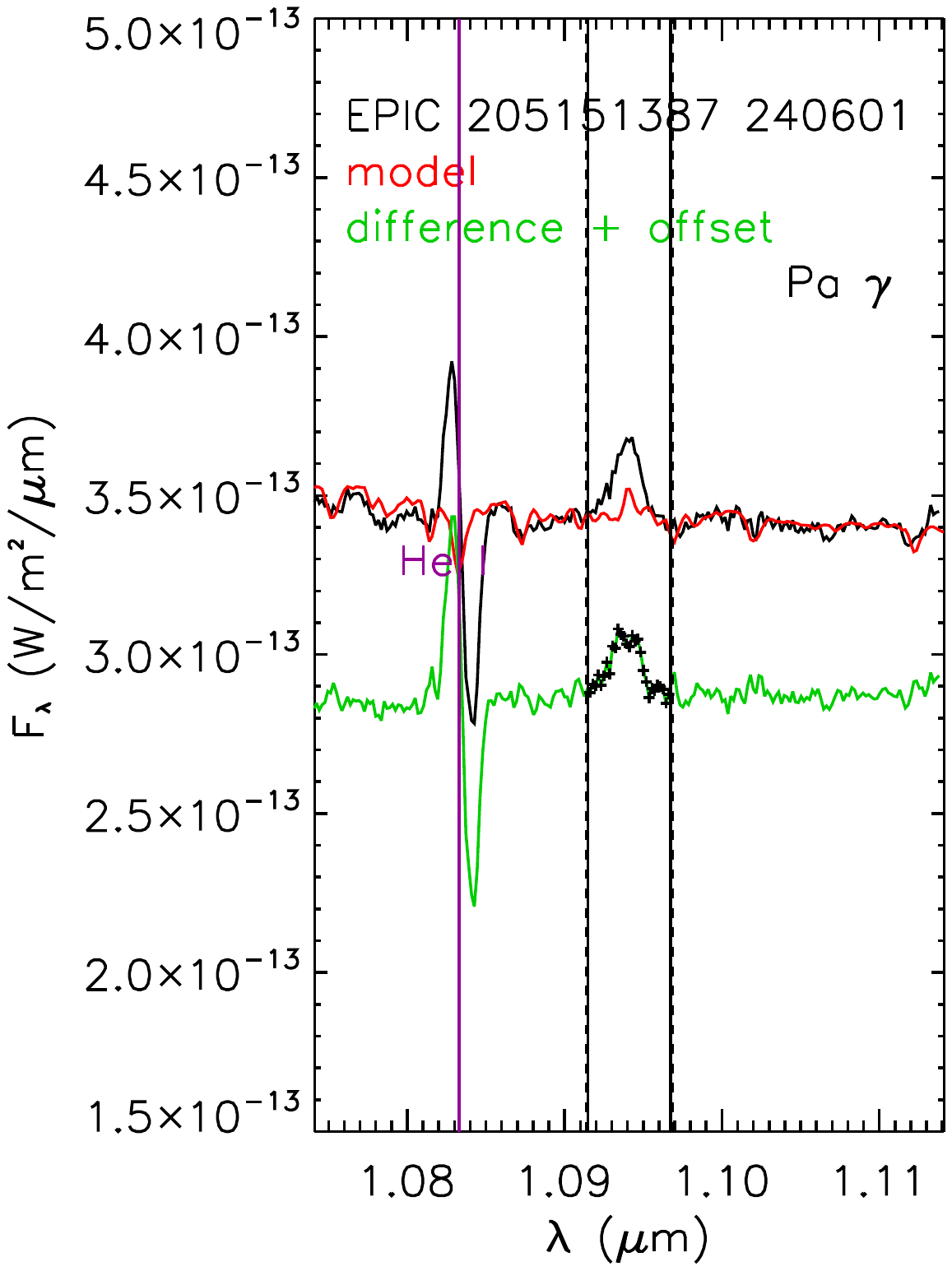}
\includegraphics[width=6.0cm, height=6.0cm]{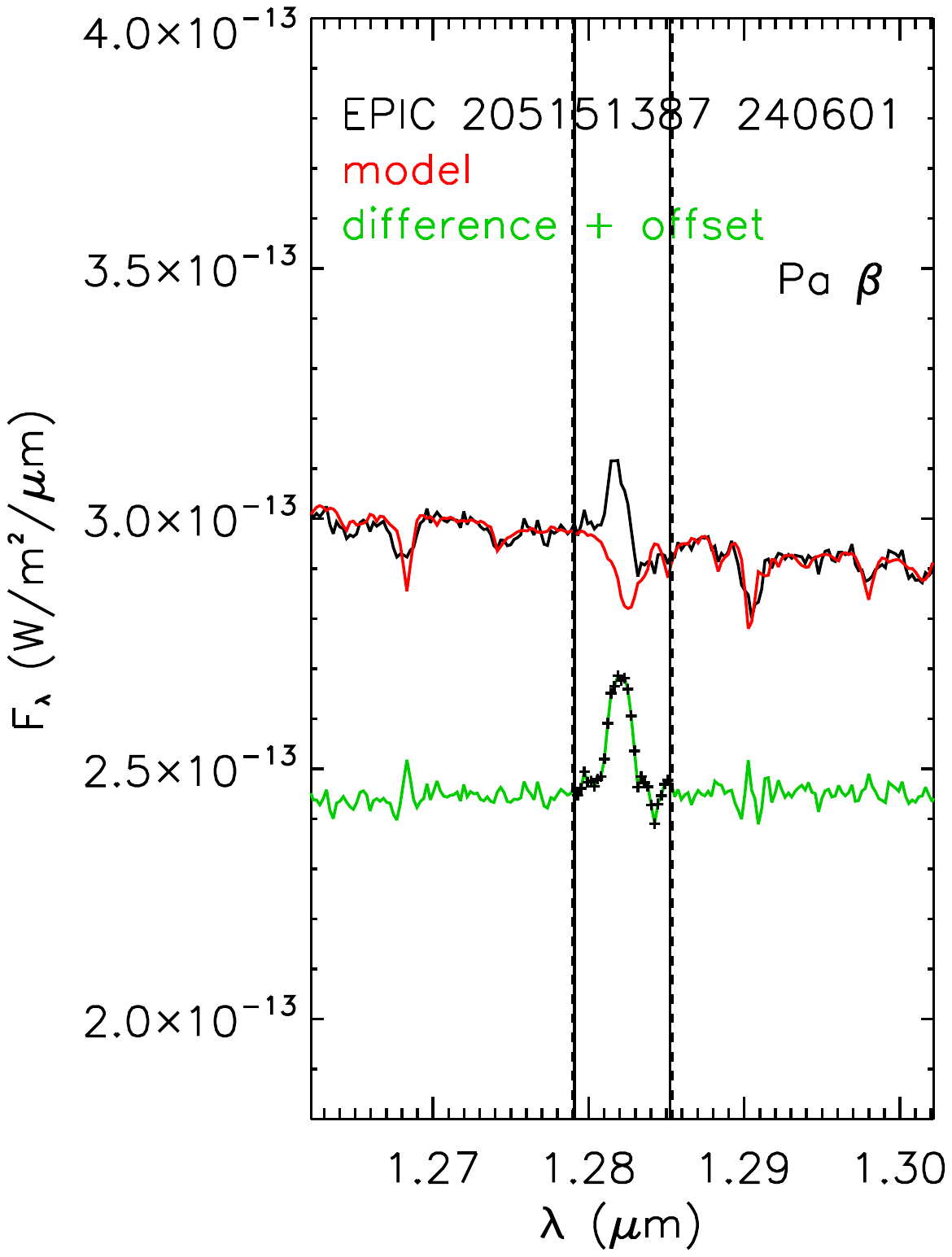}
\includegraphics[width=6.0cm, height=6.0cm]{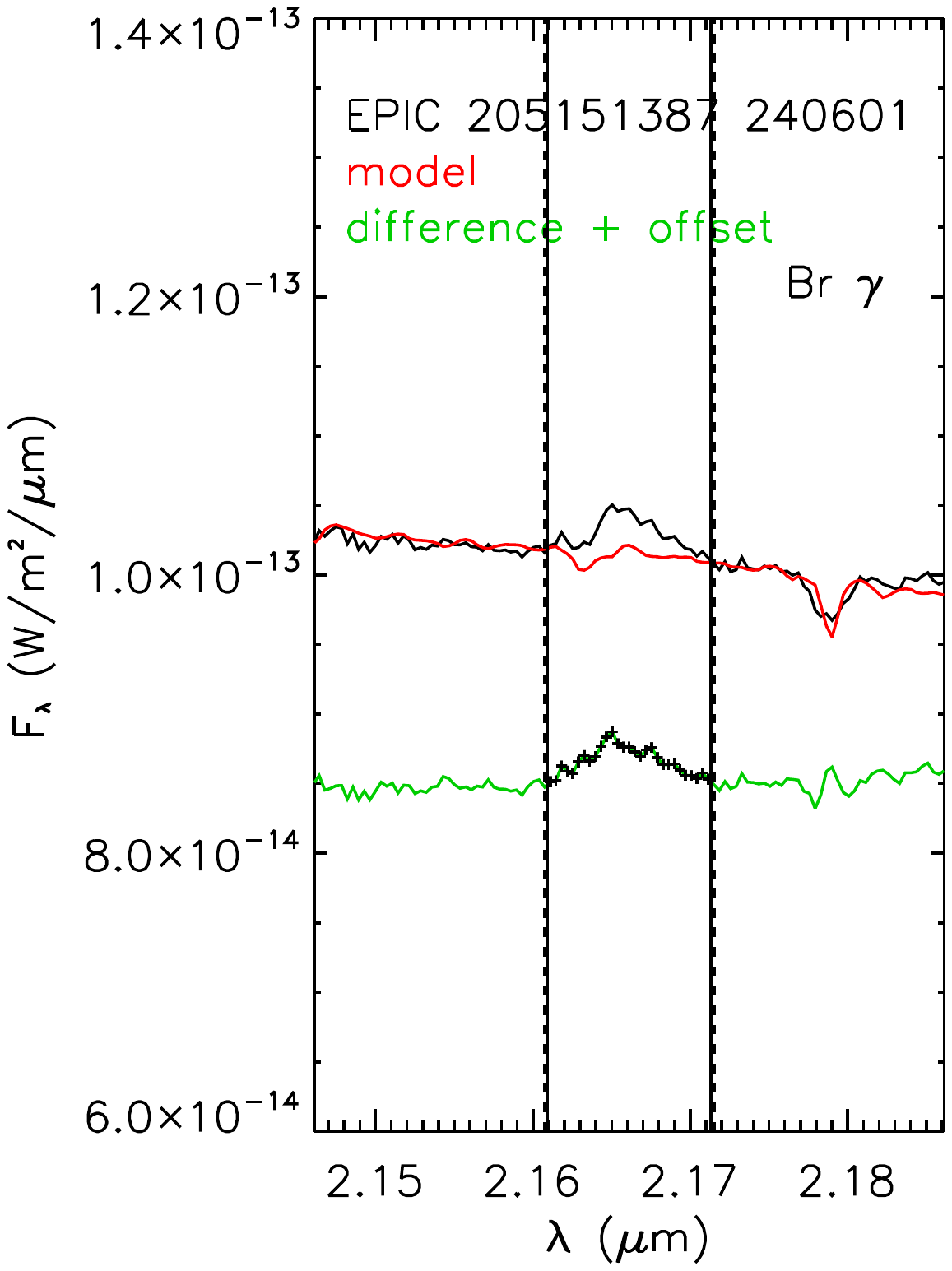}
\caption{The same as Figure A-2, except for EPIC 205151387 on 240601 UT. \label{fig:A-30}}
\end{figure}

\begin{figure}
\includegraphics[width=6.0cm, height=6.0cm]{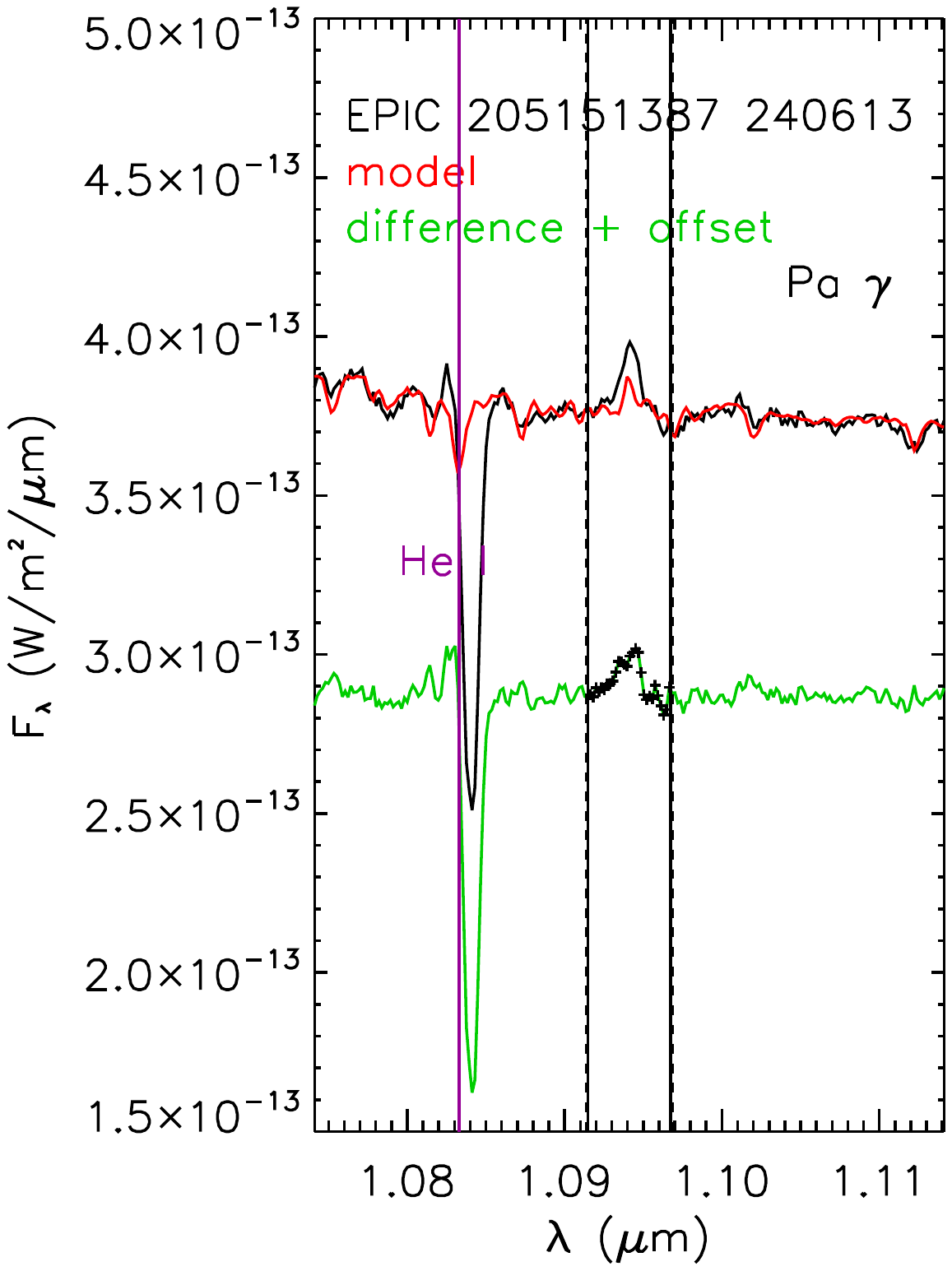}
\includegraphics[width=6.0cm, height=6.0cm]{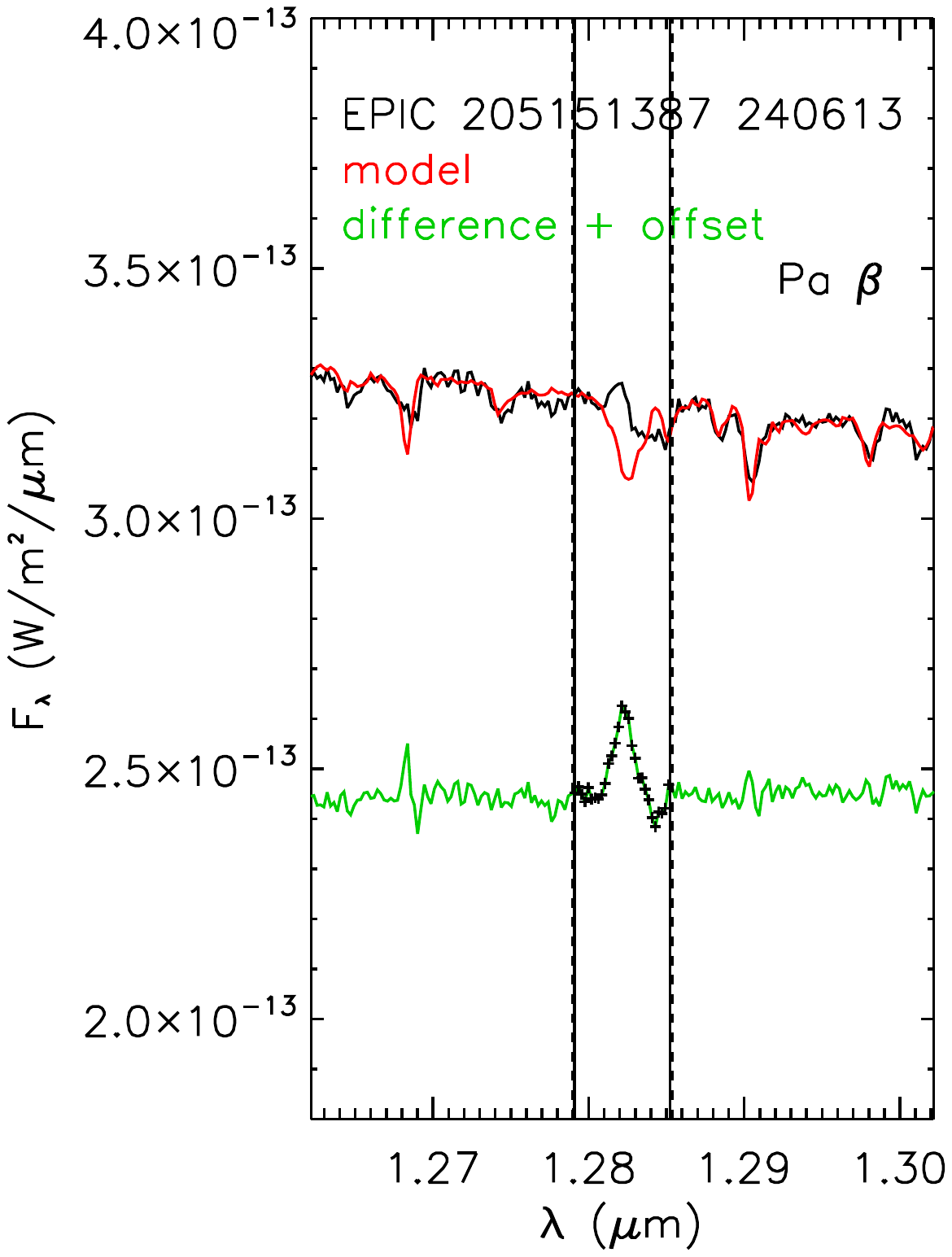}
\includegraphics[width=6.0cm, height=6.0cm]{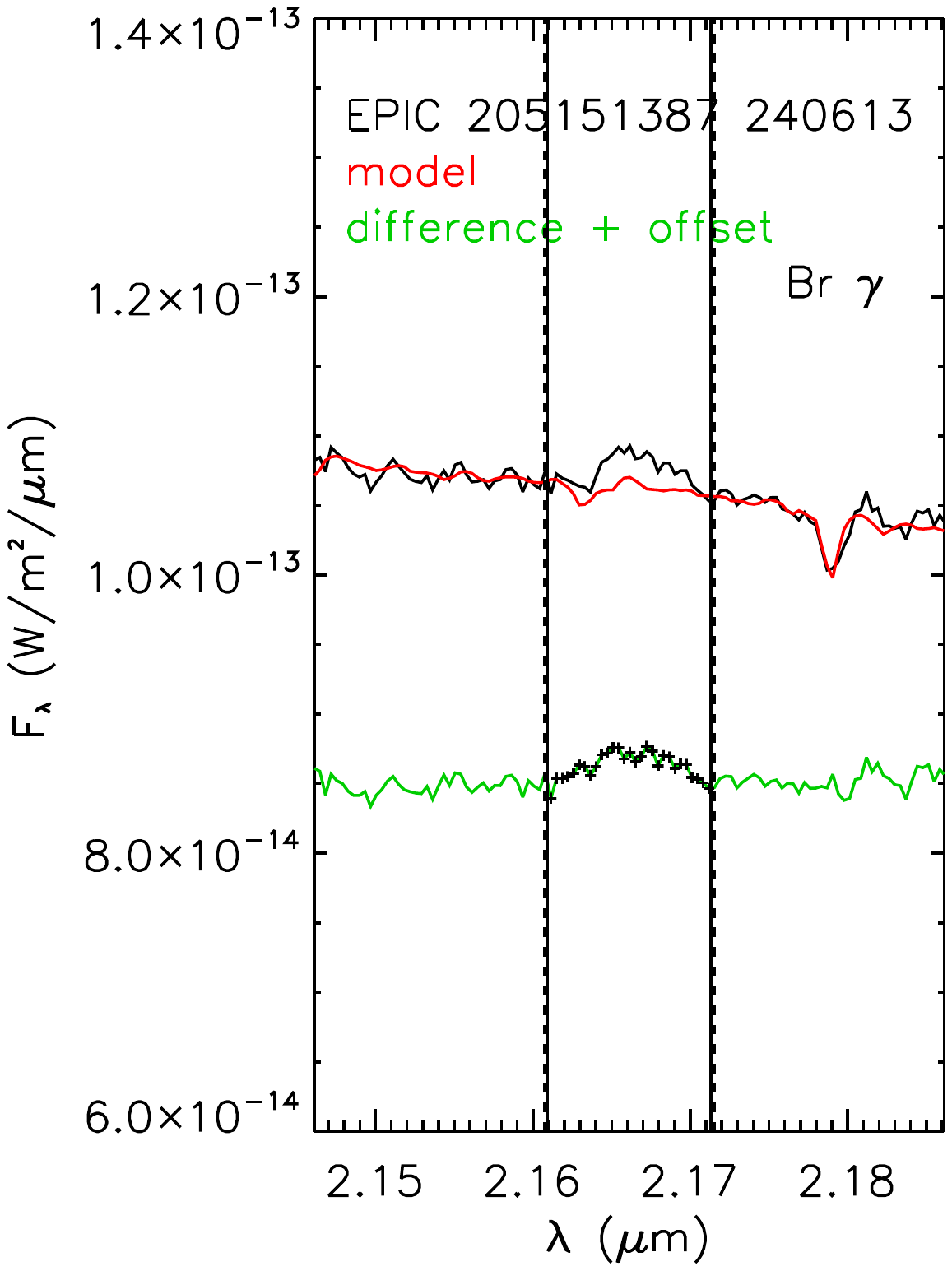}
\caption{The same as Figure A-2, except for EPIC 205151387 on 240613 UT. \label{fig:A-31}}
\end{figure}

%\clearpage

\begin{figure}
\includegraphics[width=6.0cm, height=6.0cm]{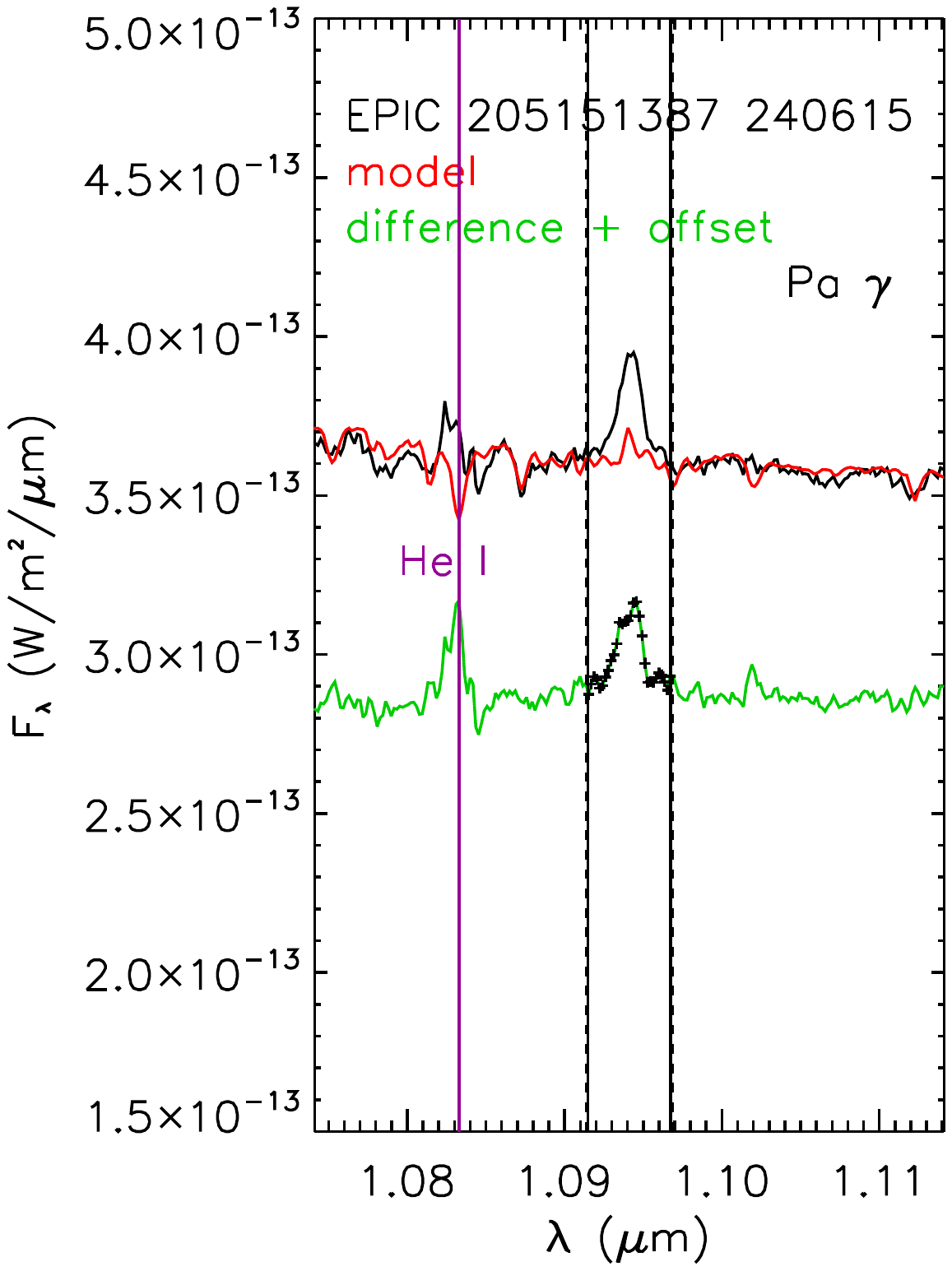}
\includegraphics[width=6.0cm, height=6.0cm]{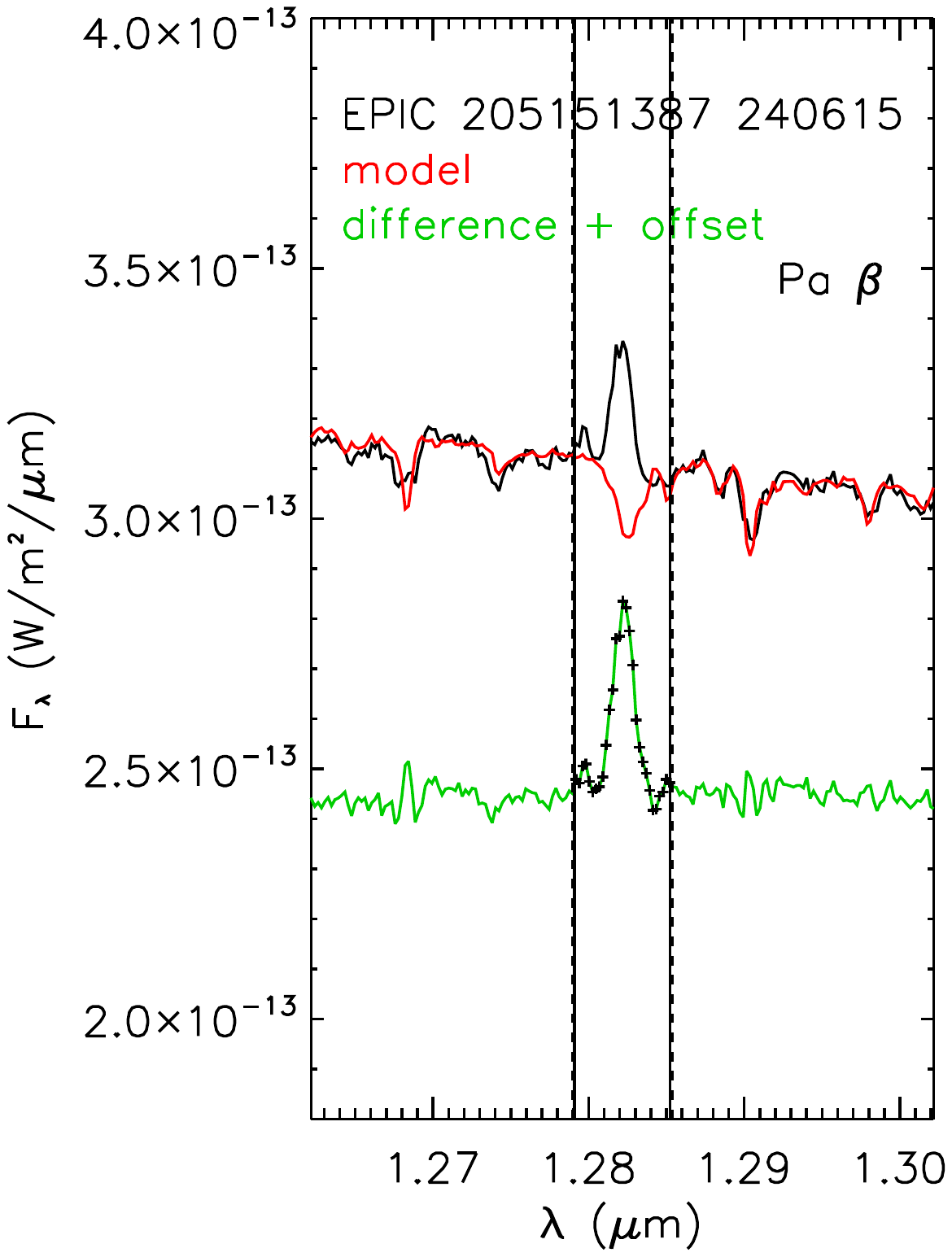}
\includegraphics[width=6.0cm, height=6.0cm]{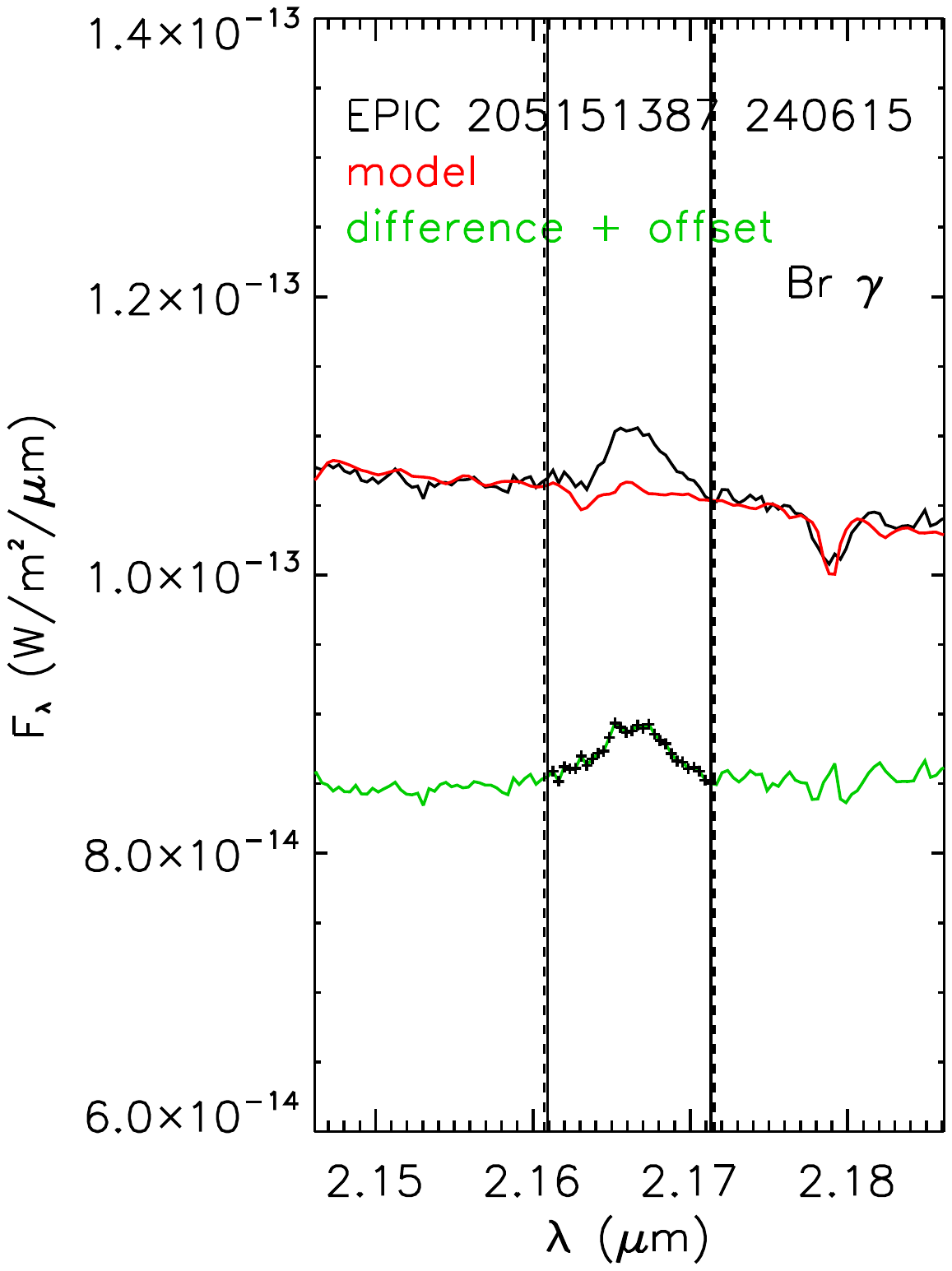}
\caption{The same as Figure A-2, except for EPIC 205151387 on 240615 UT. \label{fig:A-32}}
\end{figure}

\begin{figure}
\includegraphics[width=6.0cm, height=6.0cm]{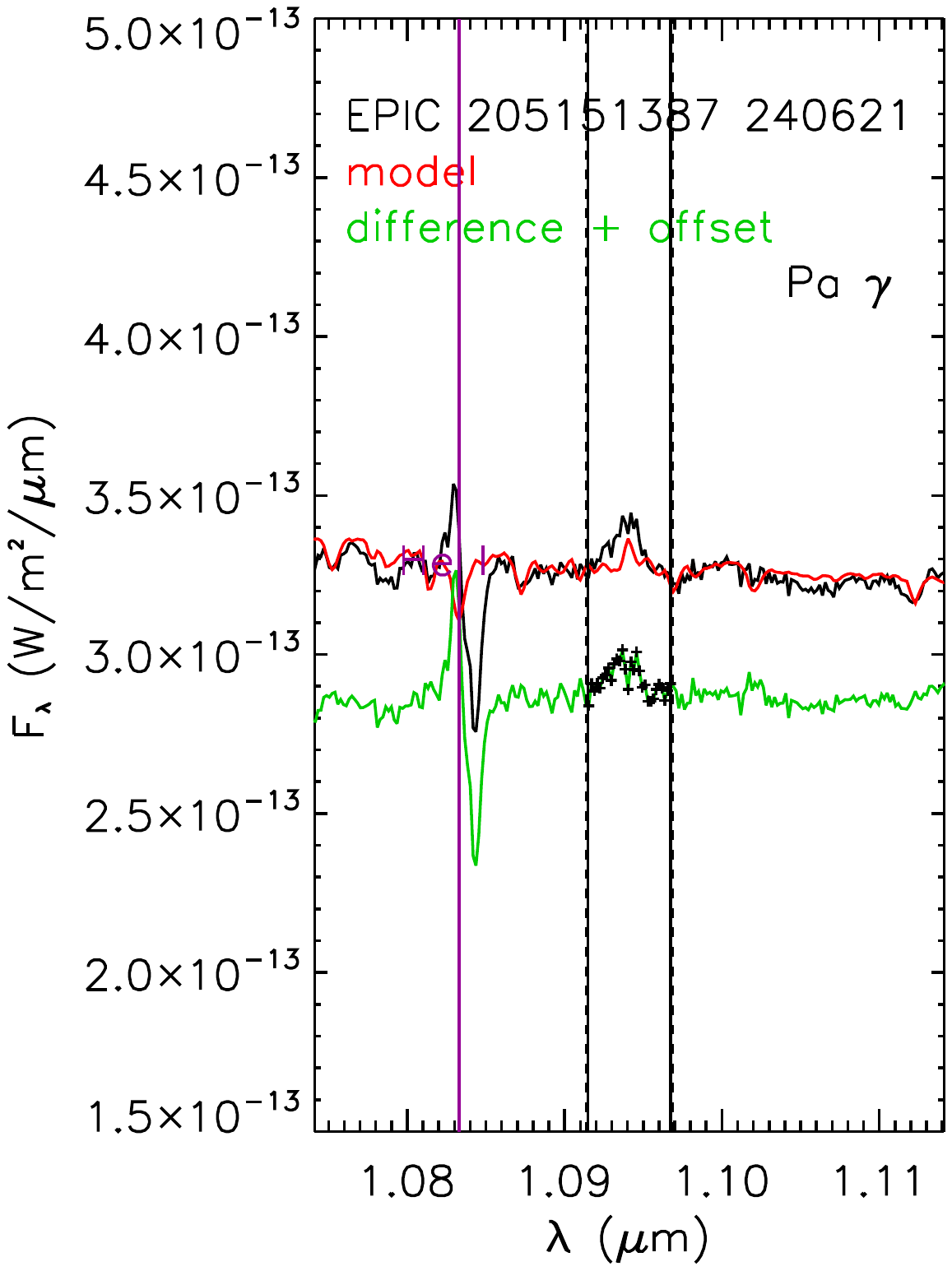}
\includegraphics[width=6.0cm, height=6.0cm]{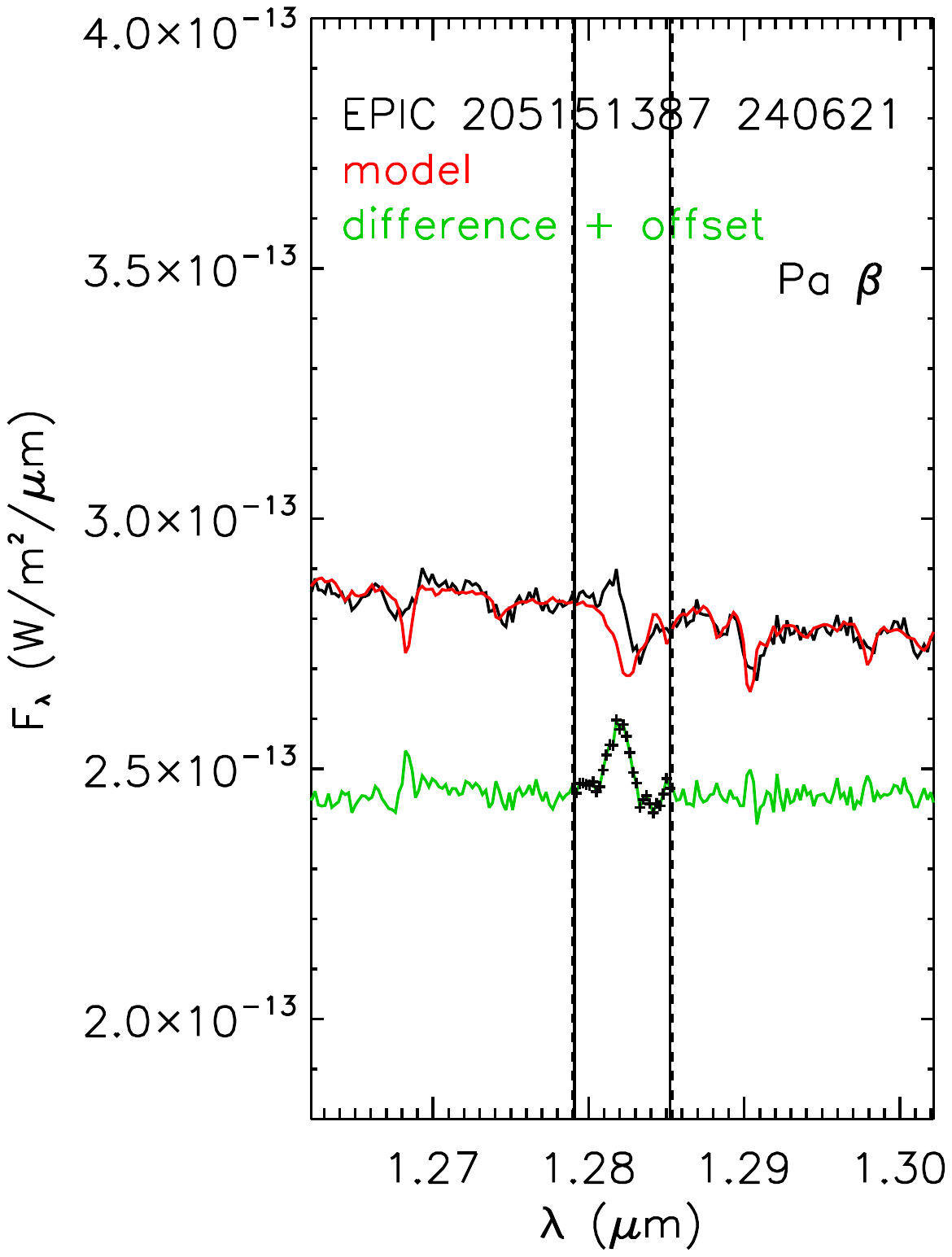}
\includegraphics[width=6.0cm, height=6.0cm]{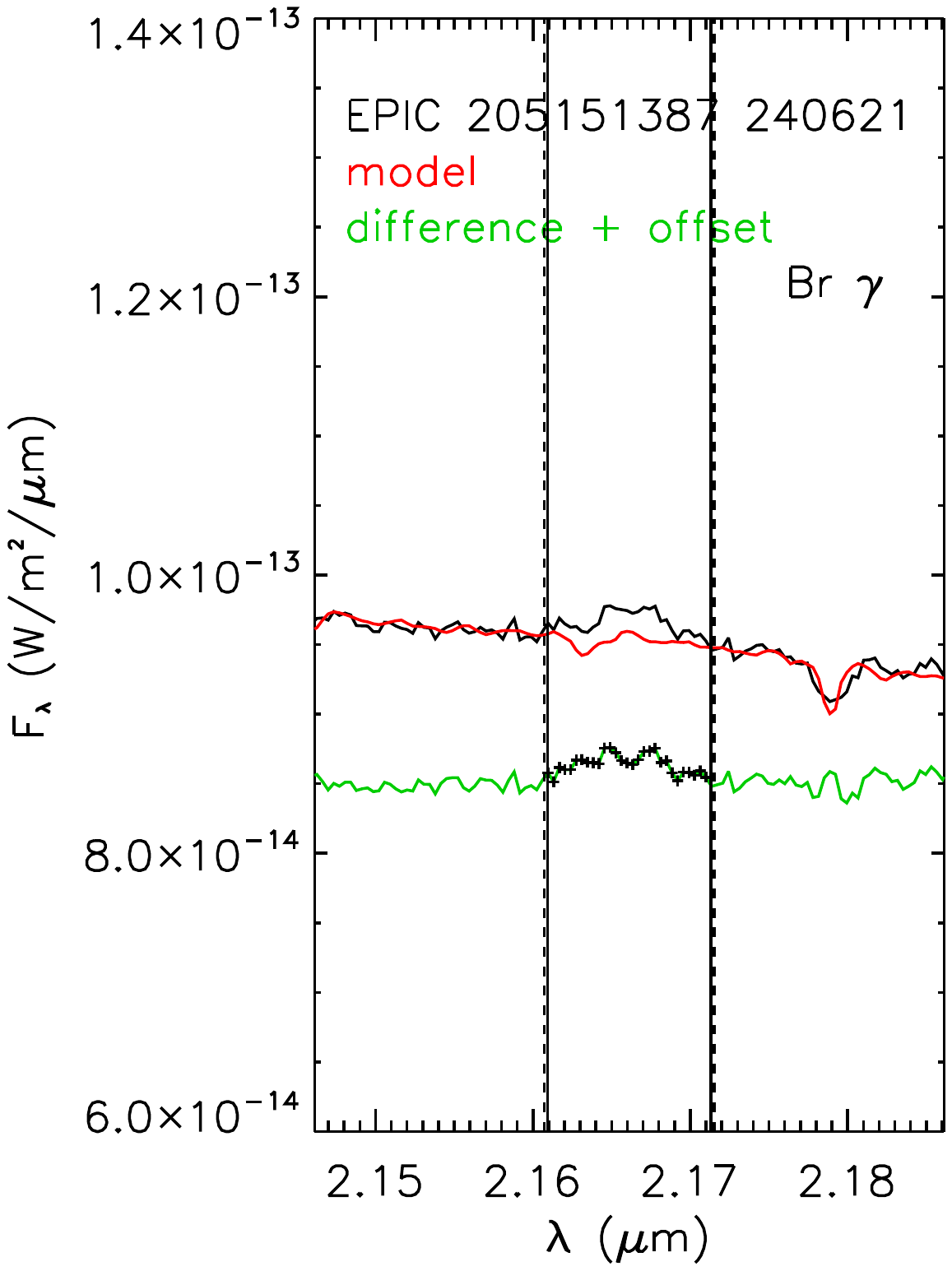}
\caption{The same as Figure A-2, except for EPIC 205151387 on 240621 UT. \label{fig:A-33}}
\end{figure}

\begin{figure}
\includegraphics[width=6.0cm, height=6.0cm]{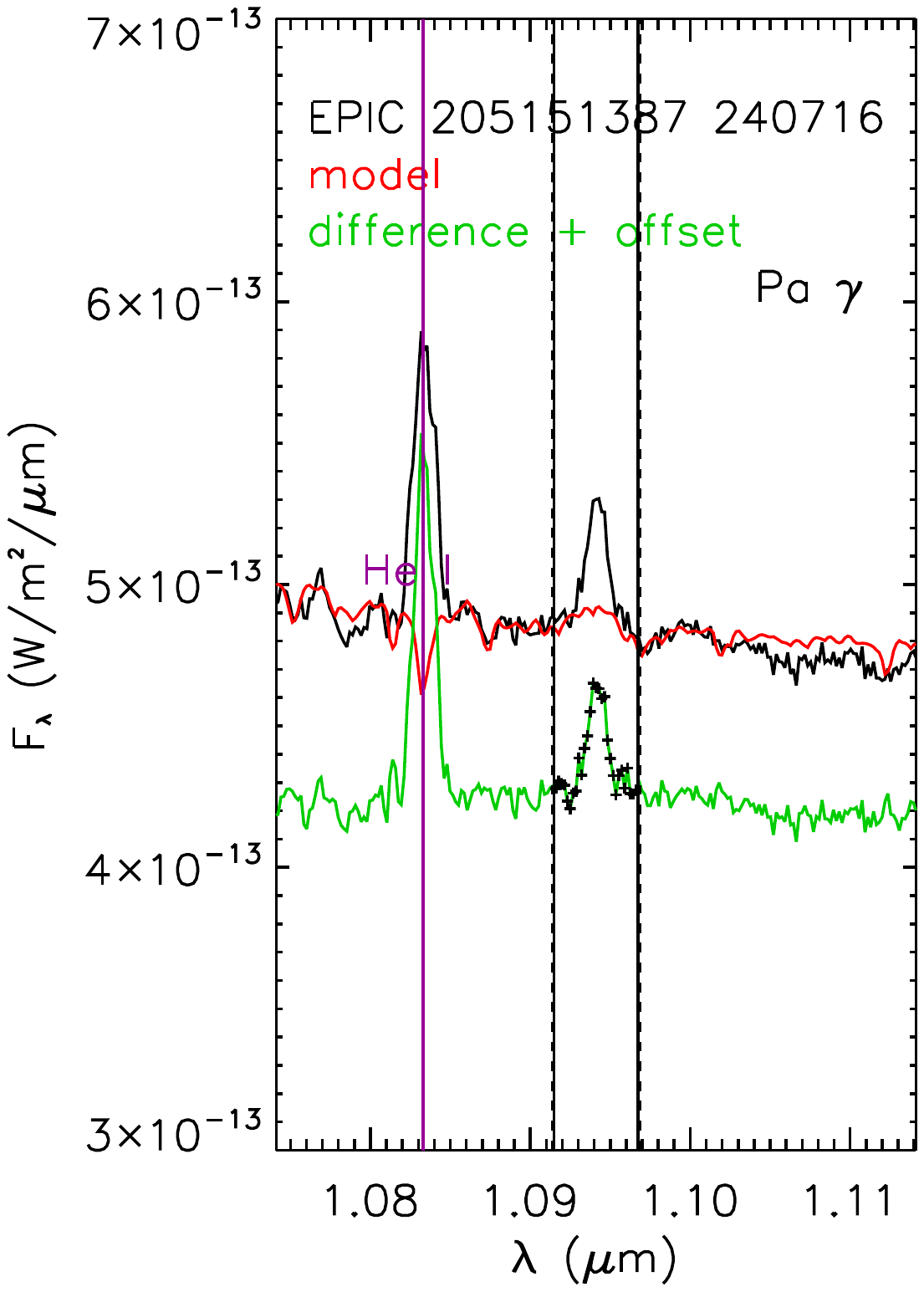}
\includegraphics[width=6.0cm, height=6.0cm]{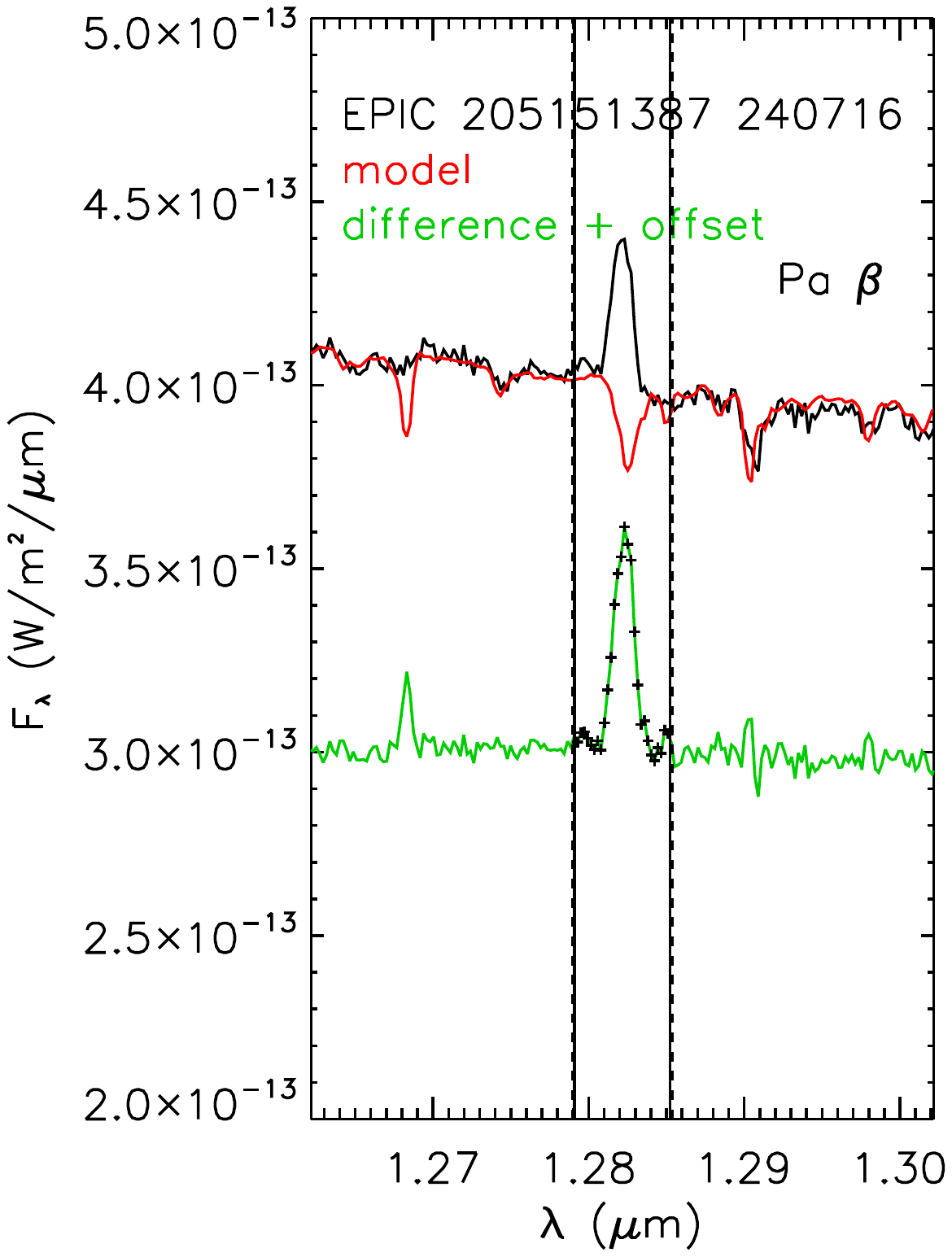}
\includegraphics[width=6.0cm, height=6.0cm]{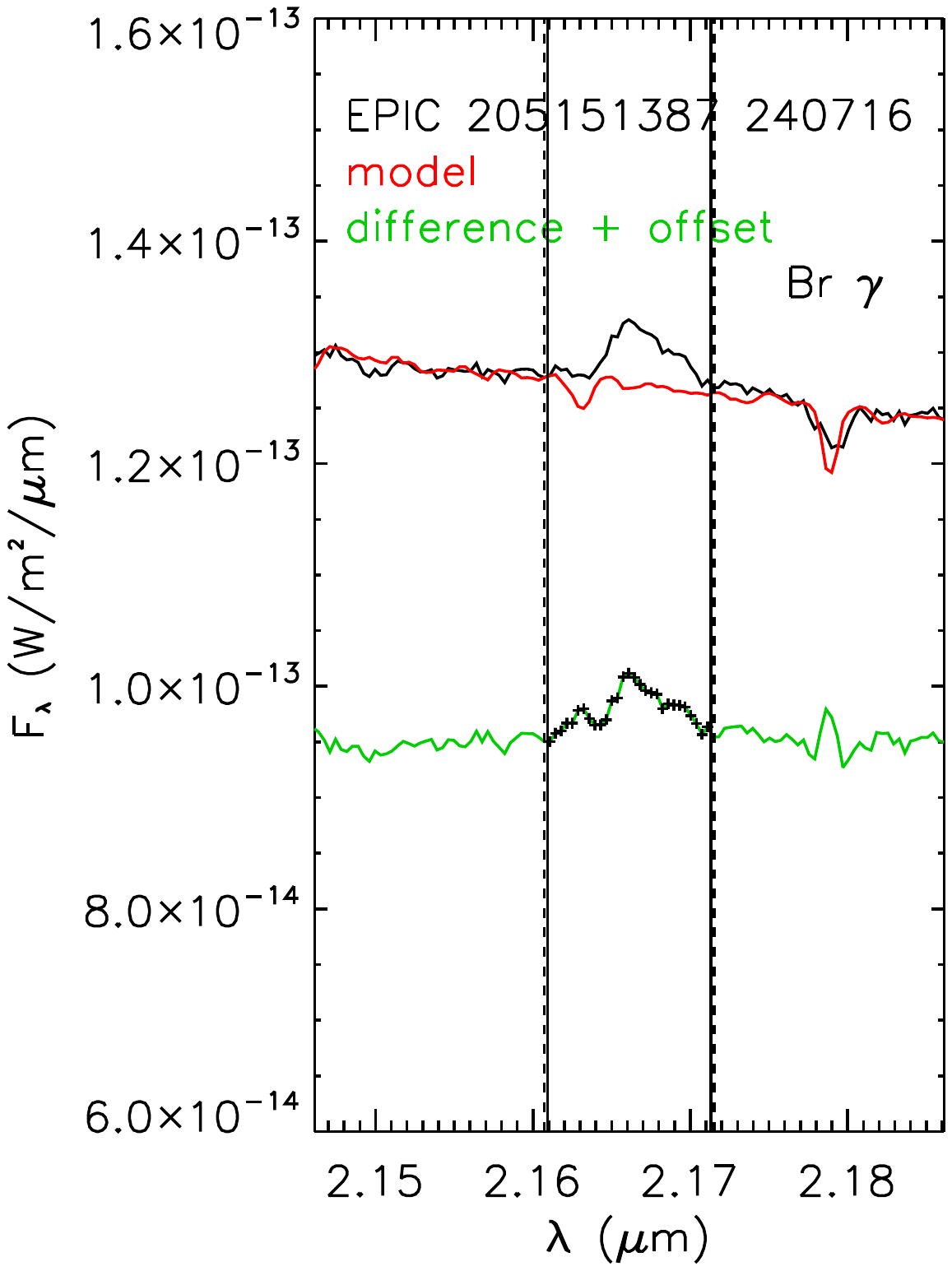}
\caption{The same as Figure A-2, except for EPIC 205151387 on 240716 UT. \label{fig:A-34}}
\end{figure}

\begin{figure}
\includegraphics[width=6.0cm, height=6.0cm]{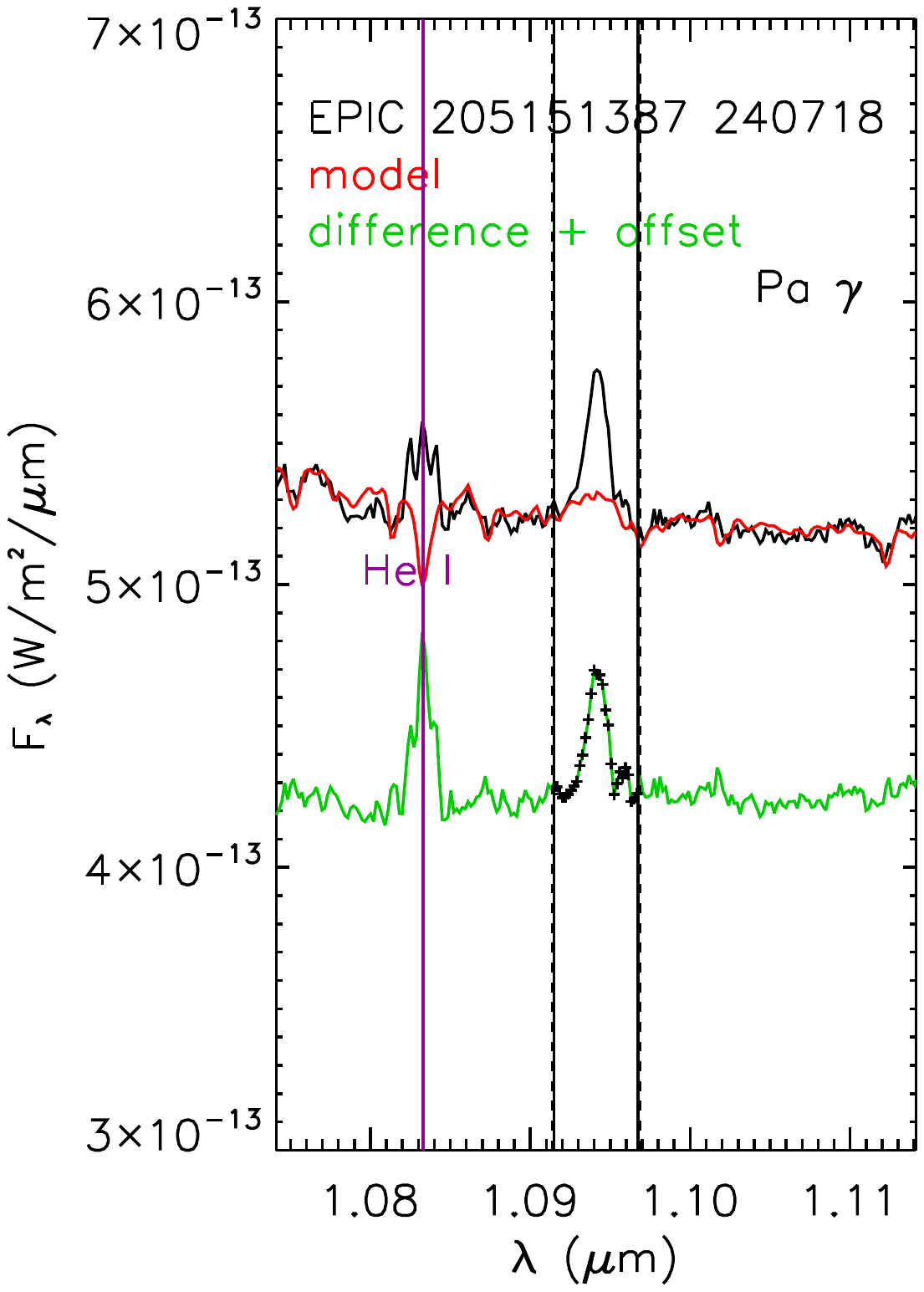}
\includegraphics[width=6.0cm, height=6.0cm]{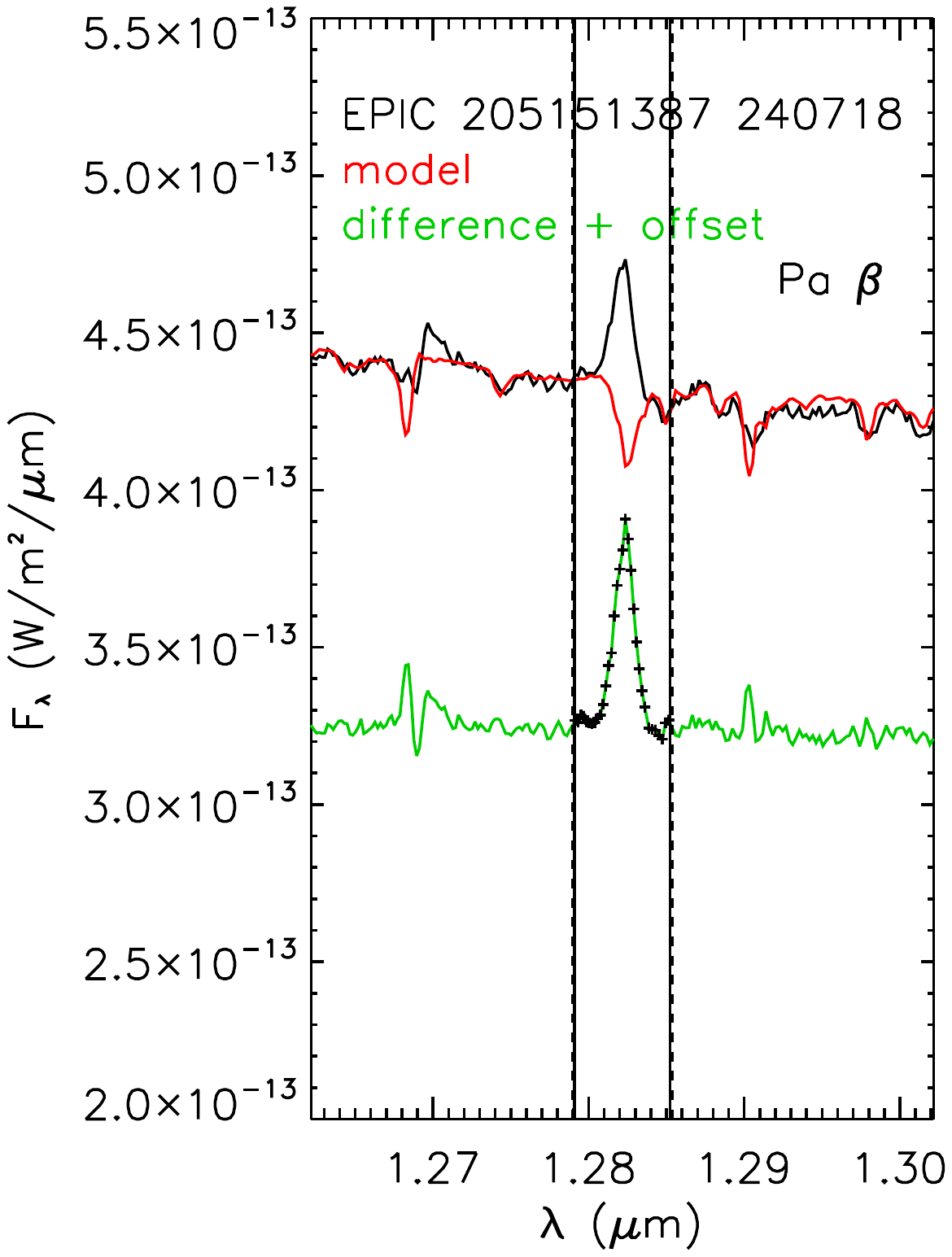}
\includegraphics[width=6.0cm, height=6.0cm]{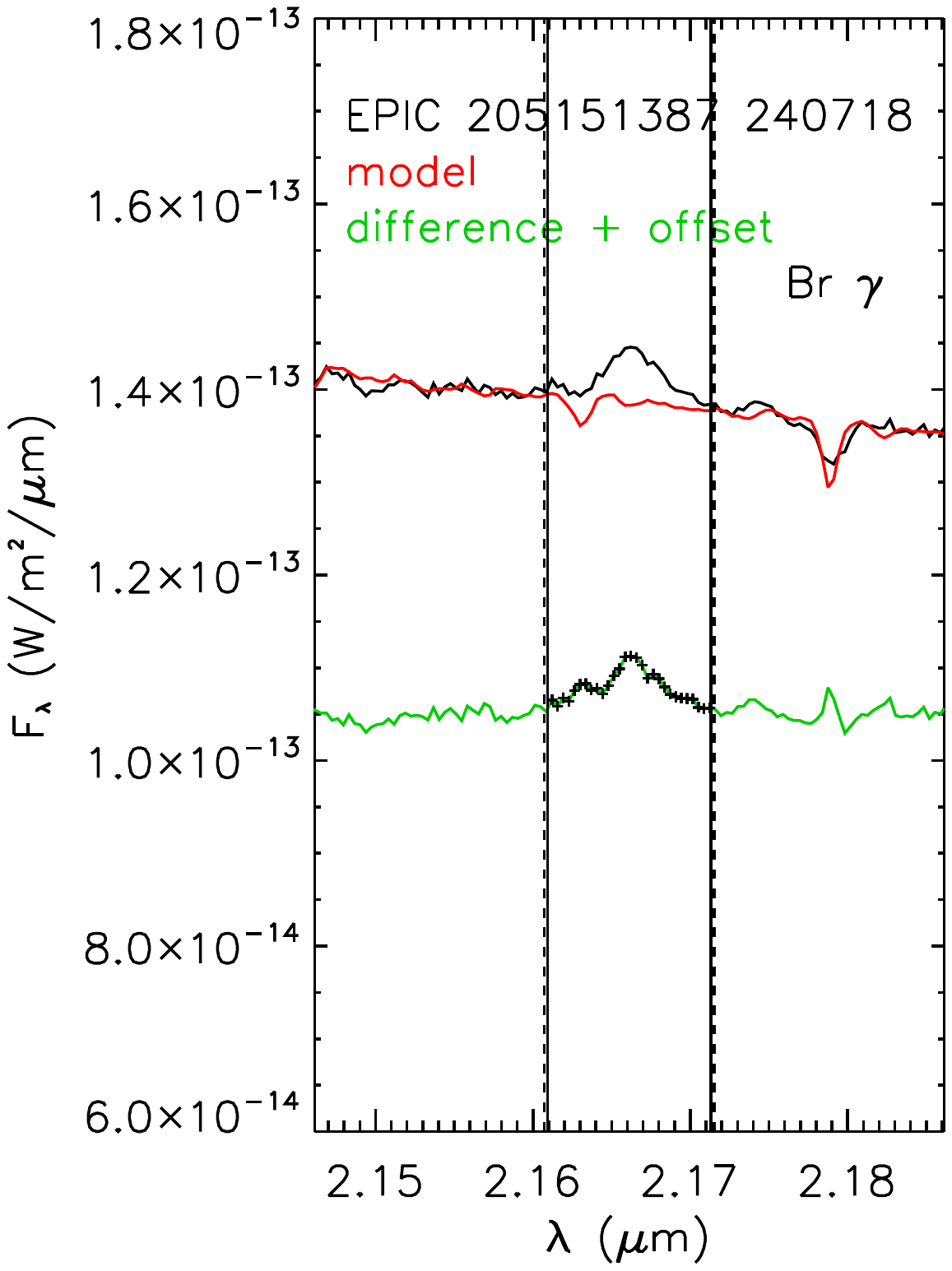}
\caption{The same as Figure A-2, except for EPIC 205151387 on 240718 UT. \label{fig:A-35}}
\end{figure}

\begin{figure}
\includegraphics[width=6.0cm, height=6.0cm]{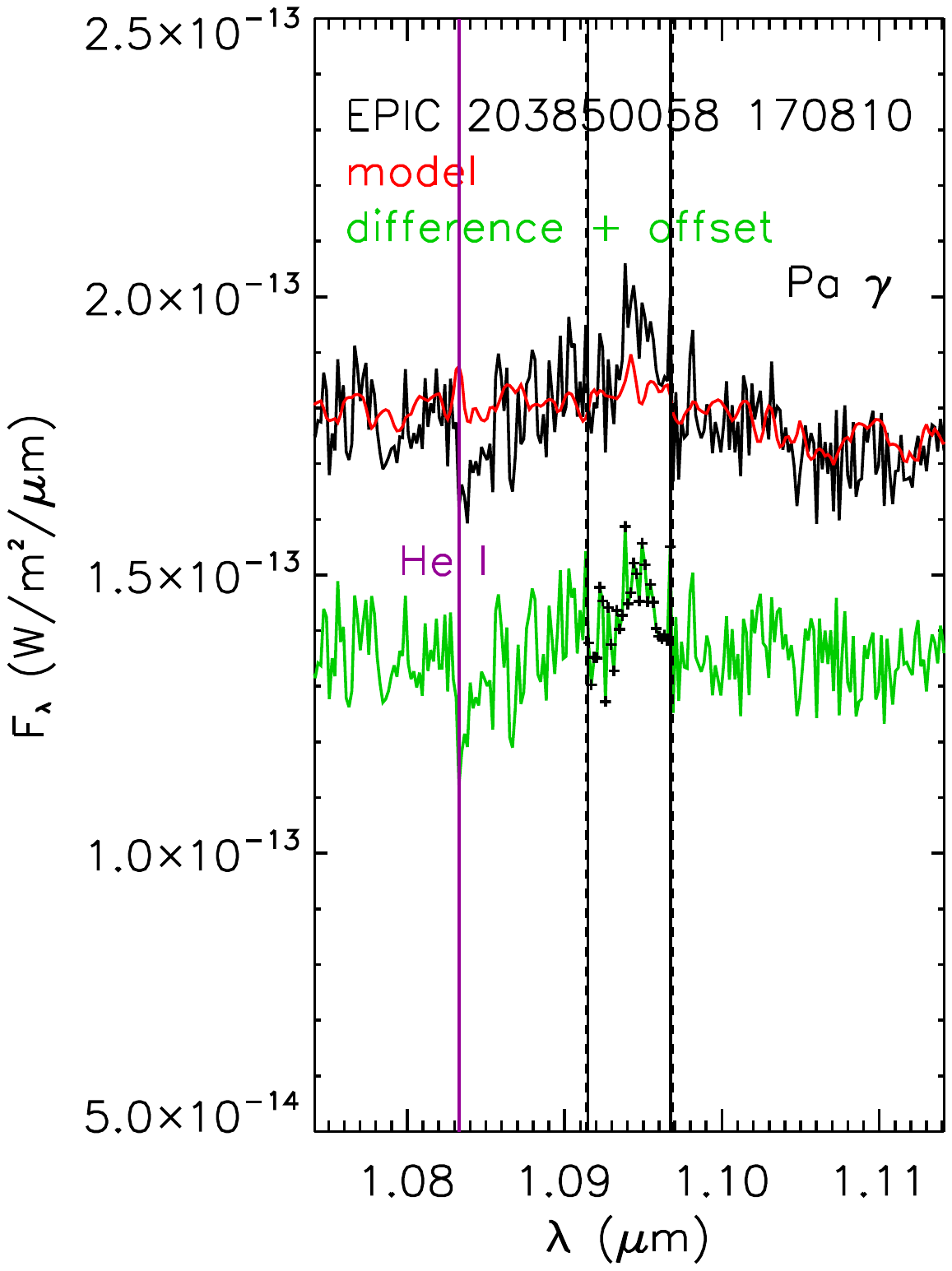}
\includegraphics[width=6.0cm, height=6.0cm]{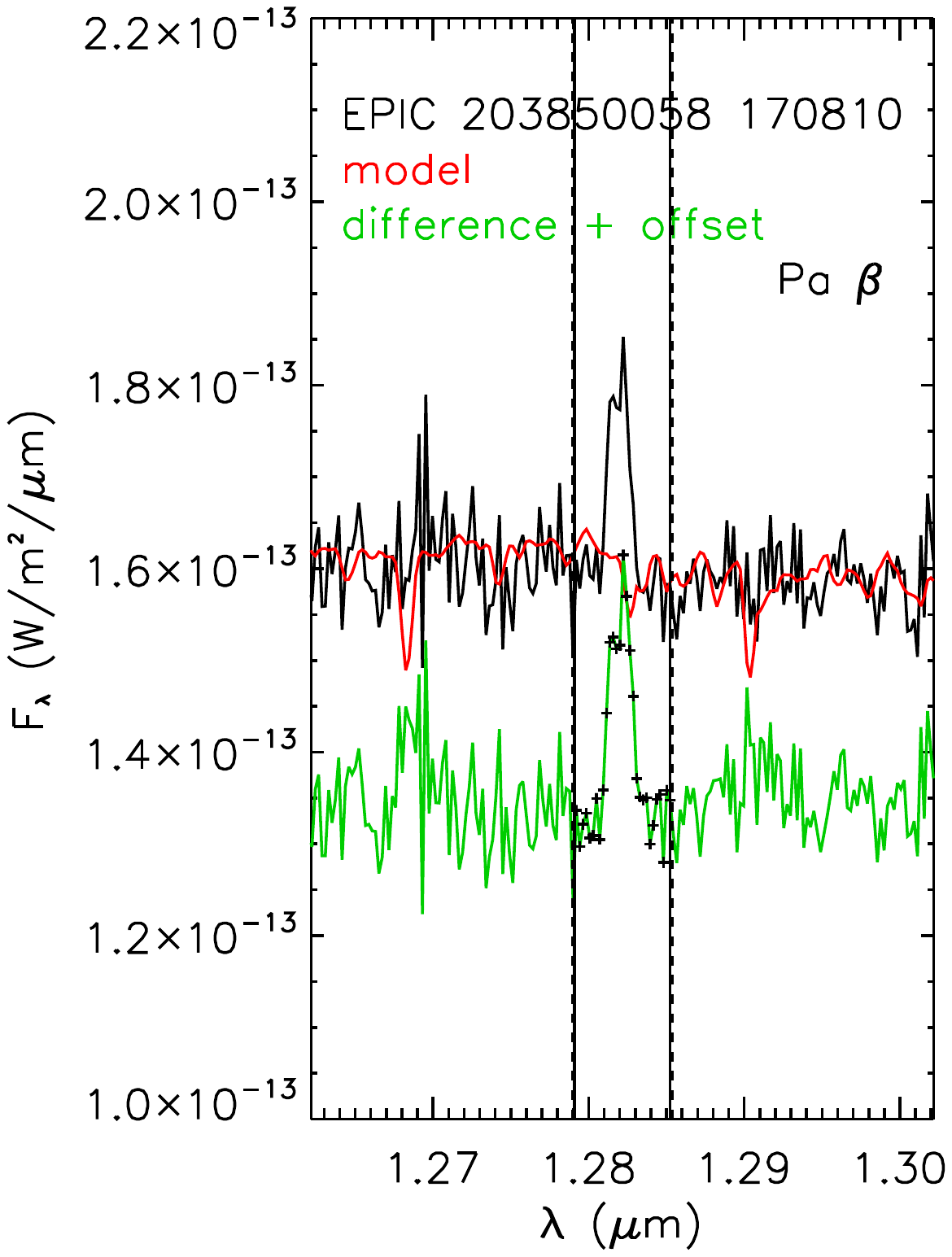}
\includegraphics[width=6.0cm, height=6.0cm]{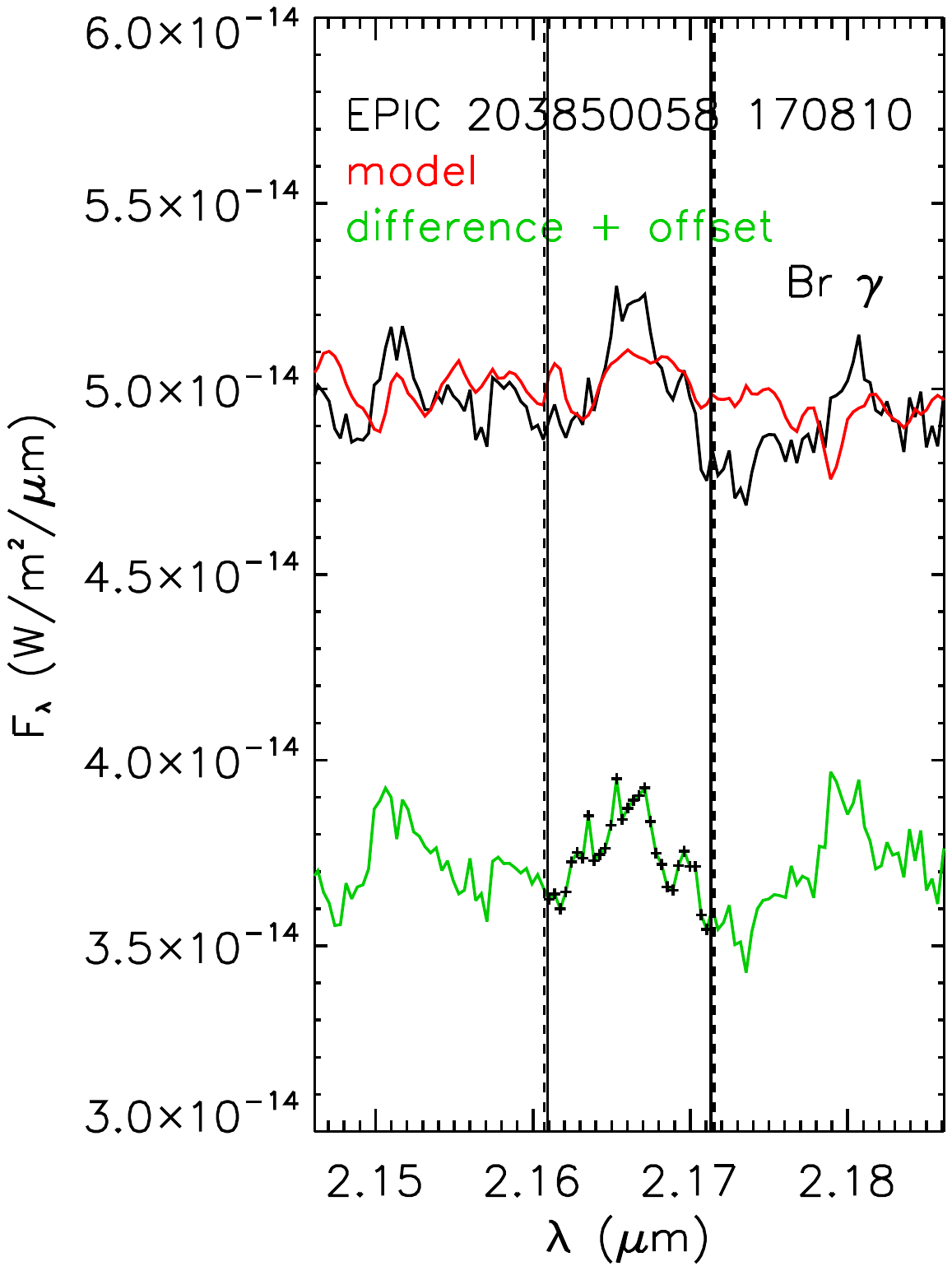}
\caption{The same as Figure A-2, except for EPIC 203850058 on 170810 UT. \label{fig:A-36}}
\end{figure}

\clearpage

\begin{figure}
\includegraphics[width=6.0cm, height=6.0cm]{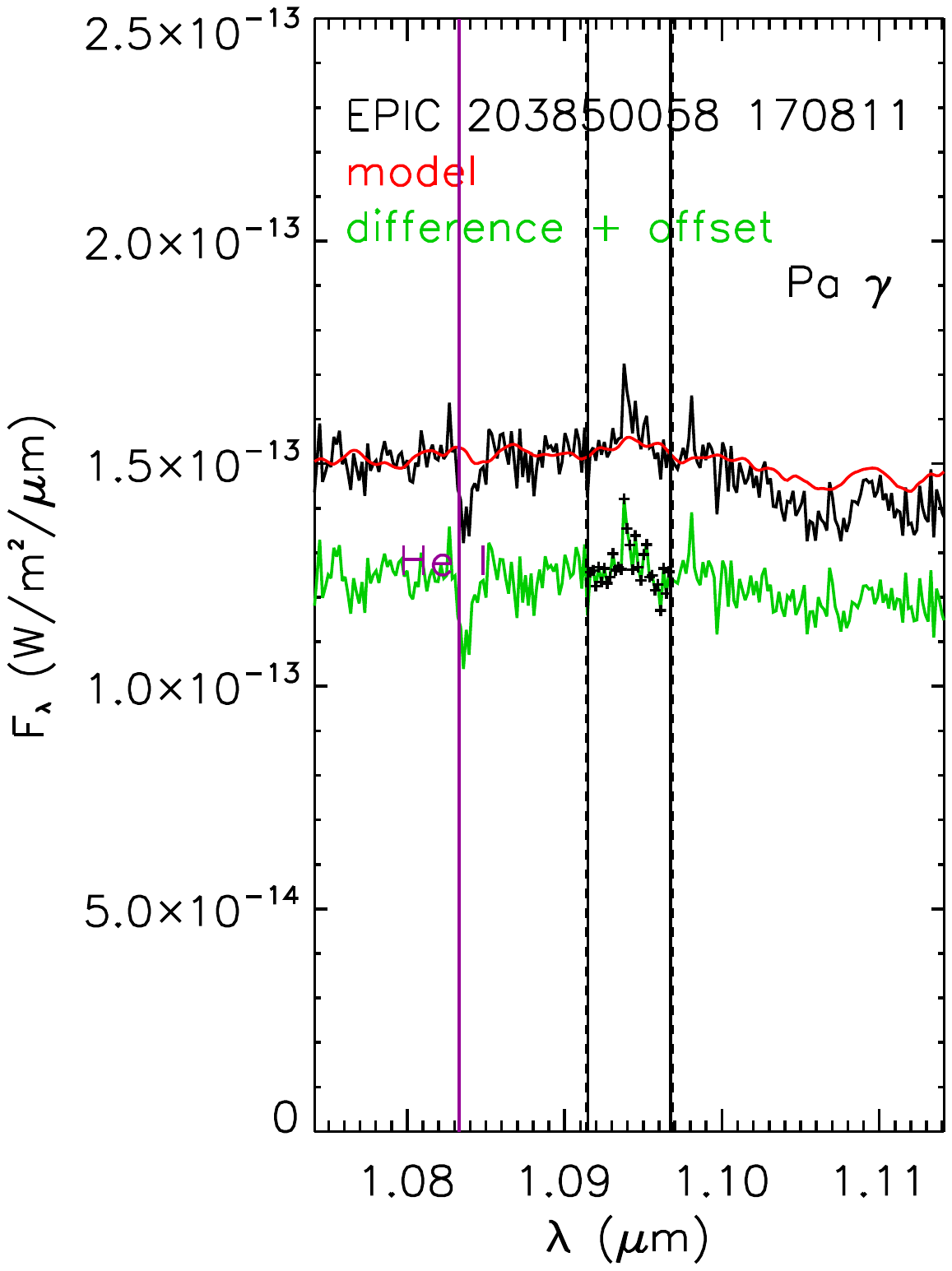}
\includegraphics[width=6.0cm, height=6.0cm]{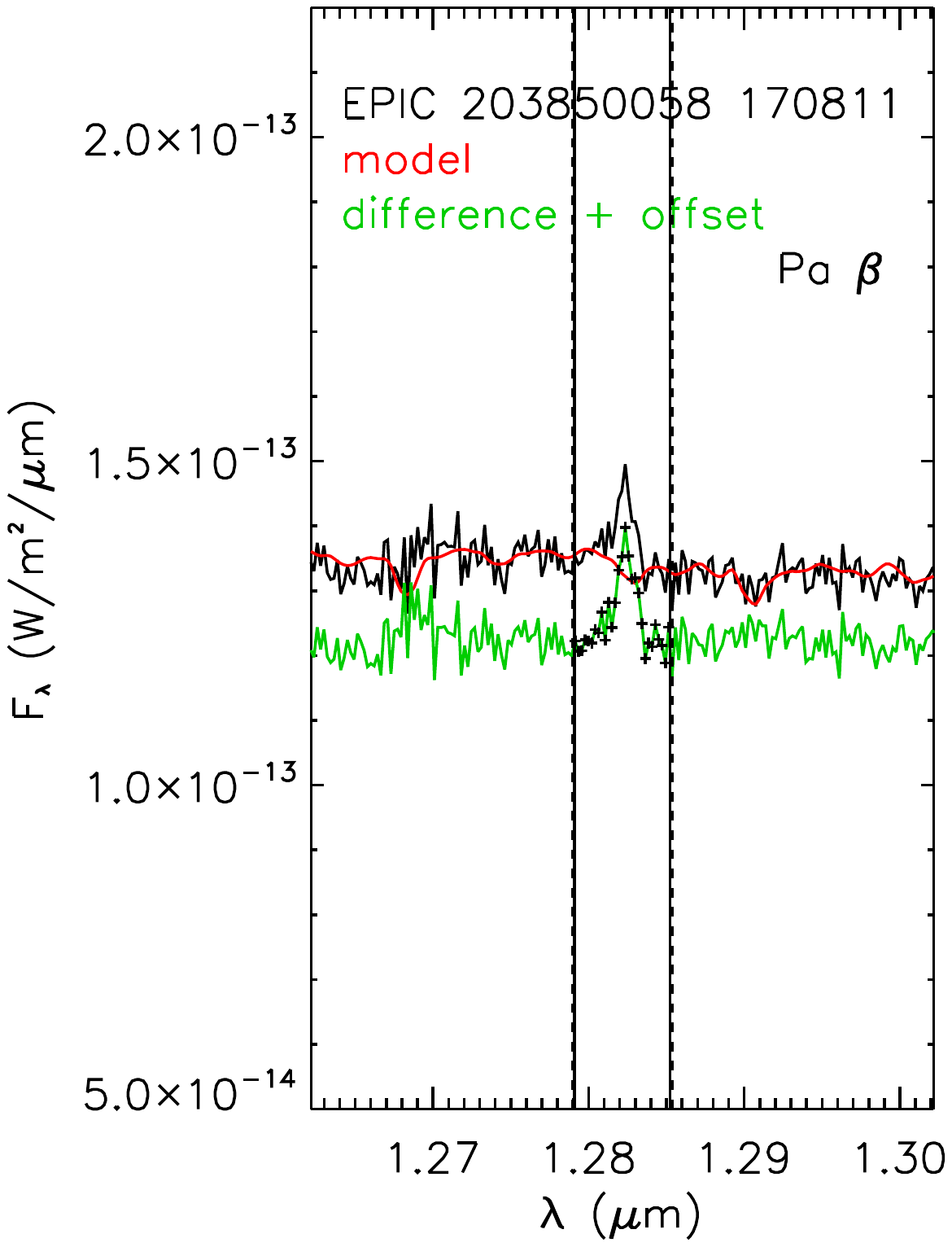}
\includegraphics[width=6.0cm, height=6.0cm]{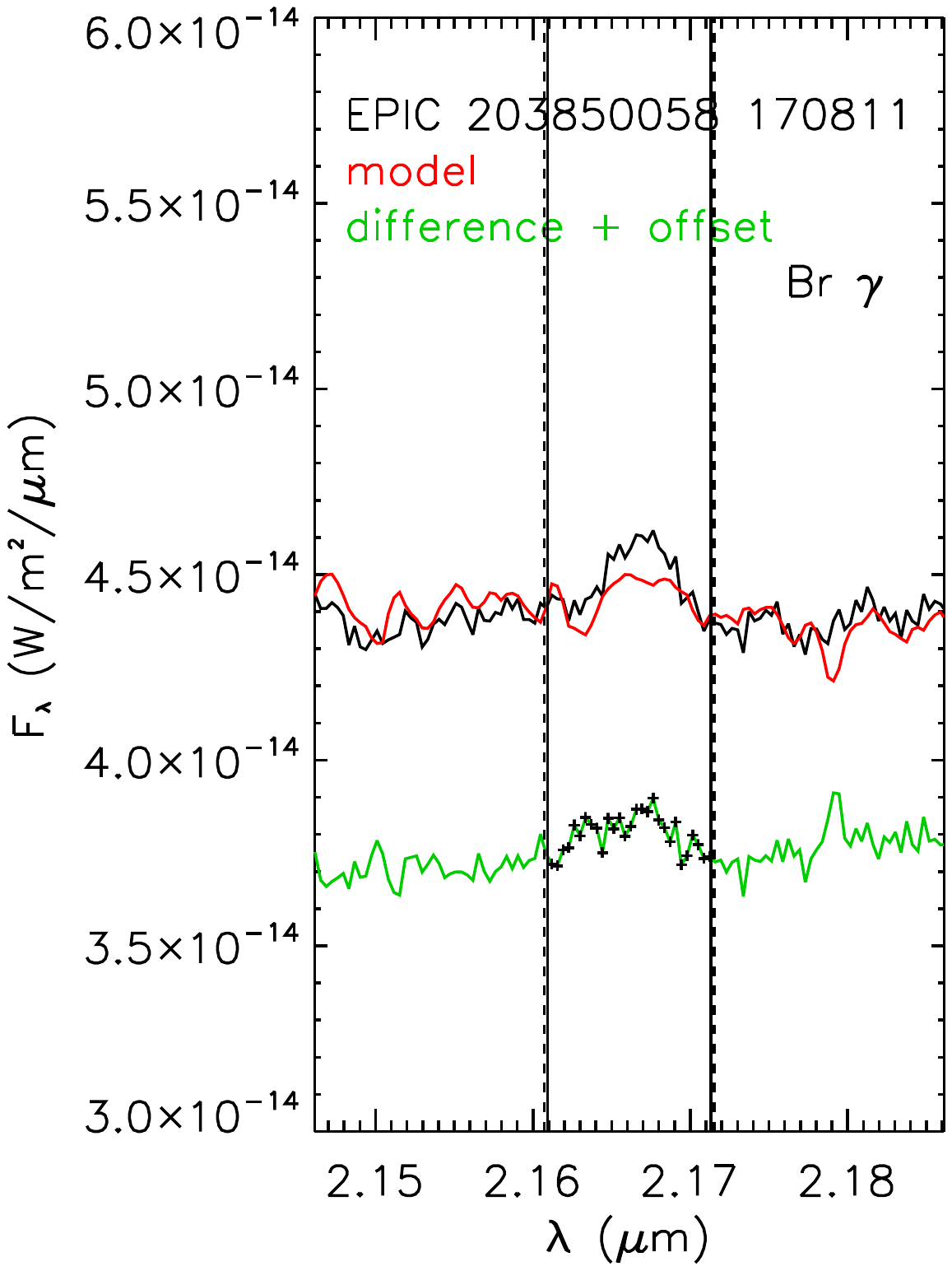}
\caption{The same as Figure A-2, except for EPIC 203850058 on 170811 UT. \label{fig:A-37}}
\end{figure}

\begin{figure}
\includegraphics[width=6.0cm, height=6.0cm]{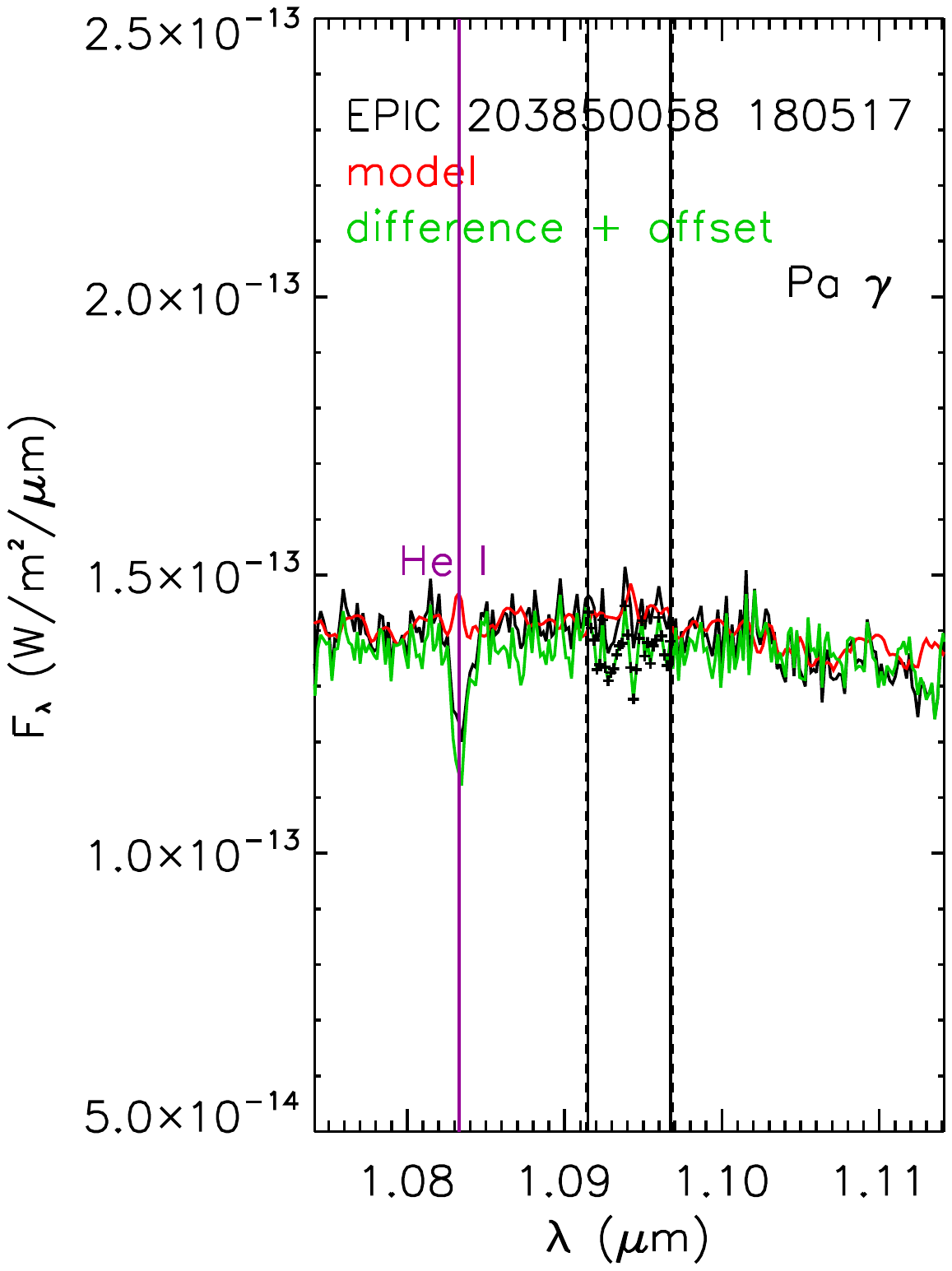}
\includegraphics[width=6.0cm, height=6.0cm]{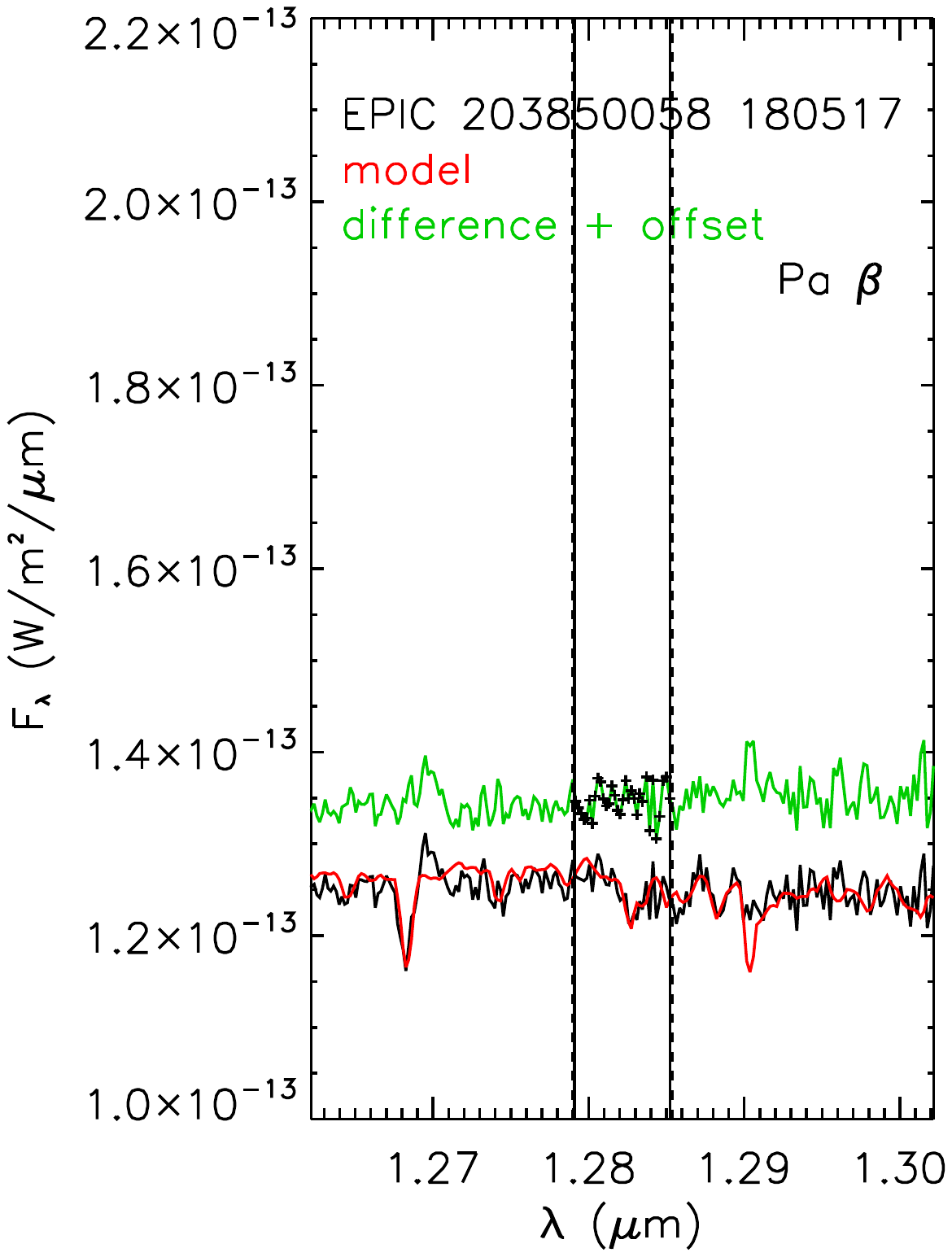}
\includegraphics[width=6.0cm, height=6.0cm]{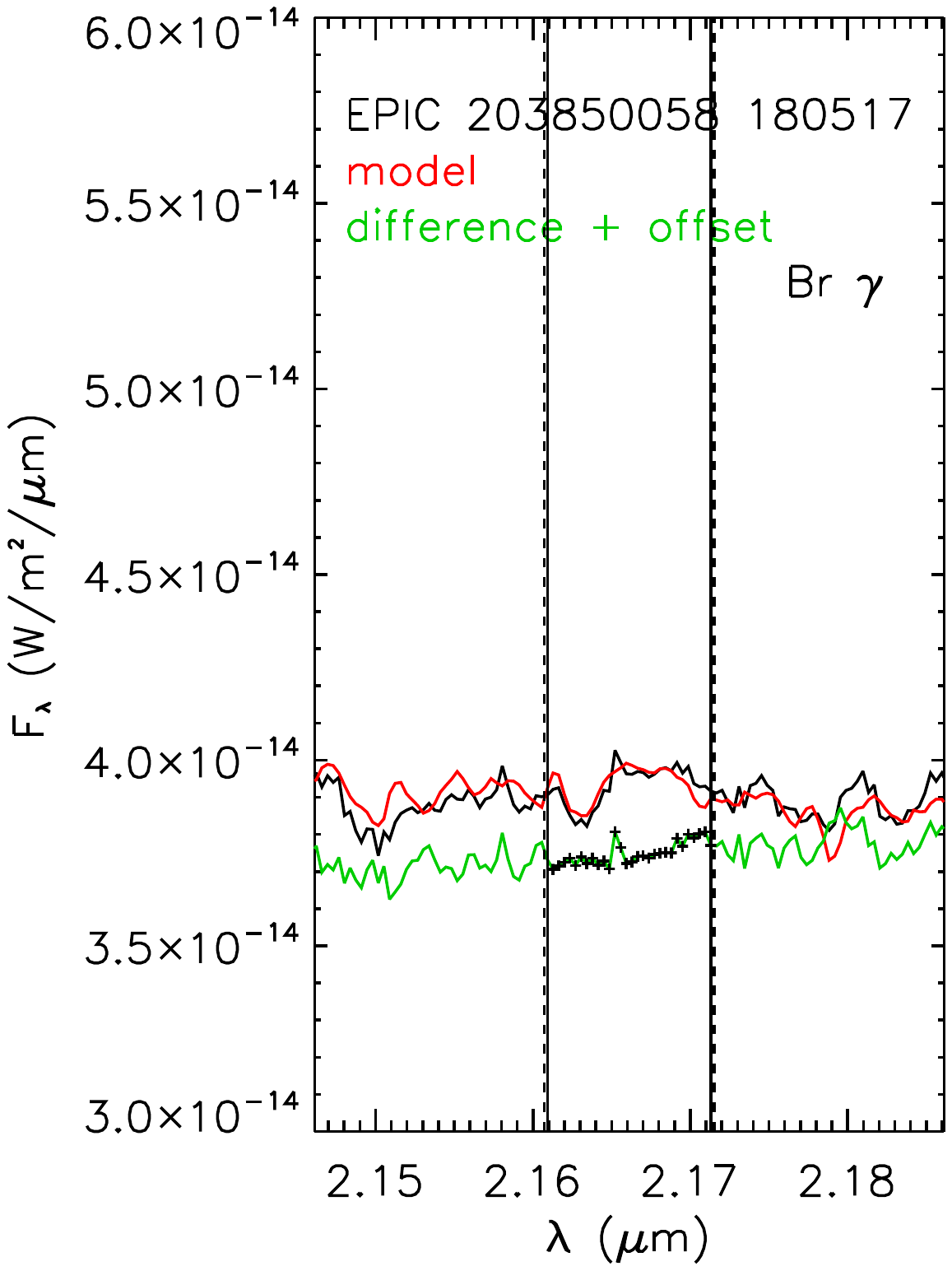}
\caption{The same as Figure A-2, except for EPIC 203850058 on 180517 UT. \label{fig:A-38}}
\end{figure}

\begin{figure}
\includegraphics[width=6.0cm, height=6.0cm]{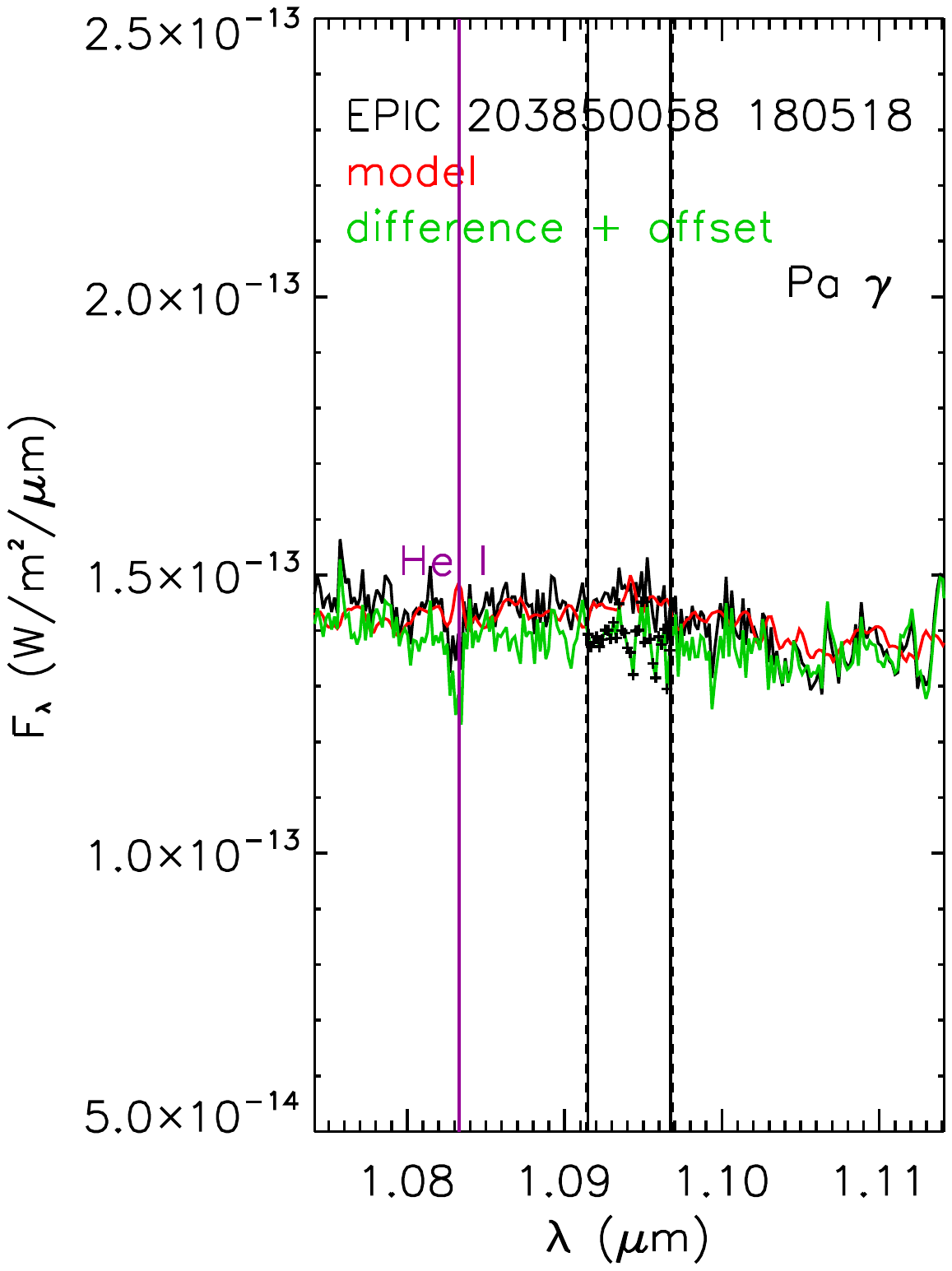}
\includegraphics[width=6.0cm, height=6.0cm]{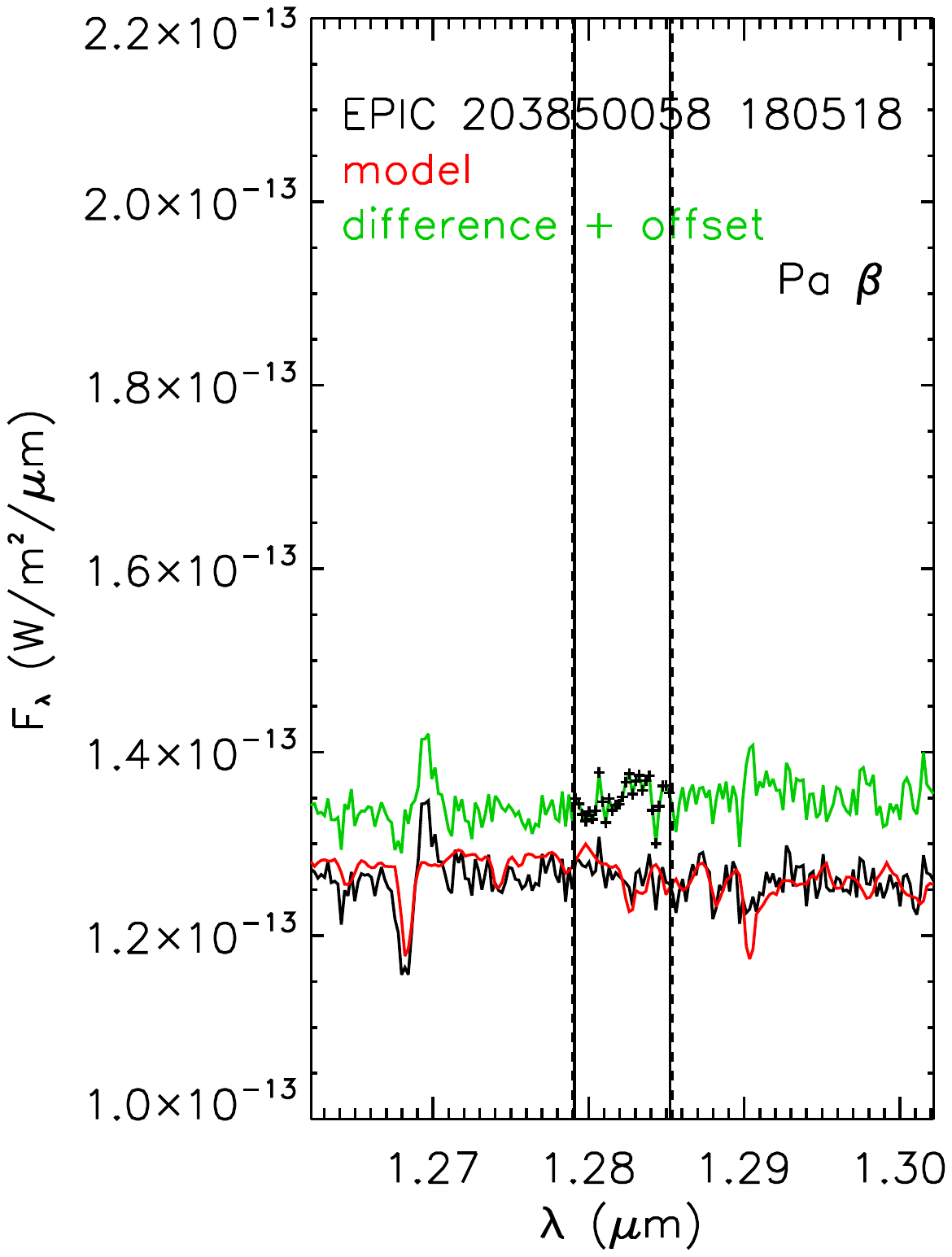}
\includegraphics[width=6.0cm, height=6.0cm]{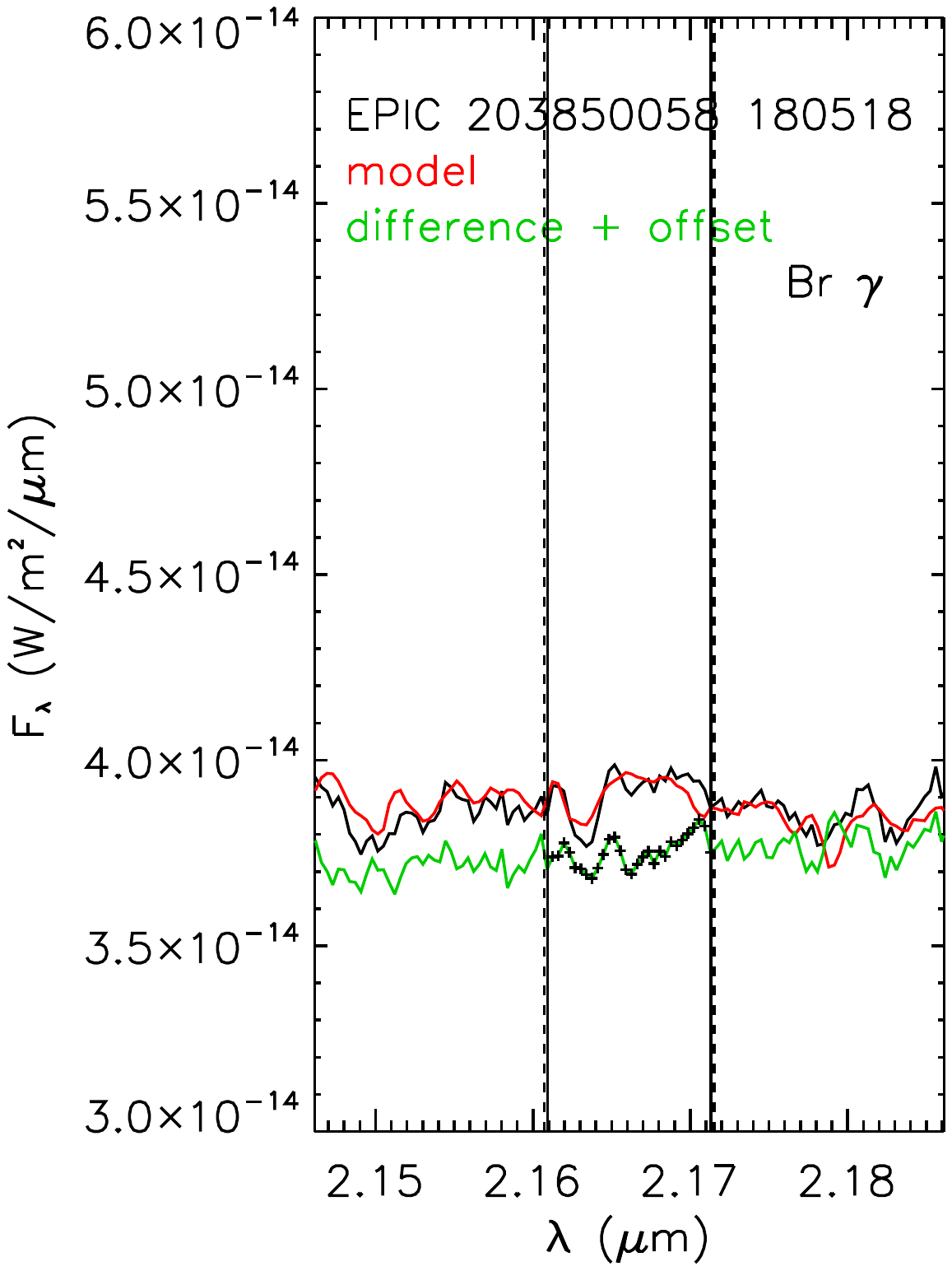}
\caption{The same as Figure A-2, except for EPIC 203850058 on 180518 UT. \label{fig:A-39}}
\end{figure}

\clearpage

\begin{figure}
\includegraphics[width=6.0cm, height=6.0cm]{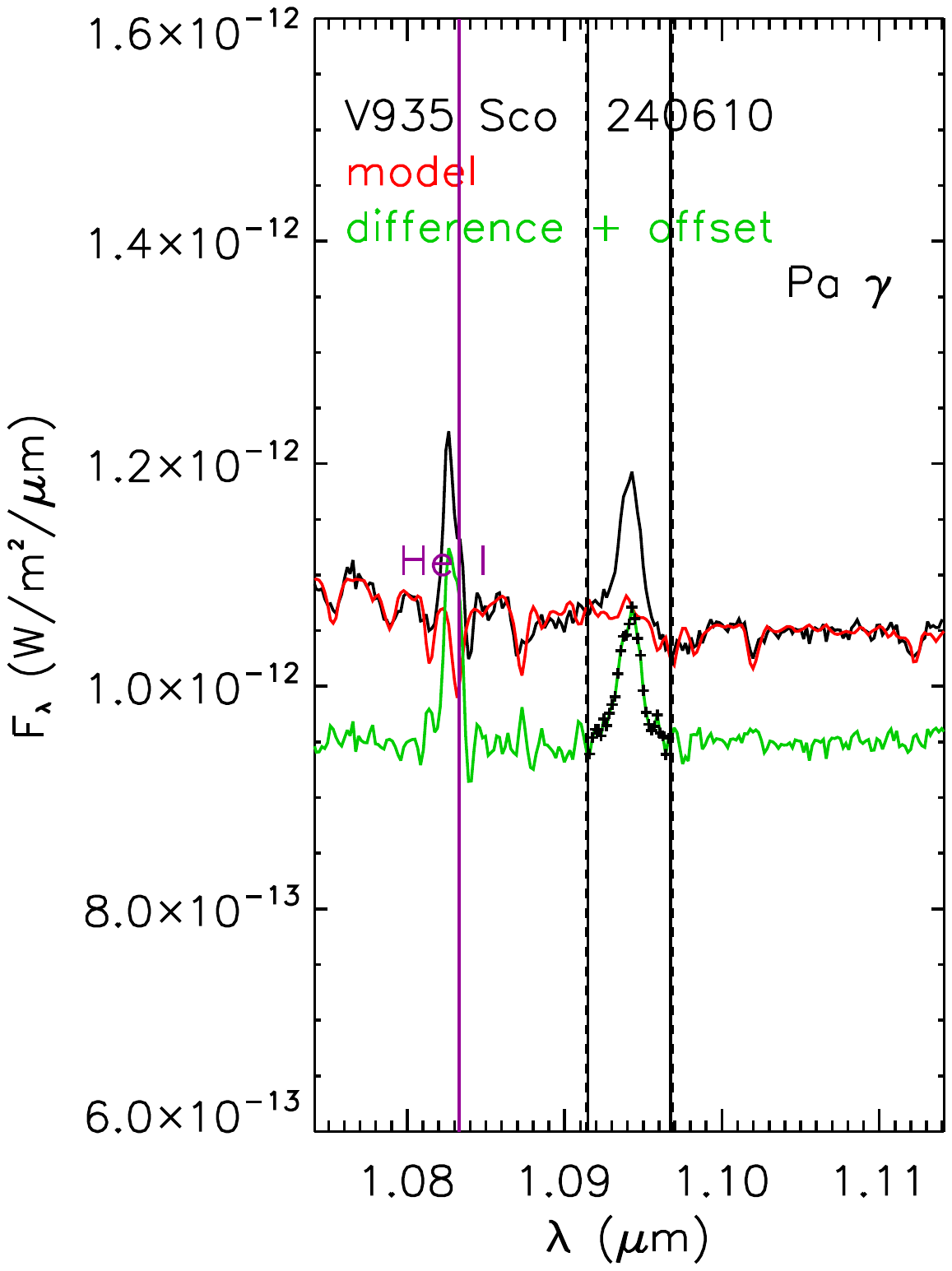}
\includegraphics[width=6.0cm, height=6.0cm]{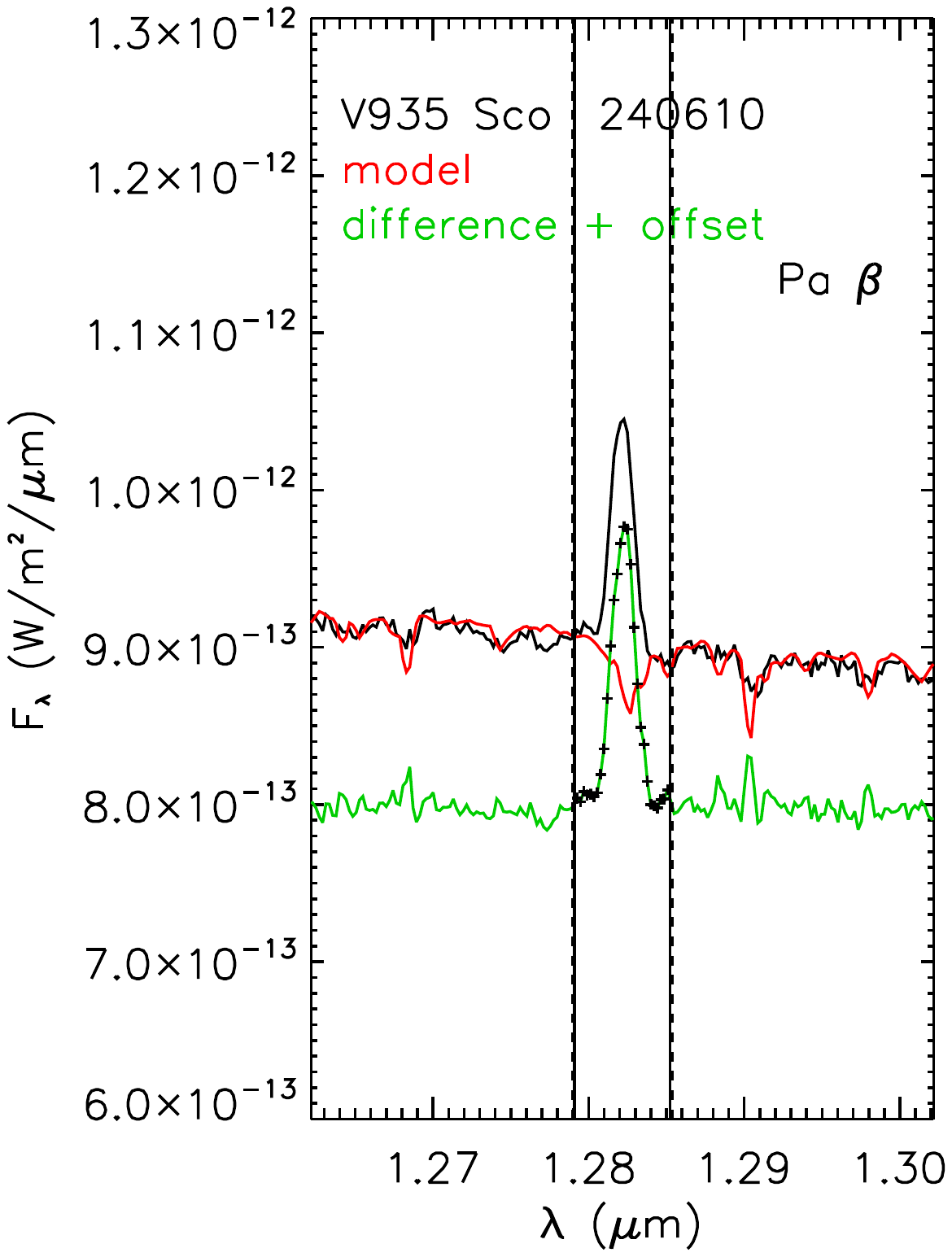}
\includegraphics[width=6.0cm, height=6.0cm]{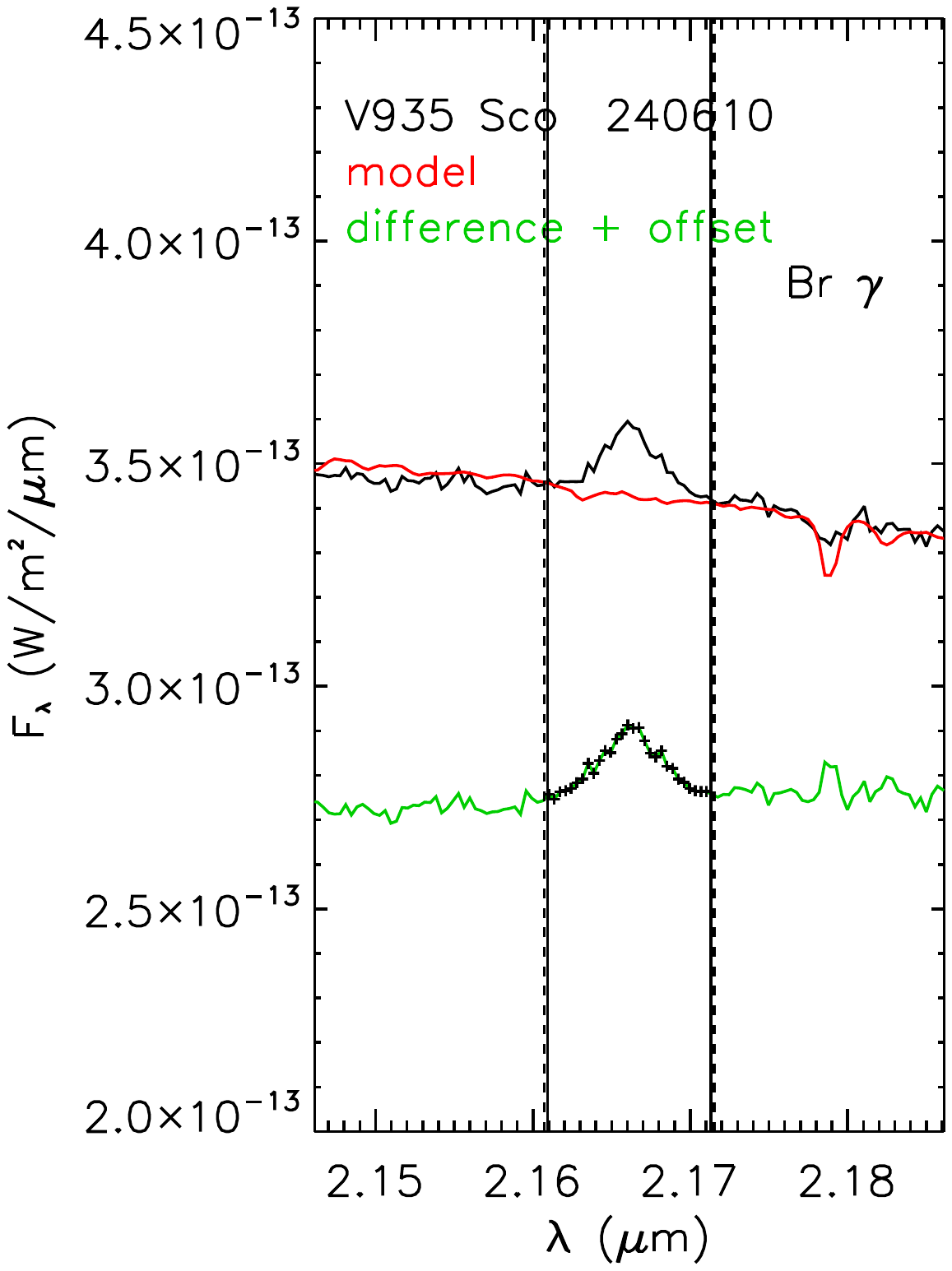}
\caption{The same as Figure A-2, except for V935 Sco on 240610 UT. \label{fig:A-42}}
\end{figure}

\begin{figure}
\includegraphics[width=6.0cm, height=6.0cm]{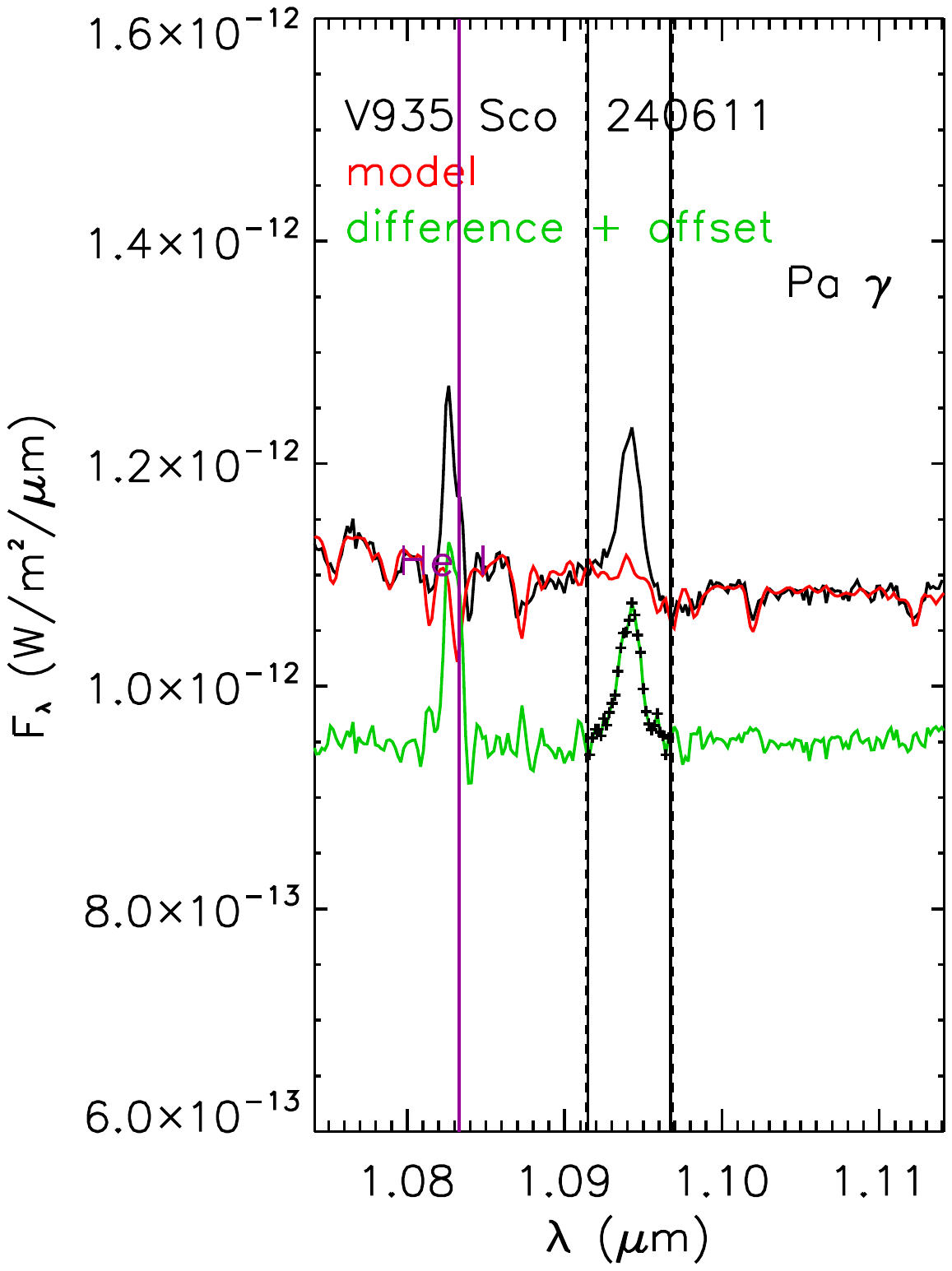}
\includegraphics[width=6.0cm, height=6.0cm]{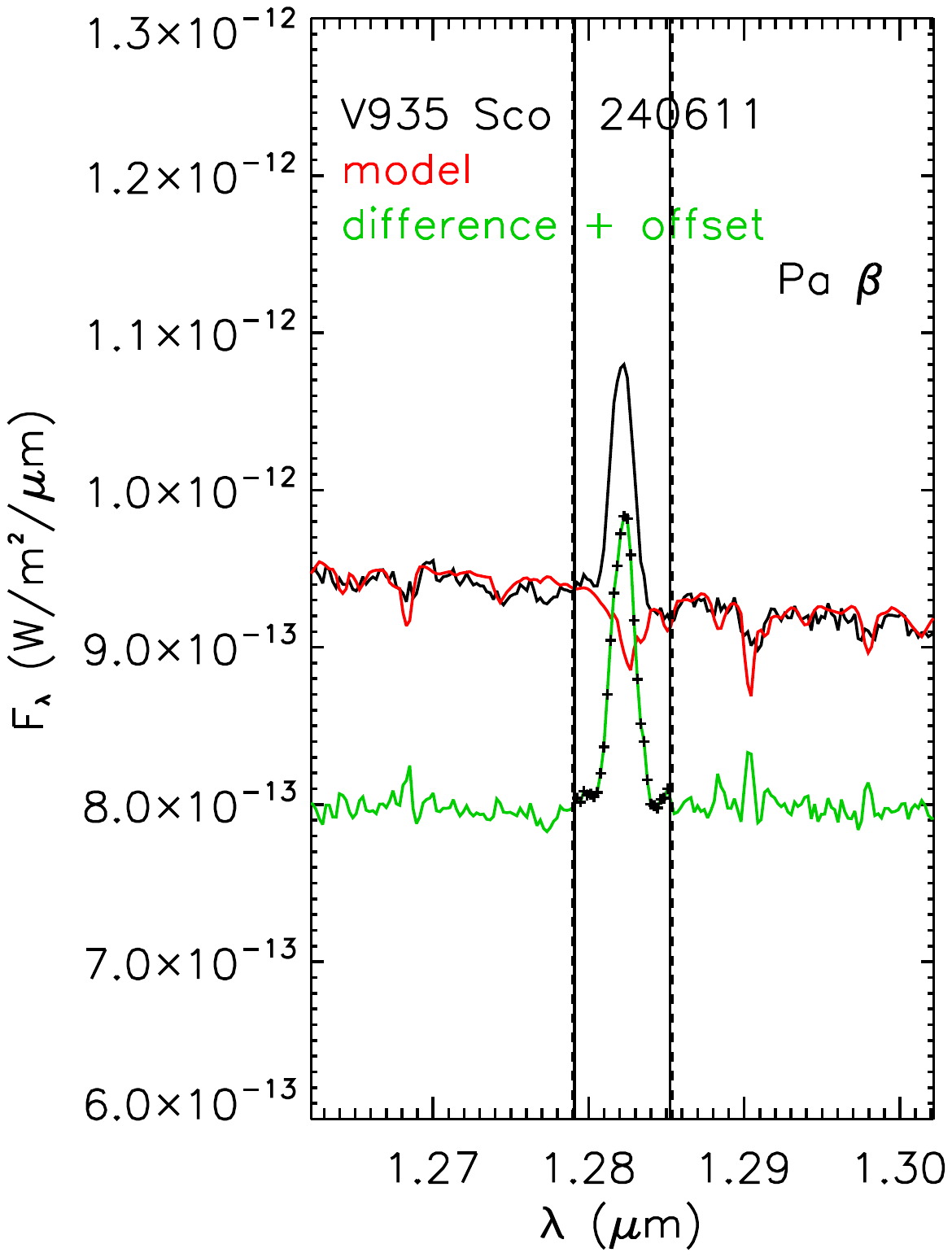}
\includegraphics[width=6.0cm, height=6.0cm]{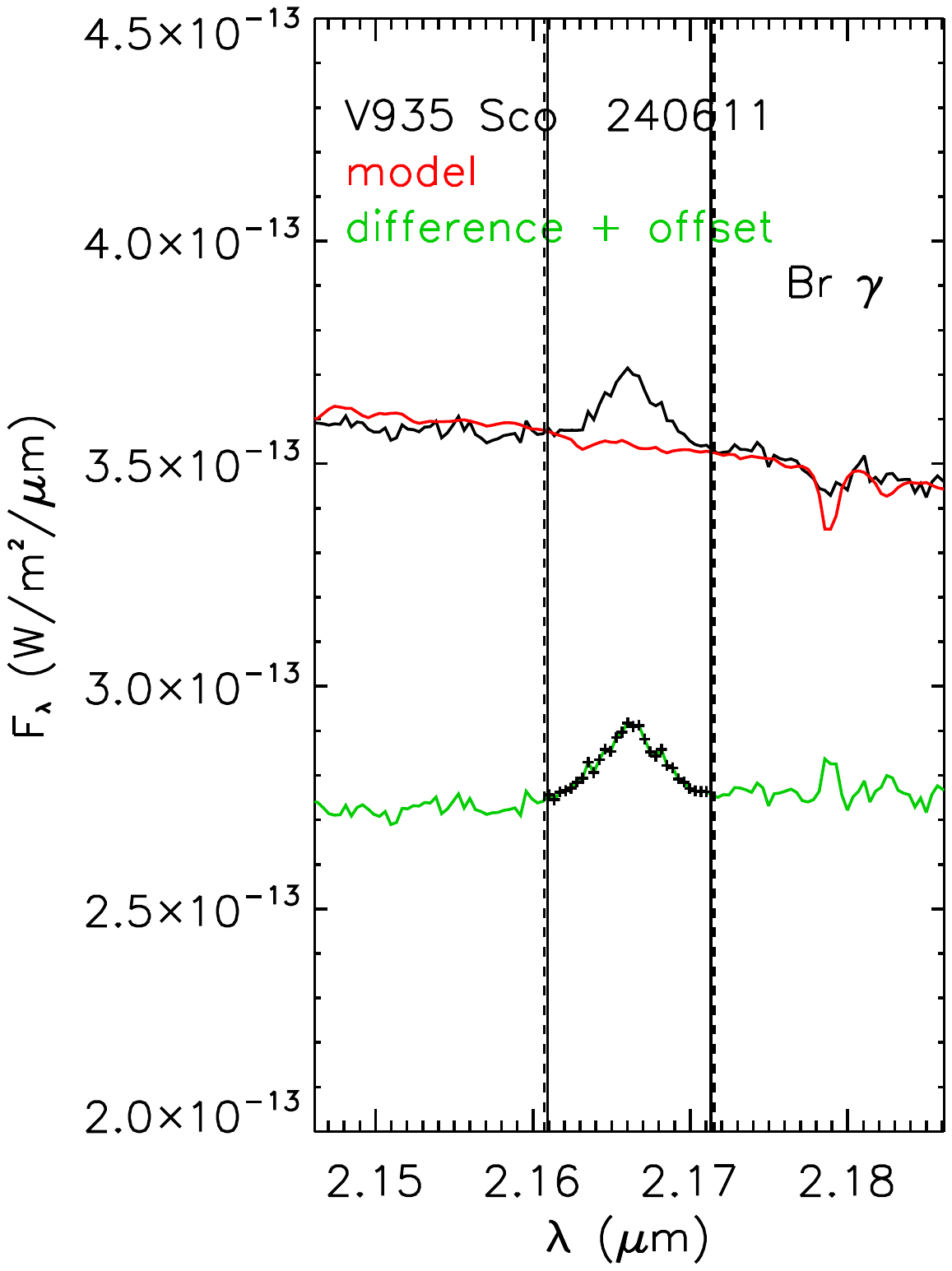}
\caption{The same as Figure A-2, except for V935 Sco on 240611UT. \label{fig:A-43}}
\end{figure}

\begin{figure}
\includegraphics[width=6.0cm, height=6.0cm]{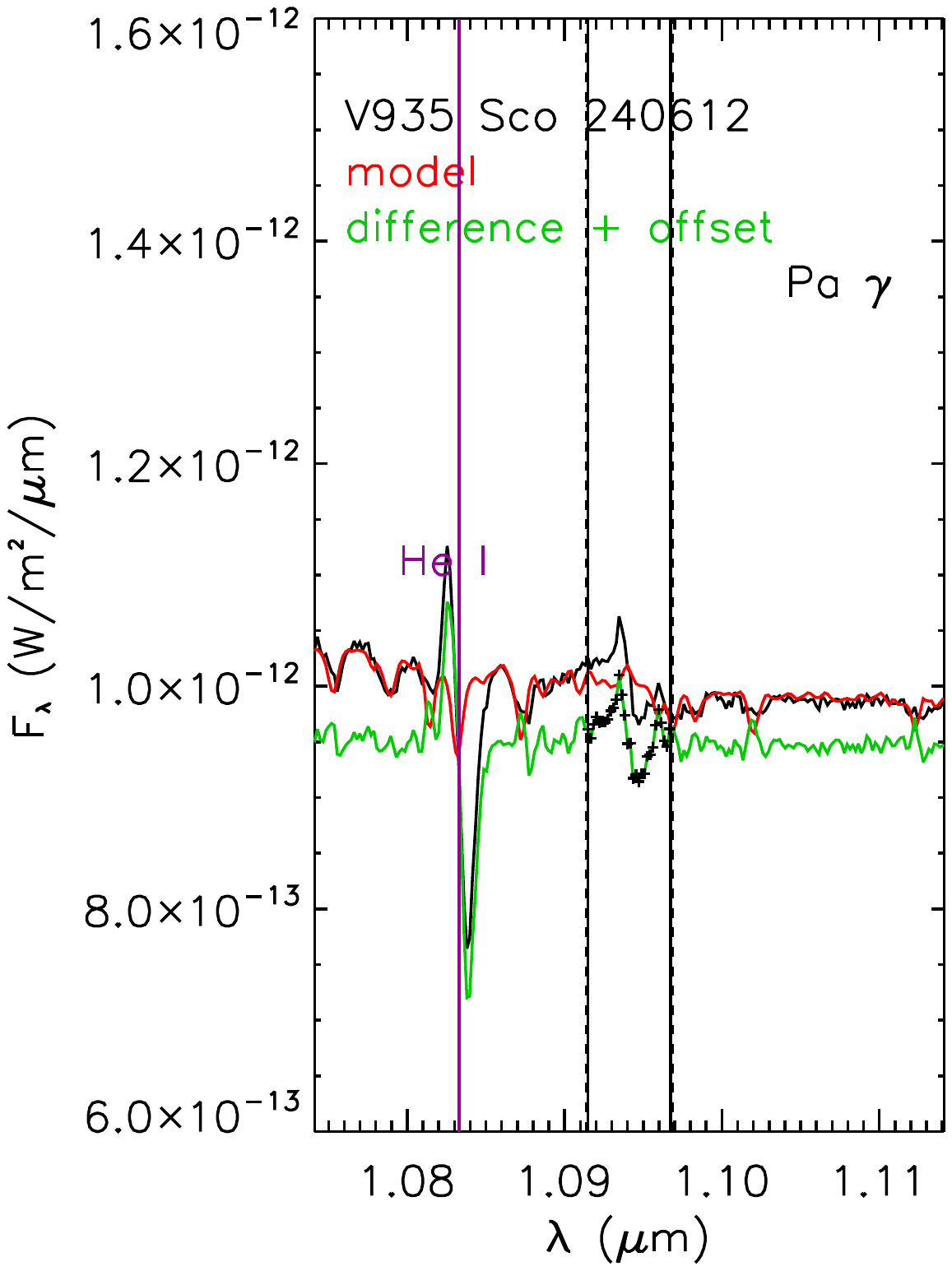}
\includegraphics[width=6.0cm, height=6.0cm]{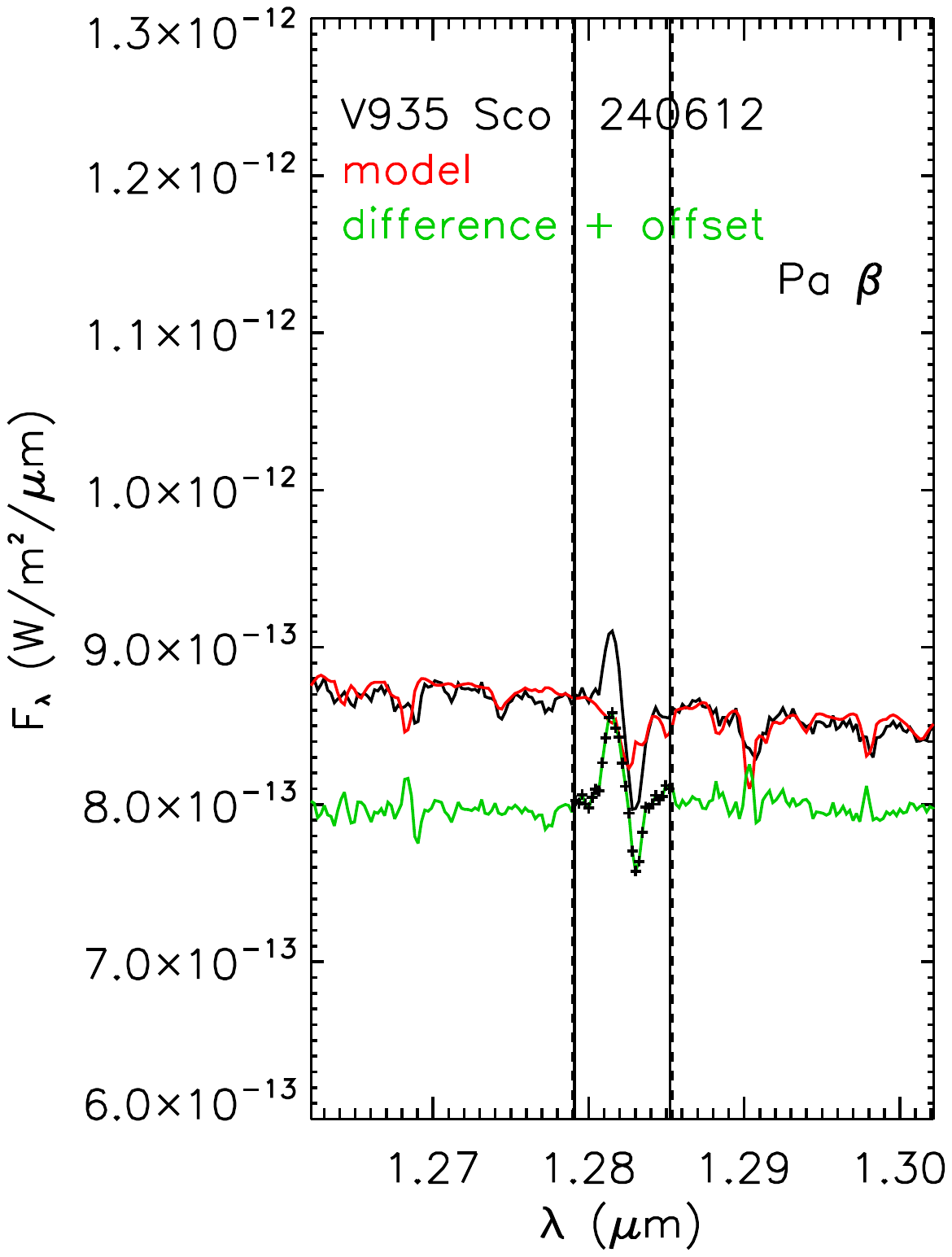}
\includegraphics[width=6.0cm, height=6.0cm]{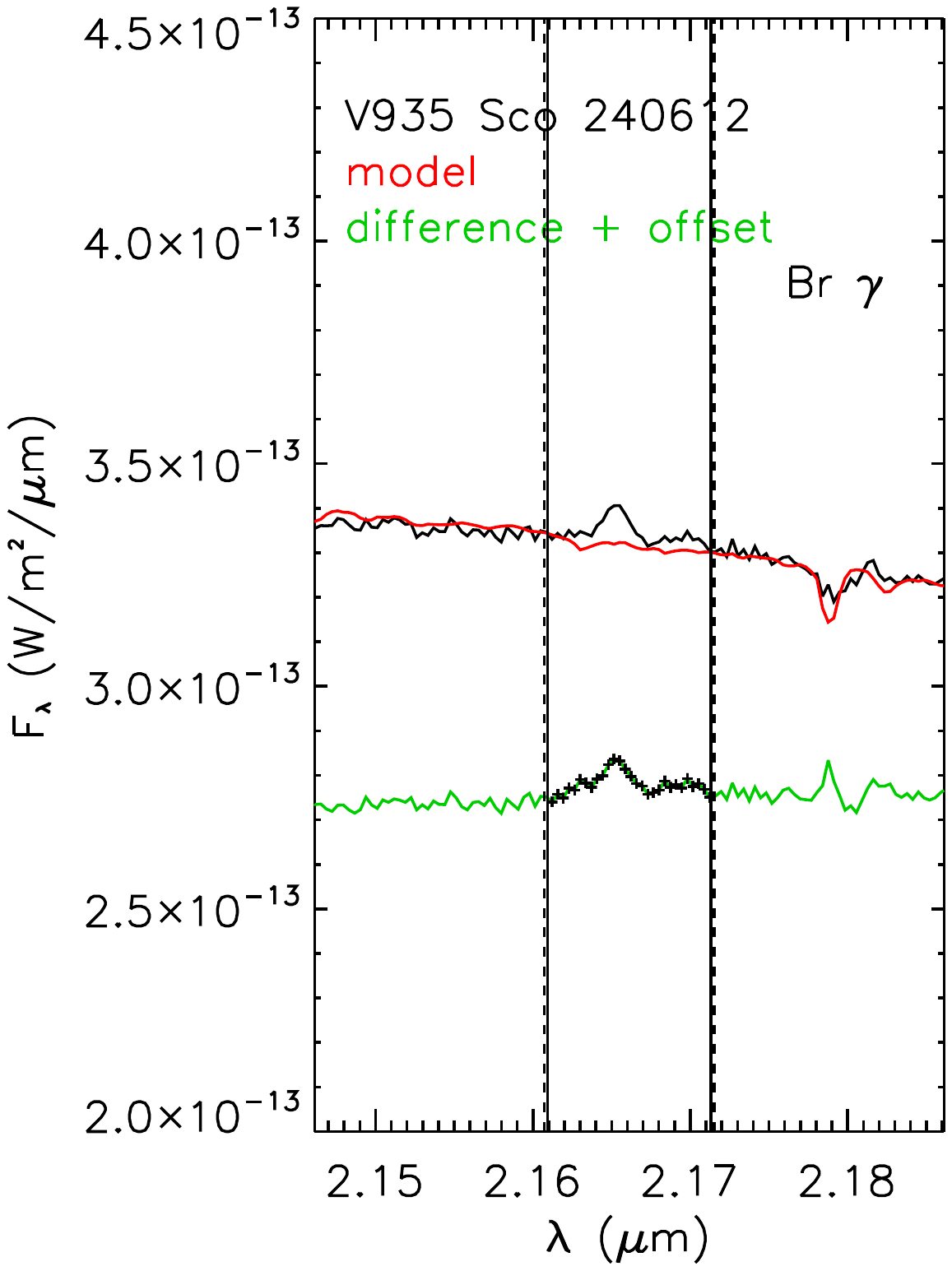} 
\caption{The same as Figure A-2, except for V935 Sco on 240612 UT. \label{fig:A-44}}
\end{figure}

\begin{figure}
\includegraphics[width=6.0cm, height=6.0cm]{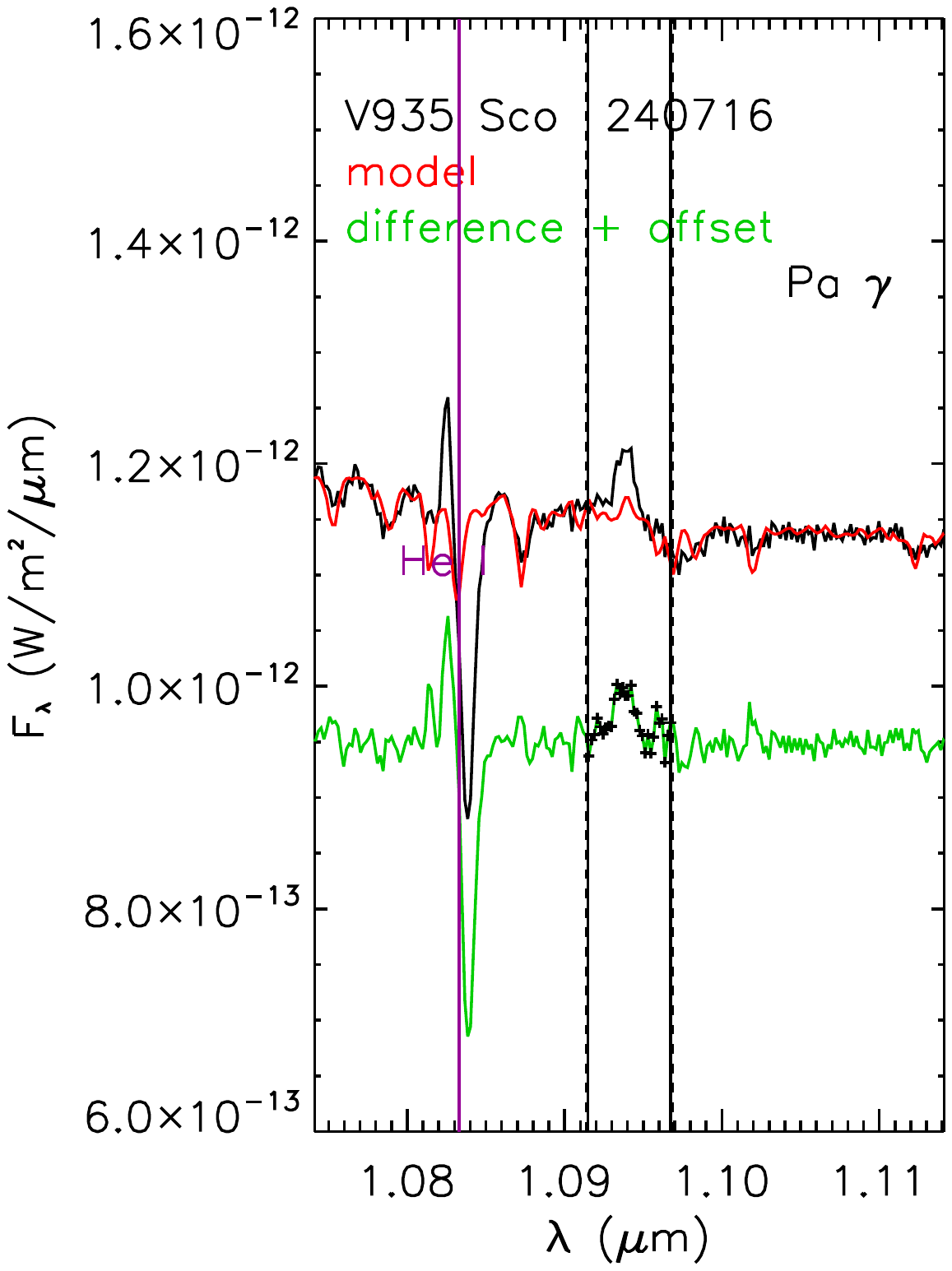}
\includegraphics[width=6.0cm, height=6.0cm]{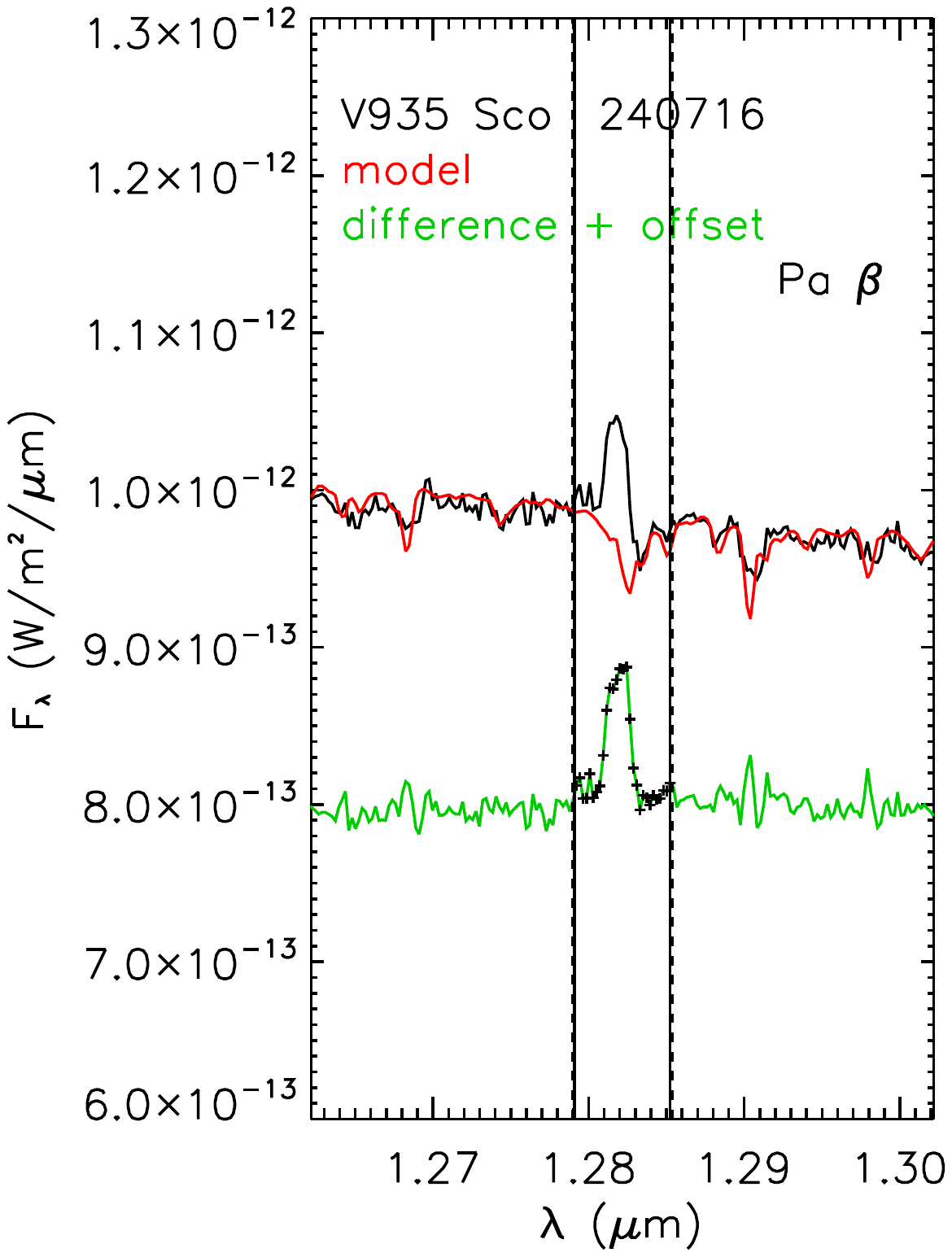}
\includegraphics[width=6.0cm, height=6.0cm]{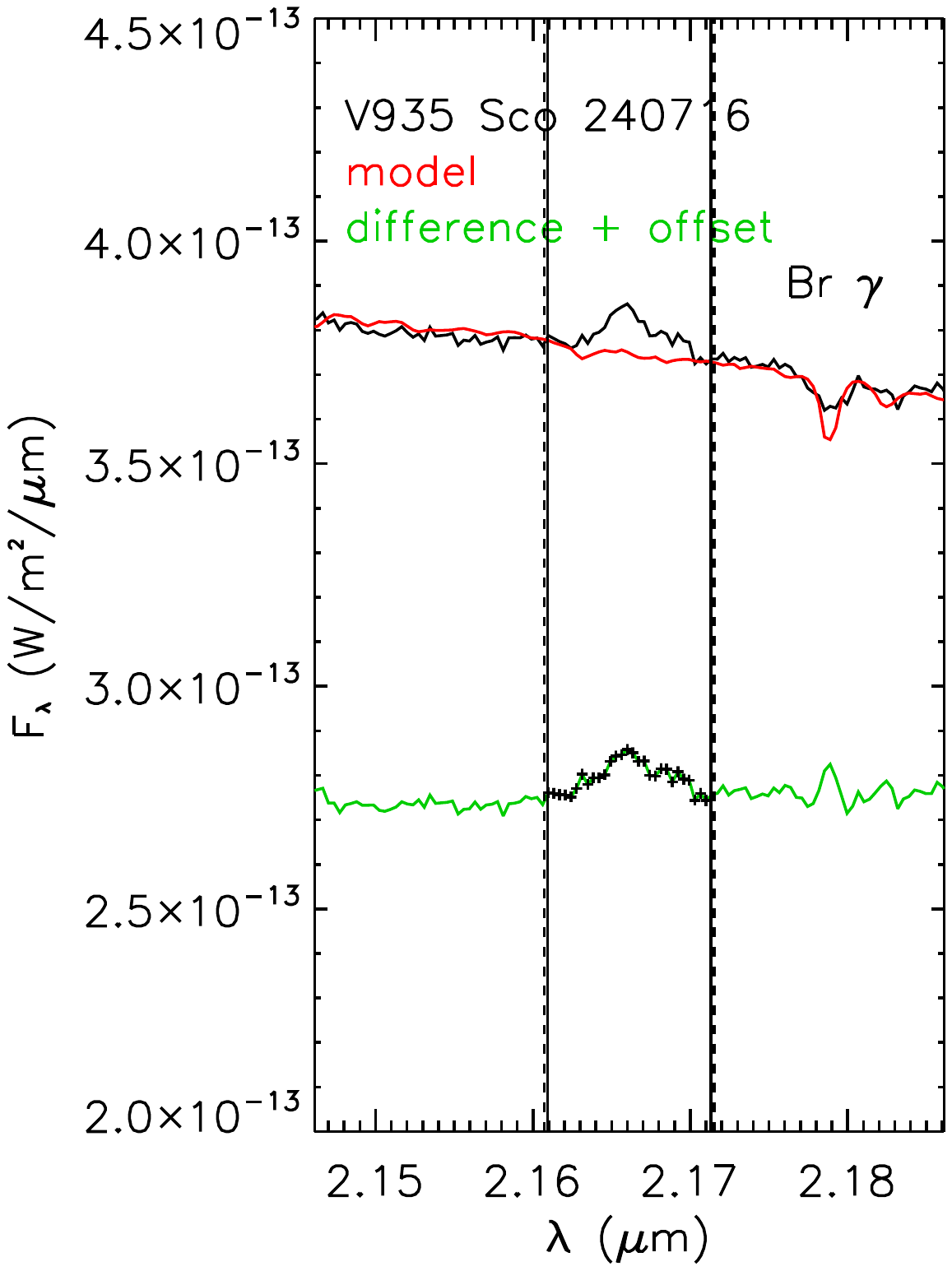}
\caption{The same as Figure A-2, except for V935 Sco on 240716 UT.  \label{fig:A-45}}
\end{figure}

\begin{figure}
\includegraphics[width=6.0cm, height=6.0cm]{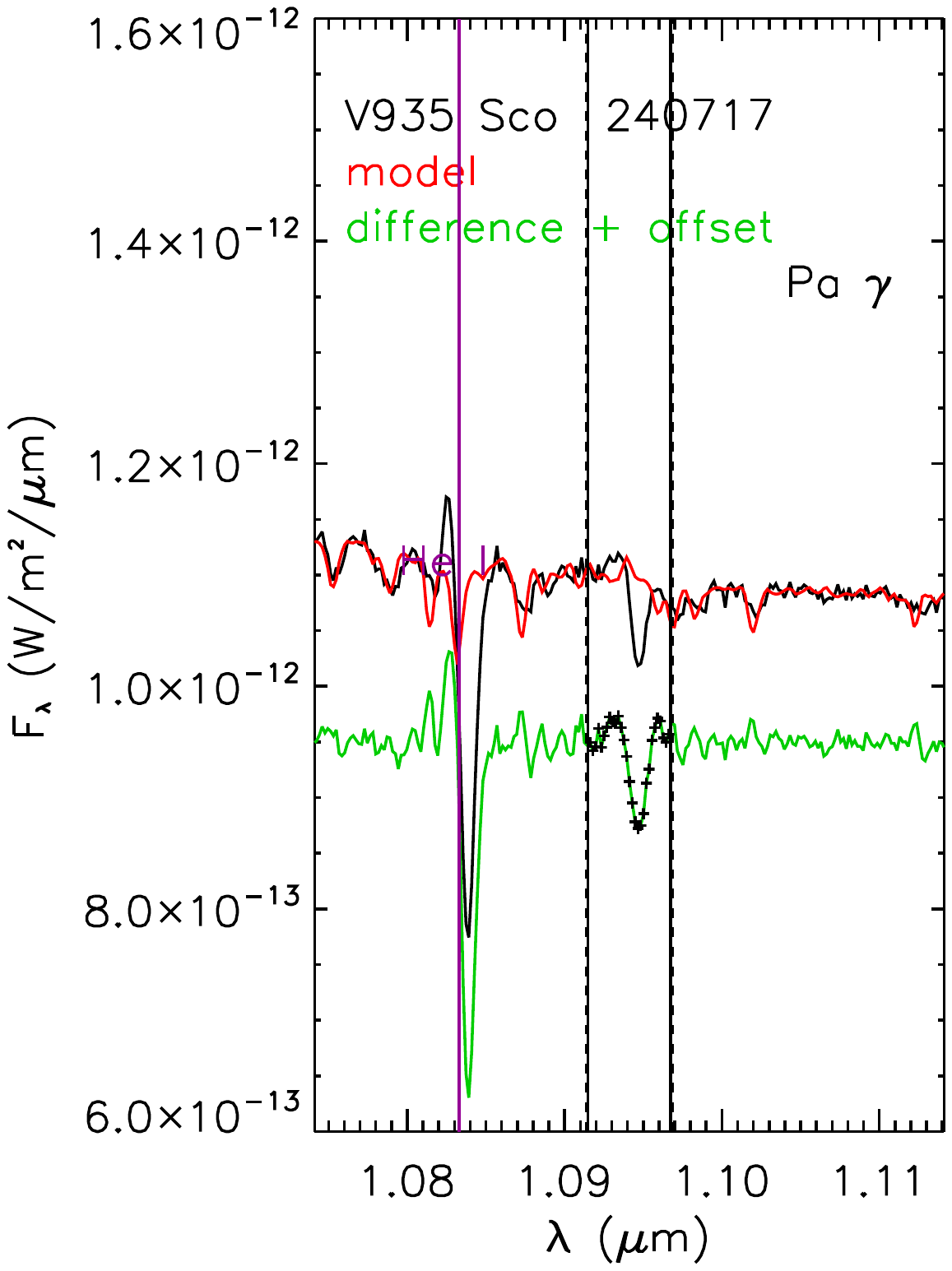}
\includegraphics[width=6.0cm, height=6.0cm]{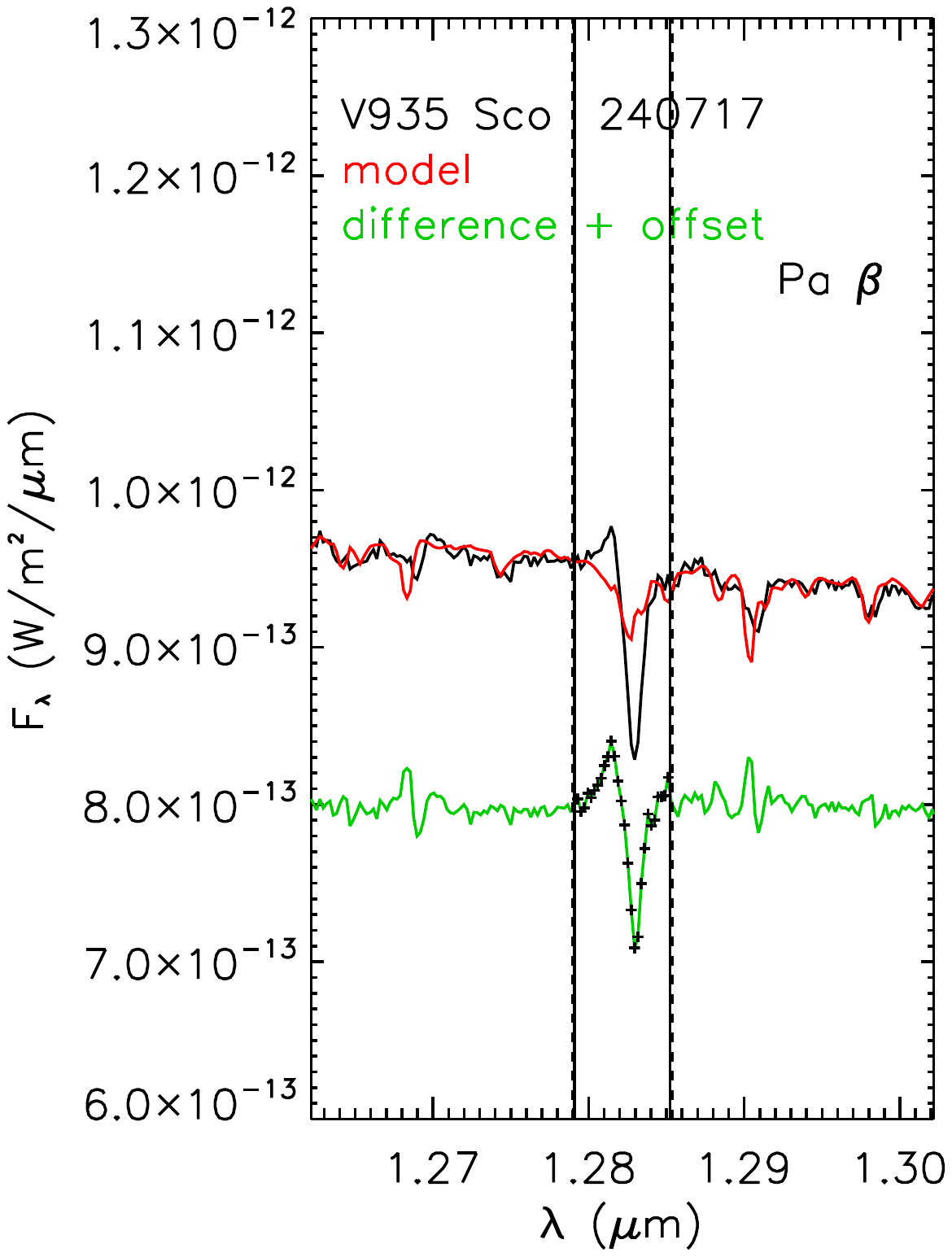}
\includegraphics[width=6.0cm, height=6.0cm]{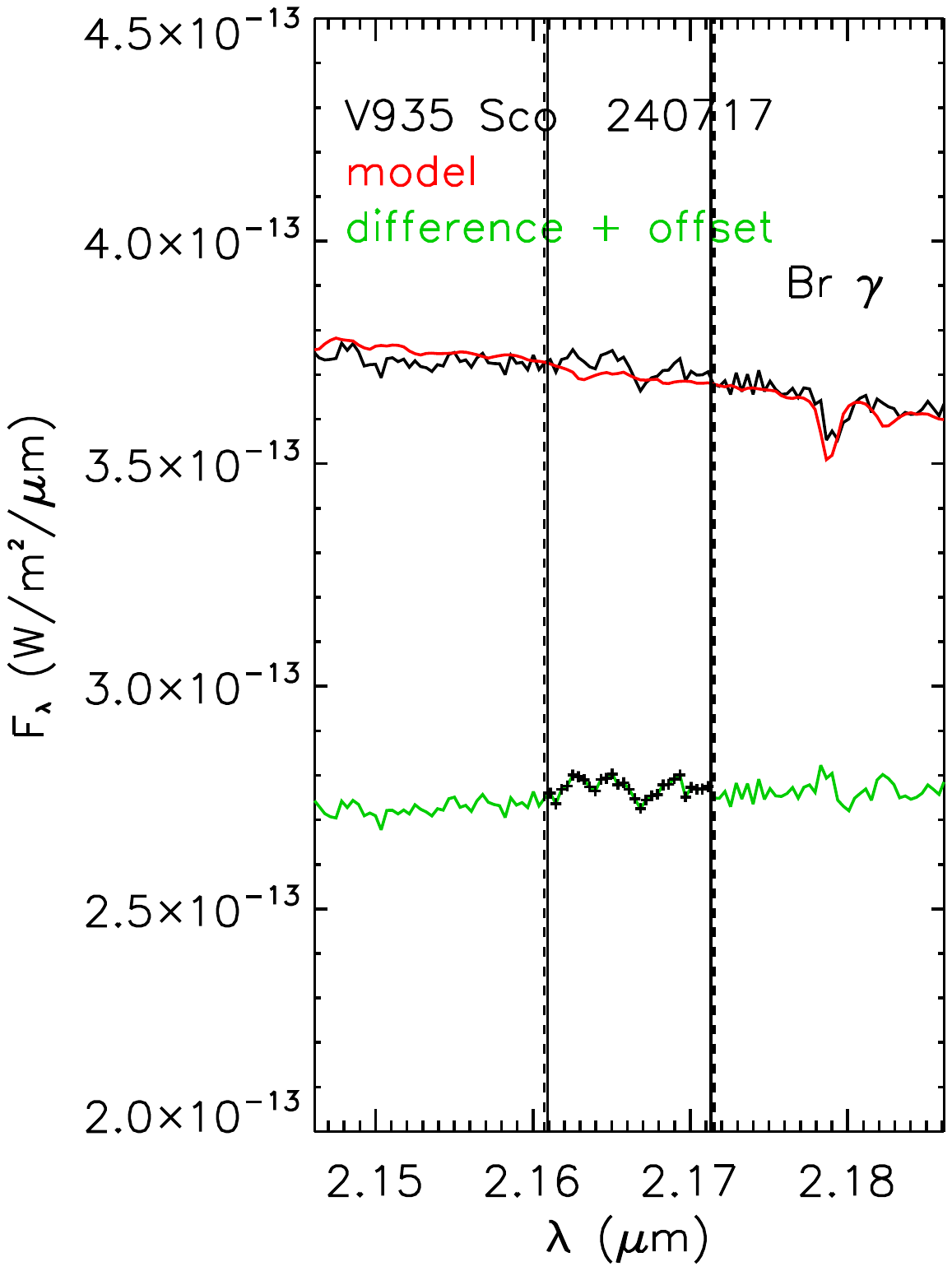}
\caption{The same as Figure A-2, except for V935 Sco on 240717 UT. \label{fig:A-46}}
\end{figure}

\begin{figure}
\includegraphics[width=6.0cm, height=6.0cm]{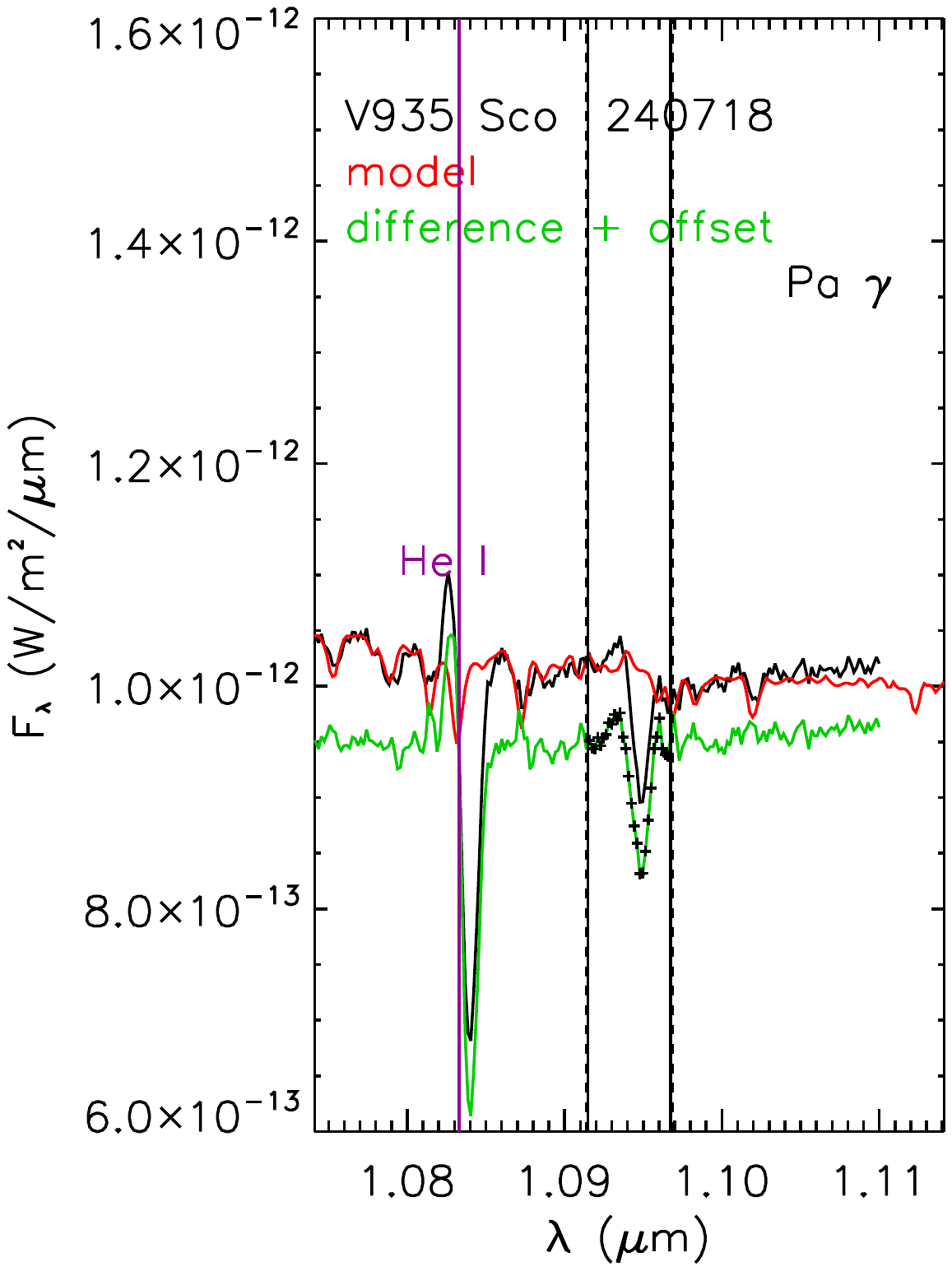}
\includegraphics[width=6.0cm, height=6.0cm]{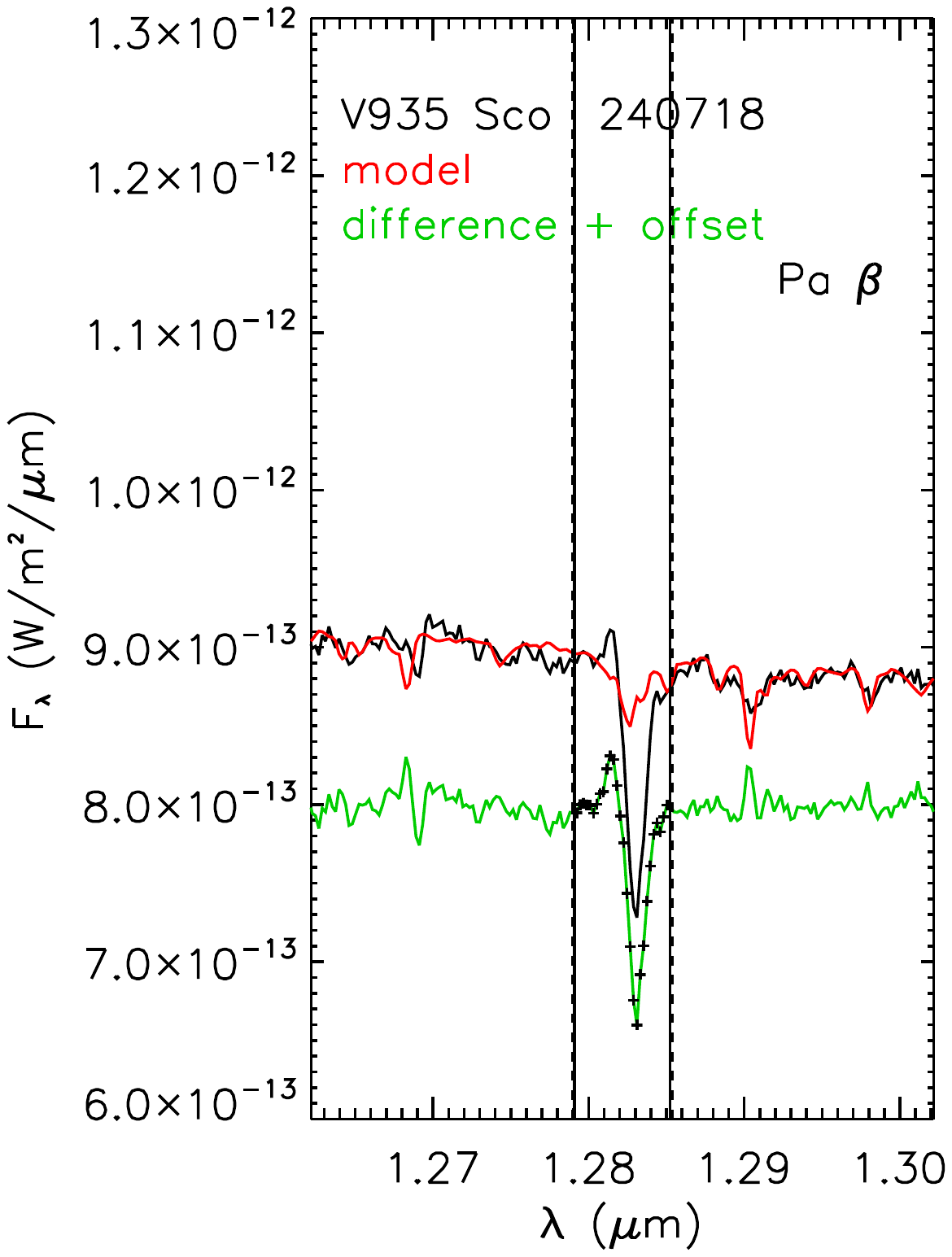}
\includegraphics[width=6.0cm, height=6.0cm]{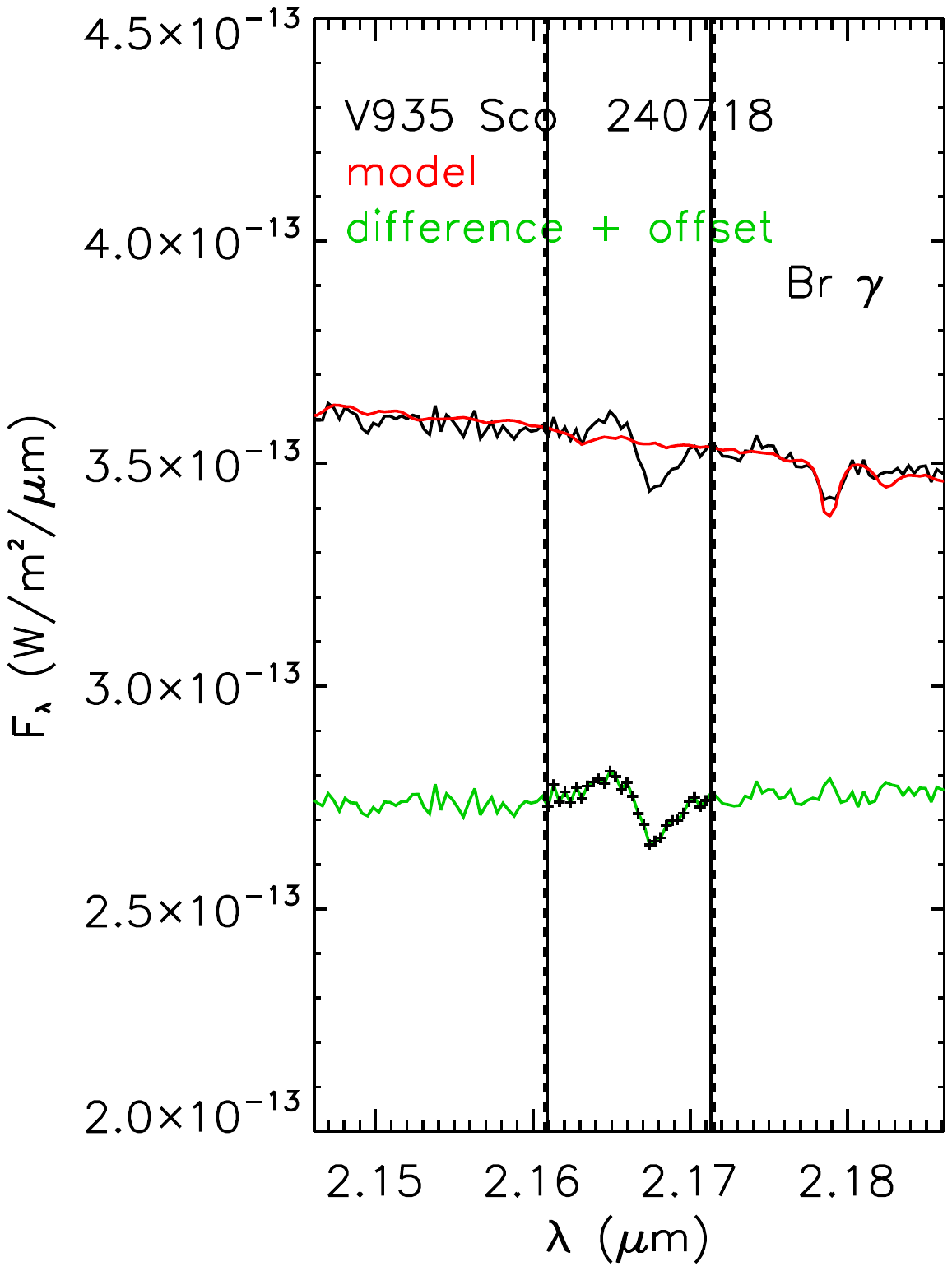}
\caption{The same as Figure A-2, except for V935 Sco on 240719 UT. \label{fig:A-47}}
\end{figure}

\begin{figure}
\includegraphics[width=6.0cm, height=6.0cm]{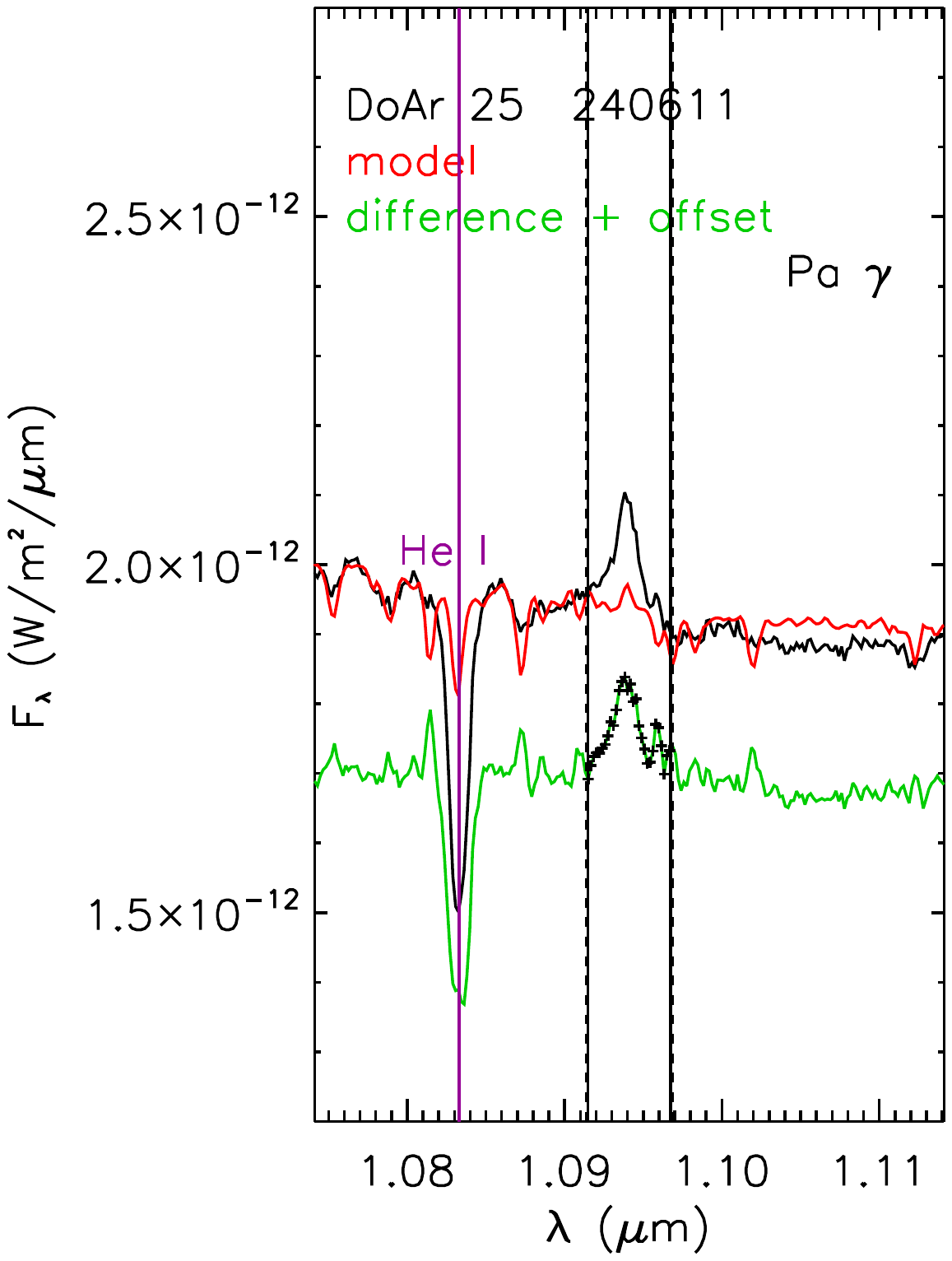}
\includegraphics[width=6.0cm, height=6.0cm]{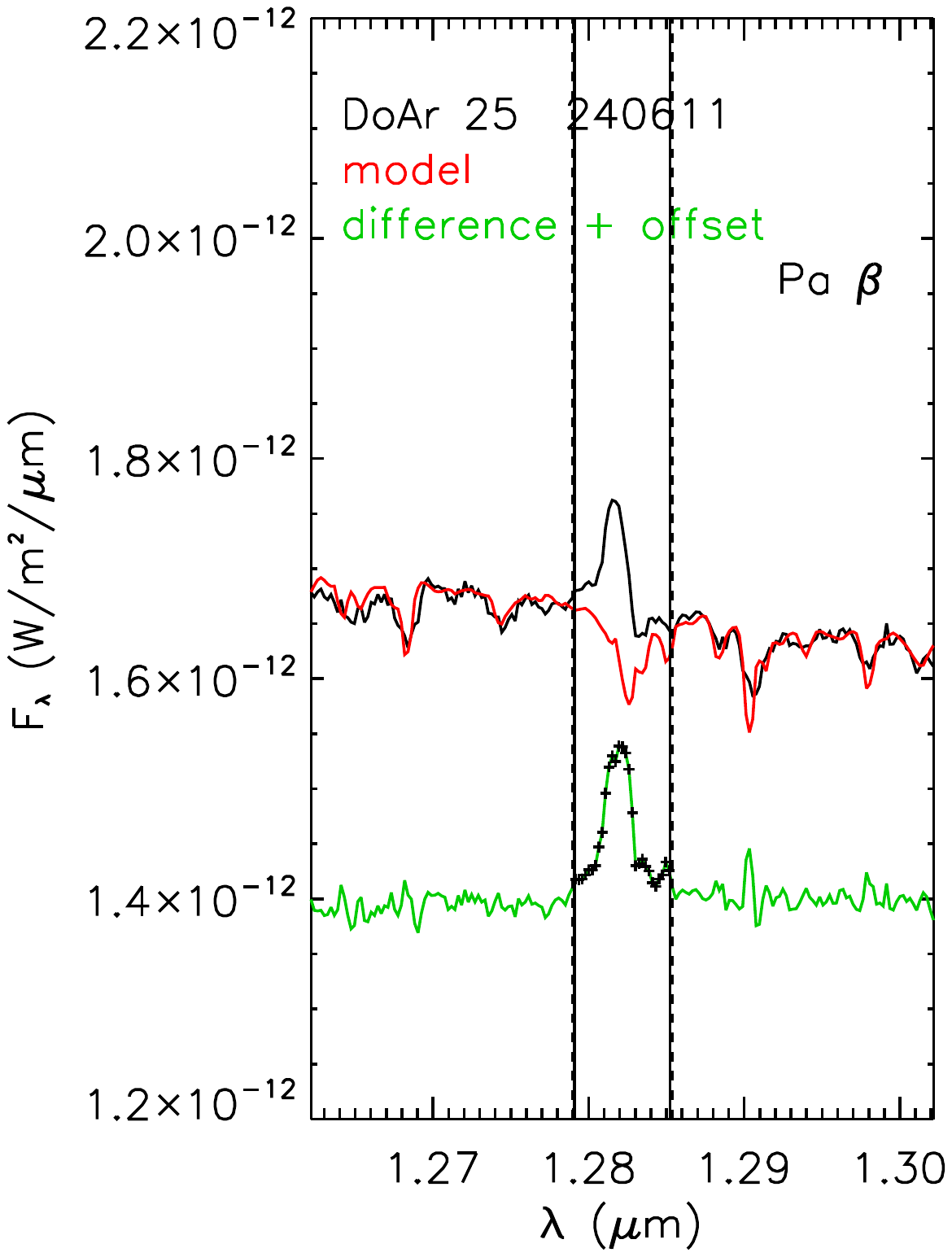}
\includegraphics[width=6.0cm, height=6.0cm]{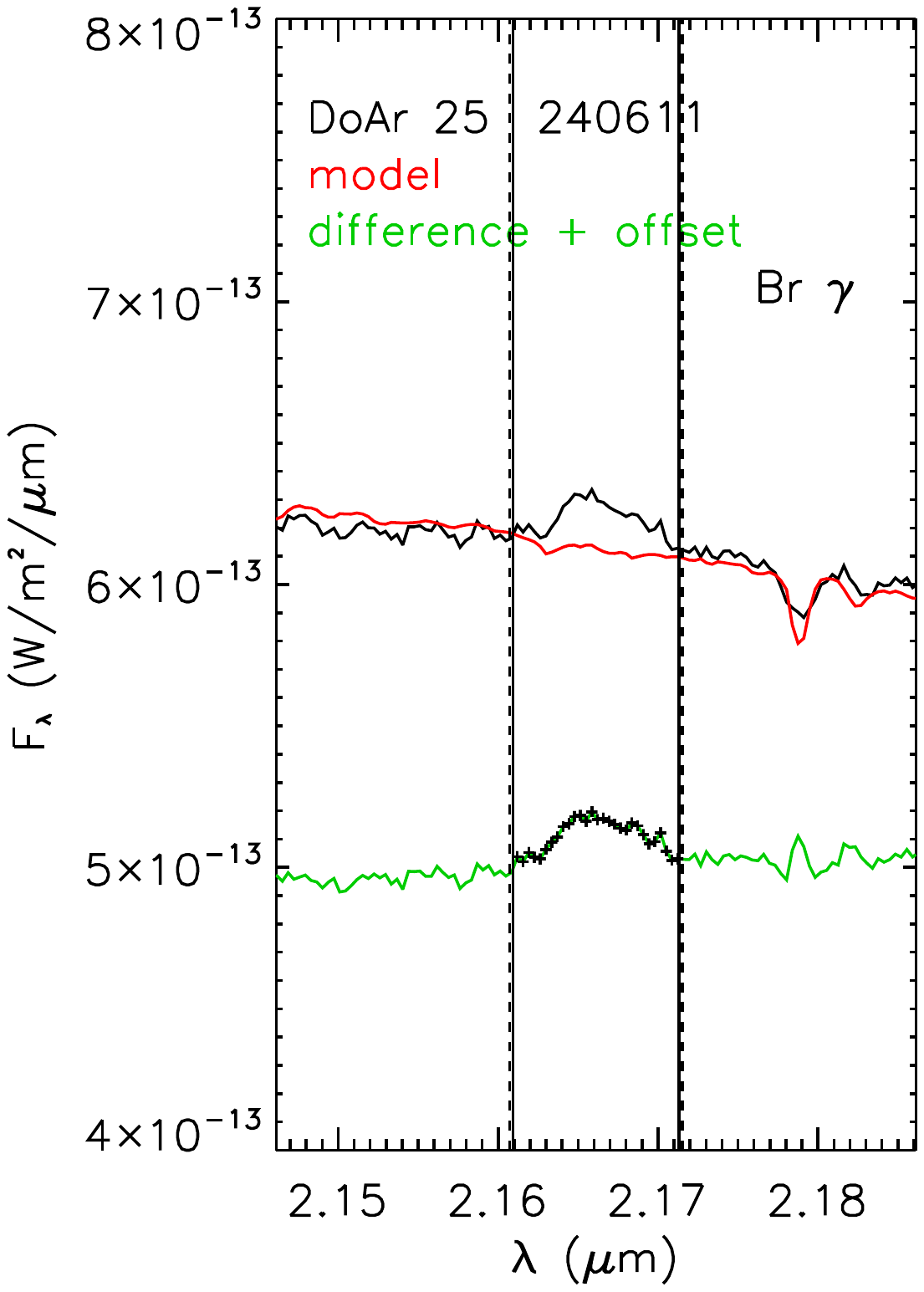}
\caption{The same as Figure A-2, except for DoAr 25 on 240611 UT. \label{fig:A-40}}
\end{figure}

\begin{figure}
\includegraphics[width=6.0cm, height=6.0cm]{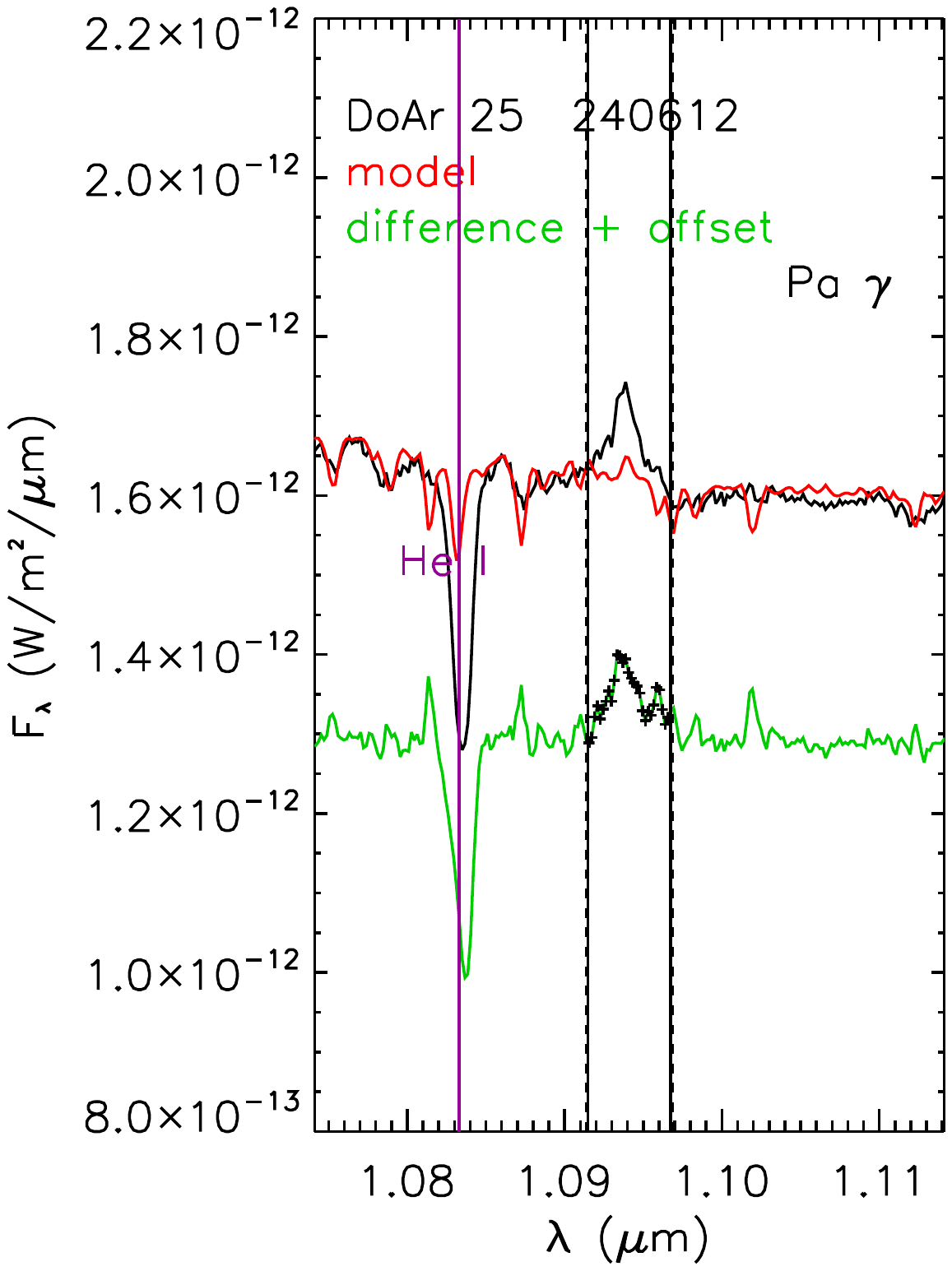}
\includegraphics[width=6.0cm, height=6.0cm]{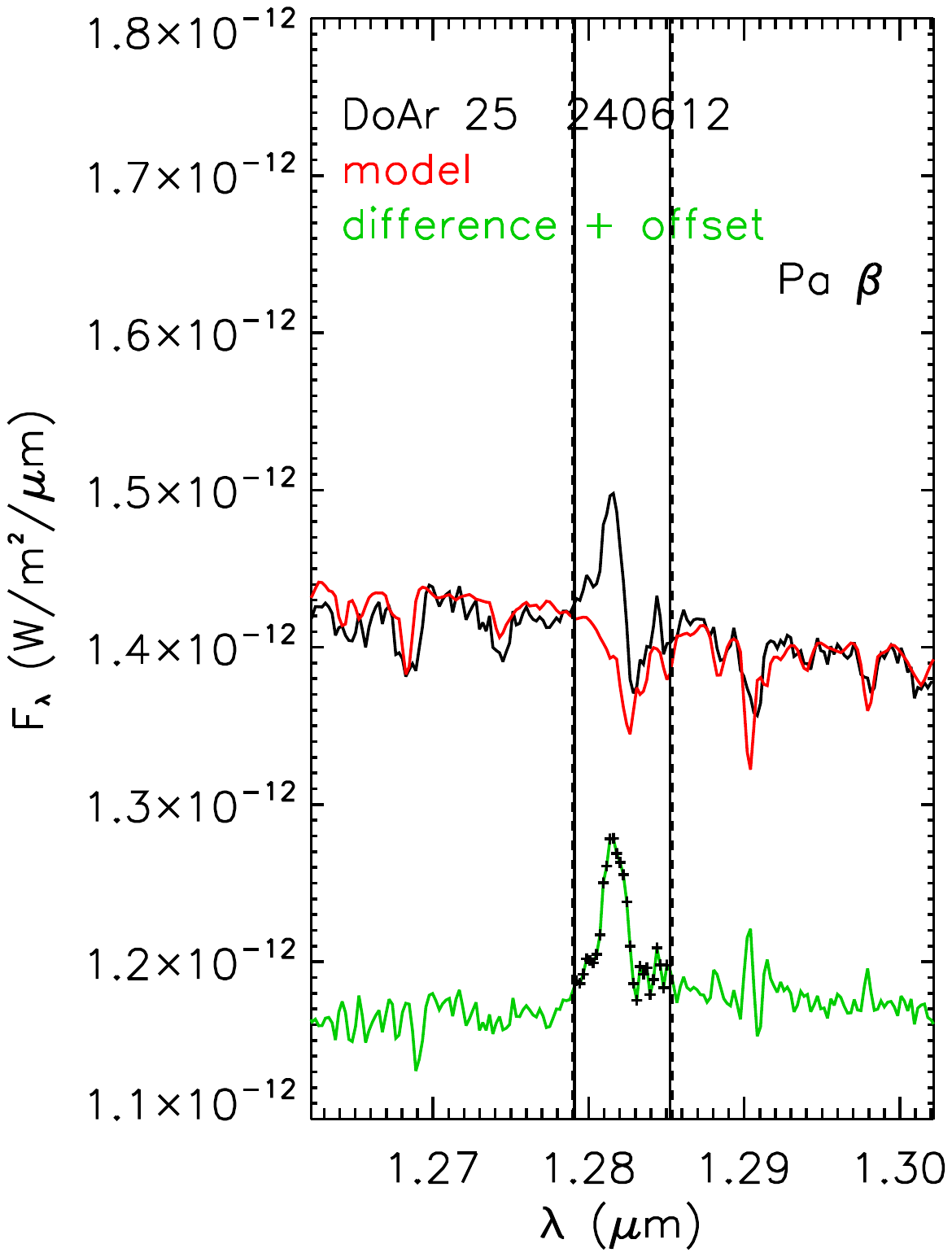}
\includegraphics[width=6.0cm, height=6.0cm]{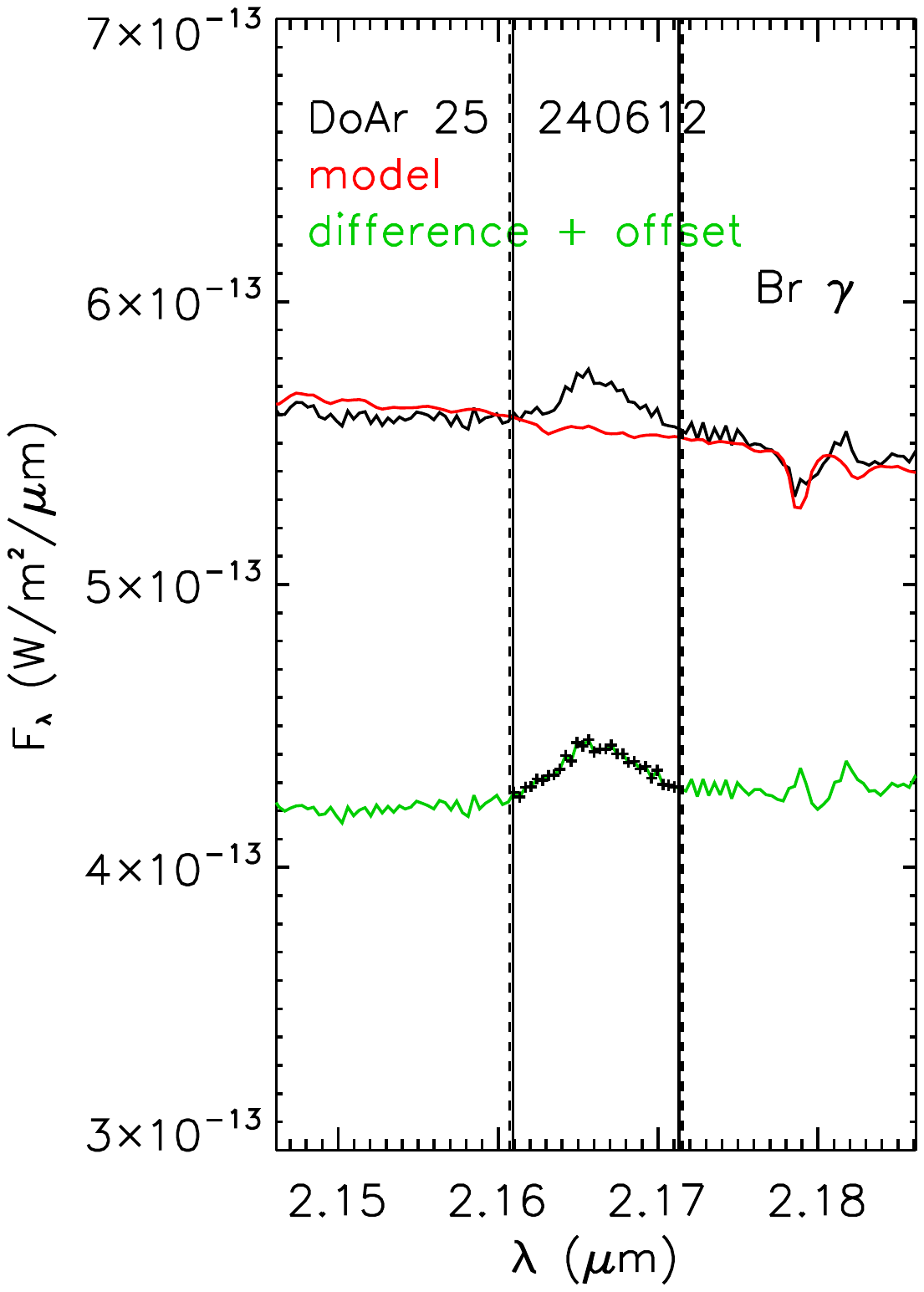}
\caption{The same as Figure A-2, except for DoAr 25 on 240612 UT. \label{fig:A-41}}
\end{figure}

%\clearpage

\end{document}